\documentclass{aa}
\usepackage{mhchem}
\usepackage{url}
\usepackage{lscape} 
\usepackage[warning,np]{numprint}
\npdecimalsign{\ensuremath{.}}
\npthousandsep{\,}
\npproductsign{\times}
\npfourdigitnosep
\usepackage{xcolor}

\newcommand{\change}[1]{#1}
\newcommand{\toprule}{\hline\hline}
\newcommand{\midrule}{\hline}
\newcommand{\bottomrule}{\hline}
\renewcommand{\cite}{\citep}
\newcommand{\textcite}{\citet}
\graphicspath{{./figures/}}

\newcommand{\myfootnotemark}[1]{\protect\renewcommand*{\thefootnote}{\alph{footnote}}\protect\footnotemark[#1]}

\begin{document}

\title{\change{Photodissociation and photoionisation of atoms and molecules of astrophysical interest}}
\author{A.~N. Heays\thanks{Current contact: Observatoire de Paris, LERMA, UMR 8112 du CNRS, 92195 Meudon, France.}
  \and A.~D.~Bosman \and E.~F.~van~Dishoeck}
\institute{Leiden Observatory, Leiden University, P.O. Box 9513, 2300 RA Leiden, The Netherlands}

\abstract{
  \change{
A new collection of photodissociation and photoionisation cross sections for 102 atoms and molecules of astrochemical interest has been assembled, along with a brief review of the basic physical processes involved.
These have been used to calculate dissociation and ionisation rates, with uncertainties, in a standard ultraviolet interstellar radiation field (ISRF) and for other wavelength-dependent radiation fields, including cool stellar and solar radiation, Lyman-$\alpha$ dominated radiation, and a cosmic-ray induced ultraviolet flux. The new ISRF rates generally agree within 30\% with our previous compilations, with a few notable exceptions. Comparison with other databases such as PHIDRATES is made. 
The reduction of rates in shielded regions was calculated as a function of dust, molecular and atomic hydrogen, atomic C, and self-shielding column densities. 
The relative importance of these shielding types depends on the atom or molecule in question and the assumed dust optical properties. 
All of the new data are publicly available from the Leiden photodissociation and ionisation database.

Sensitivity of the calculated rates to variation of temperature and isotope, and uncertainties in measured or calculated cross sections, are tested and discussed.
Tests were conducted on the new rates with an interstellar-cloud chemical model, and find general agreement (within a factor of two) in abundances obtained with the previous iteration of the Leiden database assuming an ISRF, and order-of-magnitude variations assuming various kinds of stellar radiation.
The newly parameterised dust-shielding factors makes a factor-of-two difference to many atomic and molecular abundances relative to parameters currently in the UDfA and KIDA astrochemical reaction databases.
The newly-calculated cosmic-ray induced photodissociation and ionisation rates differ from current standard values up to a factor of 5.
Under high temperature and cosmic-ray-flux conditions the new rates alter the equilibrium abundances of abundant dark cloud abundances by up to a factor of two.
The partial cross sections for \ce{H2O} and \ce{NH3} photodissociation forming OH, O, \ce{NH2} and NH are also evaluated and lead to radiation-field-dependent branching ratios.
}
}
\maketitle

\section{Introduction}
\label{sec:introduction}

Ultraviolet (UV) photons play a critical role in interstellar and
circumstellar chemistry. The realisation that photodissociation and
photoionisation processes control the abundances of atoms and small
molecules in diffuse interstellar clouds dates back nearly a century
\cite{eddington1928,kramers1946,bates1951}. Similarly,
photodissociation of parent species by UV radiation from the Sun
has long been known to explain the existence of small molecules in
cometary comae \cite{Haser57,Crovisier97}. Nowadays,
photodissociation processes are found to be important for modelling the chemistry of nearly every type of astrophysical region, from the edges of
dense clouds close to bright young stars to the surface layers of
protoplanetary disks, envelopes around evolved stars and giant
molecular clouds on galactic scales
\cite[e.g.,][]{Glassgold96,hollenbach1999,tielens2013,van_dishoeck2006,glover2012}.
Such clouds of gas and dust in which photodissociation is the dominant
molecular destruction path are termed Photodissociation or
photon-dominated regions (PDRs), although the term PDRs
originally referred mostly to high density regions close to bright O and B stars
such as found in Orion \cite{tielens1985b}.

The abundant UV photons in these regions photodissociate and
photoionise the main hydrogen, carbon, oxygen and nitrogen-containing
species, controlling the $\ce{H+}\rightarrow\ce{H}\rightarrow\ce{H2}$, $\ce{C+}\rightarrow\ce{C}\rightarrow\ce{CO}$, $\ce{O}\rightarrow\ce{O2}$ and $\ce{N}\rightarrow\ce{N2}$ transitions
\cite{tielens1985b,van_dishoeck1988b,li2013}.  Photoprocesses thus
affect the abundance of the main cooling species in the interstellar
medium, and they also generate chemically-reactive ions and radicals,
opening pathways to the formation of larger species
\cite{Sternberg95,lee1996,jansen1996,li2014a}.  The gas-phase
abundance of more complex molecules formed in this way is
simultaneously limited by their own photodestruction
\cite{teyssier2004,van_hemert2008,guzman2014}.  The photoionisation of
atoms and molecules also leads to a significant speed up of PDR
chemistry due to the enhanced reaction rates of ions compared with
neutral species \cite{tielens2013,van_dishoeck2014}.

The quantitative modelling of chemical evolution in clouds, envelopes
and disks is a prerequisite for the full interpretation of
observations of their emitting molecular lines and dust continuum.
Such models consider many physical regimes
\cite[e.g.,][]{le_petit2006,walsh2013} and involve many classes of
chemical reactions \cite{wakelam2012,mcelroy2013}.  By quantitatively
constraining the rates of photoprocesses, as is done in this paper,
other chemical and physical parameters processes affecting
observations can be more reliably determined.

The fundamental quantities governing photodissociation and ionisation
are the wavelength-dependent flux of incident UV radiation, discussed
in Sect.~\ref{sec:radiation fields}, and the wavelength-dependent
photoabsorption, photodissociation, and photoionisation cross sections
of each atom or molecule, introduced in Sects.~\ref{sec:cross sections} and \ref{sec:summary of cross sections}.
Historically, the complete and unabridged specification of these
quantities contained too much information to be included in
astrochemical models, and is actually in many cases unnecessary given
the scale of uncertainties in observations and other model parameters.
Tabulated pre-integrations of the full wavelength dependence into a
process rate (or lifetime) for different species in different kinds of
UV-irradiated environments are useful to speed up modelling.  We
calculated such rates in Sect.~\ref{sec:photo rates}.  Such tabulations
must necessarily include the column-density-dependent effect of
radiation shielding by dust, H and H$_2$ inside interstellar and
circumstellar clouds.  The wavelength dependence of such shielding is
frequently presented by a simplified parameterisation and is discussed
further in Sect.~\ref{sec:shielding functions}.

Astrochemical models can also use the full molecular and atomic
cross sections as functions of wavelength, and consider the
dissociation of species and shielding by H and H$_2$ line-by-line, to
compute the photodestruction of molecules as functions of depth into a
PDR
\cite[e.g.,][]{van_dishoeck1988b,viala1988,jansen1995,le_petit2006,Woitke09,walsh2013,li2013}.
Furthermore, astrochemical programs that employ simplified rates for photodestruction may require precomputing many of these when exploring, for example,
a range of possible dust grain ultraviolet extinction properties
\cite[e.g.,][]{van_dishoeck2006,roellig2013b}.  Fundamental atomic and molecular cross
sections such as those presented here are then required.

Even deep inside dark clouds well shielded from external radiation, a
weak UV flux is maintained. This is induced by the interaction of cosmic rays
with hydrogen.  The resulting spectrum is highly structured
\cite{prasad1983,gredel1987} and incorporation of this process into
astrochemical models also benefits from a reduction of the full
wavelength dependence into a conveniently tabulated rate.  The most-recent
tabulation of these rates is by \textcite{gredel1989}. Since that
time there have been updates for many of the photodissociation cross
sections of astrophysically relevant molecules. Here we update these
rates in Sect.~\ref{sec:cosmic rays}.

In Sect.~\ref{sec:discussion}, we discuss the potential variability of our
collected cross sections and calculated rates given their dependence on: interstellar dust optical properties, temperature,  spectrally unresolved cross sections, and isotopic substitution.  We also make a special case of studying distinct fragment branching ratios from the photodissociation of \ce{H2O} and \ce{NH3}, and assess
the significance of our new rates by means of a physically simple but
chemically complex toy astrochemical model.

All cross sections and calculated rates are available from the Leiden
Observatory database of ``photodissociation and photoionisation of
astrophysically relevant molecules'',\footnote{\url{www.strw.leidenuniv.nl/~ewine/photo}} and any future updates will
be available there.  Some of these cross sections are carried over
from the previous iteration of the Leiden database
\cite{van_dishoeck1988,van_dishoeck2006}; many species are
updated where new experimental or theoretical data has become
available, especially using the MPI Mainz UV/Vis database.\footnote{\url{satellite.mpic.de/spectral_atlas}} The
list of molecules in the database has been extended by new additions
of complex-organic species that have recently been detected in the
interstellar medium.\footnote{A community supported list of
  interstellar molecules:
  \url{wikipedia.org/wiki/List_of_interstellar}\url{_and_circumstellar_molecules}}

\section{Radiation fields}
\label{sec:radiation fields}

The photodissociation or photoionisation rate (molec./atom$^{-1}$\,s$^{-1}$) of a molecule (or atom) exposed to an ultraviolet radiation field is
\begin{equation}
  \label{eq:kpd}
 k = \int \sigma (\lambda) I(\lambda)\text{d}\lambda,
\end{equation}
where $\sigma(\lambda)$ is the appropriate photodissociation or
photoionisation cross section, to be discussed in Sect.~\ref{sec:cross
  sections}, and $I(\lambda)$ is the \change{photon-based} radiation intensity summed over all incidence angles.
\change{A photon-counting intensity was used for calculations in this paper because of the discrete nature of photodestruction events, but is directly related to the volumetric radiation energy density according to $U(\lambda)=h I(\lambda)/\lambda$ where $h$ is Planck's constant.
An angularly-differential radiation intensity may be appropriate if the incident radiation is non-isotropic.
}
The integration limits in Eq.~(\ref{eq:kpd}) are defined by
the wavelength range corresponding to the nonzero photodissociation or
ionisation cross section and radiation intensity.

\begin{figure*}
  \centering
  \includegraphics{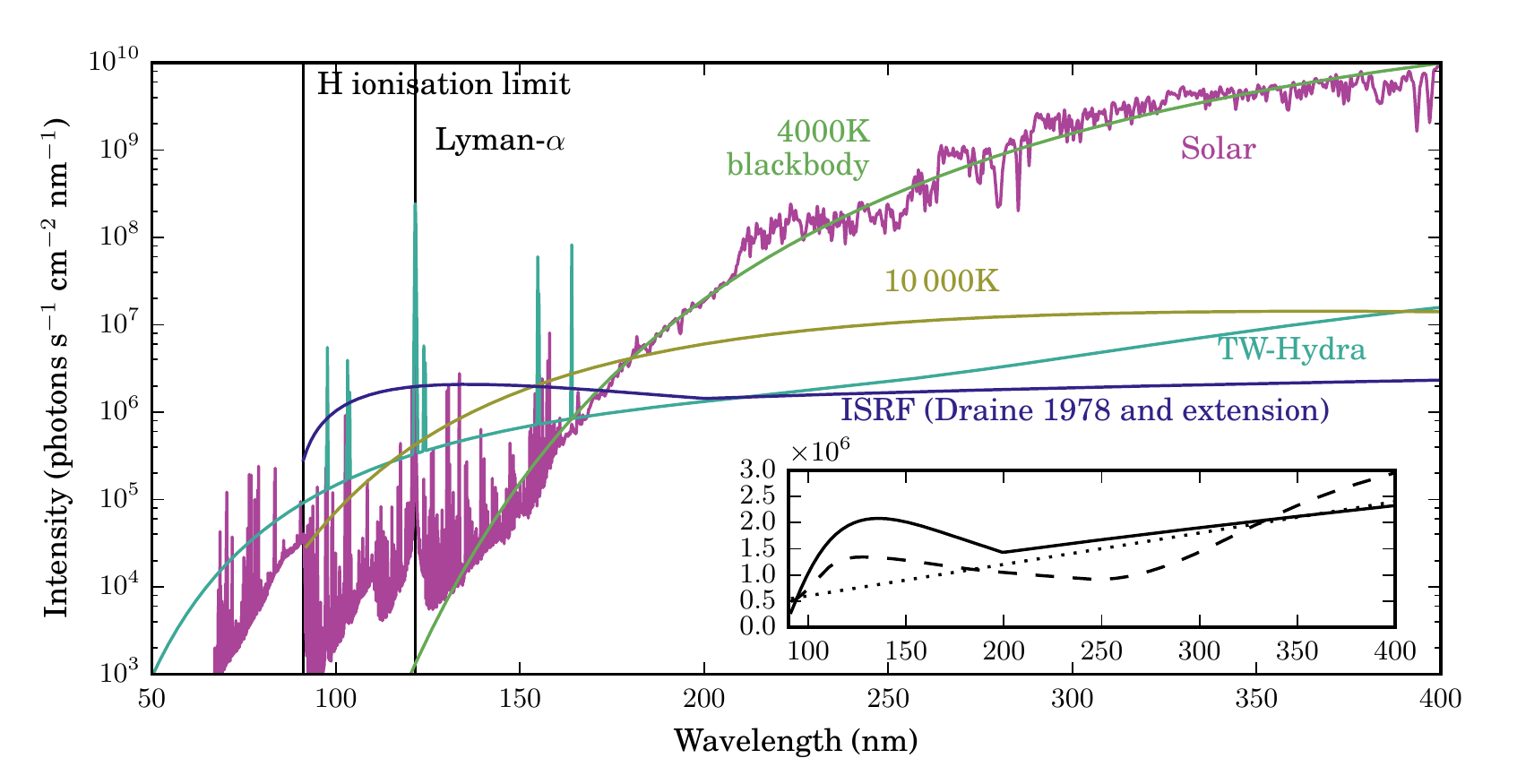}
  \caption{Wavelength dependence of some astrophysically-relevant ultraviolet radiation fields. \change{Inset: Radiation intensity in the solar neighbourhood estimated by \textcite{draine1978} (solid, modified according to \textcite{van_dishoeck1982a}), \textcite{mathis1983} (dashed), and  \textcite{Habing68} (dotted).}}
  \label{fig:radiation fields}
\end{figure*}

The average intensity of the interstellar radiation field (ISRF) can
be estimated from the number and distribution of hot stars in the
Galaxy, combined with a model for the dust distribution and its
extinction of the stellar radiation
\cite{Habing68,draine1978,mathis1983,parravano2003}.   The various estimates of
the mean UV energy density at a typical point in the local
galaxy agree to within a factor of two. Variations in this energy
density of a factor between two and three are expected throughout the galactic plane
and on time scales of a few Gyr, as massive O and B star clusters form
and die. In addition, the intensity ratio of short-wavelength photons capable of dissociating H$_2$, CO and N$_2$ and ionising atomic C ($\lambda<110$\,nm) and the broader far-ultraviolet range ($91.2<\lambda<200$\,nm) may
vary by a factor of two in location and time \cite{parravano2003}.

The wavelength dependence UV intensity as defined by \textcite{draine1978} is often adopted in
astrochemical models, and given by the formula
\begin{multline}
  \label{eq:draine field}
  I(\lambda) =   3.2028 \times 10^{13} \lambda^{-3} - 5.1542 \times 10^{15} \lambda^{-4} \\ + 2.0546 \times 10^{17} \lambda^{-5} ,
\end{multline}
where the wavelength, $\lambda$, has units of nm and the radiation intensity,
$I$, has units of photons\,cm$^{-2}$\,s$^{-1}$\,nm$^{-1}$.  This formula was intended for application within the 91.2 to 200\,nm wavelength range.
\change{
  An angularly-differential Draine field, $I(\lambda)/4\pi$, has units of photons\,cm$^{-2}$\,s$^{-1}$\,nm$^{-1}$\,sr$^{-1}$; and a scaled version of the radiation intensity may be adopted, $\chi I(\lambda)$, to describe regions with greater or lesser UV flux than the mean intensity defined by Draine.
}

The form of Eq.~(\ref{eq:draine field}) is shown in
Fig.~\ref{fig:radiation fields} and is reminiscent of a 20\,000\,K
black-body radiation field (B-type star) with some excess at shorter
wavelengths.  There is assumed to be zero flux shortwards of $91.2$\,nm
due to the ionisation continuum of atomic H that populates the
interstellar medium with a high column density for all sight lines.
An extension proposed by \textcite{van_dishoeck1982a} simulates the interstellar flux at
longer wavelengths than considered by the Draine model, and fits a
range of observed intensities between 200 and 2000\,nm to within about
50\%. 
This extension is given by the formula: \begin{equation} \label{eq:draine mod van_dishoeck1982}
  I(\lambda) = \np{3.67e4}\lambda^{0.7}\ ;\  \lambda > \np[nm]{200}.
\end{equation}
We combine the full wavelength range of the \textcite{draine1978} and
\textcite{van_dishoeck1982a} fields into a ``standard'' ISRF for the following calculations of
photodissociation and ionisation rates.

The energy intensity of the Draine field integrated between 91.2 and
200\,nm is,
\begin{equation}
  \label{eq:draine_normalisation}
  \int_{91.2}^{200} \frac{hcI(\lambda)}{\lambda}\,\text{d}\lambda =  \change{\np[W\,m^{-2}]{2.6e-6}},
\end{equation}
where $h$ is Planck's constant, and $c$ the speed of light. This
integrated value is a factor of 1.7 higher than the integrated flux of
the \textcite{Habing68} field, which is taken as the reference with
scaling factor $G_0$ in some models \cite{tielens1985b}. Thus, the
standard Draine field has $G_0$=1.7.

\change{An independent estimate of the Galactic radiation field is made by \textcite{mathis1983}, and its magnitude and wavelength dependence for the case of \np[kpc]{10} Galactocentric distance (the local Galaxy) is compared in Fig.~\ref{fig:radiation fields} with the ISRF standard we adopted.  
The Mathis et al. UV flux is generally about 35\% weaker, and photodissociation rates will be similarly reduced for all atoms and molecules, apart from those that are photodestroyed at wavelengths longer than 300\,nm, at which point the Mathis radiation becomes stronger than our standard ISRF.}

The ultraviolet field near to a star is dominated by its black
body radiation and atomic emission or absorption lines, principally
the H\,I Lyman-$\alpha$ emission line at 121.6\,nm.  \change{We
model several such radiation fields as
pure black-body emitters in the following calculations}.
Special attention to the Lyman-$\alpha$
emission spectrum is warranted because of the known high intensity of
this feature in some astrophysical situations, including fast shocks
\cite{Neufeld89a}, the active Sun \cite{lammer2012}, and young stars
\cite{valenti2000,yang2012}. Indeed, around some T-Tauri
stars, up to 90\% of the total far-ultraviolet flux is emitted in the
Lyman-$\alpha$ band \cite{bergin03uv,schindhelm2012b}.  Also, the
propagation of Lyman-$\alpha$ radiation into a disk is significantly
enhanced by scattering from the disk surface \cite{bethell2011}, where
a 121.6\,nm photon absorbed by an H atom will be ultimately re-radiated in a random direction, including into the disk.  Thus, we also
treat a pure Lyman-$\alpha$ line in our calculations.  A \np[km
s^{-1}]{200} Doppler broadening is added to the Lorentzian natural
linewidth of the Lyman-$\alpha$ transition.  This broadening is a
typical value from the observationally-constrained photospheric
emission of a sample of T-Tauri stars \cite{france2014}.

In reality, stellar spectra are not black bodies but contain many
emission or absorption lines \cite[e.g.,][]{ardila2002a,ardila2002b,leitherer2010}.
As an example of a structured stellar flux, we consider a combination
of continuum and atomic emission simulating the photosphere of the
classical T-Tauri star TW-Hydra, as deduced from UV telescope
observations \cite{france2014}.  This
observationally-derived spectrum is extrapolated to shorter and longer wavelengths
using the derived black body and accretion-induced short wavelength
excess, respectively, proposed by \textcite{nomura2005}.  This
includes an additional nonblack-body ultraviolet excess due to the
accretion of material onto the still-forming star.

The solar ultraviolet flux is measured directly in the series of
SOHO-SUMER observations \cite{curdt2001} for $\lambda<160$\,nm and
also by the UARS SOLSTICE mission \cite{woods1996}, including longer
wavelengths.  We adopt a spectrum compiled from these two data sets
corresponding to a quiet period in the Sun's radiance.  The activity
level of the Sun makes little difference for $\lambda > \np[nm]{160}$
but can induce variation of a factor of two or more at
shorter wavelengths, including enhanced Lyman-$\alpha$ radiation.
More detailed studies of the dependence of molecular photodissociation
rates on solar activity are made by
\textcite{huebner1992} and \textcite{huebner2015}.

All stellar radiation fields were normalised to match the energy
intensity of the Draine field integrated between 91.2 and 200\,nm, that is,
\change{\np[W\,m^{-2}]{2.6e-6}}.
The photodissociation and ionisation rates calculated hereafter due to
exposure of molecules and atoms to these radiation field should
subsequently be scaled to match the flux in an astrophysical
environment, which may differ by multiple orders of magnitude.
Our normalisation scheme is selected to
emphasise the wavelength-dependent effects induced by substituting
radiation fields.
A scale factor of 37\,700 should be used to increase
the solar photodissociation and photoionisation rates calculated here
to values appropriate for the approximate solar intensity at 1\,au,
assuming an integrated solar flux between 91.2 and 200\,nm of
0.098\,W\,m$^{-2}$.

For the cases of the solar and TW-Hydra radiation fields, intensity at wavelengths shorter than the ionisation threshold of
atomic H, 91.2\,nm, is included.  This is certainly appropriate for studies of
planetary atmospheres and cometary comae in the H-deprived solar
system.  
There are also several known cases of highly-evolved, hydrogen
gas-poor debris disks supporting some amount of CO
\cite{mathews2014}.  The origin of this gas is unknown but may arise
from evaporation of solids in collisions of planetesimals, allowing
for relatively low amounts of gas-phase hydrogen relative to other
species and the free transmission of short-wavelength radiation
\cite{dent2014}.

A cosmic-ray induced UV-emission spectrum is taken from the calculations of
\textcite{gredel1989}. The
energetic electrons produced from cosmic-ray ionisation of hydrogen
excite H$_2$ into excited electronic states. Spontaneous emission
back to the electronic ground state produces a rich spectrum of UV
lines, from  91.2 to 170\,nm, as well as a weak
continuum between 150 and 170\,nm. The precise spectral details depend on the
initial population of H$_2$ ro-vibrational levels and the
ortho-to-para ratio of H$_2$. Usually H$_2$ is assumed to be in its vibrational and rotational ground state in the cold interiors of dark clouds.

\section{Cross sections}
\label{sec:cross sections}

\subsection{General properties}
\label{sec:cross section general properties}
The critical data needed to describe gas-phase molecular or atomic
photoprocesses is the wavelength-dependent photoabsorption cross
section, $\sigma(\lambda)$ .  This differential quantity describes the
expected rate of photoabsorption per spectral unit of an isolated
molecule or atom, ABC, in a photon-intensity normalised radiation field, bringing it into an excited
electronic state ABC$^*$, and (oddly)
has dimension of area. The optical depth of the absorption at a
certain wavelength is given by $\tau=N\times \sigma$.
Thus, a cloud of molecules with cross section $\sigma(\lambda)=\np[cm^{2}]{e-17}$ and column density $N=\np[cm^{-2}]{e17}$ has an optical depth of 1, and a $1/e$ probability of absorbing a photon with wavelength $\lambda$.

A photo-excited molecule ABC$^*$ may decay by several channels, and the probability of each of them needs to be taken into account.
This includes dissociation (e.g., forming \ce{A + BC}), ionisation (\ce{ABC+ + e-}), or non-destructive emission (\ce{ABC + photon}).
Their respective partial photodissociation, photoionisation, and photoemission cross sections $\sigma^\text{d}(\lambda)$, $\sigma^\text{i}(\lambda)$, and $\sigma^\text{e}(\lambda)$, are the product of the photoabsorption cross section and a decay probability, $\eta^\text{d}(\lambda)$, $\eta^\text{i}(\lambda)$, and $\eta^\text{e}(\lambda)$; respectively.
We generally neglected further division of the photoabsorption cross section into decay channels leading to distinct dissociation products (e.g., A + BC, AB + C, or A + B + C) or dissociative-ionisation fragments (e.g., \ce{ABC+} or \ce{AB$^+$ + C}) because of limited branching-ratio data in the literature, although this is a very relevant issue for chemical models.
In general,
multiple fragments are energetically possible and participate
distinctly in ongoing chemistry, for example, \ce{CH4} dissociating to form
significant amounts of \ce{CH3} and \ce{CH2} in Titan's atmosphere
\cite{romanzin2005}, or H$_2$O dissociating into OH + H or O + H$_2$,
with a wavelength-dependent relative likelihood. 
As an exception, in Sect.~\ref{sec:H2O NH3 branching} we undertake to characterise the photodissociation branching of \ce{H2O} into OH and H products, and \ce{NH3} into \ce{NH2} and NH.

\begin{figure*}
  \centering
  \includegraphics[width=0.9\textwidth]{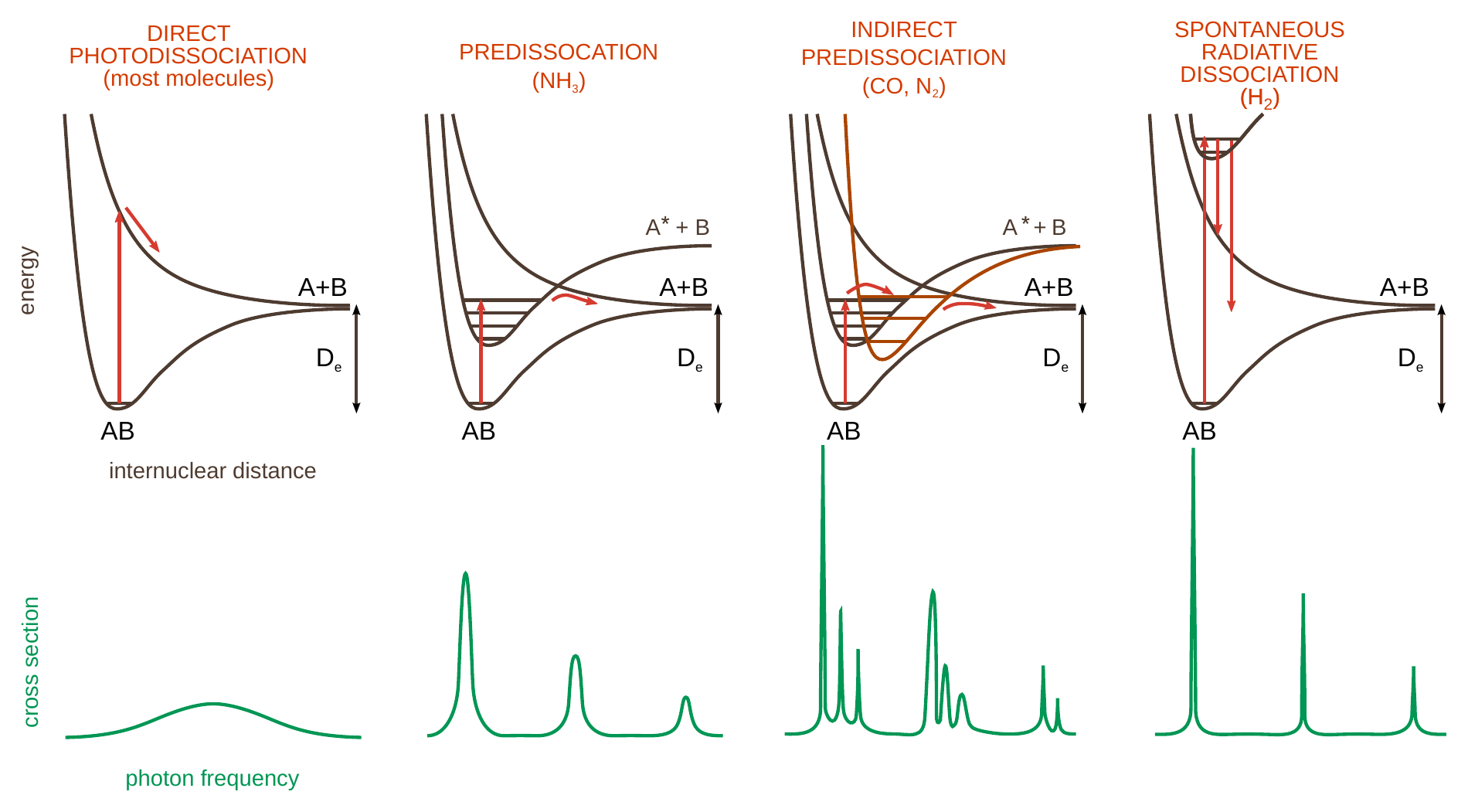}
  \caption{Schematic cross sections for photodissociation and their associated dynamical pathway (arrows) through ground and excited state potential-energy curves. For polyatomic molecules these curves represent a cross section through a multidimensional energy surface. We note that the integrated values of the various cross sections may be similar, leading to orders-or-magnitude greater peak magnitudes for indirect mechanisms. Modified from \textcite{van_dishoeck_visser2015}.}
  \label{fig:photodissociation types}
\end{figure*}

\begin{figure}[t]
  \centering
  \includegraphics[width=\linewidth]{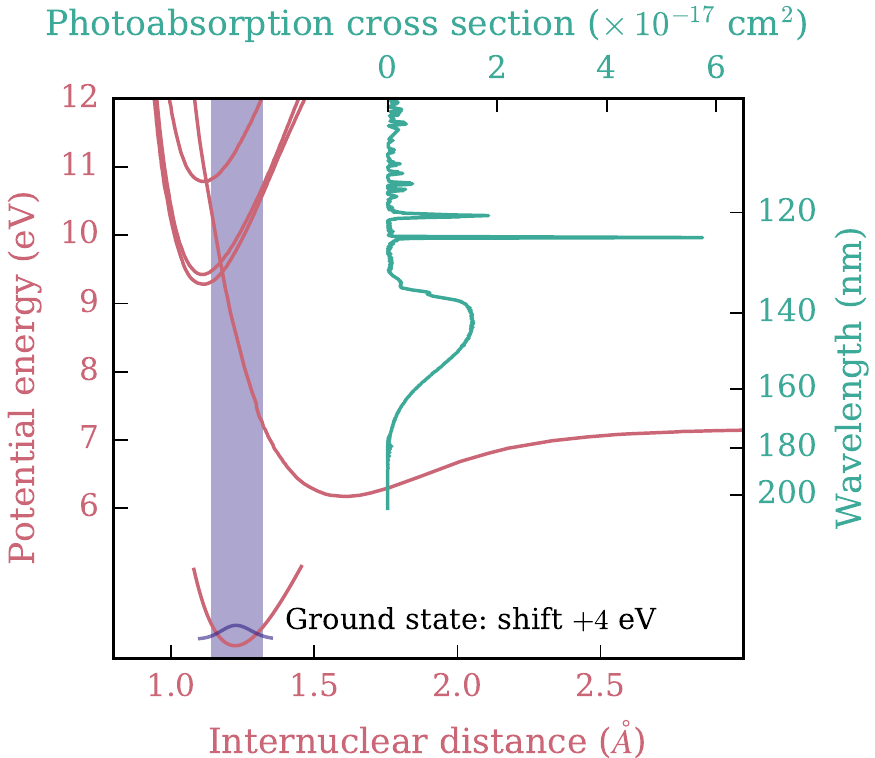}
  \caption{Potential energy curves for the ground and excited state of
    \ce{O2} (red curves) \cite{guberman1977,lewis_etal1998,lewis_etal2001b}, and the \ce{O2}
    photoabsorption section from Sect.~\ref{sec:O2} (blue curve). Shown
    on equivalent energy and wavelength scales. The energy scale is
    relative to the minimum of the ground-state electronic potential
    curve, shown here shifted upwards by 4\,eV.
    The shaded area shows the vertical excitation region.}
  \label{fig:O2_PECs}
\end{figure}

The wavelength dependence of a molecular cross section can be schematically associated with the structure of its electronically-excited states and categorised by its dissociation mechanism.
These mechanisms are depicted in Fig.~\ref{fig:photodissociation types} by potential-energy curves.
For small molecules absorption into an excited state whose potential is
repulsive along 1 or more nuclear coordinates results in 100\%-efficient direct dissociation of the molecule on sub-picosecond time scales  (see Sect. \ref{sect:theory} and Fig.~\ref{fig:O2_PECs} for a description and example of
potential energy curves). The corresponding cross section has a broad wavelength
distribution, covering several nm decades and 
peaking at the energy corresponding to vertical excitation from the ground-state
equilibrium nuclear distance to the excited repulsive curve of AB$^*$. Typical peak values range from a few
$\times 10^{-18}$ to a few $\times10^{-17}$\,cm$^{2}$.

In contrast, the cross sections for the predissociation and indirect predissociation  processes are highly
structured, consisting of sharp peaks at discrete wavelengths. 
In these
cases, the initial absorption occurs into a bound
excited electronic state, which subsequently interacts non-radiatively
with a nearby repulsive electronic state. The predissociation rate, and inversely-proportional linewidth, depends strongly on the details of this interaction and may vary from level to level, particularly in the indirect case where further intermediate states are involved.
\change{A non-unity dissociation probability will result from competitive rates for predissociation, $k^{\rm pre}$, and spontaneous emission, $A$; so that $\eta^\text{d}=k^{\rm pre}/(k^{\rm pre} + A)$.
An excited molecule decaying by emission may follow multiple competing pathways involving multiple photons of different wavelengths in a de-excitation cascade through excited and ground electronic states, and result in a super-thermal population of ground state rotational and vibrational levels. 
Only the total emissive decay rate, $A$, is considered in this paper.
}
CO and N$_2$ are the best known astrophysical examples of molecules for which predissociation is dominant. 

The fourth process is spontaneous radiative dissociation, in that the
an excited bound state radiates back into the vibrational
continuum of a lower state with a line-dependent probability. For
H$_2$, this is the dominant photodissociation pathway
\cite{stecher1967}, but not for any other interstellar molecule.
Peak cross sections for discrete lines may reach $10^{-14}$\,cm$^{2}$ over a width of $< 0.1$ nm.

Even though the peak cross sections may differ greatly for the various dissociation mechanisms depicted in Fig.~\ref{fig:photodissociation types}, the integrated cross sections $\int \sigma (\lambda) \text{d}\lambda$ are often comparable.
Further discussion and details of these phenomena may be found in \textcite{van_dishoeck1988} and \textcite{van_dishoeck_visser2015}

As a real example, Fig.~\ref{fig:O2_PECs} illustrates that both direct continuous and discrete dissociation channels are
available for a molecule like O$_2$. 
The appearance of continuum absorption between 180 and 130\,nm in Fig.~\ref{fig:O2_PECs} is consistent with an upward projection of the ground-state vibrational wavefunction to its intersection with the lowest-energy unbound excited state.
The line absorption at shorter wavelengths occurs through the predissociation of multiple bound states above 9.2\,eV.

Large molecules such as polycyclic aromatic hydrocarbons (PAHs) are
much more stable against photodissociation than the small species
considered here because the absorptions are followed by non-radiative
decay to the ground state (so-called internal conversion) from which
there is only a small probability that the molecule finds a path to
dissociation. In clouds exposed to very intense UV radiation, such as
protoplanetary disks or near active galactic nuclei,
photodissociation may however become significant on astronomical
time scales, and small PAHs (less than about 50 carbon atoms) cannot
survive. Photodissociation of these large molecules was not taken into
account for this database but is discussed most recently by
\textcite{lepage2003} and \textcite{visser2007} \cite[see also summary
in][]{van_dishoeck_visser2015}. New experimental data on the
photofragmentation and ionisation probabilities of PAHs is becoming available
\cite{zhen_junfeng2015,zhen_junfeng2016}.

In general, the key characteristics of a photoabsorption cross section are:
\begin{itemize}
  \item The long-wavelength dissociation threshold: Usually this is given by the dissociation energy of the ground electronic state. The cross section at this threshold is often orders of magnitude smaller than at shorter wavelengths. However, radiation intensity decreases rapidly with shortening wavelength for radiation fields dominated by cool stars, so even a low cross section near threshold can dominate the photodissociation rate.
  \item The ionisation threshold: This affects the relative importance of photodissociation and photoionisation. No ionisation will occur in most astrophysical environments if this threshold occurs at wavelengths shorter than \np[nm]{91.2}.
  \item The wavelengths of absorption lines: Maxima in the cross section can influence the total absorption rate if they correspond to emission lines, such as occur in the simulated TW-Hydra radiation field, or in cosmic-ray induced radiation.
  \item The cross section corresponding to the Lyman-$\alpha$ emission line at 121.3\,nm, that can singularly dominate the ultraviolet flux.
  \item The characteristic width of absorption features: the precise linewidths, that can range from $0.001$ to several $10$s of nm has a strong effect on their ability to self-shield  (Sect.~\ref{sec:shielding functions}).
\end{itemize}

Further background information and descriptions of the collected data sources and cross sections for all updated and new species relative to \textcite{van_dishoeck2006} are given in Sect. 4.
In the following subsections, more background information on experimental and theoretical determinations of cross sections is given, since this is relevant for assessing the inherent uncertainty of the various data we used, and motivating our choices of adopted cross sections.

\subsection{Experimental cross sections}
\label{sec:exp cross sections}

Photoabsorption cross sections are most frequently recorded directly,
by observing the transmission of an ultraviolet
continuum through a gas sample, with radiation generated by discharge lamps
\cite[e.g.,][]{ogawa1975,dehmer1976} or synchrotrons
\cite[e.g.,][]{yoshino1996,cheng2011}, and dispersed by diffraction
gratings or interferometry \cite{yoshino2006}.  Laser-generated
ultraviolet radiation is sometimes used in photoabsorption experiments
and provides the highest spectral resolution
\cite{gao2013,niu2014}, but generally cannot be tuned over a large
wavelength range or provide controllable intensity.  Special
techniques are required to record ultraviolet photoabsorption spectra
for wavelengths shorter than 105\,nm due to the lack of transmitting
material for use as windows or beam splitters.
\change{For example, utilising frequency-multiplied lasers \cite[e.g.,][]{ubachs2005,stark1999}, synchrotron radiation sources \cite[e.g.,][]{yoshino2006} and, recently, the vacuum-ultraviolet Fourier transform spectrometer at the SOLEIL synchrotron \cite{de_oliveira2011,eidelsberg2012}, or occasionally the interstellar laboratory \cite[e.g.,][]{federman2001}.}

The interpretation of experimental photoabsorption spectra is
generally straight-forward except where the instrumental spectral
resolution is insufficient to resolve detailed structure of molecules
with non-continuum absorption (c.f., \ce{N2} as opposed to \ce{CH4} in Sect.~\ref{sec:summary of cross sections}).
In this case, care needs to be taken not to underestimate the
integrated cross section, potentially by more than an order of
magnitude \cite{hudson_carter1968}.  Due to this issue, for some
molecules (e.g., \ce{H2}, \ce{N2}, and \ce{CO}) the recorded
absorption spectra must be analysed line-by-line and the true cross
section reconstructed without the limitation of experimental
broadening \cite{eidelsberg1992,heays2011b,glass-maujean2013a}.

Another difficulty concerns the calibration of absolute cross section
values, which rely on precise knowledge of the absorbing sample gas
column density and distribution of the ground-state rovibrational
population.  Neither quantity is generally diagnosable in
photoabsorption experiments involving transient radical species.  The
uncertainty of directly-measured photoabsorption cross sections is
usually between 10 to 20\% for stable species, and typically a factor of 2 to
5 for the case of radicals, if it is known at all.

A photoabsorption cross section can also be estimated at
relatively-low resolution by electron-energy-loss spectroscopy
\cite[e.g.,][]{chan1992,heays2012}, where monoenergetic electrons are
scattered from a low density of molecules and their final energy
spectrum mimics the resonant energy structure of the scatterer.  The
correspondence of photoabsorption to the energy loss of scattered
electrons relies on the incident beam being sufficiently energetic and
the energy-loss spectrum being recorded at small scattering angles
\cite{inokuti1971}.  The lack of spectral resolution in this kind of
experiment does not lead to an underestimate of unresolved features, in contrast to direct photoabsorption measurements, because of the linear relation between
cross section and the signal from the analysed electrons.
Electron-energy-loss cross sections can be recorded for energy-losses
spanning the entire photoabsorbing wavelength range and then
absolutely calibrated according to the Thomas-Reiche-Kuhn sum rule
\cite{backx1976} without detailed knowledge of the sample column
density.  These kinds of experiments typically yield an uncertainty of
30\% or better but can not resolve the detailed wavelength structure
of many molecules. They provide a useful comparison to benchmark the
accuracy of higher-spectral-resolution direct absorption cross sections.

The absorption of a photon with energy greater than the ionisation
energy of a molecule can produce charged fragments.  The resultant
photoions and photoelectrons can be experimentally manipulated with
electric fields and detected with high-efficiency, possibly
simultaneously \cite[e.g.,][]{backx1976,holland1993,edvardsson1999}.
When all fragments are simultaneously detected, it is possible to
spectroscopically examine the initial neutral species and produced
ions.  The simultaneous recording of photoion or photoelectron, and
photoabsorption spectra provides a direct measurement of the fraction
of excited molecules that decay via ionisation versus dissociation.
Most molecules are ionised with near-100\% efficiency by photons more than
about 2\,eV above their ionisation thresholds. 

The branching to different dissociative-ionisation channels follows
from the discrimination of photofragments with different
charge-to-mass ratios.  Less commonly, experimenters count neutral
photofragments \cite[e.g.,][]{morley1992,walter_etal1993,gao2013} or
detect their fluorescence following dissociation into excited states
\cite[e.g.,][]{lee1984,biehl1994}.  The emission of photoexcited molecules has
is sometimes recorded \cite[e.g.,][]{jonas1990,heays2014b} and
provides further information on the decay branching of excited states.

\subsection{Theoretical cross sections}
\label{sect:theory}

\subsubsection{General considerations}

Quantum-chemical calculations of the excited electronic states of
atoms and small molecules can be used to simulate photoabsorption,
dissociation, and ionisation cross sections, and are particularly
useful for species that are difficult to measure in the laboratory,
such as radicals and ions \cite[see][for
  reviews]{kirby1988,van_dishoeck1988,van_dishoeck_visser2015}.

For molecules, such calculations require knowledge of the ground
state, one or more excited states, and the transition dipole moment
connecting them.  Ground and excited states are frequently summarised
by potential energy curves describing the electrostatic interaction
energy of the electrons as a function of the nuclear configuration.
Some of these are plotted in Fig.~\ref{fig:O2_PECs} for the
1-dimensional case of a diatomic molecule, \ce{O2}.  These states
are labelled by their symmetry and a numerical label increasing with excitation
energy. For example, the 2$ ^1\Sigma^-$ state denotes the second state
of $^1\Sigma^-$ symmetry. If this state has been observed
experimentally, it often also has an alphabetic label, with the
letters; A, B, C, \dots; mostly increasing with excitation energy. For polyatomic
molecules, the notation becomes $\tilde A$, $\tilde B$,  $\tilde C$ \dots.

The ground state potential energy of a stable molecule must form a
well, leading to a quantised spectrum of bound states with increasing
vibrational excitation.  Electronically-excited states may be bound or
repulsive, that is, possess no minimum energy (see
Figs. \ref{fig:photodissociation types} and \ref{fig:O2_PECs}).  As discussed above, this distinction
dramatically affects the structure of the resultant photoabsorption
spectrum.  The example in Fig.~\ref{fig:O2_PECs} reproduces the most
important ultraviolet-excited states of \ce{O2}
\cite{lewis_etal1998,lewis_etal2001b} alongside its photoabsorption
cross section.
The strength of the cross section into each excited state
depends on its specific transition moment with the ground state, and
the size of the overlap of ground and excited vibrational
wavefunctions.  Within the Born-Oppenheimer approximation, this second
factor requires a separate calculation considering the movement
(vibration) of nuclei in a precomputed potential-energy environment.

\change{The effects of nonzero molecular rotation are not usually explicitly included in ab initio cross section calculations, but can be simulated by assigning standard rotational-line strength factors \cite{larsson1983} and assuming a population distribution of ground state rotational levels.
These factors are not always accurate if centrifugal effects significantly alter the vibrational overlap of ground and excited states or the dissociation efficiency \cite[e.g,.][]{lewis_etal2005b,heays2011b}.
}

The spectral width of absorption features is characteristic of the
lifetime of the excited state.  The 135 to 180\,nm absorption of
\ce{O2} is rapidly followed by dissociative decay into O atoms, after
less than 1\,ps.  The bound states at higher energy survive longer,
but still dissociate because of a second-order interaction induced by
the shown curve-crossing with the dissociative state
\cite{lewis_etal2002}.  States that take sufficiently long to
dissociate, greater than typical Einstein $A$ coefficients of about
1\,ns, will have time to decay radiatively by spontaneous emission.
Strong interactions lead to dissociation rates faster than $10^{12}$
s$^{-1}$, implying a 100\% dissociation efficiency. Detailed studies of
the time evolution of nuclear motion may then provide an estimate of
the dissociation branching ratio of photoexcited states
\cite[e.g.,][]{van_dishoeck1984b,kroes1997b,lewis_etal2002,heays2011b}.

The intrinsic linewidths of absorption features are given
by the inverse of the sum of the predissociation and spontaneous decay
rates, $1/(k^\textrm{pre} + A)$. A predissociation rate $k^\textrm{pre}$ as large as
$10^{11}$ s$^{-1}$ corresponds to a linewidth of \np[nm\,FWHM]{5e-4} (full-width half-maximum) at a
wavelength of 100 nm. In velocity units, this amounts to 1\,km\,s$^{-1}$, which is comparable or less than the typical turbulent Doppler broadening of an interstellar clouds.  Intrinsic
widths seen in experimental data can vary greatly, from Doppler-broadening dominated (e.g., \ce{N2}) to greater than 1\,nm
(e.g., \ce{NH3} and \ce{C2H2}), obscuring all rotational structure when
strongly predissociated.  Such accurate knowledge of absorption line
profiles is however only needed (i) to determine overlap with specific
lines that dominate the radiation field in some astrophysical
environments such as Lyman-$\alpha$; (ii) to compute optical depth
and self-shielding capacity.

For polyatomic molecules, the calculation of multidimensional
excited-state potential-energy surfaces including all degrees of freedom, and subsequent nuclear
dynamics on those surfaces, becomes computationally prohibitive.
Moreover, such detail is often not
needed to compute accurate photodestruction rates since the necessary
absorption strengths are largely determined by one or a few excited
states and, for cold molecules, the relevant nuclear motion only probes a small region of
coordinate space around the ground state equilibrium geometry.
Therefore, a simpler alternative is to only compute vertical excitation energies and
transition dipole moments defined at the equilibrium geometry, and assume a dissociation probability for
the excited state.
This reduces the photoabsorption cross section of an entire electronic transition to a single wavelength, whereas the real cross section may be very broad.
This approximation is quite sufficient for the case of photodissociation in a continuum-like radiation field, for example, the ISRF.

Our database includes vertical-excitation cross sections computed for a number of molecules and summarised in \textcite{van_dishoeck1988}, \textcite{van_dishoeck2006} and \textcite{van_hemert2008}, based on our work and that of other groups \cite[e.g.,][]{kirby1988,roueff2014}.  
These results are based on high-level configuration interaction calculations (see \textcite{van_dishoeck_visser2015} for a top level overview of such calculations).  
In the latest calculations by \textcite{van_hemert2008}, up to 9 electronic states per symmetry are considered, including diffuse (Rydberg) states.  
For the lower-energy states, comparisons with independent calculations and experiments indicate that the deduced excitation energies are accurate to better than 0.3\,eV and that oscillator strengths connecting them to the ground state agree within 30\% or better.  
For the higher states, typically the 5th root and higher per symmetry, the accuracy decreases because many states and orbitals can mix.  
Such calculations still provide a good indication of the location of those states and their combined strengths, typically within a factor of 2.

Only states above the ground-state dissociation limit and below the ionisation potential of the molecule need be taken into account for photodissociation calculations.  
The dissociation efficiency, $\eta_d$, of all calculated excited states in this range and presented here is assumed to be unity, that is, they are purely repulsive and dissociate directly, or have resonant levels and decay by predissociation (exceptions are H$_2$, CO and N$_2$ for which level-specific probabilities are available).
For larger molecules (i.e., three or more atoms) dissociation rates assuming unity efficiency should be regarded as upper limits, given that internal conversion to a lower (dissociative) electronic state is usually much more rapid than radiative decay, because of their high density of states \cite[e.g.,][]{leger1989,jochims1994}.
Above the ionisation potential, all absorption is assumed to lead to photoionisation (dissociative or not). 
Also, only states lying below the Lyman limit of 13.6\,eV are included.

Even after computing a full potential-energy surface the wavelengths and absorption oscillator strengths of known bound vibrational levels, their predissociation lifetimes and widths may still be unknown.
Additionally, the real photoabsorption cross section into a bound vibrational level may involve multiple rotational transitions, effectively increasing the width of its photoabsorption envelope.
We assumed a Gaussian profile to encompass these phenomena for theoretical predissociated levels used in our cross section database, and uniformly assumed a width of 1\,nm\,FWHM, where our following calculation of interstellar photodissociation rates is  not sensitive to the precise value of this width.

The accuracy of cross sections derived from ab initio calculations can be remarkably high, within 20\% or better, for  diatomic molecules \cite[e.g., OH][]{van_dishoeck1983,van_dishoeck1984} and sometimes for polyatomic cases (e.g., \ce{H2O} in Sect.~\ref{sec:H2O}).
The wavelengths of absorption lines exciting predissociated bound levels may be significantly in error where non-Born-Oppenheimer interactions shift energy levels and redistribute oscillator strengths between excited states \cite[e.g.,][]{van_dishoeck1984b}. 
However, inaccuracies introduced by these effects are much reduced in the calculation of interstellar photodissociation rates that average over many states (e.g., \ce{C3H} in Sects.~\ref{sec:l-C3H} and \ref{sec:c-C3H}).
The largest uncertainty in ab initio photodissociation cross sections then arises, in most cases, from inaccurately-calculated or neglected states lying close to the ionisation threshold, which are numerous and difficult to calculate or measure. 

Empirical corrections can resolve some of the uncertainty of theoretical cross sections, either by shifting absorption features to their experimentally known wavelengths, or adjusting the underlying excited state potential-energy surfaces to produced cross sections in better agreement with experiment \cite[e.g.,][]{heays2014b}.
For a few molecules in our database, we added a guessed wavelength and integrated cross section to approximate the influence of neglected high-lying states, with an associated order-of-magnitude uncertainty (in general, these additions contribute a small amount to the overall cross section and its uncertainty).
For reference, inclusion of a hypothetical state at 9\,eV with an oscillator strength of 0.1 would increase the ISRF photodissociation rates by $3.5\times 10^{-10}$\,s$^{-1}$.
In general, no corrections were made for possibly-neglected states above the ionisation limit and below 13.6\,eV.
This is because the lowest Rydberg members are generally computed explicitly, and the oscillator strengths of higher Rydberg states converging to the ionisation threshold decrease roughly as 1/$n^3$ ($n$ is the principal quantum number) and do not contribute much.

For all theoretical cross sections in our database, a minimum photodissociation cross section of \np[cm^2]{5e-20} was assumed  between the dissociation threshold and Lyman-limit at 91.2\,nm.
This weak continuum negligibly increases the integrated cross section but ensures a low but nonzero cross section overlaps the strong emission lines present in some interstellar radiation fields.

\subsubsection{Atomic photoionisation}

Atomic photoionisation cross sections have long been an object of
theoretical study due to their influence on the interpretation of
spectroscopic observations of astrophysical plasmas in ionised
interstellar gas as found around stars, active galactic nuclei, and
elsewhere \cite{seaton1951,osterbrock1979,ferland2003,tielens2013}.  
We used theoretical cross sections here for the photoionisation of some neutral atoms.  These are generally the
result of $R$-matrix calculations 
\cite{seaton1985,mendoza1996,mclaughlin2001,zatsarinny2013}, and produce continuum cross
sections that are generally accurate to within 20\%.
The specification of resonant structure evident in most atomic cross
sections presents more difficulty for this method, although the uncertainties are
diminished for photoionisation rates calculated following integration over many resonances.

\subsection{Cross section databases}

There are various public databases of photoabsorption, dissociation,
and ionisation cross sections, and some data from these were
incorporated into our assessment of molecular and atomic cross
sections.  A comprehensive set of laboratory photoabsorption cross
sections and a smaller amount of data concerning photofragment
branching ratios is contained in the MPI Mainz UV/VIS Spectral Atlas.\footnote{\url{satellite.mpic.de/spectral_atlas}}
Earlier
compilations are given by
\textcite{calvert1966,okabe1978,lee1984,gallagher1988,ashfold2006}.  The TOPbase\footnote{\url{cdsweb.u-strasbg.fr/topbase/topbase.html}} database of photoionisation cross sections
\cite{mendoza1996} includes $R$-matrix
calculations for many atoms, including their highly-charged states.  A
collation of molecular and atomic cross sections from multiple sources
is contained in the PHIDRATES database\footnote{\url{phidrates.space.swri.edu}}
\cite{huebner1992,huebner2015} as well as
calculations of their photodissociation and photoionisation rates in
the ISRF and solar radiation fields.  Our compilation differs somewhat
from \textcite{huebner2015} for molecules in common, due to different
choices of cross section data and a larger focus on highly excited
electronic states in our work that are more important for the ISRF
than for the solar radiation field.

More specialised databases containing cross sections of astrochemical
interest are the MOLAT and SESAM databases of vacuum-ultraviolet (VUV)
spectroscopy\footnote{\url{molat.obspm.fr} and \url{sesam.obspm.fr}}, including CO, \ce{H2}, and
\ce{N2}; the Harvard CfA VUV database\footnote{\url{www.cfa.harvard.edu/amp/ampdata/cfamols.html}}
including primary data on many small molecules including wavelengths
as short as 80\,nm.; and the UGA Opacity Project database\footnote{\url{www.physast.uga.edu/ugamop/index.html}}.
The VAMDC virtual portal\footnote{\url{portal.vamdc.org}} integrates some of these data.

    \onecolumn
    \begin{landscape}
    \begin{centering}
    \renewcommand*{\thefootnote}{\alph{footnote}}
    \begin{longtable}{ccccccccl}
    \caption{Some summary properties of atomic and molecular cross sections. \label{tab:cross_section_properties}}\\
    \hline\hline
    {} 
    & Thresh. 
    & \multicolumn{2}{c}{Threshold (nm)\footnotemark[2]} 
    & \multicolumn{2}{c}{Ly-$\alpha$ cross section (cm$^2$)\footnotemark[3]} 
    & \multicolumn{2}{c}{Uncertainty\footnotemark[4]} 
    & References\footnotemark[5] 
    \\
    Species 
    & Dissoc. prod.\footnotemark[1] 
    & Dissoc. 
    & Ionis. 
    & Dissoc. 
    & Ionis. 
    & ISRF 
    & Ly-$\alpha$ 
    & Cross sec./Dissoc. threshold/Ionis. threshold
    \\
    \hline
    \endfirsthead
    \caption{continued.}\\
    \hline\hline
    {} 
    & Thresh. 
    & \multicolumn{2}{c}{Threshold (nm)\footnotemark[2]} 
    & \multicolumn{2}{c}{Ly-$\alpha$ cross section (cm$^2$)\footnotemark[3]} 
    & \multicolumn{2}{c}{Uncertainty\footnotemark[4]} 
    & References\footnotemark[5] 
    \\
    Species 
    & Dissoc. prod.\footnotemark[1] 
    & Dissoc. 
    & Ionis. 
    & Dissoc. 
    & Ionis. 
    & ISRF 
    & Ly-$\alpha$ 
    & Cross sec./Dissoc. threshold/Ionis. threshold
    \\
    \hline
    \endhead
    \hline
    \endfoot
\ce{H} & -- & -- & 91.2 & 0 & 0 & A & -- & {\scriptsize Sect.~\ref{sec:H}/--/\textcite{kelly1987_partI}} \\
\ce{Li} & -- & -- & 230 & 0 & \np{1.33e-18} & A & A & {\scriptsize \textcite{huebner2015}/--/\textcite{kelly1987_partI}} \\
\ce{C} & -- & -- & 110 & 0 & 0 & B & -- & {\scriptsize Sect.~\ref{sec:C}/--/\textcite{kelly1987_partI}} \\
\ce{N} & -- & -- & 85 & 0 & 0 & A & -- & {\scriptsize Sect.~\ref{sec:N}/--/\textcite{kelly1987_partI}} \\
\ce{O} & -- & -- & 91.04 & 0 & 0 & B & -- & {\scriptsize Sect.~\ref{sec:O}/--/\textcite{kelly1987_partI}} \\
\ce{Na} & -- & -- & 241 & 0 & \np{1.07e-19} & B & B & {\scriptsize \textcite{huebner2015}/--/\textcite{kelly1987_partI}} \\
\ce{Mg} & -- & -- & 162 & 0 & \np{7.13e-20} & B & B & {\scriptsize Sect.~\ref{sec:Mg}/--/\textcite{kelly1987_partI}} \\
\ce{Al} & -- & -- & 207 & 0 & \np{8.71e-18} & B & B & {\scriptsize Sect.~\ref{sec:Al}/--/\textcite{kelly1987_partI}} \\
\ce{Si} & -- & -- & 152 & 0 & \np{3.61e-17} & B & B & {\scriptsize Sect.~\ref{sec:Si}/--/\textcite{kelly1987_partI}} \\
\ce{P} & -- & -- & 118 & 0 & 0 & C & -- & {\scriptsize Sect.~\ref{sec:P}/--/\textcite{kelly1987_partI}} \\
\ce{S} & -- & -- & 120 & 0 & 0 & A & -- & {\scriptsize Sect.~\ref{sec:S}/--/\textcite{kelly1987_partI}} \\
\ce{Cl} & -- & -- & 96 & 0 & 0 & B & -- & {\scriptsize Sect.~\ref{sec:Cl}/--/\textcite{kelly1987_partI}} \\
\ce{K} & -- & -- & 286 & 0 & \np{2.22e-19} & B & B & {\scriptsize Sect.~\ref{sec:K}/--/\textcite{kelly1987_partI}} \\
\ce{Ca} & -- & -- & 203 & 0 & \np{9.13e-19} & B & B & {\scriptsize Sect.~\ref{sec:Ca}/--/\textcite{kelly1987_partI}} \\
\ce{Ti} & -- & -- & 182 & 0 & \np{5.72e-18} & C & C & {\scriptsize \textcite{reilman1979}/--/\textcite{sohl1990}} \\
\ce{Cr} & -- & -- & 183 & 0 & \np{7.35e-18} & B & B & {\scriptsize \textcite{reilman1979}/--/\textcite{kelly1987_partI}} \\
\ce{Mn} & -- & -- & 167 & 0 & \np{4.42e-20} & B & B & {\scriptsize$\left\{\begin{tabular}{l} \textcite{reilman1979,mcguire1968}/--/ \\\textcite{kelly1987_partII}\end{tabular}\right.$} \\
\ce{Fe} & -- & -- & 158 & 0 & \np{4.86e-18} & A & A & {\scriptsize Sect.~\ref{sec:Fe}/--/\textcite{kelly1987_partII}} \\
\ce{Co} & -- & -- & 158 & 0 & \np{1.77e-19} & B & B & {\scriptsize \textcite{reilman1979}/--/\textcite{kelly1987_partI}} \\
\ce{Ni} & -- & -- & 162 & 0 & \np{3.12e-19} & B & B & {\scriptsize \textcite{van_dishoeck1988}/--/\textcite{kelly1987_partII}} \\
\ce{Zn} & -- & -- & 132 & 0 & \np{5.91e-20} & B & B & {\scriptsize Sect.~\ref{sec:Zn}/--/\textcite{kelly1987_partII}} \\
\ce{Rb} & -- & -- & 297 & 0 & \np{1.49e-19} & B & B & {\scriptsize$\left\{\begin{tabular}{l} \textcite{weisheit1972,suemitsu1983}/--/ \\\textcite{sanonetti2006}\end{tabular}\right.$} \\
\ce{Ca+} & -- & -- & 104 & 0 & 0 & B & -- & {\scriptsize \textcite{black1972}/--/\textcite{kelly1987_partI}} \\
\ce{H-} & -- & -- & 1470 & 0 & \np{4.75e-18} & B & B & {\scriptsize \textcite{wishart1979}/--/\textcite{kinghorn1997}} \\
\ce{H2} & \ce{H + H} & 274 & 82 & 0 & 0 & \phantom{+}A+ & -- & {\scriptsize$\left\{\begin{tabular}{l} Sect.~\ref{sec:H2}/\textcite{huber_herzbergIV}/ \\\textcite{wolniewicz1995}\end{tabular}\right.$} \\
\ce{H2+} & \ce{H + H+} & 290 & -- & \np{6.83e-18} & 0 & A & A & {\scriptsize \textcite{dunn1968}/\textcite{babb2015}/--} \\
\ce{H3+} & \ce{H+ + H2} & 283 & -- & 0 & 0 & A & -- & {\scriptsize \textcite{kulander1978}/\textcite{cosby1988}/--} \\
\ce{CH} & \ce{C + H} & 358 & 117 & \np{5.00e-20} & 0 & C & C & {\scriptsize$\left\{\begin{tabular}{l} \textcite{van_dishoeck1987}/\textcite{huber_herzbergIV}/ \\\textcite{huber_herzbergIV}\end{tabular}\right.$} \\
\ce{CH+} & \ce{C + H+} & 303 & -- & \np{5.00e-20} & 0 & A & C & {\scriptsize \textcite{kirby1980,uzer1978}/\textcite{huber_herzbergIV}/--} \\
\ce{CH2} & \ce{C + H2} & 417 & 119 & \np{5.00e-20} & 0 & C & C & {\scriptsize$\left\{\begin{tabular}{l}\textcite{van_dishoeck1996,kroes1997b}/\\\textcite{kroes1997b}/\\\textcite{herzberg1966_polyatomic_molecules}\end{tabular}\right.$} \\
\ce{CH2+} & \ce{CH+ + H} & 204 & -- & \np{5.00e-20} & 0 & B & C & {\scriptsize Sect.~\ref{sec:CH2+}/\textcite{vane2007}/--} \\
\ce{CH3} & \ce{CH + H2} & 271 & 126 & \np{5.00e-20} & \np{5.31e-18} & B & B & {\scriptsize Sect.~\ref{sec:CH3}/\textcite{yu1984}/\textcite{schulenberg2006}} \\
\ce{CH4} & \ce{CH3 + H} & 277 & 98 & \np{1.80e-17} & 0 & \phantom{+}A+ & \phantom{+}A+ & {\scriptsize Sect.~\ref{sec:CH4}/\textcite{gans2011}/\textcite{sorenson1995}} \\
\ce{CH4+} & \ce{CH3+ + H} & 1031 & -- & \np{5.00e-20} & 0 & B & C & {\scriptsize Sect.~\ref{sec:CH4+}/\textcite{van_dishoeck1980}/--} \\
\ce{C2} & \ce{C + C} & 193 & 102 & \np{4.98e-19} & 0 & B & B & {\scriptsize$\left\{\begin{tabular}{l} \textcite{pouilly1983}/\textcite{feller2000}/ \\\textcite{huber_herzbergIV}\end{tabular}\right.$} \\
\ce{C2H} & \ce{C2      + H} & 253 & 109 & \np{3.21e-18} & 0 & B & C & {\scriptsize$\left\{\begin{tabular}{l} Sect.~\ref{sec:C2H}/\textcite{van_hemert2008}/ \\\textcite{van_hemert2008}\end{tabular}\right.$} \\
\ce{C2H2} & \ce{C2H + H} & 217 & 109 & \np{5.78e-17} & 0 & \phantom{+}A+ & B & {\scriptsize Sect.~\ref{sec:C2H2}/\textcite{wu2010}/\textcite{metzger1964}} \\
\ce{C2H4} & \ce{C2H2 + H2} & 258 & 118 & \np{2.36e-17} & 0 & C & C & {\scriptsize$\left\{\begin{tabular}{l} \textcite{zelikoff1953}/\textcite{petrank1992}/ \\\textcite{knowles1974}\end{tabular}\right.$} \\
\ce{C2H6} & \ce{C2H5 + H} & 287 & 108 & \np{2.26e-17} & 0 & \phantom{+}A+ & \phantom{+}A+ & {\scriptsize$\left\{\begin{tabular}{l} Sect.~\ref{sec:C2H6}/\textcite{bauschlicher1995}/ \\\textcite{chupka1971}\end{tabular}\right.$} \\
\ce{C3} & \ce{C2 + C} & 268 & 102 & \np{1.38e-17} & 0 & C & C & {\scriptsize Sect.~\ref{sec:C3}/\textcite{kim1997}/\textcite{benedikt2005}} \\
\ce{\textit{l}-C3H} & \ce{C3      + H} & 379 & 144 & \np{5.00e-20} & 0 & B & C & {\scriptsize$\left\{\begin{tabular}{l} Sect.~\ref{sec:l-C3H}/\textcite{van_hemert2008}/ \\\textcite{van_hemert2008}\end{tabular}\right.$} \\
\ce{\textit{c}-C3H} & \ce{C3      + H} & 289 & 129 & \np{5.00e-20} & 0 & B & C & {\scriptsize$\left\{\begin{tabular}{l} Sect.~\ref{sec:c-C3H}/\textcite{van_hemert2008}/ \\\textcite{van_hemert2008}\end{tabular}\right.$} \\
\ce{HC3H} & \ce{l-C3H     + H} & 400 & 138 & \np{5.00e-20} & 0 & B & C & {\scriptsize Sect.~\ref{sec:HC3H}/\textcite{mebel1998}/\textcite{taatjes2005}} \\
\ce{\textit{l}-C3H2} & \ce{l-C3H     + H} & 320 & 119 & \np{5.00e-20} & 0 & B & C & {\scriptsize Sect.~\ref{sec:l-C3H2}/\textcite{mebel1998}/\textcite{clauberg1992}} \\
\ce{\textit{c}-C3H2} & \ce{c-C3H     + H} & 284 & 136 & \np{5.00e-20} & 0 & B & C & {\scriptsize Sect.~\ref{sec:c-C3H2}/\textcite{mebel1998}/\textcite{clauberg1992}} \\
\ce{\textit{l}-C4} & \ce{C3      + C} & 263 & 116 & \np{5.00e-20} & 0 & B & C & {\scriptsize Sect.~\ref{sec:l-C4}/\textcite{benedikt2005}/\textcite{ortiz1993}} \\
\ce{\textit{l}-C4H} & \ce{C4      + H} & 267 & 129 & \np{5.00e-20} & 0 & B & C & {\scriptsize$\left\{\begin{tabular}{l} Sect.~\ref{sec:l-C4H}/\textcite{van_hemert2008}/ \\\textcite{van_hemert2008}\end{tabular}\right.$} \\
\ce{\textit{l}-C5H} & \ce{C5      + H} & 348 & 168 & \np{5.00e-20} & 0 & B & C & {\scriptsize$\left\{\begin{tabular}{l} Sect.~\ref{sec:l-C5H}/\textcite{van_hemert2008}/ \\\textcite{van_hemert2008}\end{tabular}\right.$} \\
\ce{OH} & \ce{O + H} & 279 & 95 & \np{4.07e-18} & 0 & A & B & {\scriptsize$\left\{\begin{tabular}{l}\textcite{van_dishoeck1983};\\\textcite{van_dishoeck1984}/\\\textcite{huber_herzbergIV}/\\\textcite{garcia2015}\end{tabular}\right.$} \\
\ce{OH+} & \ce{O+ + H} & 247 & -- & \np{2.58e-25} & 0 & A & A & {\scriptsize \textcite{saxon1986}/\textcite{levin2000}/--} \\
\ce{H2O} & \ce{H + OH} & 242 & 98 & \np{1.53e-17} & 0 & \phantom{+}A+ & \phantom{+}A+ & {\scriptsize Sect.~\ref{sec:H2O}/\textcite{mordaunt1994}/\textcite{page1988}} \\
\ce{O2} & \ce{O + O} & 242 & 103 & \np{2.14e-19} & 0 & A & B & {\scriptsize$\left\{\begin{tabular}{l} Sect.~\ref{sec:O2}/\textcite{huber_herzbergIV}/ \\\textcite{dibeler1967}\end{tabular}\right.$} \\
\ce{O2+} & \ce{O + O+} & 186 & -- & \np{2.52e-18} & 0 & B & C & {\scriptsize Sect.~\ref{sec:O2+}/\textcite{huber_herzbergIV}/--} \\
\ce{HO2} & \ce{H + O2} & 476 & 109 & \np{5.00e-20} & 0 & C & C & {\scriptsize$\left\{\begin{tabular}{l} \textcite{mcadam1987,van_dishoeck1988}/\textcite{sawyer1989}/ \\\textcite{litorja1998}\end{tabular}\right.$} \\
\ce{H2O2} & \ce{OH + OH} & 556 & 117 & \np{9.65e-18} & 0 & A & A & {\scriptsize Sect.~\ref{sec:H2O2}/\textcite{holt1948}/\textcite{litorja1998}} \\
\ce{O3} & \ce{O2 + O} & 1180 & 99 & \np{2.97e-17} & 0 & B & B & {\scriptsize$\left\{\begin{tabular}{l} Sect.~\ref{sec:O3}/\textcite{grebenshchikov2007}/ \\\textcite{willitsch2005}\end{tabular}\right.$} \\
\ce{CO} & \ce{C + O} & 110 & 88 & 0 & 0 & \phantom{+}A+ & -- & {\scriptsize$\left\{\begin{tabular}{l} Sect.~\ref{sec:CO}/\textcite{tchang-brillet_etal1992}/ \\\textcite{zhao2014a}\end{tabular}\right.$} \\
\ce{CO+} & \ce{C+ + O} & 149 & -- & \np{5.00e-20} & 0 & B & C & {\scriptsize Sect.~\ref{sec:CO+}/\textcite{huber_herzbergIV}/--} \\
\ce{CO2} & \ce{CO + O} & 227 & 90 & \np{6.54e-20} & 0 & \phantom{+}A+ & \phantom{+}A+ & {\scriptsize Sect.~\ref{sec:CO2}/\textcite{song2014}/\textcite{wang1988}} \\
\ce{HCO} & \ce{H + CO} & 2037 & 152 & \np{5.00e-20} & 0 & C & C & {\scriptsize$\left\{\begin{tabular}{l}\textcite{bruna1976,van_dishoeck1988}/\\\textcite{mourik2000}/\\\textcite{mayer1995}\end{tabular}\right.$} \\
\ce{HCO+} & \ce{H + CO+} & 145 & -- & 0 & 0 & B & -- & {\scriptsize Sect.~\ref{sec:HCO+}/\textcite{koch1995b}/--} \\
\ce{H2CO} & \ce{H2 + CO} & 361 & 114 & \np{9.43e-18} & 0 & B & B & {\scriptsize Sect.~\ref{sec:H2CO}/\textcite{hopkins2007}/\textcite{guyon1976}} \\
\ce{NH} & \ce{N + H} & 362 & 92 & \np{3.10e-19} & 0 & B & B & {\scriptsize Sect.~\ref{sec:NH}/\textcite{gibson1985}/\textcite{de_beer1991b}} \\
\ce{NH+} & \ce{N+ + H} & 281 & -- & \np{4.68e-20} & 0 & C & C & {\scriptsize \textcite{van_dishoeck1986}/\textcite{tarroni1997}/--} \\
\ce{NH2} & \ce{NH + H} & 314 & 111 & \np{3.03e-20} & 0 & C & C & {\scriptsize Sect.~\ref{sec:NH2}/\textcite{gibson1985}/\textcite{willitsch2006}} \\
\ce{NH3} & \ce{NH + H2} & 301 & 122 & \np{8.37e-18} & \np{3.11e-19} & \phantom{+}A+ & \phantom{+}A+ & {\scriptsize Sect.~\ref{sec:NH3}/\textcite{leach2005}/\textcite{herzberg1966_polyatomic_molecules}} \\
\ce{N2} & \ce{N + N} & 127 & 79 & 0 & 0 & \phantom{+}A+ & -- & {\scriptsize Sect.~\ref{sec:N2}/\textcite{roncin_etal1989}/\textcite{klynning_pages1982}} \\
\ce{NO} & \ce{N + O} & 191 & 134 & \np{4.81e-19} & \np{2.08e-18} & A & B & {\scriptsize$\left\{\begin{tabular}{l} Sect.~\ref{sec:NO}/\textcite{huber_herzbergIV}/ \\\textcite{huber_herzbergIV}\end{tabular}\right.$} \\
\ce{NO2} & \ce{NO + O} & 398 & 127 & \np{5.00e-20} & \np{1.06e-19} & B & C & {\scriptsize \textcite{lee1984}/\textcite{lee1982}/\textcite{huber_herzbergIV}} \\
\ce{N2O} & \ce{N2 + O} & 735 & 96 & \np{3.85e-18} & 0 & B & C & {\scriptsize \textcite{lee1984}/\textcite{dibeler1967b}/\textcite{herzberg1966_polyatomic_molecules}} \\
\ce{CN} & \ce{C + N} & 160 & 87 & \np{4.00e-19} & 0 & B & C & {\scriptsize Sect.~\ref{sec:CN}/\textcite{huang1992}/\textcite{berkowitz1962}} \\
\ce{HCN} & \ce{H + CN} & 243 & 91 & \np{3.40e-17} & 0 & A & A & {\scriptsize Sect.~\ref{sec:HCN}/\textcite{davis1968}/\textcite{fridh1975}} \\
\ce{HC3N} & \ce{H + C3N} & 244 & 107 & \np{1.98e-17} & 0 & A & B & {\scriptsize Sect.~\ref{sec:HC3N}/\textcite{halpern1988}/\textcite{leach2014}} \\
\ce{CH3OH} & \ce{CH3O + H} & 280 & 113 & \np{1.39e-17} & 0 & \phantom{+}A+ & \phantom{+}A+ & {\scriptsize Sect.~\ref{sec:CH3OH}/\textcite{nee1985}/\textcite{nee1985}} \\
\ce{CH3CN} & \ce{CH2CN + H} & 308 & 102 & \np{1.09e-17} & 0 & A & A & {\scriptsize Sect.~\ref{sec:CH3CN}/\textcite{schwell2008}/\textcite{nuth1982}} \\
\ce{CH3SH} & \ce{CH3S + H} & 331 & 131 & \np{2.18e-17} & \np{2.37e-17} & A & B & {\scriptsize Sect.~\ref{sec:CH3SH}/\textcite{wilson1994}/\textcite{morgan1995}} \\
\ce{CH3CHO} & \ce{HCO + CH3} & 342 & 121 & \np{2.03e-17} & \np{1.23e-18} & A & A & {\scriptsize$\left\{\begin{tabular}{l} Sect.~\ref{sec:CH3CHO}/\textcite{heazlewood2009}/ \\\textcite{knowles1974}\end{tabular}\right.$} \\
\ce{CH3NH2} & \ce{CH3NH + H} & 244 & 135 & 0 & \np{2.44e-17} & A & A & {\scriptsize Sect.~\ref{sec:CH3NH2}/\textcite{hubin-franskin2002}/\textcite{hu2002}} \\
\ce{NH2CHO} & \ce{NH3 + CO} & 208 & 121 & \np{1.82e-17} & \np{1.88e-19} & A & A & {\scriptsize Sect.~\ref{sec:NH2CHO}/\textcite{gingell1997}/\textcite{leach2010}} \\
\ce{C2H5OH} & \ce{C2H5O + H} & 208 & 119 & \np{2.56e-17} & 0 & A & A & {\scriptsize Sect.~\ref{sec:C2H5OH}/\textcite{feng2002}/\textcite{cool2005}} \\
\ce{C3H7OH} & \ce{C3H7O + H} & 208 & 123 & \np{5.16e-17} & \np{2.80e-19} & A & A & {\scriptsize$\left\{\begin{tabular}{l} Sect.~\ref{sec:C3H7OH}/\textcite{salahub1971}/ \\\textcite{cool2005}\end{tabular}\right.$} \\
\ce{SH} & \ce{S + H} & 345 & 119 & \np{8.07e-18} & 0 & B & C & {\scriptsize$\left\{\begin{tabular}{l}Sect.~\ref{sec:SH}/\\\textcite{peebles2002}/\\\textcite{hsu1994}\end{tabular}\right.$} \\
\ce{SH+} & \ce{S+ + H} & 320 & -- & \np{7.41e-20} & 0 & B & C & {\scriptsize Sect.~\ref{sec:SH+}/\textcite{gustafsson1988}/--} \\
\ce{H2S} & \ce{H + SH} & 318 & 119 & \np{5.39e-17} & 0 & A & A & {\scriptsize Sect.~\ref{sec:H2S}/\textcite{shiell2000}/\textcite{cheng_bingming1998}} \\
\ce{CS} & \ce{C + S} & 169 & 110 & \np{5.36e-17} & 0 & C & C & {\scriptsize$\left\{\begin{tabular}{l} Sect.~\ref{sec:CS}/\textcite{coppens1995}/ \\\textcite{coppens1995}\end{tabular}\right.$} \\
\ce{CS2} & \ce{CS + S} & 278 & 123 & \np{1.71e-17} & \np{1.04e-18} & A & A & {\scriptsize Sect.~\ref{sec:CS2}/\textcite{okabe1972}/\textcite{fischer1993}} \\
\ce{OCS} & \ce{CO + S} & 280 & 111 & \np{7.14e-18} & 0 & A & A & {\scriptsize Sect.~\ref{sec:OCS}/\textcite{molina1981}/\textcite{wang1988}} \\
\ce{S2} & \ce{S + S} & 283 & 133 & 0 & \np{1.80e-18} & B & C & {\scriptsize Sect.~\ref{sec:S2}/\textcite{frederix2009}/\textcite{liao1986}} \\
\ce{SO} & \ce{S + O} & 233 & 120 & \np{1.06e-16} & 0 & B & C & {\scriptsize$\left\{\begin{tabular}{l} Sect.~\ref{sec:SO}/\textcite{reddy2000}/ \\\textcite{norwood1989}\end{tabular}\right.$} \\
\ce{SO2} & \ce{O + SO} & 219 & 100 & \np{3.86e-17} & 0 & A & A & {\scriptsize Sect.~\ref{sec:SO2}/\textcite{becker1995}/\textcite{holland1995}} \\
\ce{SiH} & \ce{Si + H} & 417 & 155 & \np{5.00e-20} & 0 & B & C & {\scriptsize$\left\{\begin{tabular}{l} \textcite{lewerenz1983}/\textcite{neufeld2009}/ \\\textcite{huber_herzbergIV}\end{tabular}\right.$} \\
\ce{SiH+} & \ce{Si+ + H} & 391 & -- & \np{2.22e-17} & 0 & A & B & {\scriptsize \textcite{van_dishoeck1988}/\textcite{huber_herzbergIV}/--} \\
\ce{SiO} & \ce{Si + O} & 150 & 108 & \np{5.00e-20} & 0 & B & C & {\scriptsize Sect.~\ref{sec:SiO}/\textcite{huber_herzbergIV}/\textcite{huber_herzbergIV}} \\
\ce{HCl} & \ce{H + Cl} & 279 & 97 & \np{9.19e-19} & 0 & A & A & {\scriptsize Sect.~\ref{sec:HCl}/\textcite{huber_herzbergIV}/\textcite{weiss1970}} \\
\ce{HCl+} & \ce{H + Cl+} & 267 & -- & \np{8.44e-19} & 0 & C & C & {\scriptsize Sect.~\ref{sec:HCl+}/\textcite{hotop1975}/--} \\
\ce{NaCl} & \ce{Na + Cl} & 293 & -- & \np{5.00e-20} & 0 & C & C & {\scriptsize$\left\{\begin{tabular}{l}\textcite{silver1986,zeiri1983}/\\\textcite{huber_herzbergIV}/\\--\end{tabular}\right.$} \\
\ce{PH} & \ce{P + H} & 398 & 122 & \np{1.13e-19} & 0 & C & C & {\scriptsize$\left\{\begin{tabular}{l} \textcite{bruna1981}/\textcite{gao_yufeng2014}/ \\\textcite{berkowitz1989}\end{tabular}\right.$} \\
\ce{PH+} & \ce{P+ + H} & 369 & -- & \np{1.06e-18} & 0 & C & C & {\scriptsize \textcite{bruna1981}/\textcite{huber_herzbergIV}/--} \\
\ce{AlH} & \ce{Al + H} & 392 & 156 & \np{5.00e-20} & 0 & C & C & {\scriptsize$\left\{\begin{tabular}{l} Sect.~\ref{sec:AlH}/\textcite{baltayan1979}/ \\\textcite{marinelli1982}\end{tabular}\right.$} \\
\ce{LiH} & \ce{Li + H} & 579 & 152 & \np{5.00e-20} & 0 & A & C & {\scriptsize$\left\{\begin{tabular}{l} \textcite{kirby1978}/\textcite{neufeld2009}/ \\\textcite{carmona-novillo1996}\end{tabular}\right.$} \\
\ce{MgH} & \ce{Mg + H} & 623 & -- & \np{5.00e-20} & 0 & C & C & {\scriptsize \textcite{kirby1979b}/\textcite{neufeld2009}/--} \\
\ce{NaH} & \ce{Na + H} & 606 & 179 & \np{5.00e-20} & 0 & A & C & {\scriptsize$\left\{\begin{tabular}{l} \textcite{kirby1978}/\textcite{neufeld2009}/ \\\textcite{langhoff1985}\end{tabular}\right.$} \\
\end{longtable}
\end{centering}
\begin{minipage}{\linewidth}
    \footnotetext[1]{Photodissociation products at the threshold wavelength. We note that this is not necessarily the main product since significant (sometimes dominant) branching to other fragments is possible at shorter wavelengths. See discussion in Sect.~\ref{sec:cross section general properties}, Sect.~\ref{sec:cross sections} for selected individual molecules and Sect.~\ref{sec:H2O NH3 branching} for \ce{H2O} and \ce{NH3}.}
    \footnotetext[2]{Long-wavelength threshold for the onset of dissociation or ionisation. For dissociation, the ground-state thermodynamic limit is given unless this is known to be inaccessible by photoexcitation, in that case a shorter wavelength limit is given.}
    \footnotetext[3]{Photodissociation or ionisation cross section at the Lyman-$\alpha$ wavelength, 121.6\,nm.}
    \footnotetext[4]{Estimated uncertainty in the magnitude of the overall photodissociation cross section weighted by the ISRF, and at the Lyman-$\alpha$ wavelength in particular, rated according to the system:\\
    A+: Accurate to within 20\%.\\
    A: Accurate to within 30\%.\\
    B: Accurate to within a factor of 2.\\
    C: Accurate to within a factor of 10.}
    \footnotetext[5]{Cross section source references or subsection where these are provided, and references for the given threshold wavelengths.}
\end{minipage}
\renewcommand*{\thefootnote}{\arabic{footnote}}
\end{landscape}
\twocolumn

\section{Compiled cross sections}
\label{sec:summary of cross sections}

\begin{figure*}
  \centering
  \includegraphics{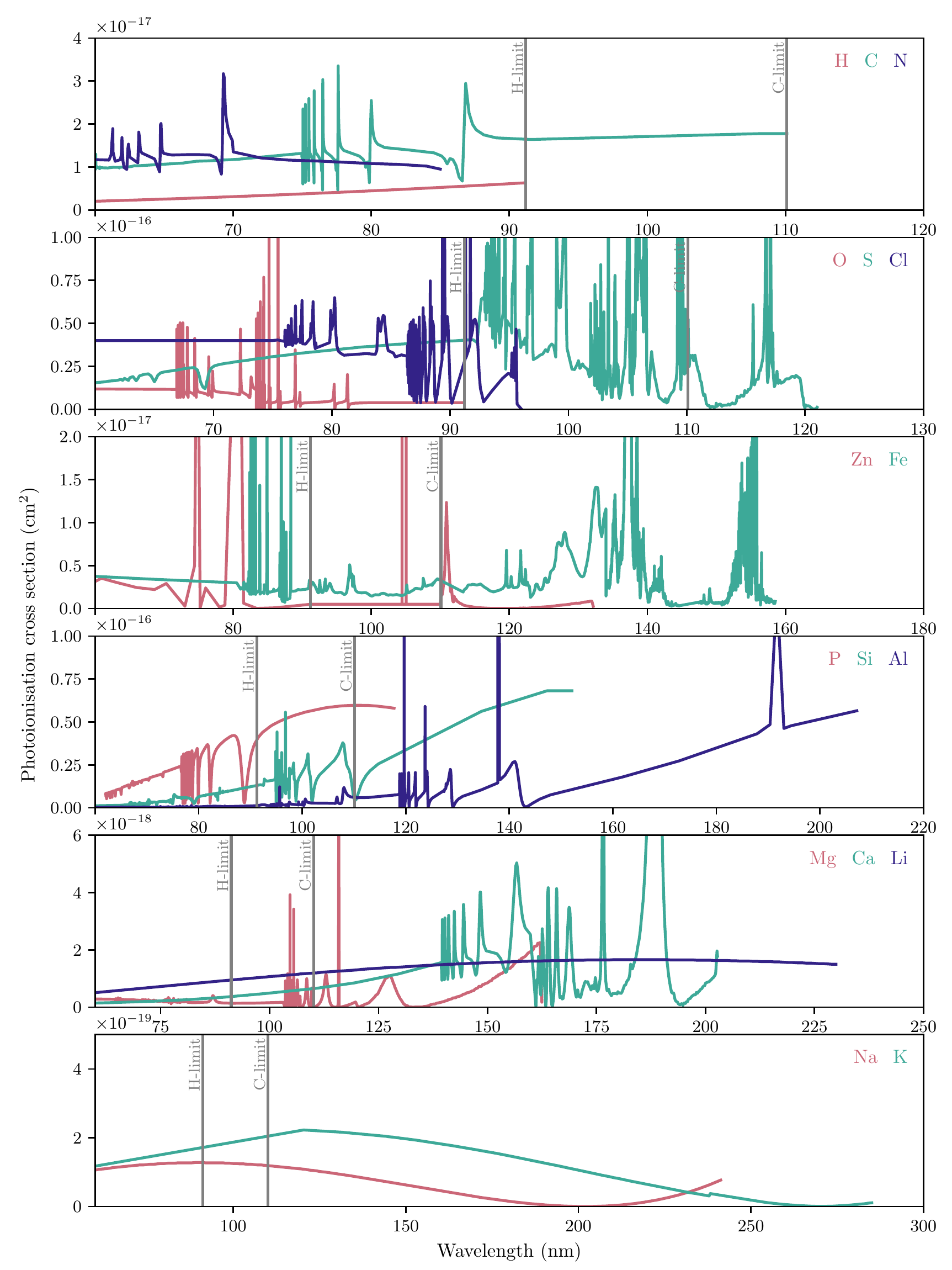}
  \caption{ Compiled atomic photoionisation cross sections. The ionisation thresholds of H and C are indicated by vertical lines.}
  \label{fig:atomic_cross_sections}
\end{figure*}

\begin{figure*}
  \centering
  \includegraphics[width=0.49\textwidth]{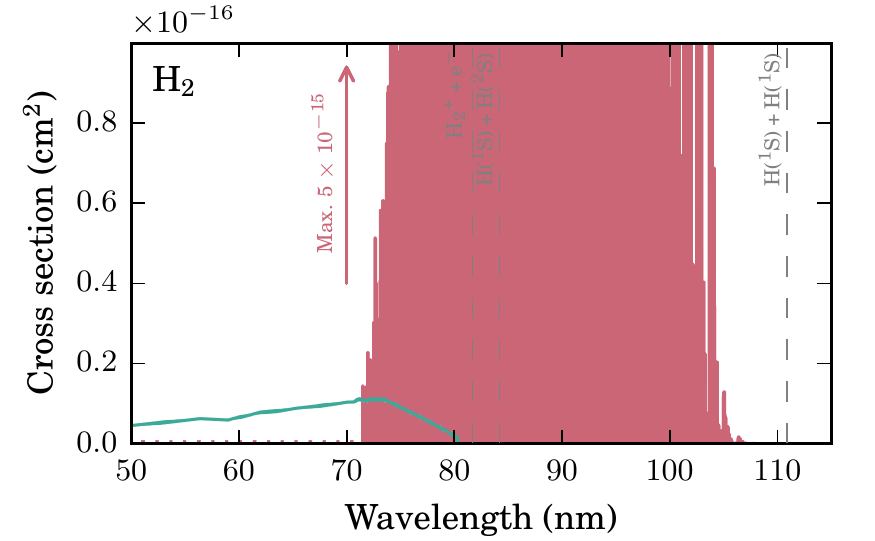}  
  \includegraphics[width=0.49\textwidth]{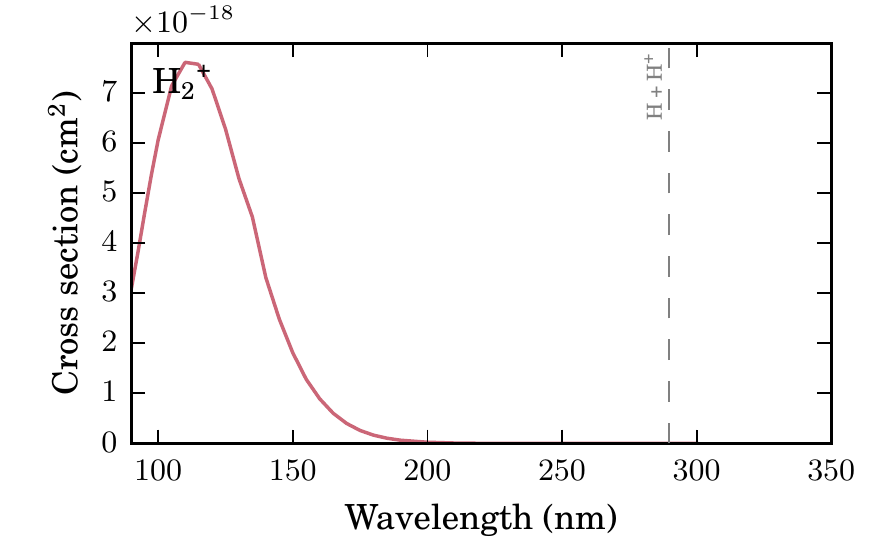} 
  \includegraphics[width=0.49\textwidth]{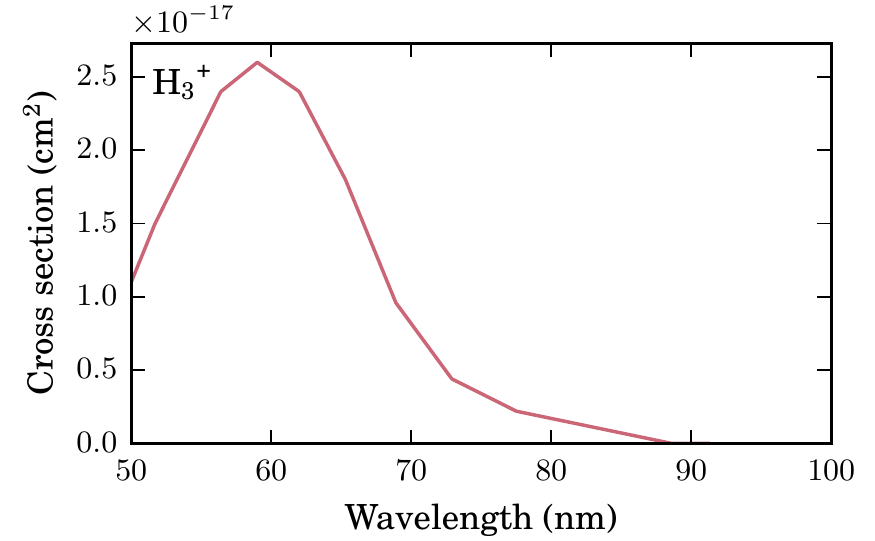} 
  \includegraphics[width=0.49\textwidth]{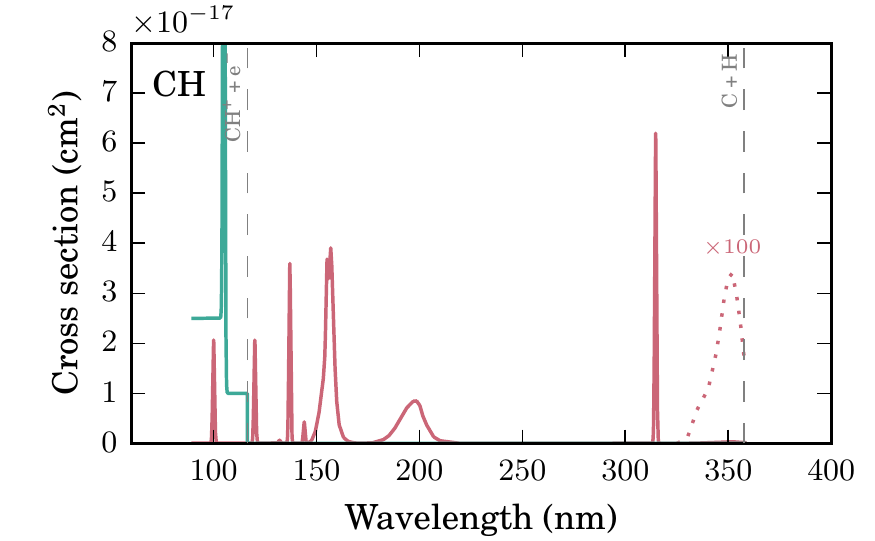}  
  \includegraphics[width=0.49\textwidth]{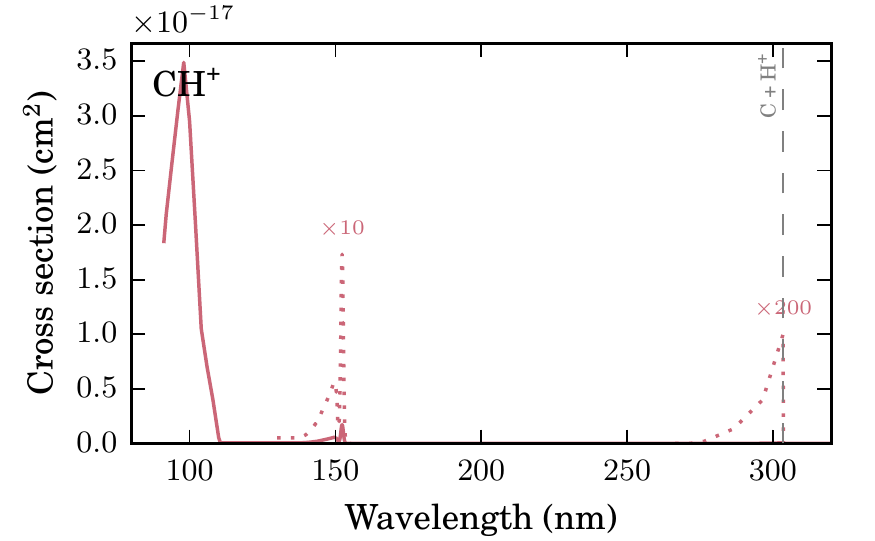} 
  \includegraphics[width=0.49\textwidth]{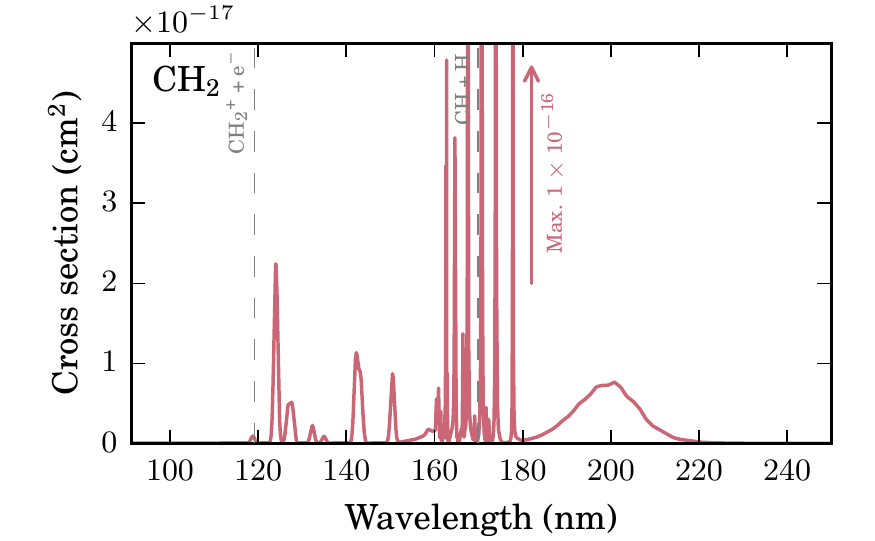} 
  \includegraphics[width=0.49\textwidth]{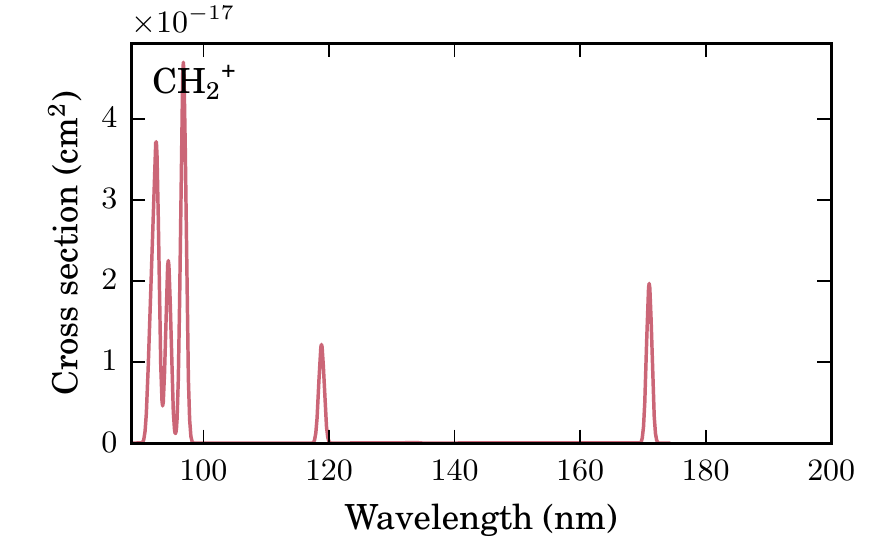}
  \includegraphics[width=0.49\textwidth]{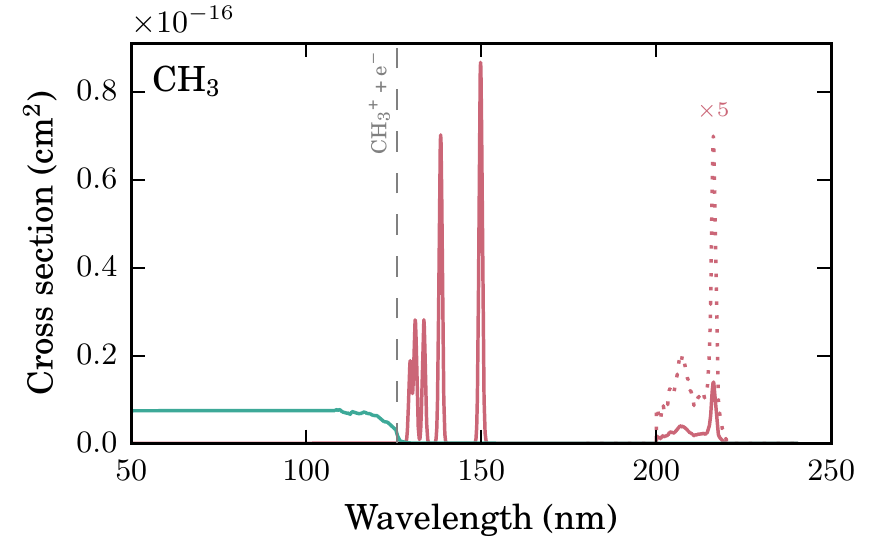} 
  \caption{Cross sections of molecules. Red: Photodissociation. Blue: Photoionisation. Some photofragmentation thresholds are also labelled.}
  \label{fig:cross sections H2}  
  \label{fig:cross sections H2+} 
  \label{fig:cross sections H3+} 
  \label{fig:cross sections CH}  
  \label{fig:cross sections CH+} 
  \label{fig:cross sections CH2} 
  \label{fig:cross sections CH2+}
  \label{fig:cross sections CH3} 
\end{figure*}

\begin{figure*}
  \centering
  \includegraphics[width=0.49\textwidth]{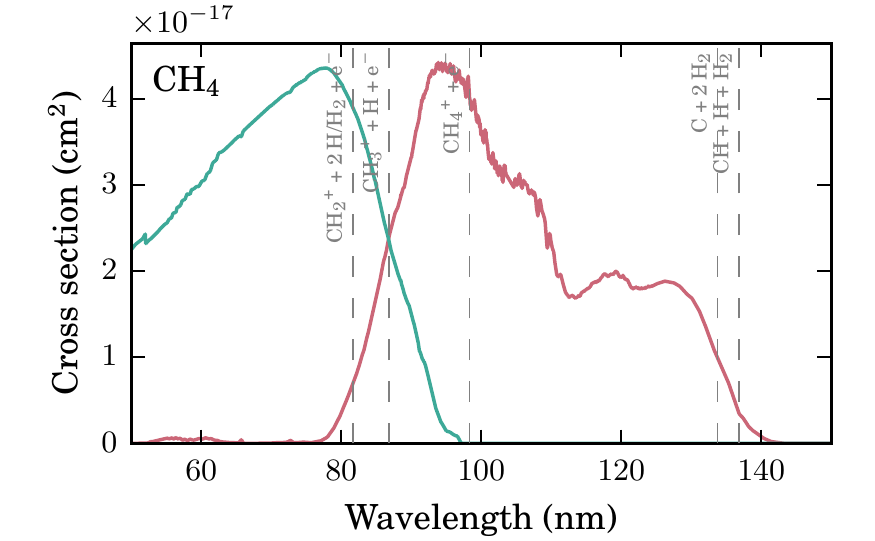} 
  \includegraphics[width=0.49\textwidth]{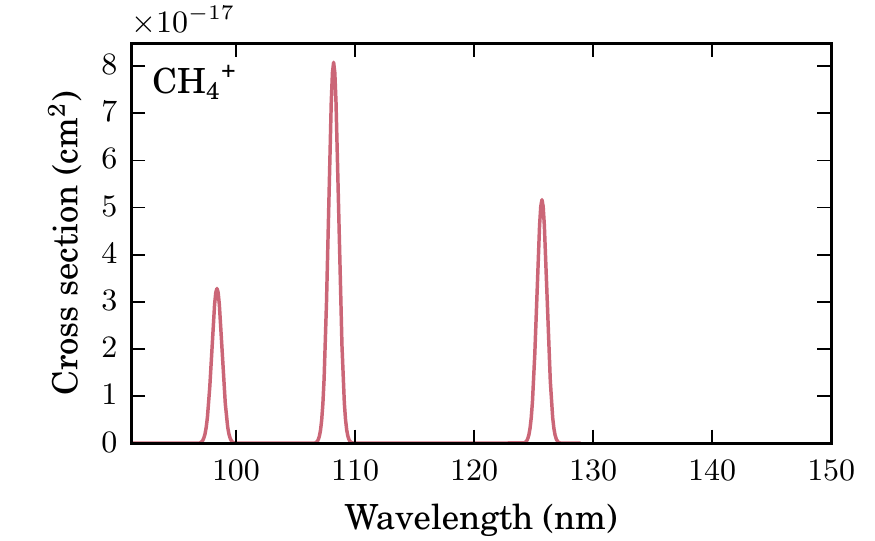}
  \includegraphics[width=0.49\textwidth]{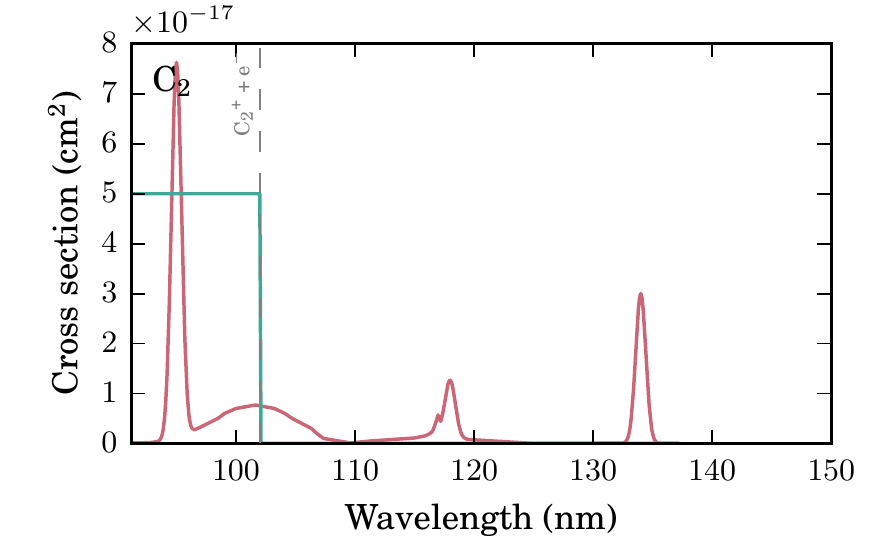}  
  \includegraphics[width=0.49\textwidth]{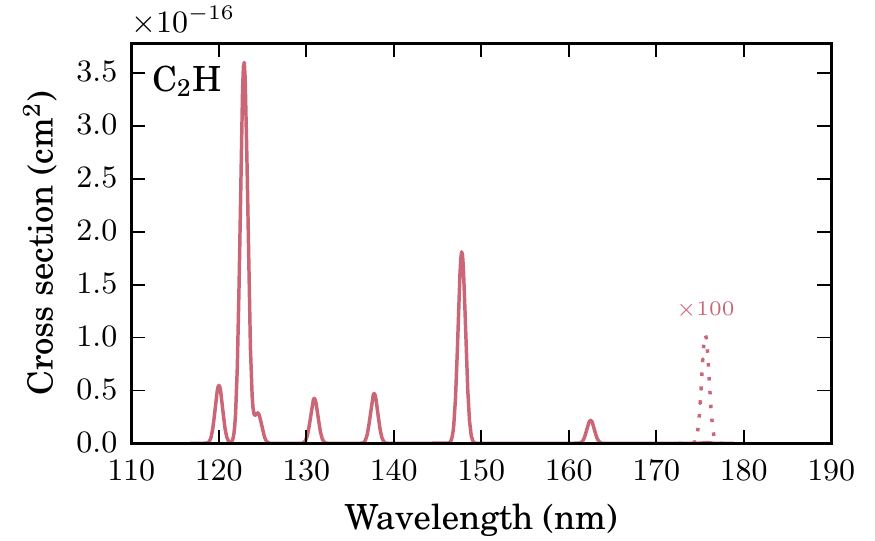} 
  \includegraphics[width=0.49\textwidth]{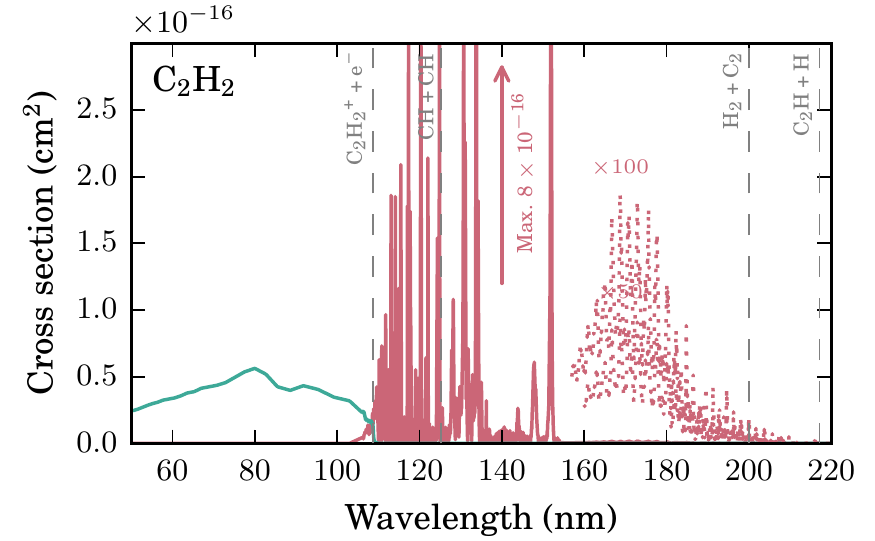}
  \includegraphics[width=0.49\textwidth]{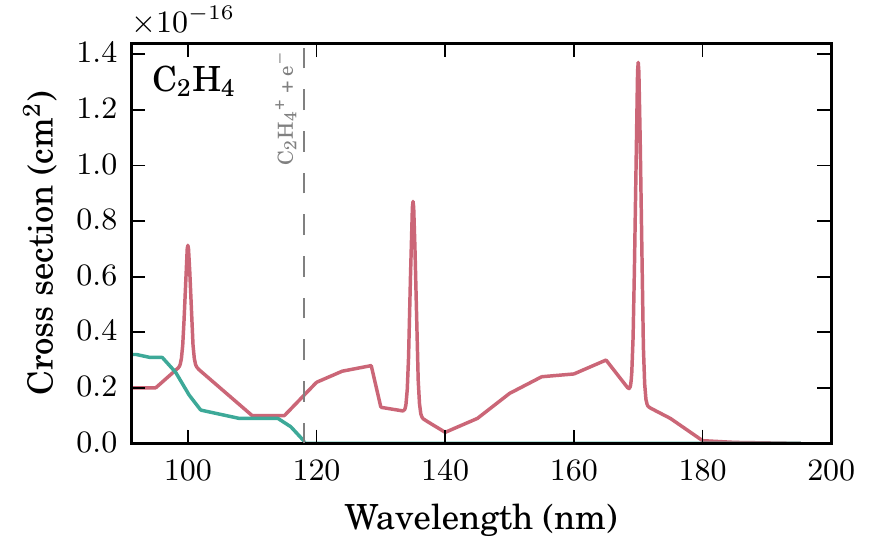}
  \includegraphics[width=0.49\textwidth]{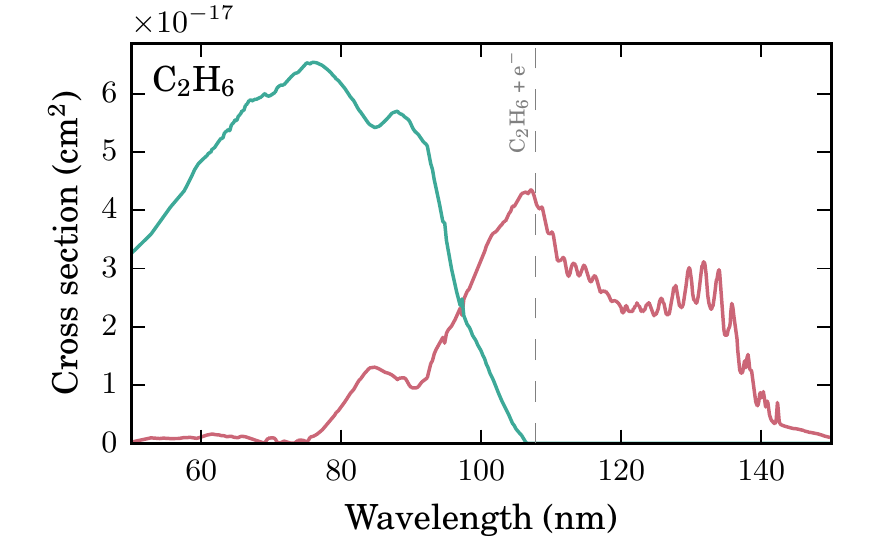}
  \includegraphics[width=0.49\textwidth]{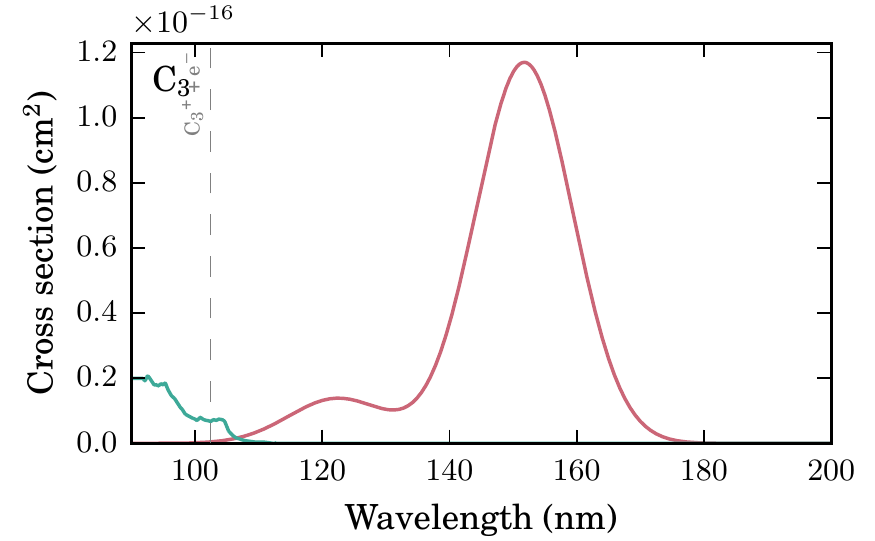}  
  \caption{Cross sections of molecules. Red: Photodissociation. Blue: Photoionisation. Some photofragmentation thresholds are also labelled.}
  \label{fig:cross sections CH4}   
  \label{fig:cross sections CH4+}  
  \label{fig:cross sections C2}    
  \label{fig:cross sections C2H}   
  \label{fig:cross sections C2H2}  
  \label{fig:cross sections C2H4}  
  \label{fig:cross sections C2H6}  
  \label{fig:cross sections C3}    
\end{figure*}

\begin{figure*}
  \centering
  \includegraphics[width=0.49\textwidth]{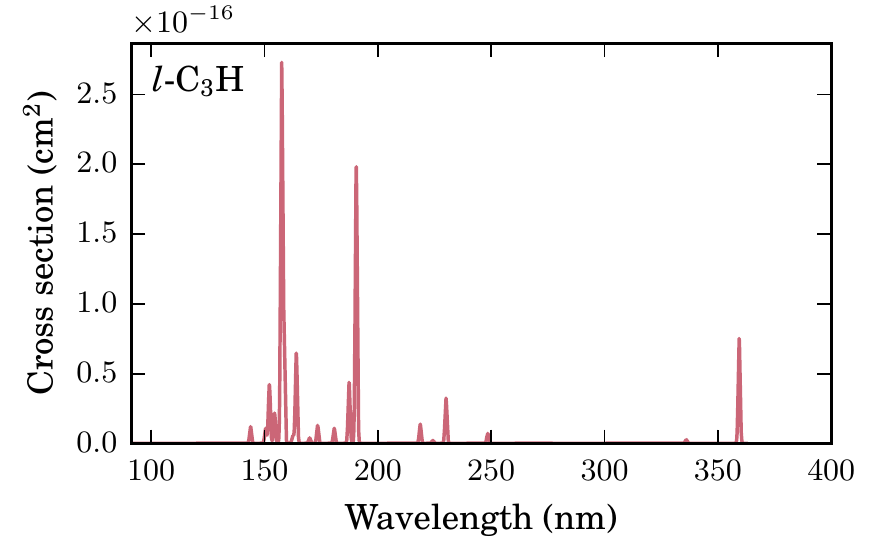} 
  \includegraphics[width=0.49\textwidth]{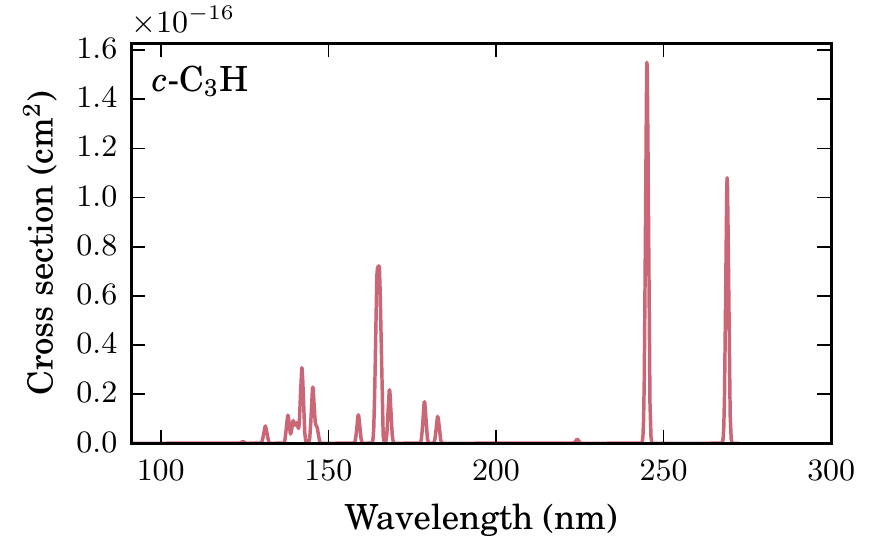} 
  \includegraphics[width=0.49\textwidth]{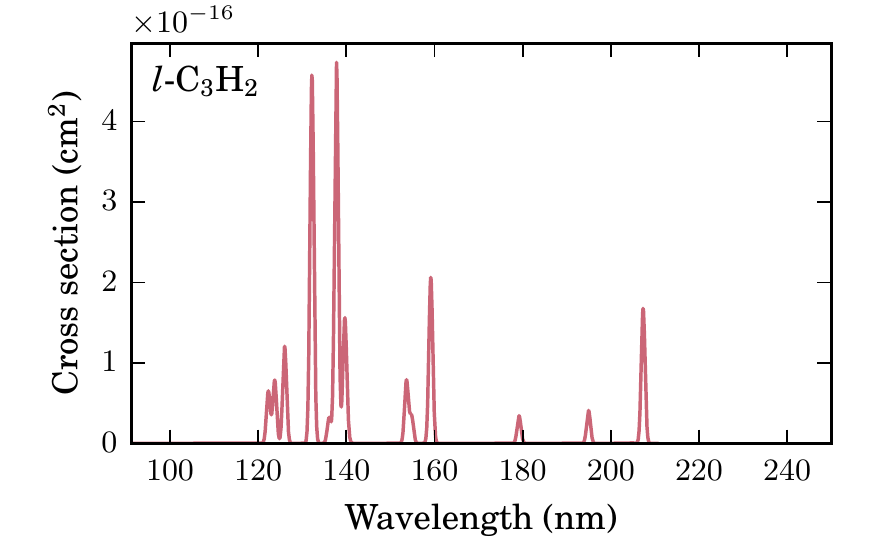}
  \includegraphics[width=0.49\textwidth]{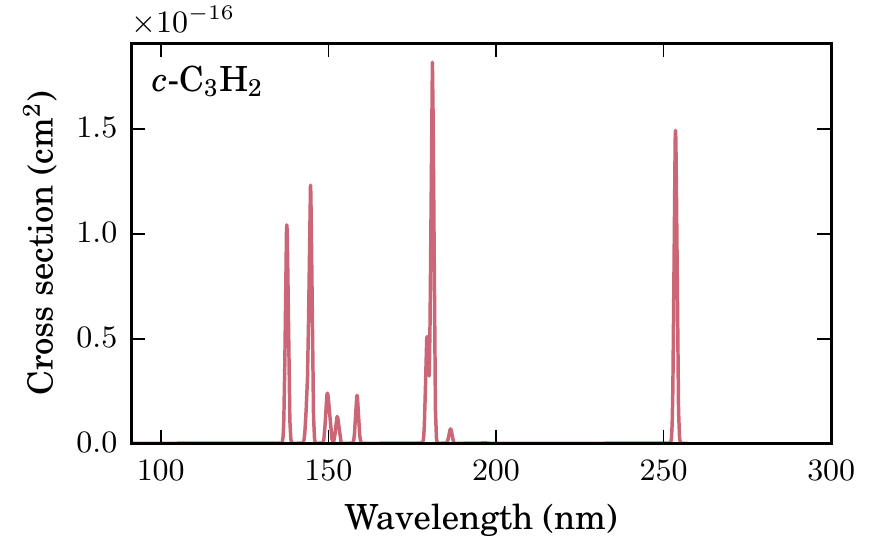}
  \includegraphics[width=0.49\textwidth]{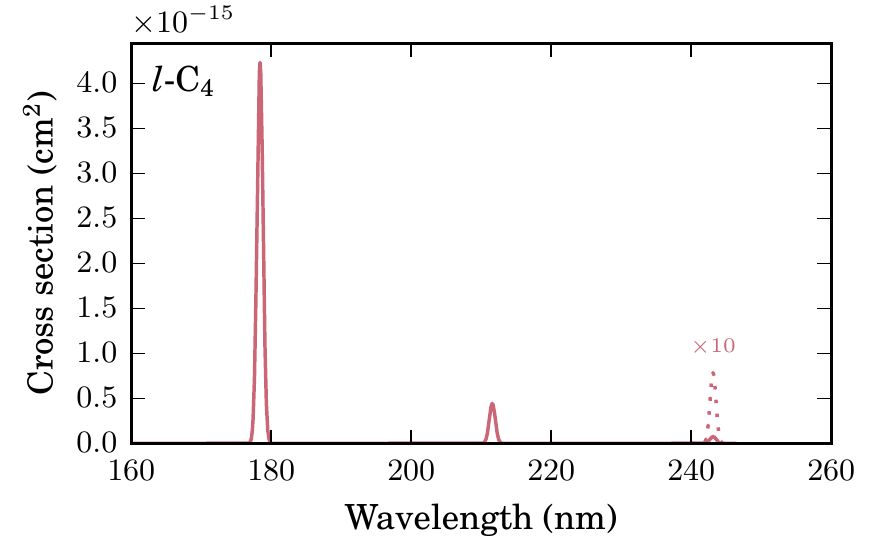}  
  \includegraphics[width=0.49\textwidth]{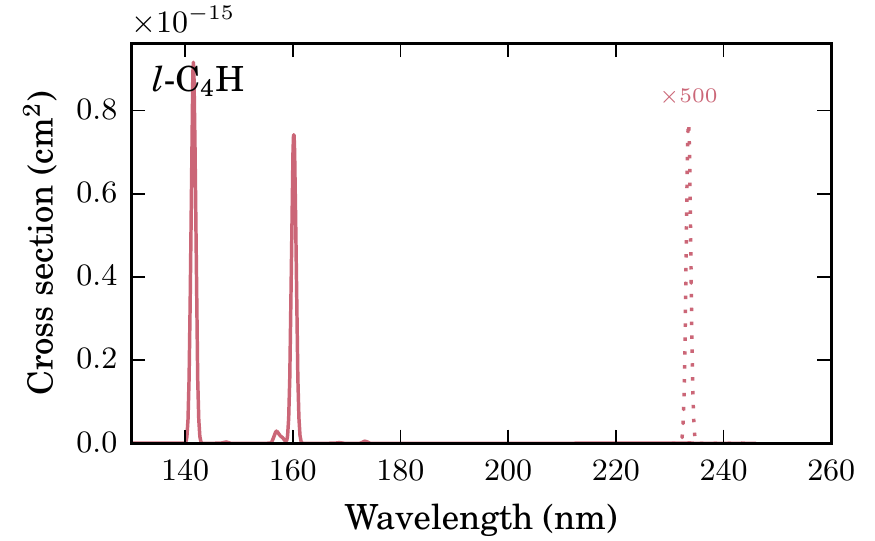} 
  \includegraphics[width=0.49\textwidth]{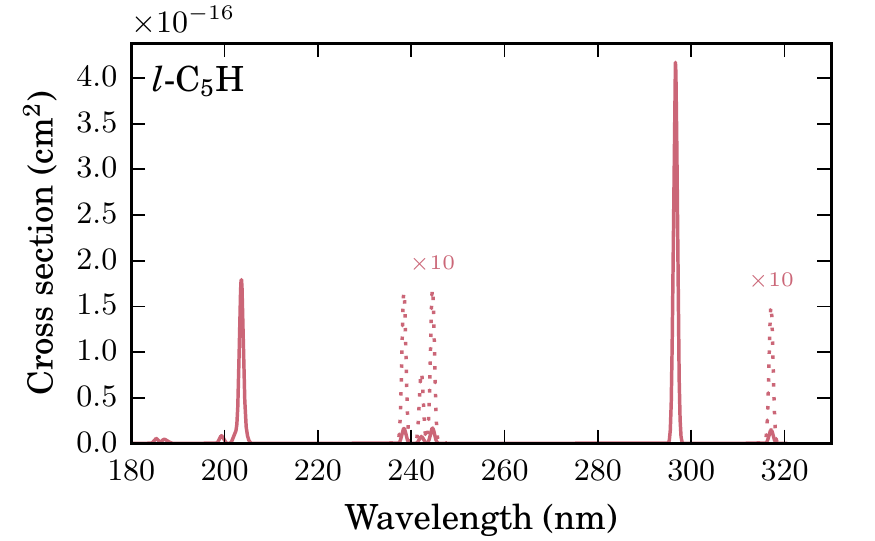} 
  \includegraphics[width=0.49\textwidth]{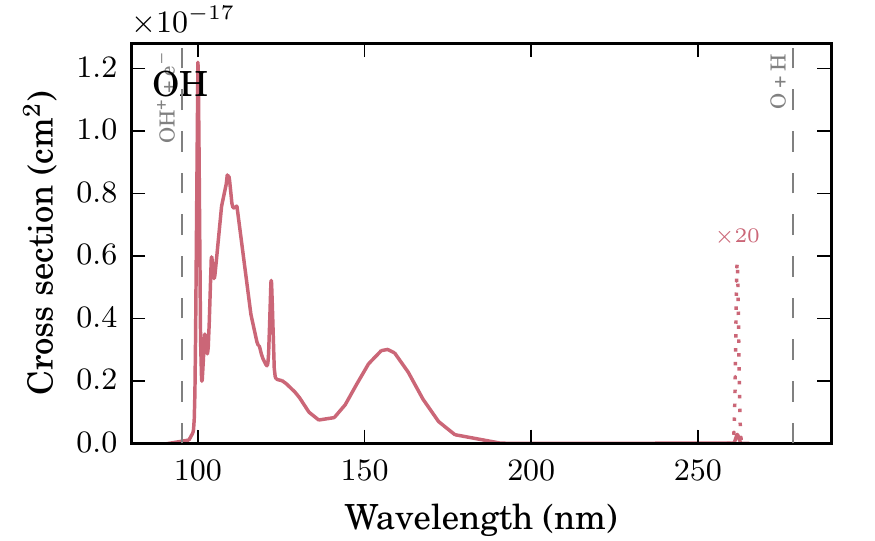}    
  \caption{Cross sections of molecules. Red: Photodissociation. Blue: Photoionisation. Some photofragmentation thresholds are also labelled.}
  \label{fig:cross sections l-C3H}   
  \label{fig:cross sections c-C3H}   
  \label{fig:cross sections l-C3H2}  
  \label{fig:cross sections c-C3H2}  
  \label{fig:cross sections l-C4}    
  \label{fig:cross sections l-C4H}   
  \label{fig:cross sections l-C5H}   
  \label{fig:cross sections OH}      
\end{figure*}

\begin{figure*}
  \centering
  \includegraphics[width=0.49\textwidth]{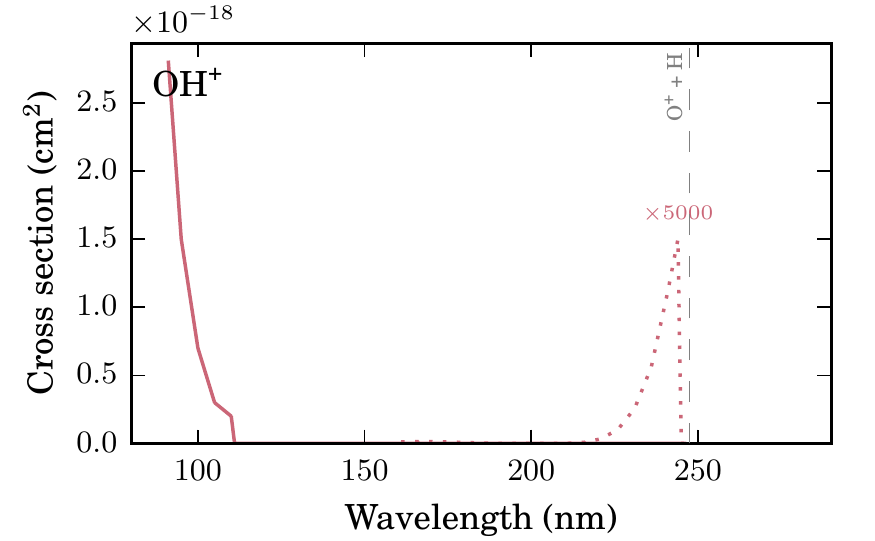} 
  \includegraphics[width=0.49\textwidth]{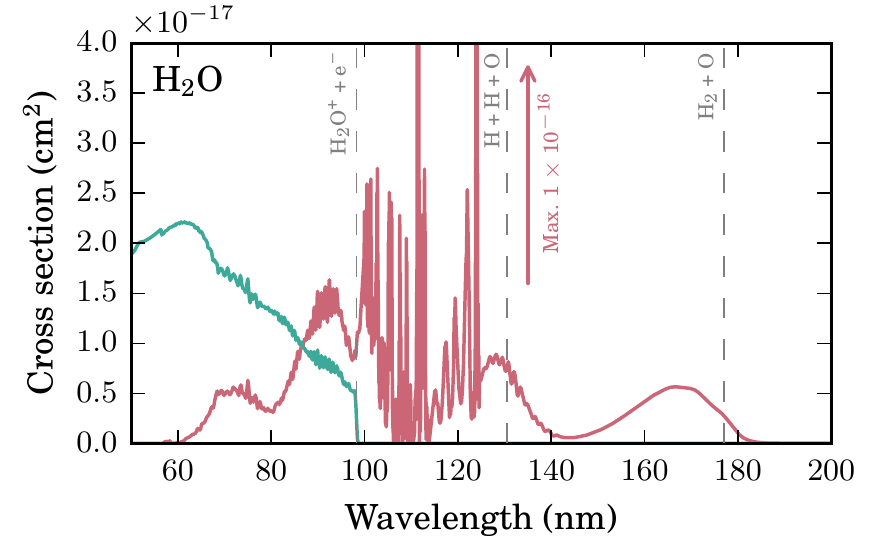} 
  \includegraphics[width=0.49\textwidth]{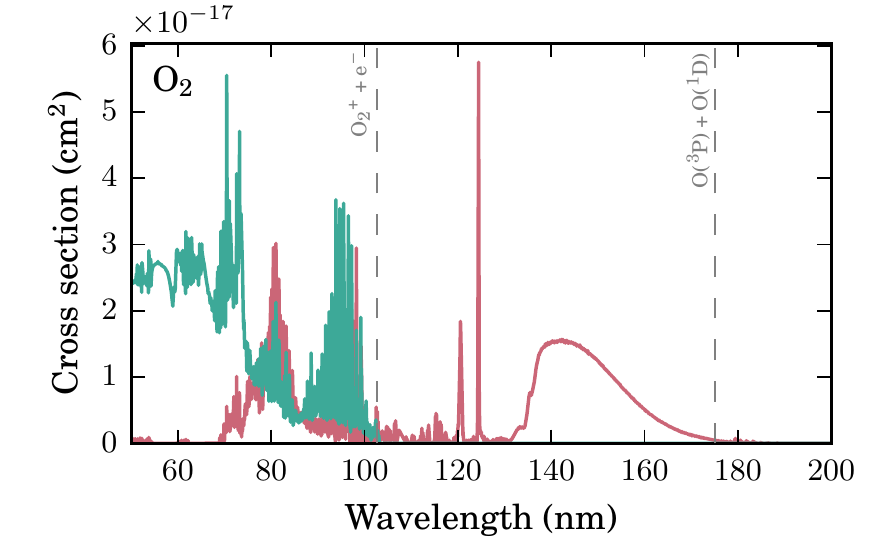}  
  \includegraphics[width=0.49\textwidth]{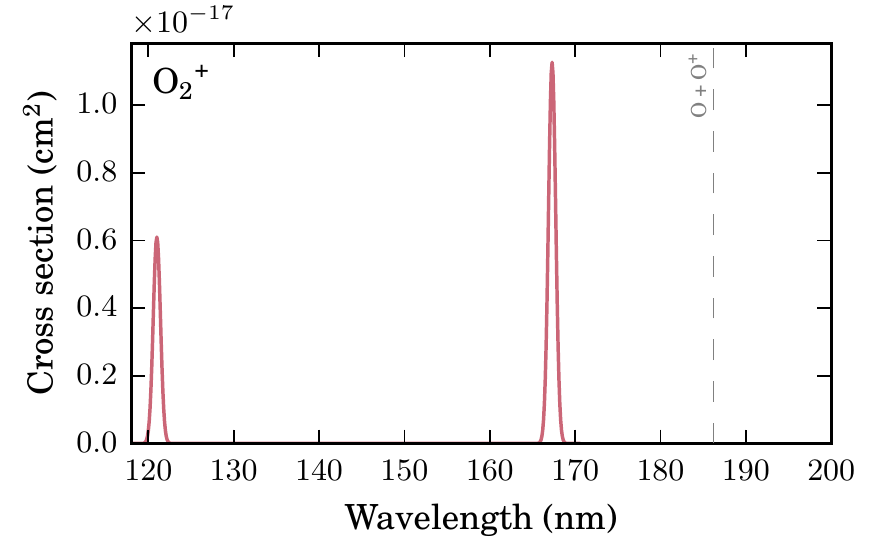} 
  \includegraphics[width=0.49\textwidth]{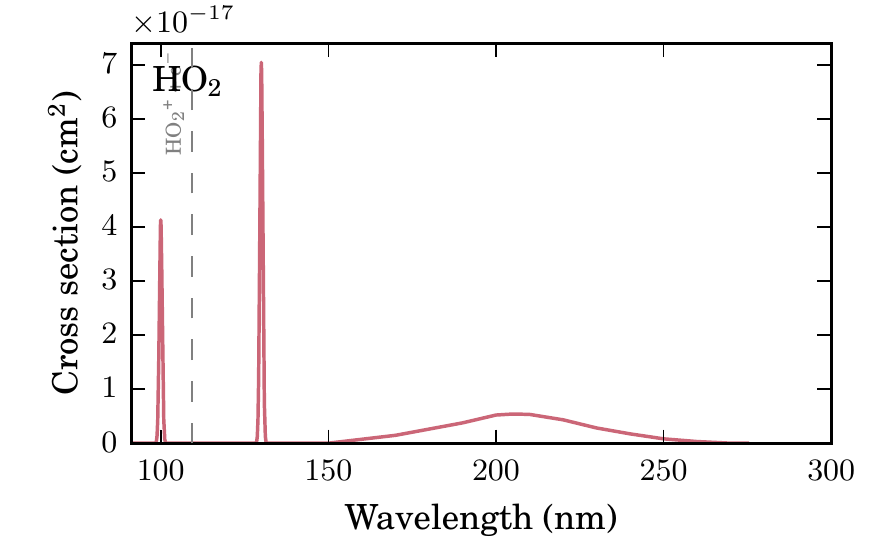} 
  \includegraphics[width=0.49\textwidth]{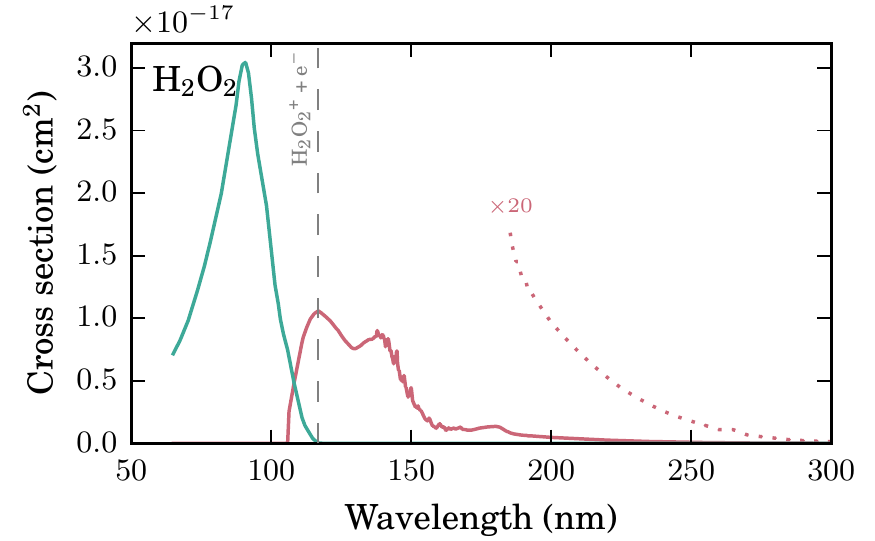}
  \includegraphics[width=0.49\textwidth]{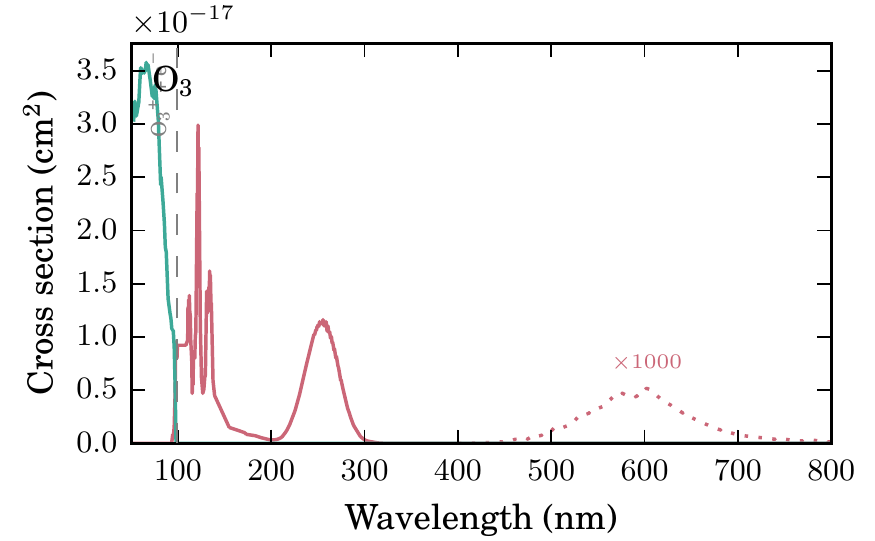}  
  \includegraphics[width=0.49\textwidth]{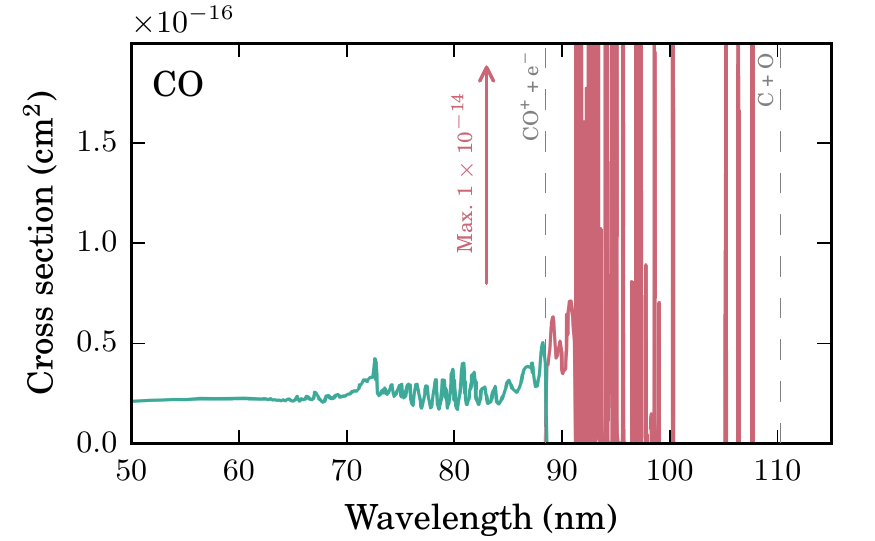}  
  \caption{Cross sections of molecules. Red: Photodissociation. Blue: Photoionisation. Some photofragmentation thresholds are also labelled.}
  \label{fig:cross sections OH+}  
  \label{fig:cross sections H2O}  
  \label{fig:cross sections O2}   
  \label{fig:cross sections O2+}  
  \label{fig:cross sections HO2}  
  \label{fig:cross sections H2O2} 
  \label{fig:cross sections O3}   
  \label{fig:cross sections CO}   
\end{figure*}

\begin{figure*}
  \centering
  \includegraphics[width=0.49\textwidth]{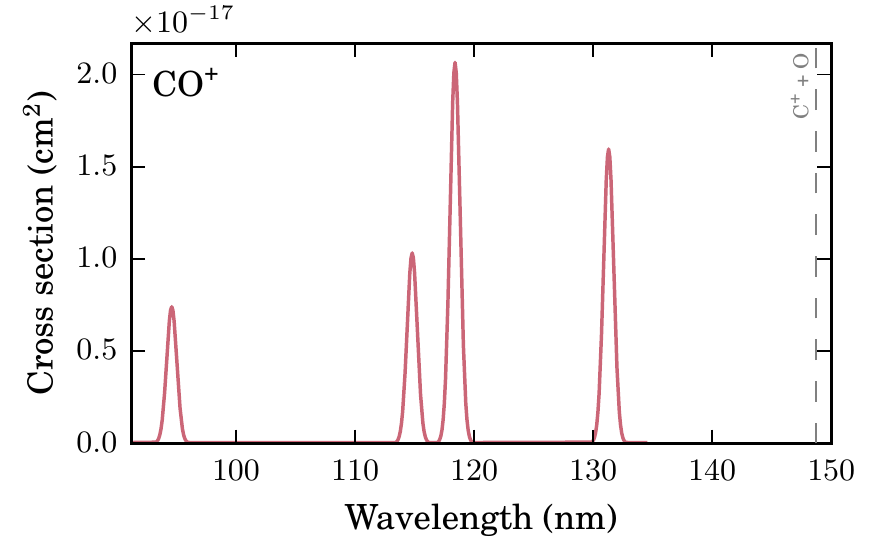} 
  \includegraphics[width=0.49\textwidth]{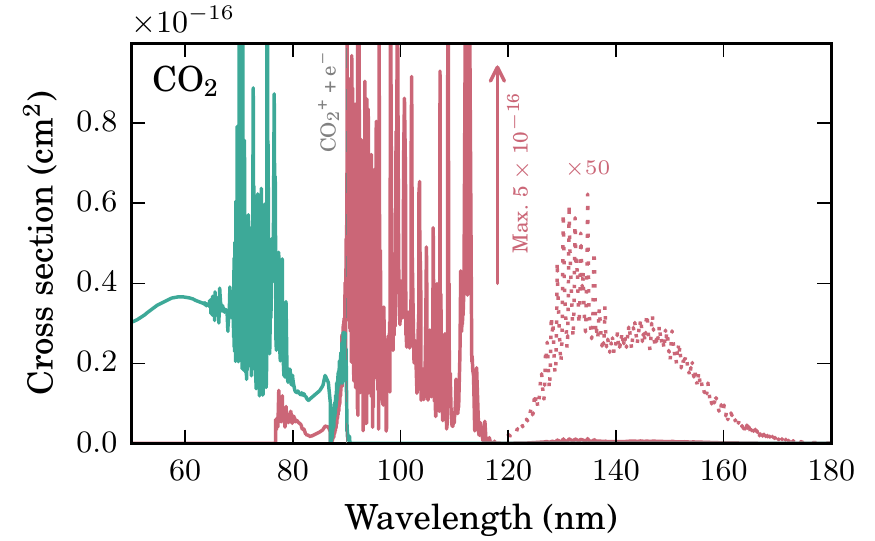} 
  \includegraphics[width=0.49\textwidth]{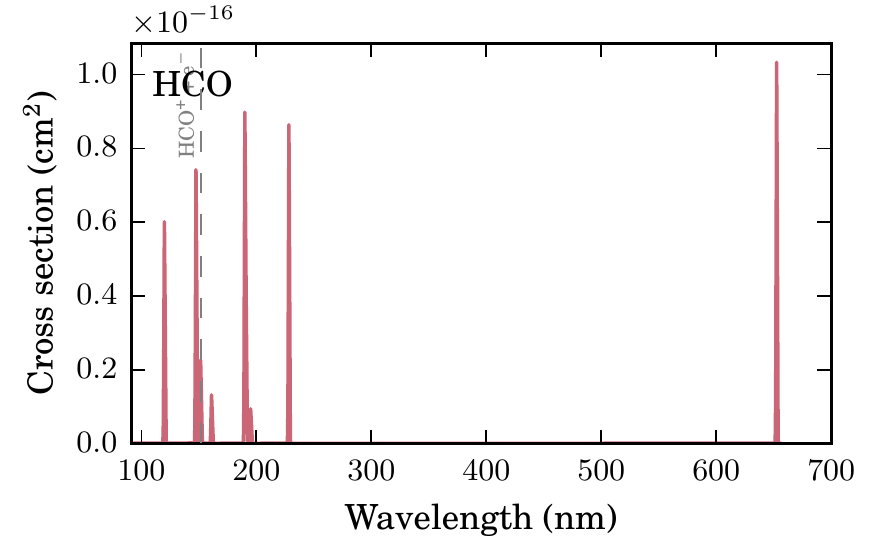} 
  \includegraphics[width=0.49\textwidth]{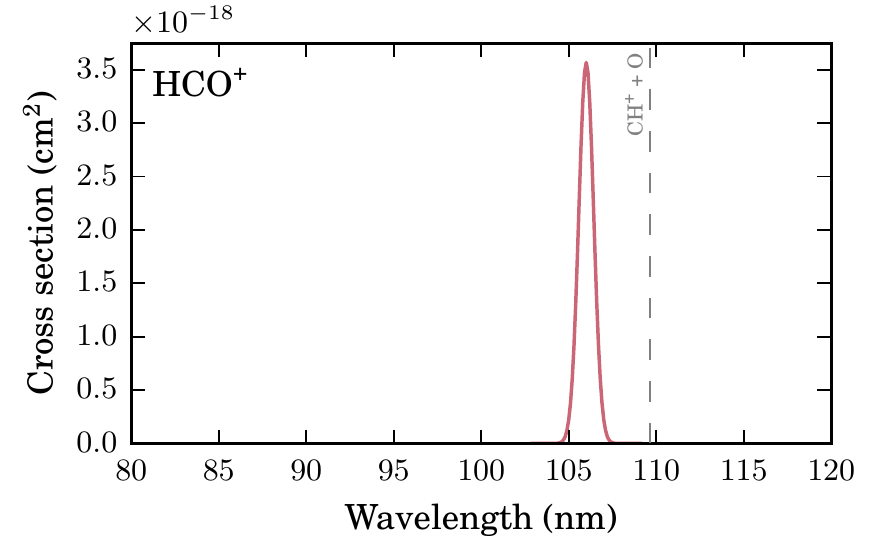}
  \includegraphics[width=0.49\textwidth]{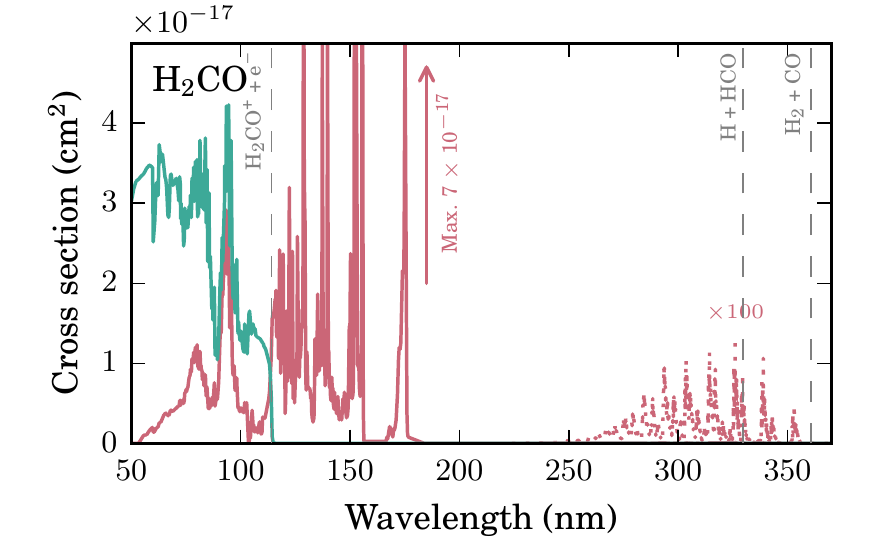}
  \includegraphics[width=0.49\textwidth]{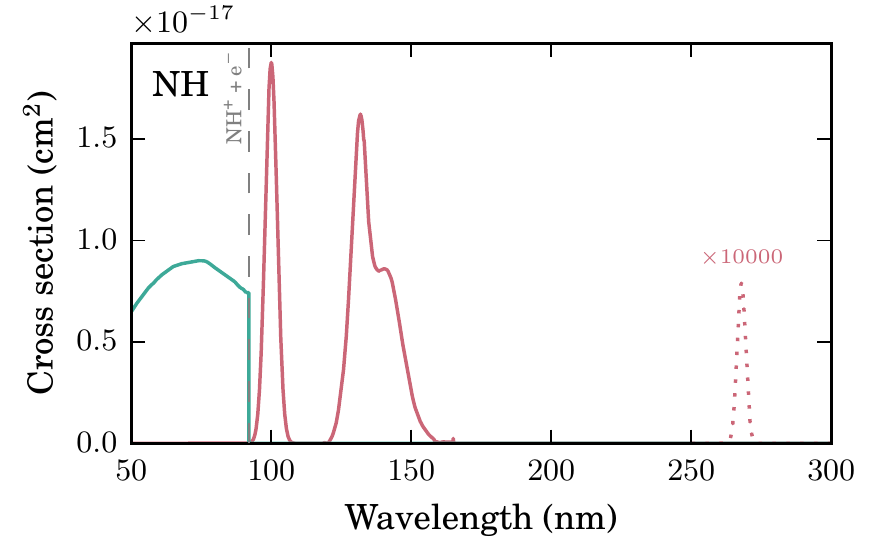}  
  \includegraphics[width=0.49\textwidth]{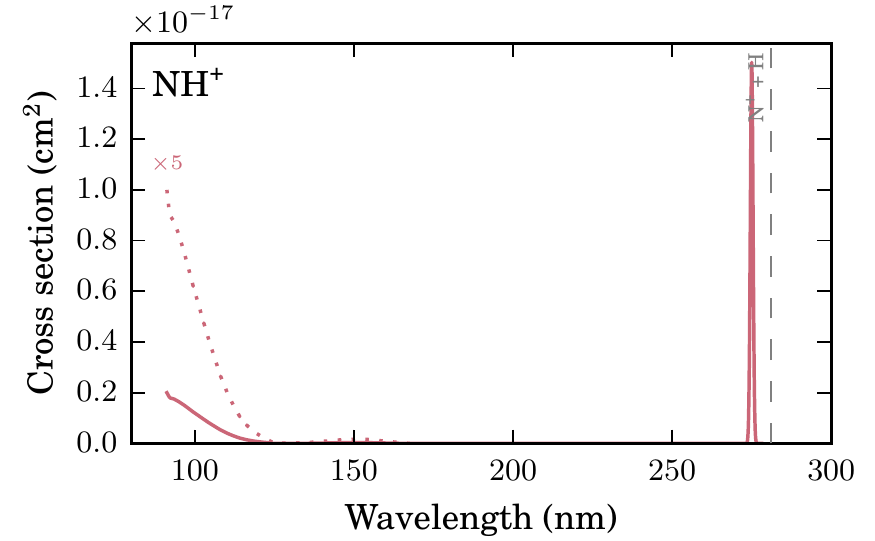} 
  \includegraphics[width=0.49\textwidth]{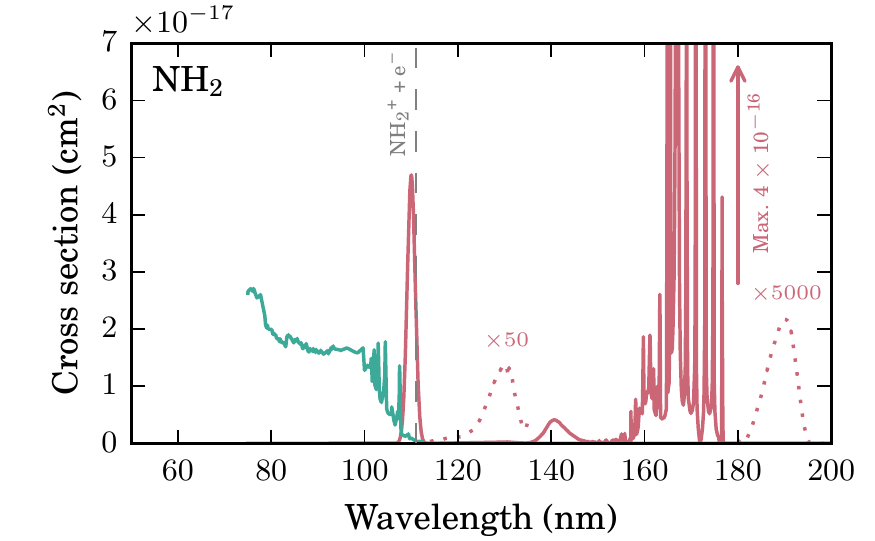} 
  \caption{Cross sections of molecules. Red: Photodissociation. Blue: Photoionisation. Some photofragmentation thresholds are also labelled.}
  \label{fig:cross sections CO+}  
  \label{fig:cross sections CO2}  
  \label{fig:cross sections HCO}  
  \label{fig:cross sections HCO+} 
  \label{fig:cross sections H2CO} 
  \label{fig:cross sections NH}   
  \label{fig:cross sections NH+}  
  \label{fig:cross sections NH2}  
\end{figure*}

\begin{figure*}
  \centering
  \includegraphics[width=0.49\textwidth]{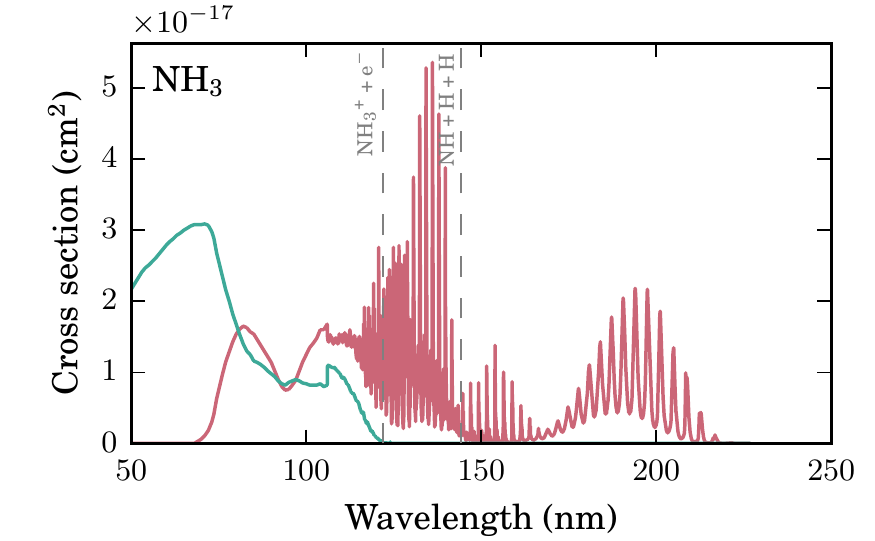} 
  \includegraphics[width=0.49\textwidth]{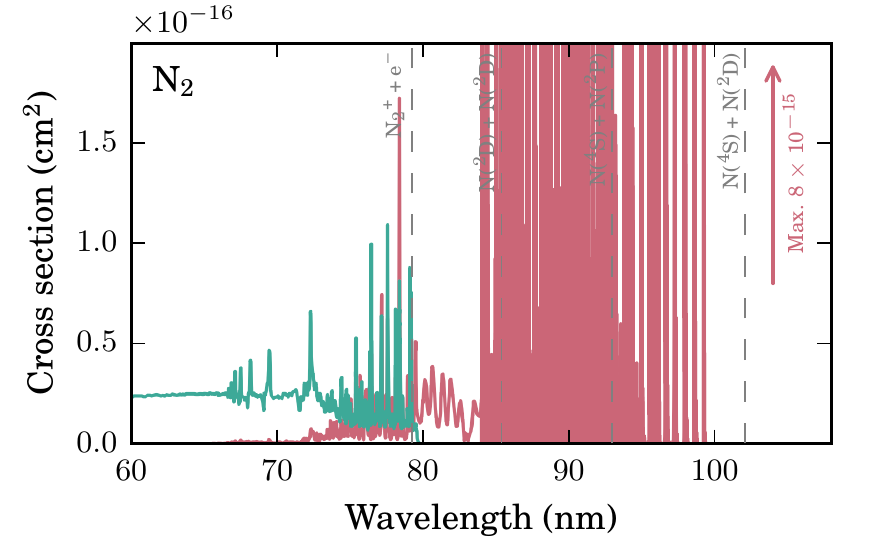}  
  \includegraphics[width=0.49\textwidth]{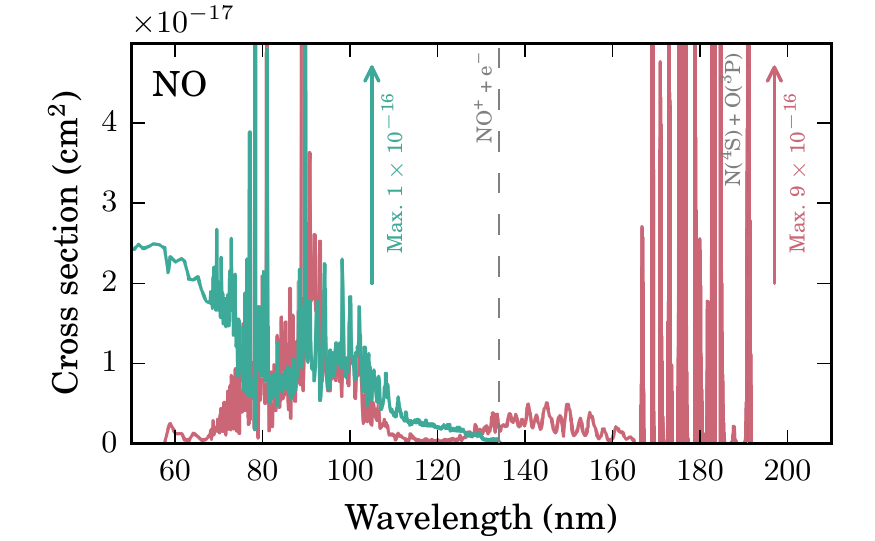}  
  \includegraphics[width=0.49\textwidth]{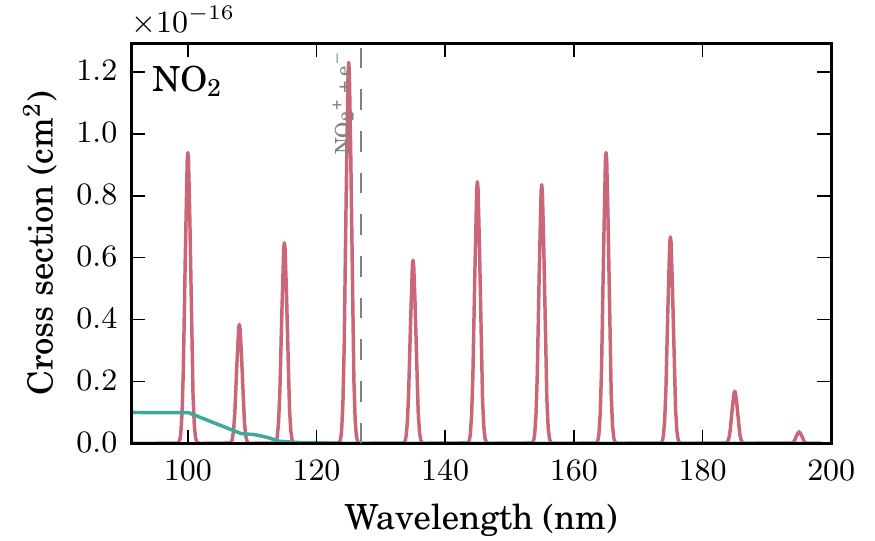} 
  \includegraphics[width=0.49\textwidth]{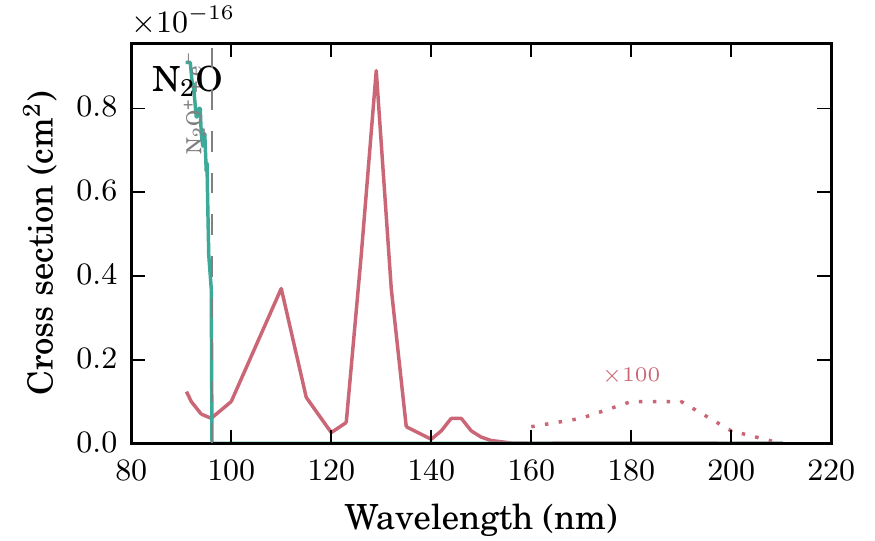} 
  \includegraphics[width=0.49\textwidth]{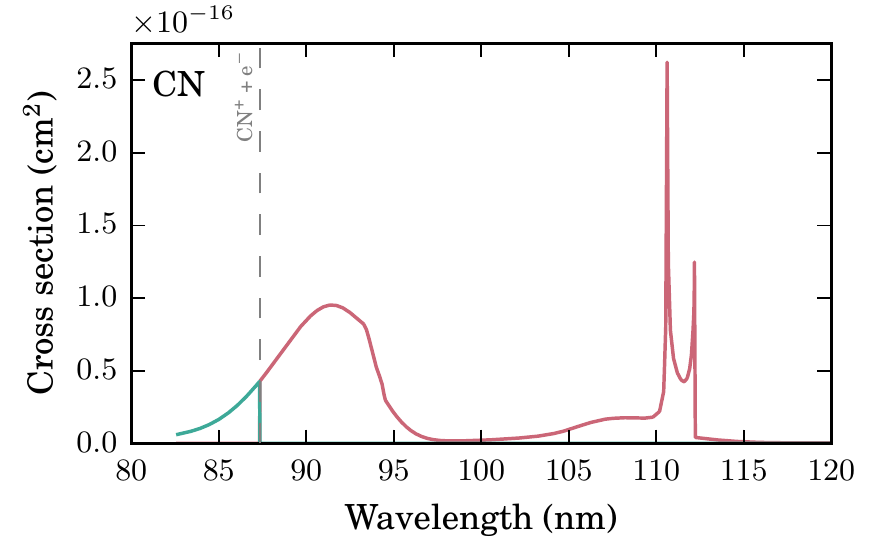}  
  \includegraphics[width=0.49\textwidth]{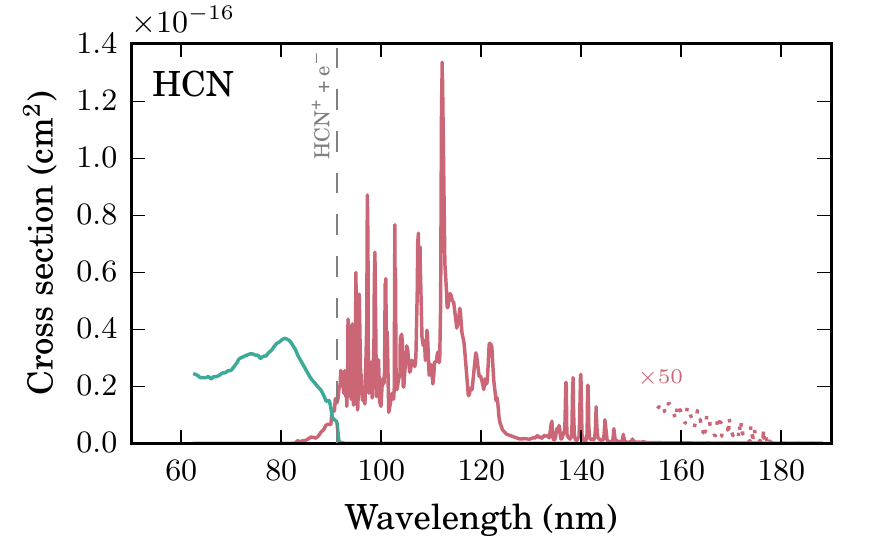} 
  \includegraphics[width=0.49\textwidth]{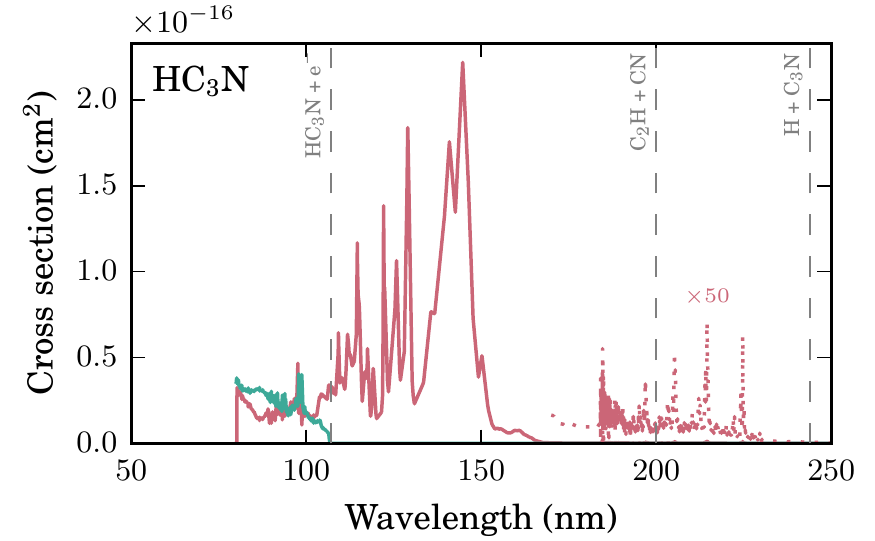}
  \caption{Cross sections of molecules. Red: Photodissociation. Blue: Photoionisation. Some photofragmentation thresholds are also labelled.}
  \label{fig:cross sections NH3} 
  \label{fig:cross sections N2}  
  \label{fig:cross sections NO}  
  \label{fig:cross sections NO2} 
  \label{fig:cross sections N2O} 
  \label{fig:cross sections CN}  
  \label{fig:cross sections HCN} 
  \label{fig:cross sections HC3N}
\end{figure*}

\begin{figure*}
  \centering
  \includegraphics[width=0.49\textwidth]{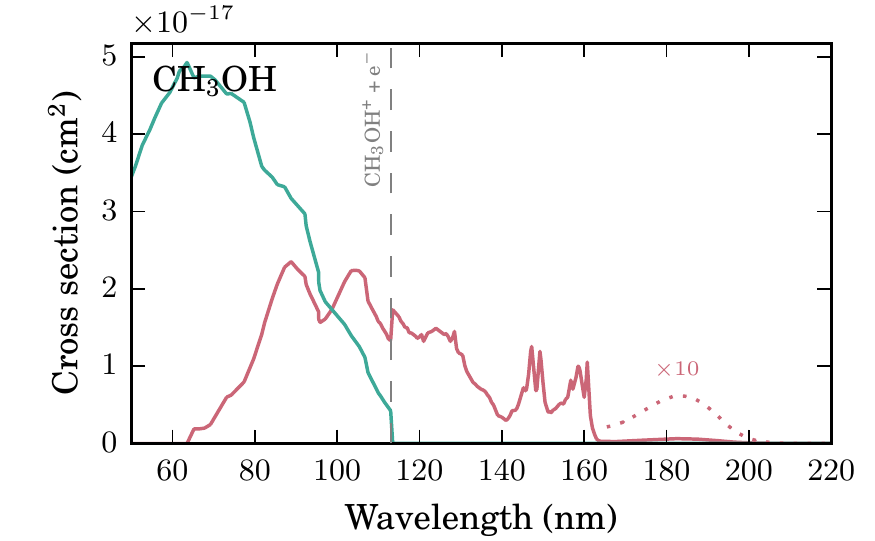} 
  \includegraphics[width=0.49\textwidth]{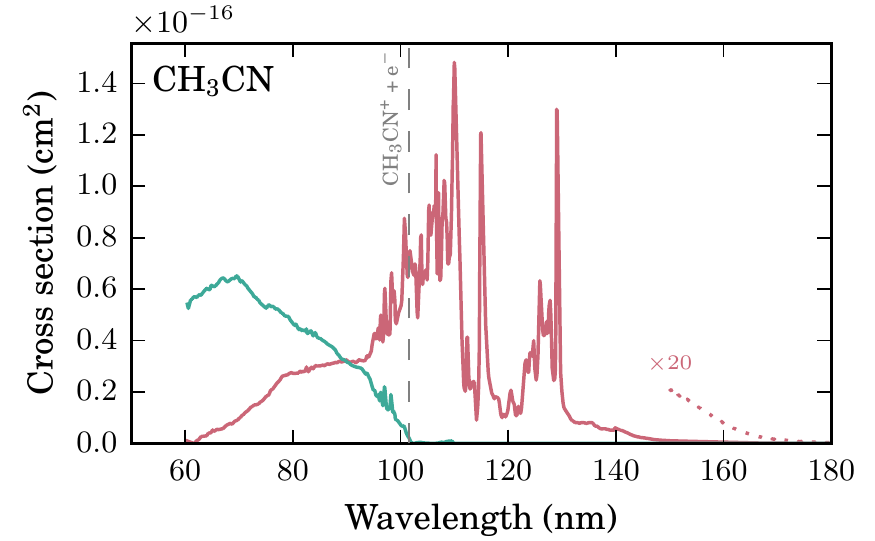} 
  \includegraphics[width=0.49\textwidth]{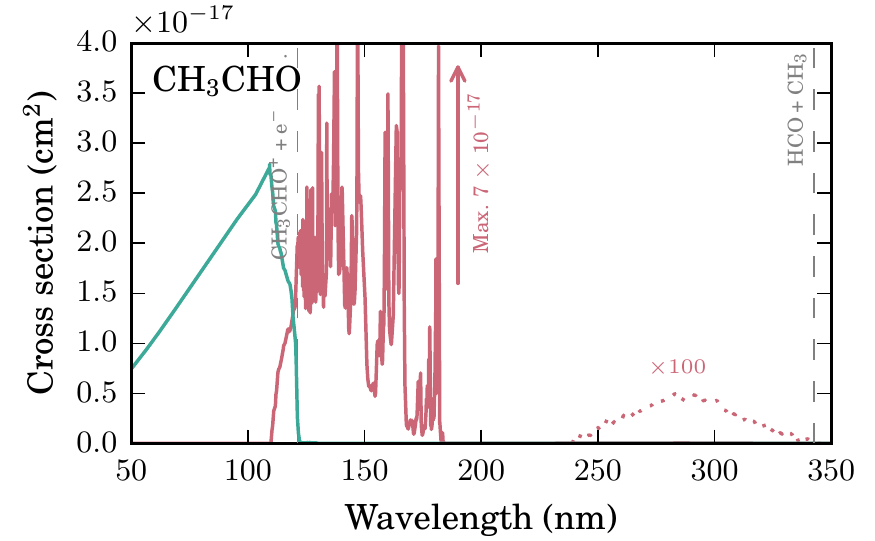}
  \includegraphics[width=0.49\textwidth]{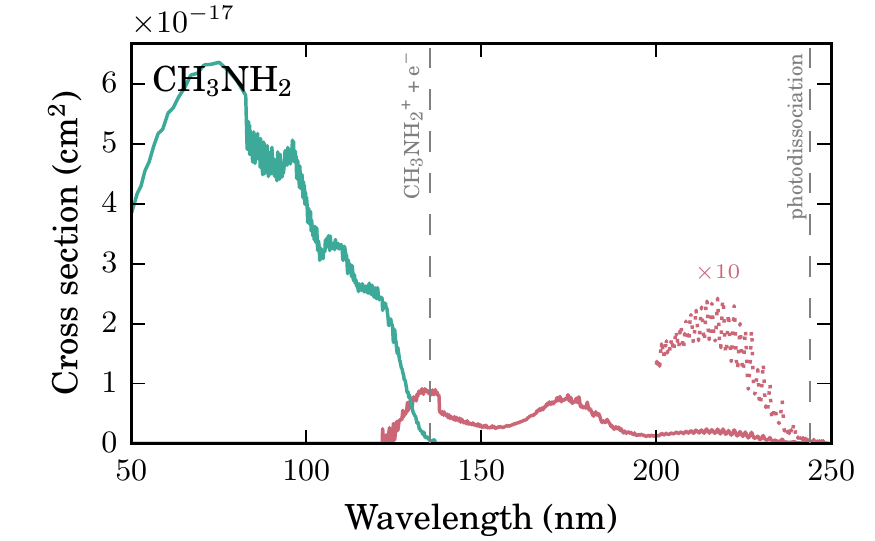}
  \includegraphics[width=0.49\textwidth]{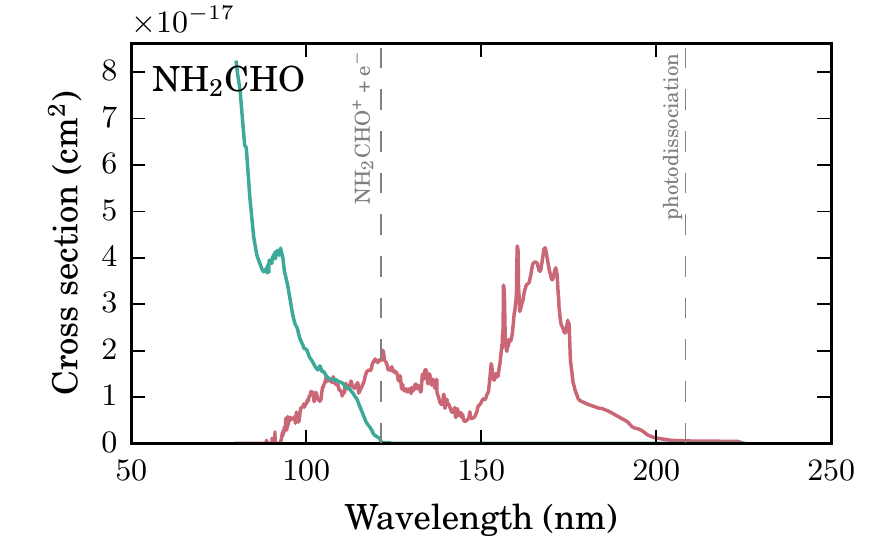}
  \includegraphics[width=0.49\textwidth]{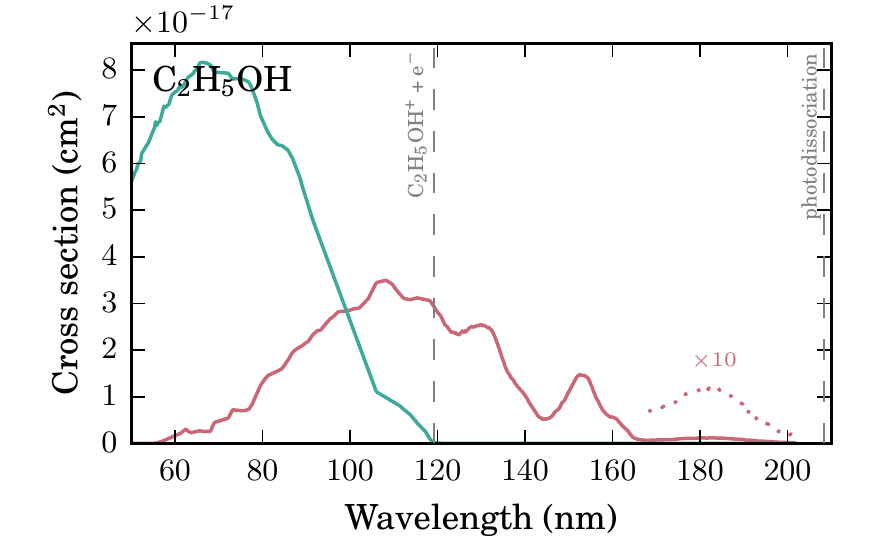}
  \includegraphics[width=0.49\textwidth]{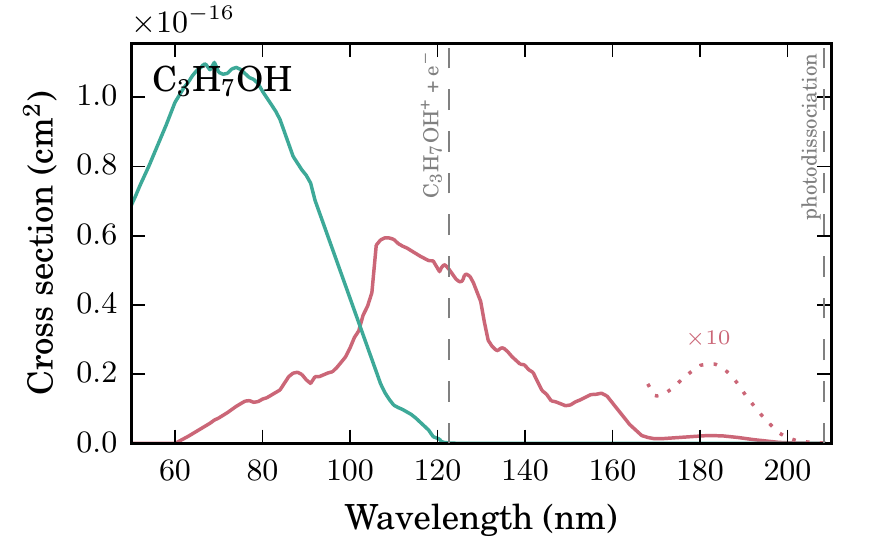}
  \includegraphics[width=0.49\textwidth]{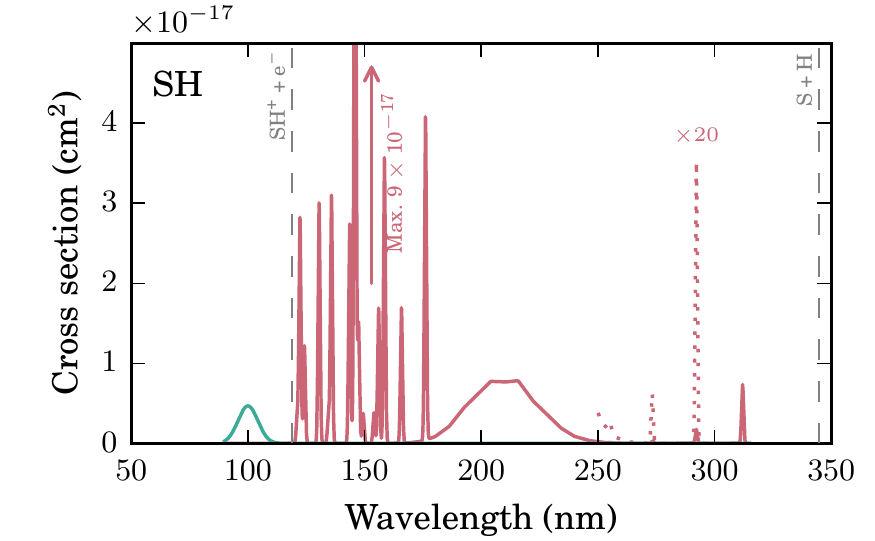}    
  \caption{Cross sections of molecules. Red: Photodissociation. Blue: Photoionisation. Some photofragmentation thresholds are also labelled.}
  \label{fig:cross sections CH3OH} 
  \label{fig:cross sections CH3CN} 
  \label{fig:cross sections CH3CHO}
  \label{fig:cross sections CH3NH2}
  \label{fig:cross sections NH2CHO}
  \label{fig:cross sections C2H5OH}
  \label{fig:cross sections C3H7OH}
  \label{fig:cross sections SH}    
\end{figure*}

\begin{figure*}
  \centering
  \includegraphics[width=0.49\textwidth]{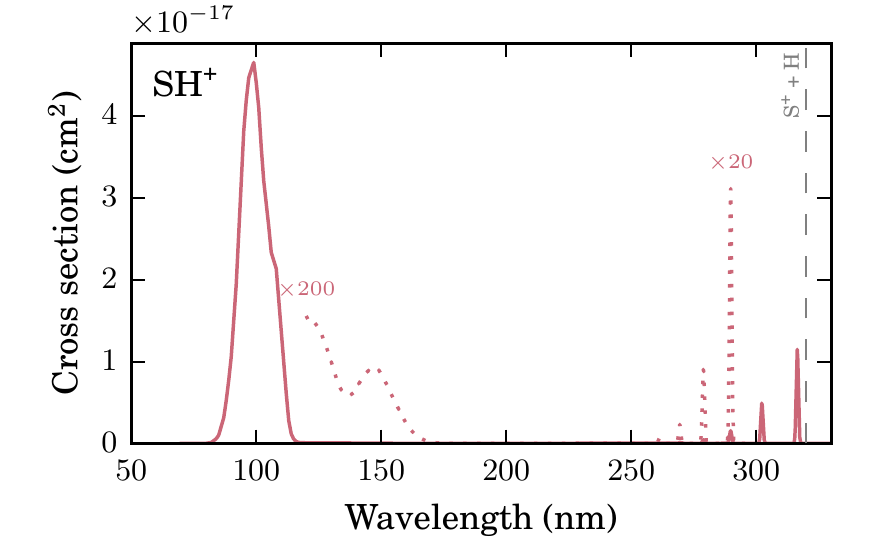}
  \includegraphics[width=0.49\textwidth]{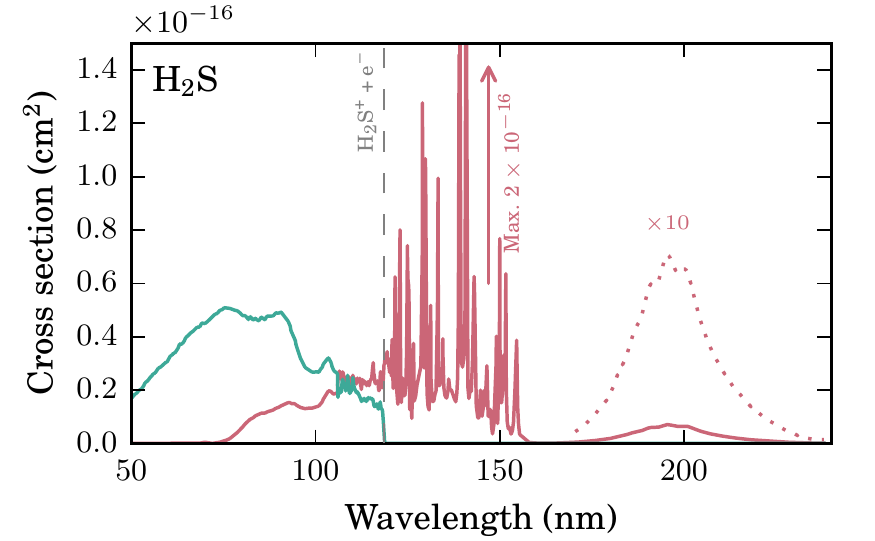}
  \includegraphics[width=0.49\textwidth]{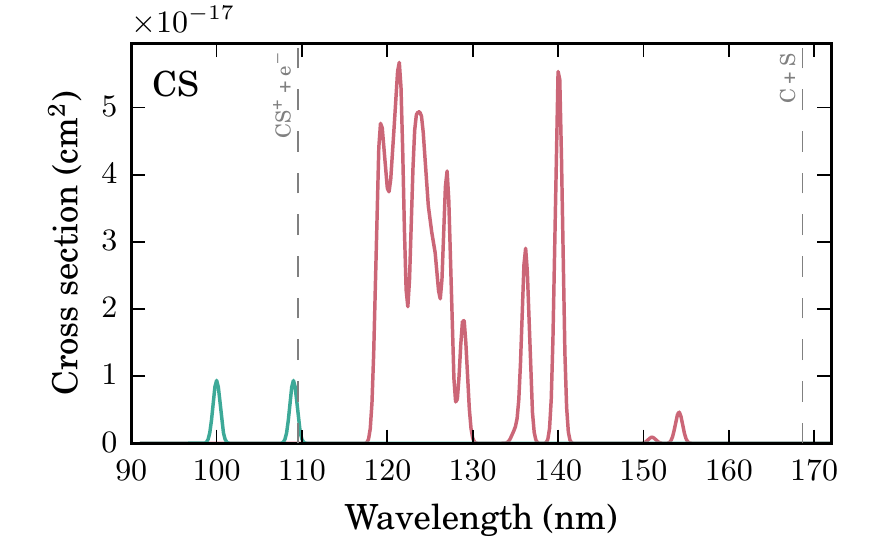}
  \includegraphics[width=0.49\textwidth]{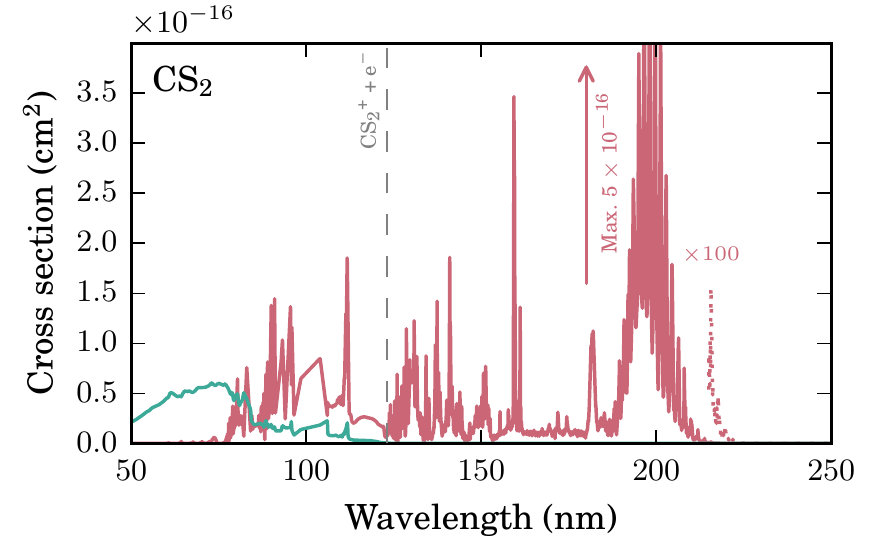}
  \includegraphics[width=0.49\textwidth]{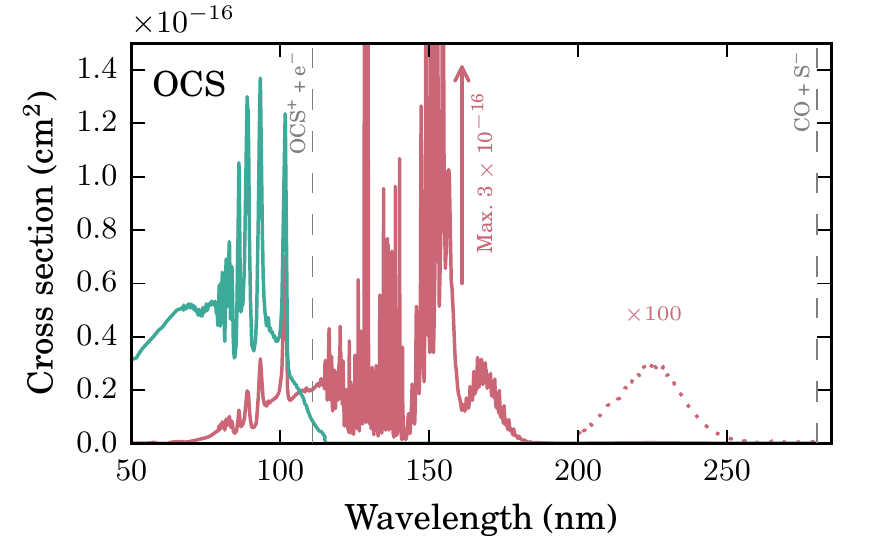}
  \includegraphics[width=0.49\textwidth]{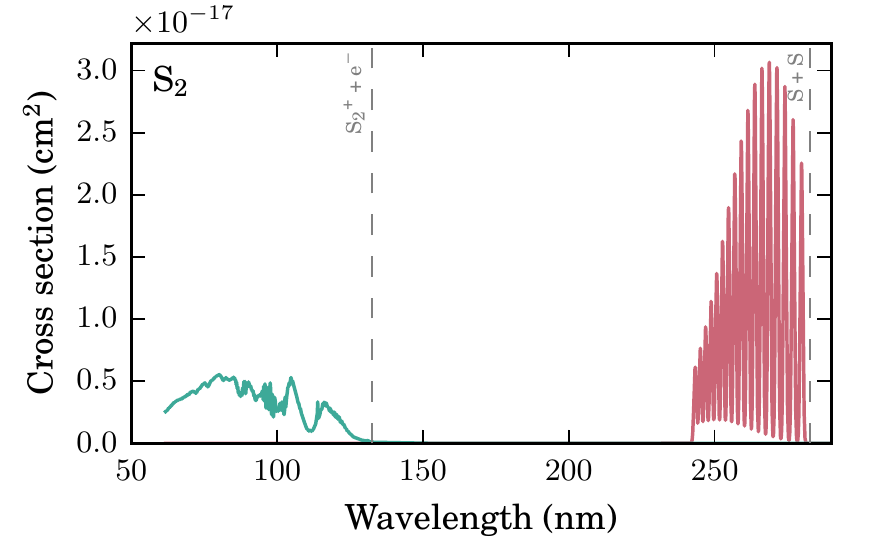}
  \includegraphics[width=0.49\textwidth]{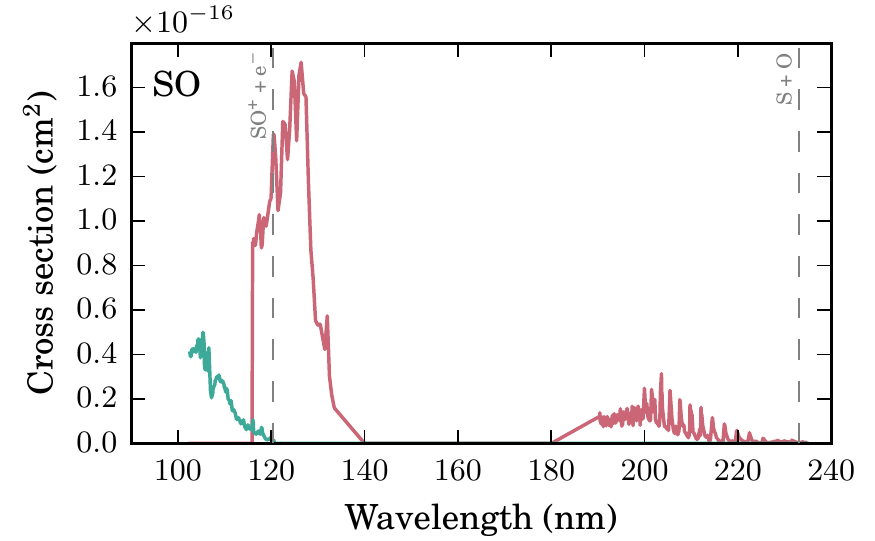}
  \includegraphics[width=0.49\textwidth]{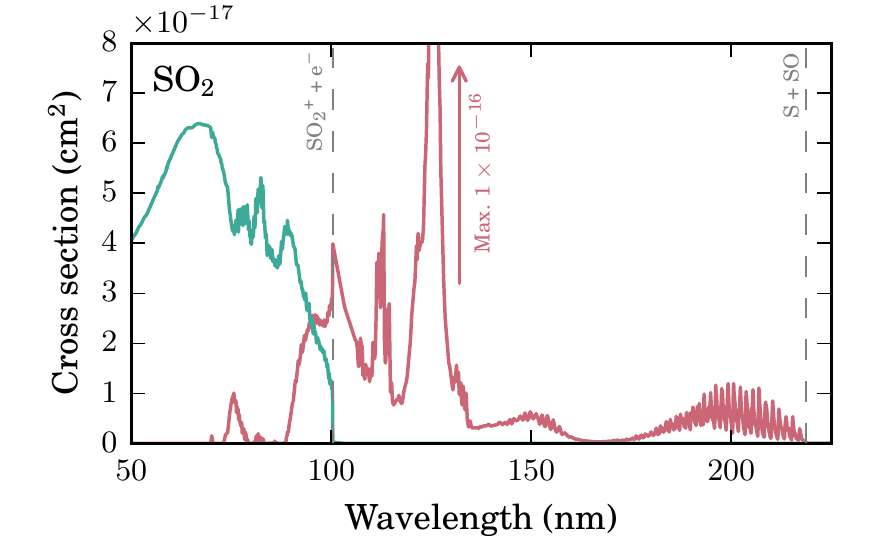}
  \caption{Cross sections of molecules. Red: Photodissociation. Blue: Photoionisation. Some photofragmentation thresholds are also labelled.}
  \label{fig:cross sections SH+}
  \label{fig:cross sections H2S}
  \label{fig:cross sections CS} 
  \label{fig:cross sections CS2}
  \label{fig:cross sections OCS}
  \label{fig:cross sections S2} 
  \label{fig:cross sections SO} 
  \label{fig:cross sections SO2}
\end{figure*}

\begin{figure*}
  \centering
  \includegraphics[width=0.49\textwidth]{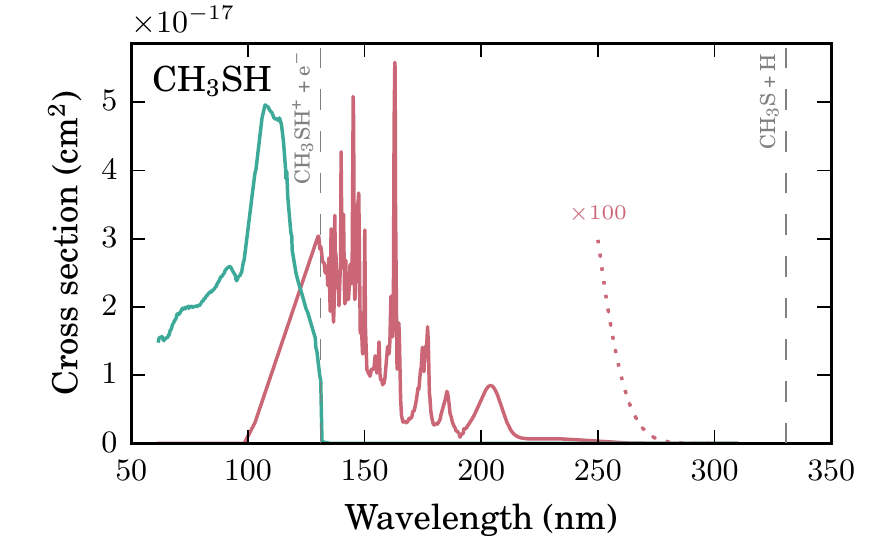}
  \includegraphics[width=0.49\textwidth]{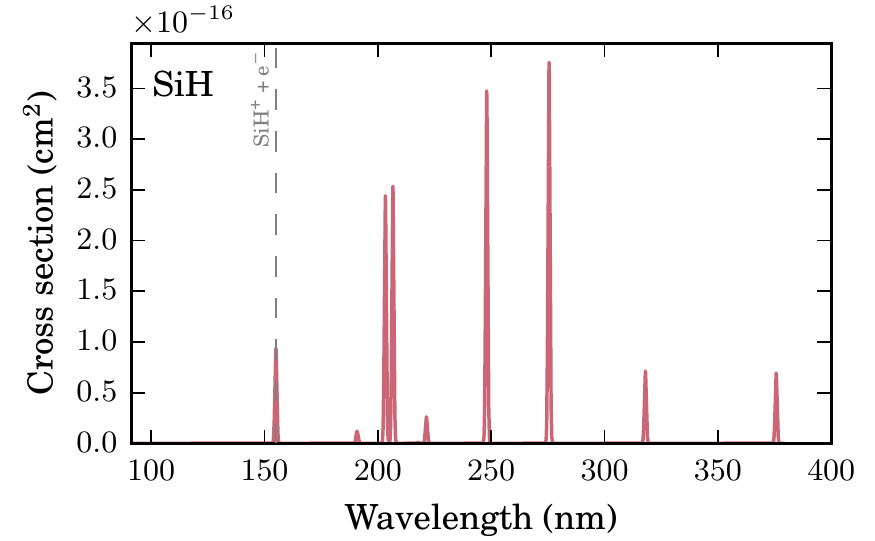}
  \includegraphics[width=0.49\textwidth]{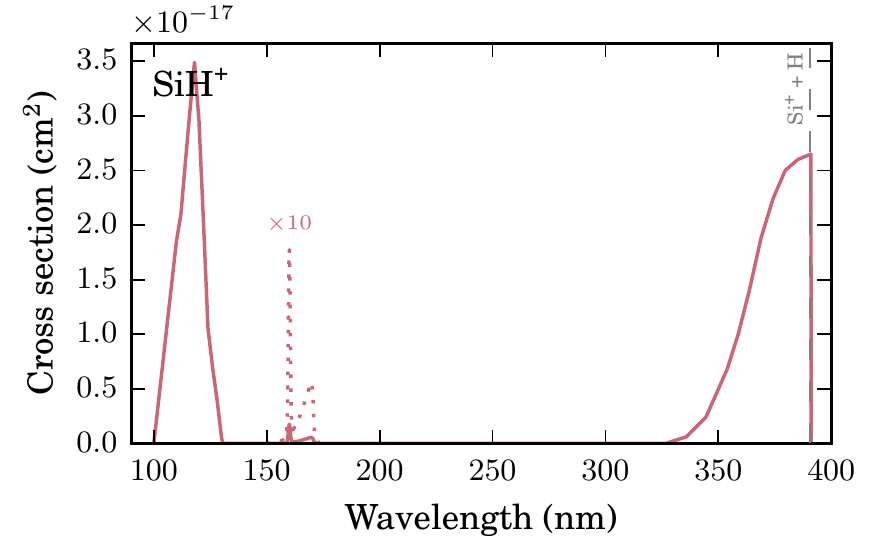}
  \includegraphics[width=0.49\textwidth]{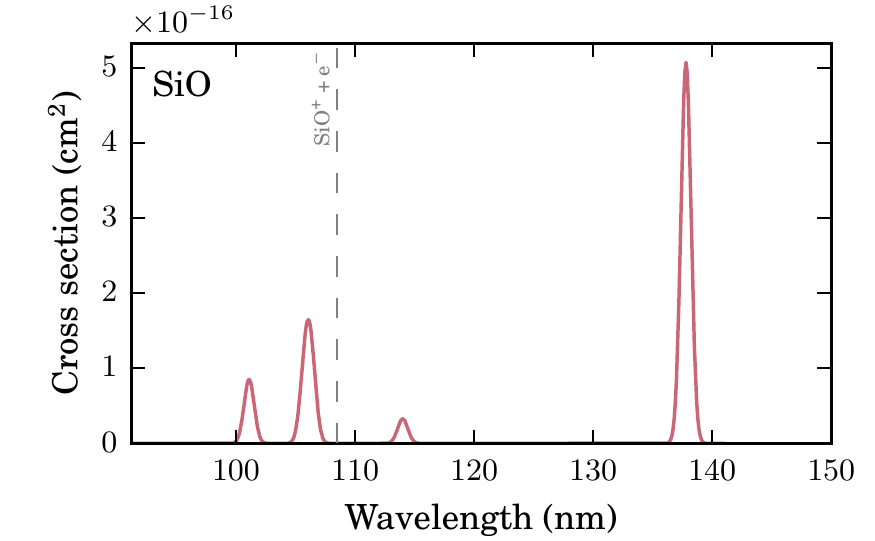} 
  \includegraphics[width=0.49\textwidth]{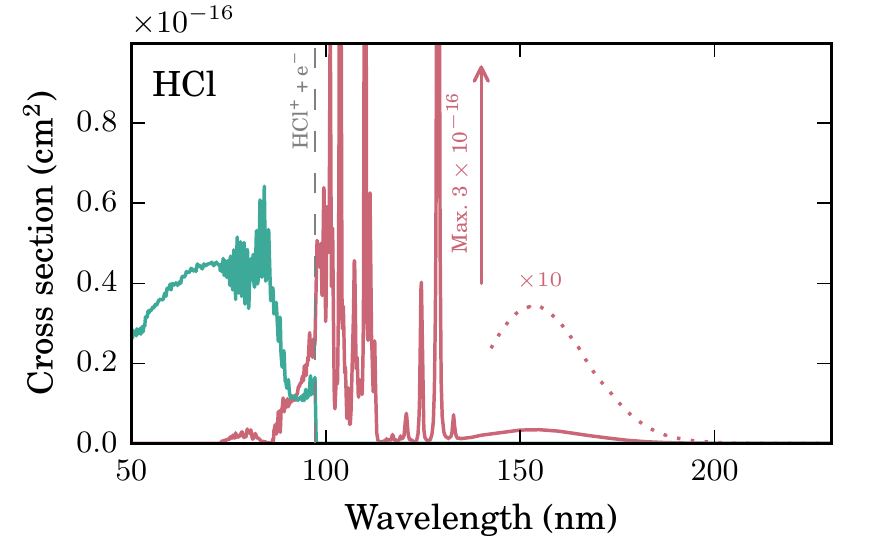} 
  \includegraphics[width=0.49\textwidth]{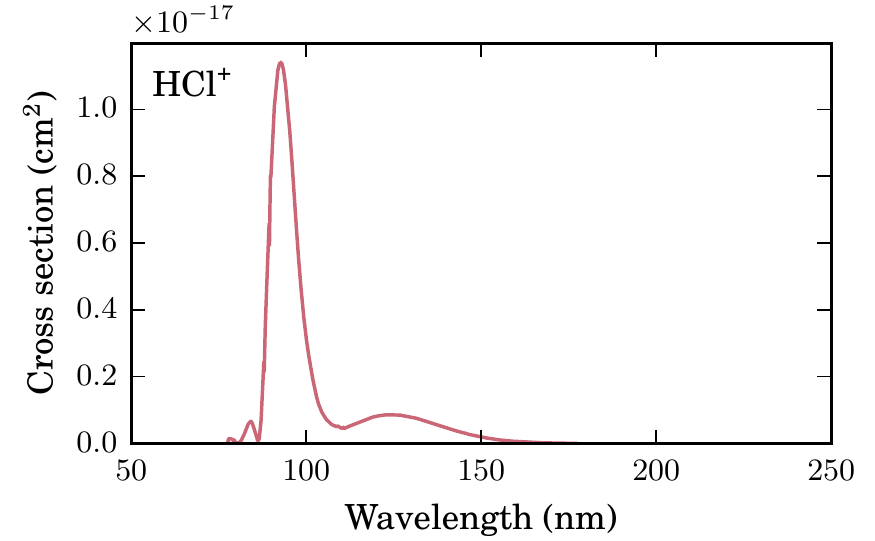}
  \includegraphics[width=0.49\textwidth]{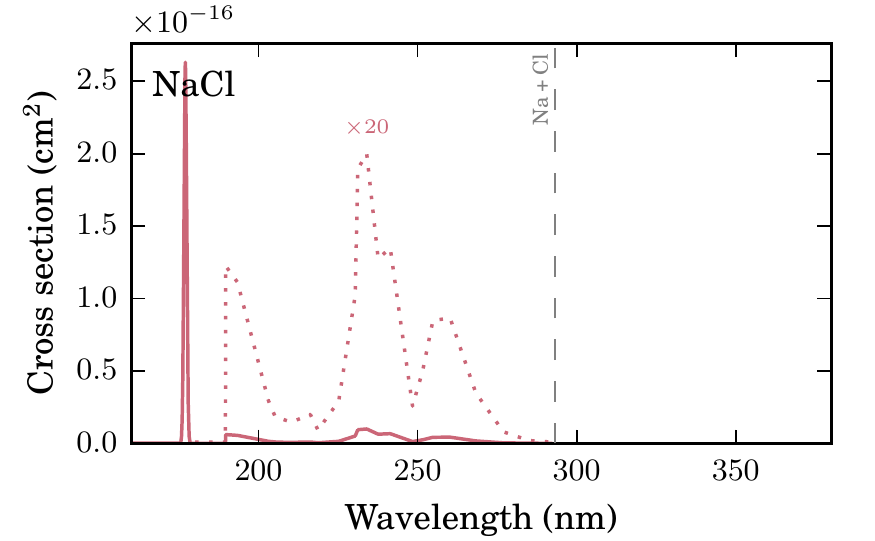}
  \includegraphics[width=0.49\textwidth]{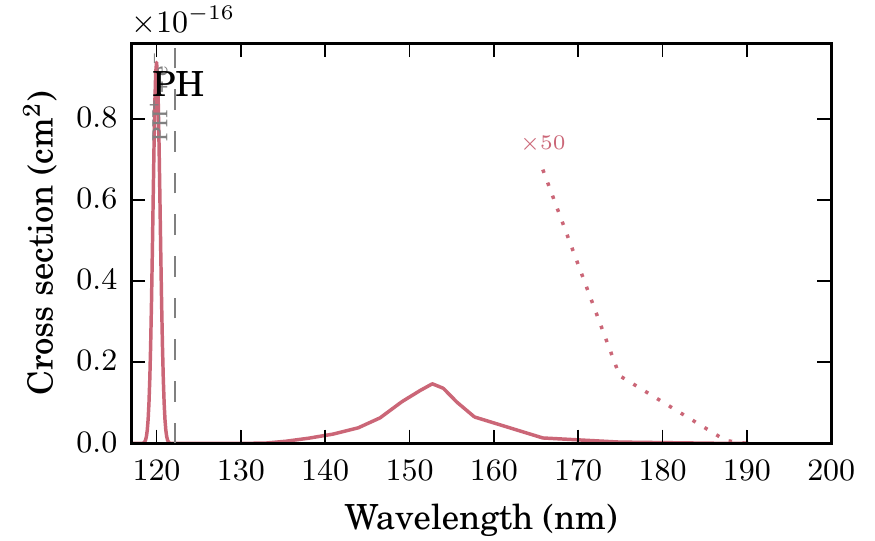}  
  \caption{Cross sections of molecules. Red: Photodissociation. Blue: Photoionisation. Some photofragmentation thresholds are also labelled.}
  \label{fig:cross sections CH3SH}
  \label{fig:cross sections SiH}
  \label{fig:cross sections SiH+} 
  \label{fig:cross sections SiO}  
  \label{fig:cross sections HCl}  
  \label{fig:cross sections HCl+} 
  \label{fig:cross sections NaCl} 
  \label{fig:cross sections PH}   
\end{figure*}

\begin{figure*}
  \centering
  \includegraphics[width=0.49\textwidth]{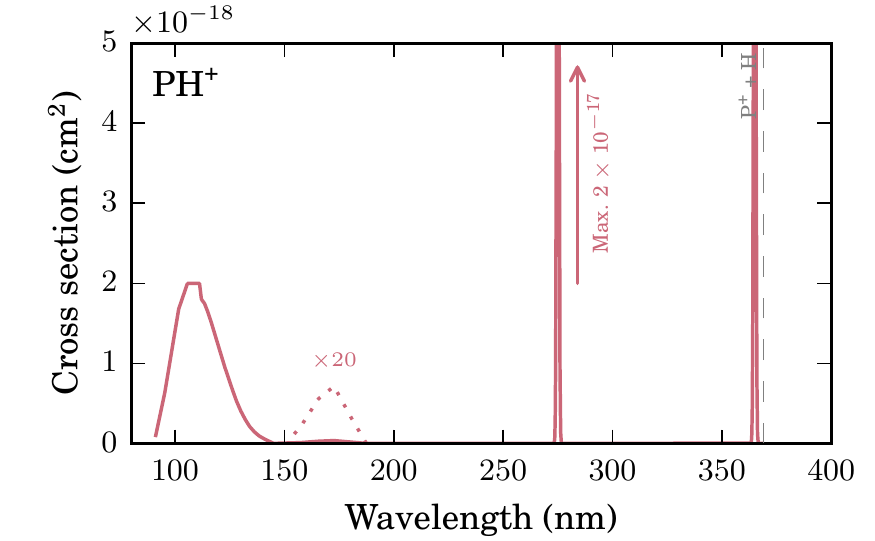} 
  \includegraphics[width=0.49\textwidth]{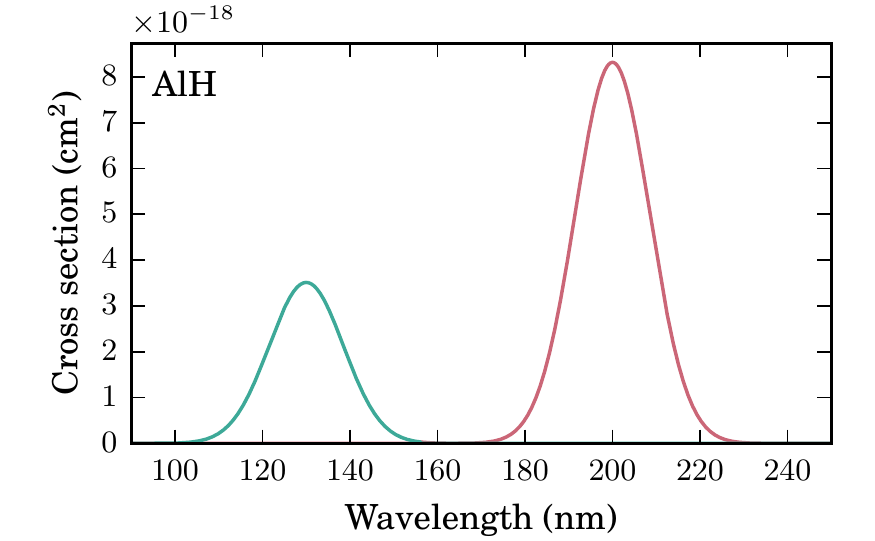} 
  \includegraphics[width=0.49\textwidth]{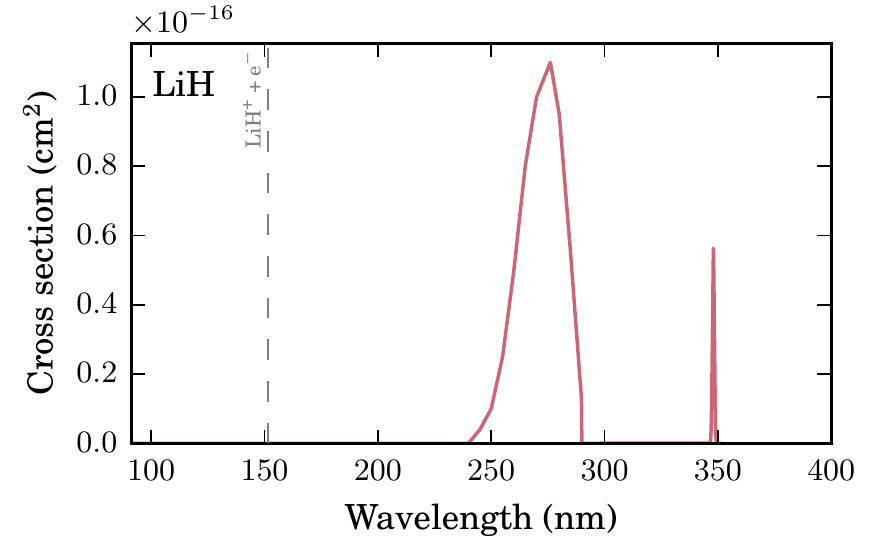}
  \includegraphics[width=0.49\textwidth]{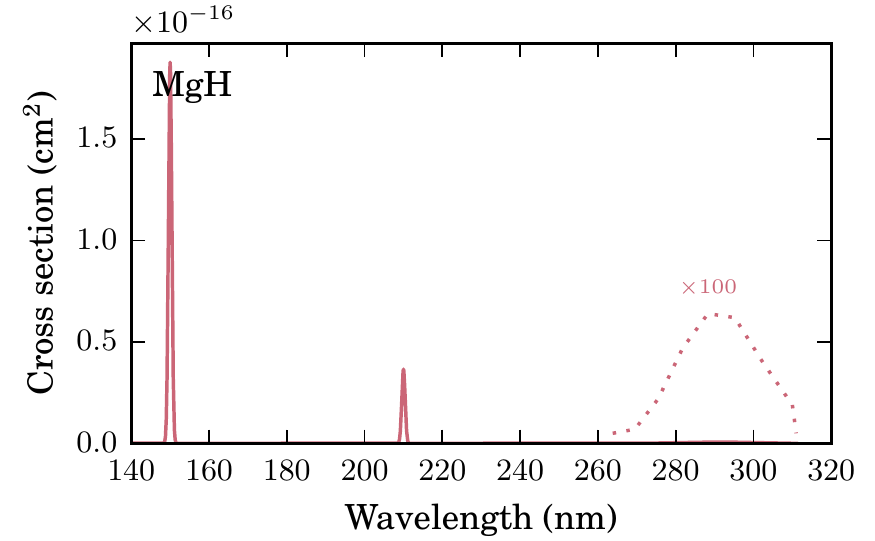}
  \caption{Cross sections of molecules. Red: Photodissociation. Blue: Photoionisation. Some photofragmentation thresholds are also labelled.}
  \label{fig:cross sections PH+}  
  \label{fig:cross sections AlH}  
  \label{fig:cross sections LiH}
  \label{fig:cross sections MgH}
\end{figure*}

In this section the cross sections of atoms and molecules in our database are presented.
All cross sections are plotted in Figs.~\ref{fig:atomic_cross_sections} to \ref{fig:cross sections MgH} and have some summarised properties listed in Table~\ref{tab:cross_section_properties}.
Complete descriptions of the source material for most cross section are given in Sects.~\ref{sec:H} to \ref{sec:AlH}.
For some species, we did not exhaustively reappraise the literature and instead give a reference to its cross section in Table \ref{tab:cross_section_properties}.
\change{There are dissociation and ionisation thresholds listed in Table~\ref{tab:cross_section_properties} for all species where these are relevant. In some cases the listed molecular dissociation limits correspond to greater photon energies (shorter wavelengths) than the dissociation energies of their ground electronic states, due to the lack of accessible excited states for photoabsorption at these energies.}

\clearpage

\subsection{Cross section uncertainties}
\label{sec:cross section uncertainties}
We assigned uncertainties to each overall molecular and atomic cross section according to estimates within their source material, where available, or based on the general accuracy of the various experimental and theoretical methods used, as discussed in Sects.~\ref{sec:exp cross sections} and \ref{sect:theory}.

We limited our estimated uncertainties to three broad categories for simplicity and in view of the ubiquity of large uncertainties in many other key parameters in astrophysical modelling.
These categories are:
\begin{itemize}
\change{\item[A\textsuperscript{+}:] accurate to within 20\%}
  \item[A:] accurate to within 30\%
  \item[B:] accurate within a factor of 2
  \item[C:] accurate within a factor of 10
\end{itemize}
For the purposes of programs requiring uncertainties in terms of log-normal standard deviations, our rating system corresponds approximately to $2\sigma$ uncertainties.

For some molecules, the compilation of data sources into a single best estimated cross section introduces clear wavelength-dependence into the cross section uncertainty, which we weighted according to the wavelength dependence of the ISRF to give the estimates in Table~\ref{tab:cross_section_properties}.
Then, a greater uncertainty at the shortest wavelengths will not contribute as much to our uncertainty estimate as near the long-wavelength threshold.

The uncertainty of photodestruction rates calculated according to Eq.~(\ref{eq:kpd}) will potentially differ for non-ISRF radiation fields.
This is most significant for the case of a Lyman-$\alpha$ dominated radiation field, where the cross section uncertainty at 121.6\,nm is most important.
For molecules with weak and uncertain continua at this wavelength, or sharply resonant structure that may be experimentally under-resolved, recorded at an inappropriate temperature, or calculated with a line position error, the uncertainty may be significantly larger than for the overall cross section.
For this reason we provide a separate uncertainty rating applicable to the Lyman-$\alpha$ wavelength.

The comparison of independent cross section measurements and calculations allows for testing their claimed uncertainties.
An example is \ce{H2O} in Sect.~\ref{sec:unresolved structure}, where we find that the ISRF photodissociation rate obtained with cross section data from four independent experiments with quite different methodologies agrees within 10\%.

\subsection{Cross section database format}

The collected cross sections, with references, are available for download from the Leiden Observatory database.\footnote{\url {www.strw.leidenuniv.nl/~ewine/photo}}
The data files are given in two alternative formats for convenient utilisation, in text files listing the continuum and line absorption of each cross section separately, and in binary format encoding the full wavelength dependence of absorption lines.

The first case provides for a concise data format and rapid
calculation of the photodissociation rates for line absorption in a
continuum field.
This is at the cost of information regarding their shapes and its overlap with a structured radiation field.
The strengths of discrete absorption lines are represented by their integrated cross section; related to
the well-known (dimensionless) oscillator strength, $f_{u\ell}$ for a transition between upper state $u$ and lower state $\ell$; and defined as follows,
\begin{align}
  \label{eq:integrated_cross_section_to_f_value}
  \sigma^\text{int} &= \int\sigma(\lambda)\,\text{d}\lambda \\
                     &=  \frac{e^2}{4 \epsilon_0 m_e c^2} \lambda_0^2 x_{\ell} f_{u\ell} \eta^d, \\
                     &=  \np{8.85e-20}\lambda_0^2 x_{\ell} f_{u\ell} \eta^d.
\end{align}
Here, the final form is appropriate for case of $\sigma$ and $\lambda$ in units of cm$^2$ and nm, respectively.
The integration domain for the cross sections is the wavelength range
of an absorption feature, whether it is an individual ro-vibrational
transition, or an entire electronic-state transition containing many
rotational-vibrational levels.  Here, $e$, $m_e$, $c$, and $\epsilon_0$ are the electron
charge and mass, the speed of light, and permittivity of free space, respectively, and  $\lambda_0$ is the central
wavelength of the transition.
Finally, $\eta^{\rm d}$ is the dissociation probability of the upper level and  $x_{\ell}$ is a ground-state fractional population.
The latter is 1 for oscillator strengths describing transitions
between entire electronic states but will depend on the ground state
excitation temperature when individual rotational or vibrational
transitions are considered.

Then, the photodissociation or ionisation rate defined in Eq.~(\ref{eq:kpd}) is substituted by the equation,
\begin{equation}
  \label{eq:kpd with lines}
  k = \int \sigma^\text{cont} (\lambda) I(\lambda)\text{d}\lambda + \sum_{i=1}^\text{all lines}I(\lambda_0)\sigma^\text{int}.
\end{equation}

As an example\, the photodissociation cross section of
the \ce{CH2} radical is available as the data file {\tt ch2.pd} with contents:
{\small
\begin{verbatim}
 CH2 P.D. cf. van Dishoeck et al. ('96)
   17
    1  1187.6  0.100E-16
    2  1227.6  0.800E-19
    3  1240.0  0.240E-15
    4  1267.7  0.460E-16
    5  1276.9  0.490E-16
         (more data)
   13  1433.3  0.540E-16
   14  1504.7  0.920E-16
   15  1585.5  0.620E-17
   16  1595.7  0.450E-18
   17  1680.0  0.200E-16
  306
 -2750.
    1  1427.6  0.100E-20
    2  1516.3  0.100E-18
    3  1555.7  0.499E-18
    4  1564.1  0.684E-18
    5  1572.6  0.870E-18
         (more data)
  302  2388.4  0.733E-21
  303  2408.2  0.707E-21
  304  2428.4  0.173E-21
  305  2448.9  0.361E-21
  306  2750.0  0.000E+00
\end{verbatim}
}

This file can be serially decoded as follows:

  \begin{description}
  \item[Line 1] Describes the contents of the data file, provides the 
main literature reference.
  \item[Line 2] Number of discrete lines through which photodissociation occurs
   for this molecule, $n_l$
  \item[Line 3 to {$\bf 3+n_l$}]\hfill\\[-3ex]
    \begin{description}
    \item[Field 1] Index of a line
    \item[Field 2] Wavelength of the line centre (\AA)
    \item[Field 3] Integrated cross section of this line  (cm$^2$\,\AA)
    \end{description}
  \item[Line {$\bf 3+n_l+1$}] Number of continuum points defined, $n_c$
  \item[Line {$\bf 3+n_l+2$}] Long wavelength threshold for the continuum data, if set to {\tt -1} or blank the last explicitly listed wavelength is used.
  \item[Line {$\bf 3+n_l+3$ to $\bf 3+n_l+3+n_c$}]\hfill\\[-3ex]
    \begin{description}
    \item[Field 1] Index of this point
    \item[Field 2] Wavelength (\AA)
    \item[Field 3] Cross section  (cm$^2$) 
    \end{description}
  \end{description}
  
All cross sections are also provided as data files in an alternative format including a full specification of their wavelength dependence.
This may require tens or hundreds of thousands of wavelength points to capture the cross
section of highly-structured molecules.  
To facilitate the handling of such large datasets we provide them in the {\tt hdf5} binary format.\footnote{\url{www.hdfgroup.org}}
For the case of \ce{CH2} with file name {\tt CH2.hdf5}, this contains a
dataset labelled {\tt README}, providing a description of the file
contents, literature references, and uncertainty estimate; a binary array labelled {\tt wavelength} in
units of nm; and multiple molecular cross sections in units of cm$^2$, labelled {\tt photoabsorption}, {\tt photodissociation}, and {\tt photoionisation}. 
The discrete lines listed with integrated cross sections in the text-formatted files are included in the continuum-only binary files with Gaussian profiles of 1\,nm\,FWHM.
Astrophysically-important rates calculated in the following are not sensitive to the precise value of this assumed width.

\subsection{Photodissociation and photoionisation of atoms and molecules}

In the following, the cross sections of
atoms and molecules are discussed and summarised. These subsections (and various tables in the paper) are ordered by chemical type, for example, atoms, hydro-carbon species, and oxygen-, nitrogen-, sulphur- and
metal-containing molecules.
A summary of dissociation and ionisation thresholds, cross sections at the Lyman-$\alpha$ wavelength, and the estimated uncertainty of these cross sections is provided in Table~\ref{tab:cross_section_properties}.
The ordering of the following subsections follows the row ordering of Table~\ref{tab:cross_section_properties}.

\subsubsection{H -- hydrogen}
\label{sec:H}
The hydrogen photoionisation continuum, 91.2\,nm and shorter, is
calculated for the TOPbase database
\cite{mendoza1996} and agrees very well with an
experimental measurement \cite{palenius1976} where they overlap.  We
simulate the longer-wavelength Lyman-series line absorption from a
list of transition wavelengths and oscillator strengths from the NIST
atomic database\footnote{\url{www.nist.gov/pml/data/asd.cfm}} \cite{kramida2010}.
We adopted a Gaussian-shaped Doppler broadening of \np[km\,s^{-1}]{1} to accompany the Lorentzian natural linewidths of the H lines in the simulated photoabsorption cross section.

\subsubsection{\ce{C} -- carbon}
\label{sec:C}
The calculated cross section for C was taken from the TOPbase database \cite{mendoza1996,nahar1997}.

\subsubsection{\ce{N} -- nitrogen}
\label{sec:N}
The long-wavelength photoionisation threshold of atomic nitrogen,
85\,nm, is shorter than the Lyman-limit, 91.2\,nm, and its
photoionisation in the interstellar medium is then suppressed.  The
cross section adopted here is taken from the absolute experimental
data of \textcite{samson1990}, apart from the resonant region between 61.8 and
71.5\,nm where the relative-magnitude high-resolution measurement of
\textcite{dehmer1974} was used, after scaling this to match the Samson
\& Angel value at 67\,nm, in a continuum region.

\subsubsection{\ce{O} -- oxygen}
\label{sec:O}
A theoretical calculation of the oxygen photoionisation cross section
\cite{cantu1988,huebner2015} was used and is in good agreement with an
absolute photoionisation yield \cite{angel1988} that lacks the spectral
resolution to reveal the majority of calculated photoionisation
resonances.

\subsubsection{Mg -- magnesium}
\label{sec:Mg}
The experimental photoabsorption cross section of \textcite{yih1998}
was adopted for the photoionisation cross section of Mg for
wavelengths between 120 and 163\,nm.  At shorter wavelengths the
calculated TOPbase cross section
\cite{mendoza1996} was used after rescaling by a
factor of 1.14 to match the integrated value of \textcite{yih1998}
where they overlap.

\subsubsection{\ce{Al} -- aluminium}
\label{sec:Al}
The calculated cross section for Al was taken from the TOPbase database \cite{mendoza1996} and is in reasonable agreement with an absolutely-calibrated experimental measurement \cite{kohl1973}.

\subsubsection{Si -- silicon}
\label{sec:Si}
There are multiple R-matrix calculations of Si photoionisation
\cite{mendoza1988,nahar1993} that are broadly in agreement.  Here, we
adopt a cross section from the TOPbase database
\cite{mendoza1996,huebner2015}.

\subsubsection{\ce{P} -- phosphorus}
\label{sec:P}
The photoionisation cross section determined by \textcite{tayal2004}
from R-matrix calculations is adopted between 62 and 118\,nm.

\subsubsection{\ce{S} -- sulphur}
\label{sec:S}
The recent R-matrix calculation of \textcite{barthel2015} is adopted for the
atomic-S photoionisation cross section for wavelengths shorter
than 93\,nm, and an experimental measurement for longer
wavelengths \cite{gibson1986}.  This non-absolute experiment includes
the correct energy location of many resonances appearing in this cross
section, including some that arise from excited fine-structure
components of the S ground state and may not be populated under all
astrophysical conditions.  The experiment was scaled to match the
calculated cross section after integrating over the range 96 to
121\,nm.

\subsubsection{\ce{Cl} -- chlorine}
\label{sec:Cl}
Two experimental Cl photoionisation cross sections are connected spanning from the \ce{Cl+} ionisation threshold (95.6\,nm) to 60\,nm.
We adopt the measurement of \textcite{cantu1988} for wavelengths longer than 75\,nm, and 
 \textcite{ruscic1983} at shorter wavelengths.

\subsubsection{K -- potassium}
\label{sec:K}
The most recent experimental measurement of the potassium
photoionisation cross section \cite{sandner1981} covers the
wavelengths region between 238 and 285\,nm.  We supplemented this
between 120 and 238\,nm with a recent R-matrix calculation
\cite{zatsarinny2010}, and adopted a linearly-decreasing cross section
at still shorter wavelengths.

\subsubsection{\ce{Ca} -- calcium}
\label{sec:Ca}
A recent high-resolution measurement of the calcium photoabsorption cross section is measured by \textcite{yih1998} from 160\,nm to the ionisation threshold at 202\,nm, and adopted here as a photoionisation cross section.
This was combined with a further measurement that better resolves the two dominant resonant features at 176.5 and 188\,nm \cite{carter1971}, over the regions 176.3 to 176.8\,nm and 187.1 to 189.6\,nm.
The R-matrix calculation from the TOPbase database \cite{mendoza1996} was used for wavelengths 
shorter than 160\,nm.

\subsubsection{Fe -- iron}
\label{sec:Fe}

Multiple measurements were combined into a single Fe photoionisation
cross section file.  An absolute cross section measured at 154\,nm
\cite{lombardi1978} was used to scale the wavelength-dependent
relative photoionisation cross section of \textcite{hansen1977}.  In
turn, this was used to calibrate the measurements of
\textcite{reed2009} and \textcite{tondello1975} by comparing
integrated values over their overlapping ranges.  For the case of
\textcite{reed2009} separate calibration factors were used for the
ranges 125 to 145\,nm and 145 to 158\,nm to account for an apparent
wavelength dependence of laser power in their experiment.  For
  wavelengths shorter than those covered by the various experiments
  the R-matrix calculation of \textcite{bautista1997} was used, and at
  the shortest wavelengths the fitting formulate of
  \textcite{verner1996}.  The relevant scaling factors and wavelength
  ranges are listed in Table~\ref{tab:Fe data sources}.

\begin{table}
  \caption{References and wavelength ranges of concatenated Fe cross sections.}
  \label{tab:Fe data sources}
  \footnotesize
  \centering
  \renewcommand*{\thefootnote}{\alph{footnote}}
  \begin{tabular}{ccl}
    \toprule
    Wavelength (nm) & Scaling (cm$^2$)\footnotemark[1] & Reference \\
    \midrule
    $<80$           & --                                & \textcite{verner1996}  \\
    ~80 -- ~88 & --                                & \textcite{bautista1997}\\
    ~88 -- 123 & \np{2.62e-18}                     & \textcite{tondello1975}\\
    123 -- 134 & \np{1.88e-18}                     & \textcite{hansen1977}  \\
    134 -- 146 & \np{2.01e-18}                     & \textcite{reed2009}    \\
    146 -- 158 & \np{2.71e-18}                     & \textcite{reed2009}    \\
    \bottomrule
  \end{tabular}
  \begin{minipage}{\linewidth}
  \footnotetext[1]{The amount by which the arbitrary units of the various photoion yields were scaled to provide a cross section.}
  \end{minipage}
\end{table}

\subsubsection{Zn -- zinc}
\label{sec:Zn}
There are two measurements of the Zn photoabsorption cross section
between 110\,nm and the photoionisation threshold at 132\,nm
\cite{marr1969,yih1998} that disagree by a factor of two.  We used an
average of these two measurements where they coincide, the measurement
of \textcite{marr1969} between 91 and 110\,nm and the
relative-photoionisation cross section of \textcite{harrison1969} at
shorter wavelengths, after scaling its integrated magnitude to match
the continuum level of an R-matrix calculation \cite{huebner2015}.

\subsubsection{\ce{H2} -- hydrogen}
\label{sec:H2}

Molecular hydrogen has a line-dominated spectrum beginning at 112\,nm
with absorption into the Lyman ($B\,{}^1\Sigma^+_u\leftarrow X\,{}^1\Sigma^+_g$) and Werner ($C\,{}^1\Pi_u\leftarrow X\,{}^1\Sigma^+_g$) bands, and retains its
resonant character even beyond the ionisation threshold at 80.3\,nm.
The excited rovibrational lines and their tendency to partially
predissociate or autoionise is well studied experimentally and
theoretically by the group of Abgrall and Glass-Maujean et al.
\cite{abgrall_etal1993,abgrall_etal1993b,abgrall_etal1993c,abgrall_etal1994,abgrall_roueff2006,abgrall2000,glass-maujean2010,glass-maujean2013b,glass-maujean2013c,glass-maujean2013a},
culminating in a detailed experimental and theoretical database of
absorption and emission lines,\footnote{\url{molat.obspm.fr}} as well as by
other groups
\cite[e.g.,][]{dehmer1976,chan1992,samson1994,jonin2000,philip2004b,hollenstein2006,dickenson2014,holland2014}. 

Spontaneous emission from photoexcited $B\,{}^1\Sigma^+_u$ and $C\,{}^1\Pi_u$ states into ground state vibrational levels produces a spectrum of vacuum- and far-ultraviolet emission lines.
Emission into the ground state continuum is also possible and provides a dissociation mechanism at wavelengths below the direct-photodissociation threshold of 84\,nm \cite{field1966,stecher1967,abgrall1997,abgrall1999}.
This spontaneous radiative dissociation mechanism is experimentally and theoretically verified \cite{dalgarno1970,stephens1972}.
The overall  \ce{H2} dissociation efficiency assuming several of the radiation fields discussed in Sect.~\ref{sec:radiation fields} is listed in Table~\ref{tab:H2 photodissociation efficiency} and varies between 5 and 15\% for continuum interstellar radiation fields.
This efficiency is somewhat larger (up to 25\%) for the solar and simulated TW-Hydra radiation fields because these include flux shorter than the Lyman limit at 91.2\,nm and the dissociation fraction of the \ce{H2} cross section increases with decreasing wavelength.

\begin{table}
  \caption{\ce{H2} integrated photodissociation efficiency assuming several radiation field types.\myfootnotemark{1}}
  \label{tab:H2 photodissociation efficiency}
  \centering
  \begin{tabular}[c]{rc}
    \toprule
    Radiation field type    & Efficiency (fraction)\\
    \midrule
    ISRF                    & 0.12           \\
    4000\,K black body       & 0.05           \\
    \np{10000}\,K black body & 0.11           \\
    \np{20000}\,K black body & 0.13           \\
    Solar                   & 0.25           \\
    TW-Hydra                & 0.16           \\
    \bottomrule
  \end{tabular}
  \begin{minipage}{\linewidth}
    \footnotetext[1]{Assuming an ortho-\ce{H2}:para-\ce{H2} ratio of 0:1.}
  \end{minipage}
\end{table}

\begin{figure}
  \centering
  \includegraphics{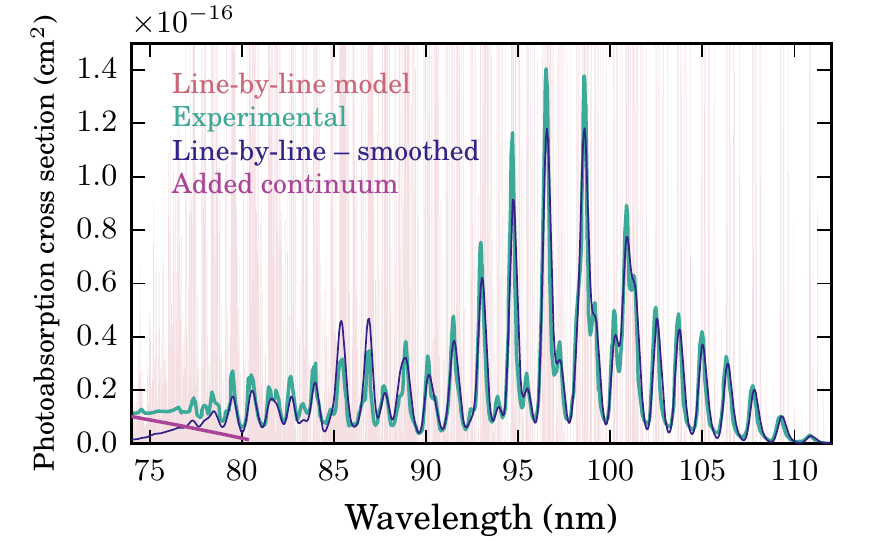}
  \label{fig:H2_cross_section_breakdown}
  \caption{\ce{H2} photoabsorption cross section. The under-resolved experimental measurement of \cite{chan1992}, and a model spectrum before and after smoothing to the experimental resolution. Also shown is a correction to the model accounting for neglected excitation mechanisms.}
\end{figure}

Measured and calculated transition frequencies, strengths, and predissociation efficiencies of individual rovibrational transitions from  the database of Abgrall and Glass-Maujean et
  al. \change{were employed} to reconstruct the temperature-dependent photoabsorption cross
sections for \ce{H2} line-by-line, including excitation to $B\,{}^1\Sigma^+_u$, $B'\,{}^1\Sigma^+_u$,
$C\,{}^1\Pi_u$, and $D\,{}^1\Pi_u$ excited valence states, and many $np\pi$ and $np\sigma$
Rydberg levels.  The completeness of this database for wavelengths
longer than 80\,nm is evidenced by the agreement within 10\% over this
region of its integrated cross section with the low-resolution
electron-energy-loss-derived absolute cross section of
\textcite{chan1992}.  
This comparison is shown in Fig.~\ref{fig:H2_cross_section_breakdown} after smoothing the line-by-line model spectrum to the under resolved experimental spectrum.
The deficit of the model cross section relative to \textcite{chan1992} for wavelengths shorter than 80\,nm was assumed to arise from direct ionisation to \ce{H2+} or
autoionising highly-excited \ce{H2} levels that are neglected in the
line-by-line model.
This deficit is then added to the line-by-line
model forming a continuum shortwards of 80\,nm. The cross section of \textcite{backx1976} was
adopted at the shortest wavelengths, $<62$\,nm.  The use of a line-absorption model permits calculation of
cross sections with a range of ground-state excitation states and
Doppler profiles.

\begin{table}
  \caption{\ce{H2} photodissociation rate in the interstellar  radiation field of \textcite{draine1978}.}
  \label{tab:H2 photodissociation rates}
  \centering
  \footnotesize
  \begin{tabular}[c]{cc}
    \toprule
    & Rate (s$^{-1}$)\\
    \midrule
    Present result assuming ortho:para=0:1 & \np{5.1e-11}\\
    Present result assuming ortho:para=3:1 & \np{5.7e-11}\\
    \textcite{sternberg2014} & \np{5.8e-11} \\
    \bottomrule
  \end{tabular}
\end{table}
There is a great deal of previous work done on \ce{H2}
photodissociation due to its importance to the balance of atomic and
molecular hydrogen in the Universe.  This subject is well reviewed and
state-of-the-art calculations made in \textcite{sternberg2014} and
references therein.
The correctness of our simulated \ce{H2} photodissociation cross
section is verified by comparing an ISRF dissociation rate with the
calculations of \textcite{sternberg2014}, listed in Table~\ref{tab:H2
  photodissociation rates}.
Two calculations are made, for an ortho:para ratio of \ce{H2} ground state levels of 3:1 and 0:1.
The increased rate for the 3:1 case is due to increased population of
rotationally-excited \ce{H2} levels that are more likely to decay
dissociatively.

\subsubsection{\ce{H3+} -- trihydrogen cation}
\label{sec:H3+}
The electronic excitation of cold \ce{H3+} has not been measured in the laboratory, due to the difficult of cooling the highly-symmetric radical ions that must be formed in situ, although its photodissociation from excited ground state rotational-vibrational levels is studied in some detail \cite[e.g.,][]{carrington1984}.
Calculations of its excited states find no allowed transitions with wavelength longer than about 70\,nm \cite{talbi1988}, well above its 283\,nm ground state dissociation threshold \cite{kulander1978}.
Equilibrium potential-energies and transition moments for the ground state accessible by photoabsorption are calculated by \textcite{talbi1988} and van~Dishoeck (unpublished), leading to very similar cross sections. 
The cross section of van~Dishoeck was adopted for our photodissociation database.

\subsubsection{\ce{CH2+} -- methylene cation}
\label{sec:CH2+}
According to the calculations of \textcite{theodorakopoulos1991} this ion has many dipole-allowed excited states below 13.6 eV. The 1, 2 and 3$^2$B$_2$, 2, 3 and 4 $^2$A$_1$ and 2$^2$B$_1$ states were included in our cross section with $f$=0.008, 0.0001, 0.02, 0.01, 0.06, 0.05 and 0.03,
respectively, and $\eta^\text{d}=1$ assumed.

\subsubsection{\ce{CH3} -- methyl radical}
\label{sec:CH3}
\begin{table}
  \caption{References and wavelength ranges of concatenated \ce{CH3} cross sections.}
  \label{tab:CH3 data sources}
  \begin{minipage}[c]{1.0\linewidth}
    \centering
    \begin{tabular}{cl}
      \toprule
      Wavelength (nm) & Reference \\
      \midrule
      200 -- 220 & \textcite{cameron2002}\\
      126 -- 200 & \parbox{0.6\linewidth}{\textcite{herzberg1956} and assumed continuum}\\
      108 -- 126 & \textcite{gans2010}\\
     $ <$ 108   & \parbox{0.6\linewidth}{Extrapolation of \textcite{gans2010}}\\
      \bottomrule    
    \end{tabular}
  \end{minipage}
\end{table}

Several \ce{CH3} absorption bands are photographically observed
between 120 and 230\,nm by \textcite{herzberg1956} revealing
predissociation-broadened bands with widths of around 1\,nm, and very
roughly estimated strengths.  Higher-resolution absolutely-calibrated
photoabsorption measurements of the longest-wavelength absorption
features \cite{cameron2002,khamaganov2007} (200\,nm to the
photoabsorption threshold at 220\,nm) allow for a rough calibration of
the other photographic features based on their apparent saturation.

An absolute photoionisation cross section is recorded in the neighbourhood of the 126\,nm ionisation threshold \cite{gans2010}, which was extrapolated here to shorter wavelengths.
A summary of cross sectional data sources is given in Table~\ref{tab:CH3 data sources}.

\subsubsection{\ce{CH4} -- methane}
\label{sec:CH4}

\ce{CH4} photoabsorbs significantly at wavelengths shorter than
140\,nm with an unstructured cross section that peaks at 90\,nm,
indicating mostly direct photodissociation.
Here,  the
experimental cross section of \textcite{kameta2002} between 52 and
124\,nm was used, and the slightly-lower resolution data of \textcite{au1993}
outside this range.  The cross sections of \textcite{kameta2002}
was scaled down by a factor or 0.95 in order to agree with other more
reliably calibrated overlapping measurements \cite{au1993,lee2009}.

Many neutral and ionised fragments are observed following
\ce{CH4} photodissociation and these data are well summarised
elsewhere
\cite{backx1975b,huebner1992,gans2011,blitz2012,huebner2015}.  Here,
 the wavelength-dependent branching ratios of \textcite{kameta2002}
were used to decompose photoabsorption into ionisation and dissociation cross
sections, assuming zero branching to fluorescence.  

\subsubsection{\ce{CH4+} -- methane cation}
\label{sec:CH4+}
This ion is subject to detailed theoretical
studies indicating several dissociative excited states \cite{van_dishoeck1980}.
Oscillator strengths into the 2, 3 $^2$A$_1$ and 2$^2$B$_1$ states
are taken to be $f$=0.04, 0.04 and 0.08, respectively, with $\eta^\text{d}$=1.

\subsubsection{\ce{C2H} -- ethynyl radical}
\label{sec:C2H}

Absorption into the 5 $^2\Sigma^+$ state lying around 10\,eV dominates
the interstellar photodissociation of this molecule. The higher
$^2\Pi$ states in the 8.5 to 10.5 eV range can also contribute
significantly as far as the C$_2$H ionization potential of 11.4 eV.
The oscillator strengths listed in Table 1 of the calculations of
\textcite{van_hemert2008} are used with $\eta^\text{d}$=1.

\subsubsection{\ce{C2H2} -- acetylene}
\label{sec:C2H2}
\begin{table}
  \caption{References and wavelength ranges of concatenated \ce{C2H2} cross sections.}
  \label{tab:C2H2 data sources}
  \begin{minipage}[c]{1.0\linewidth}
    \centering
    \begin{tabular}{cl}
      \toprule
      Wavelength (nm) & Reference \\
      \midrule
  ~~~6 -- 106   &  \textcite{cooper1995}     \\
  106 -- 110 &  \textcite{xia1991}      \\
  110 -- 154 &  \textcite{cheng2011}      \\
  154 -- 210 &  \textcite{wu1989}         \\
  210 -- 300 &  \textcite{vattulainen1997}\\
      \bottomrule    
    \end{tabular}
  \end{minipage}
\end{table}

The \ce{C2H2} photoabsorption cross section is compiled from a
collection of measurements
\cite{wu1989,xia1991,cooper1995,vattulainen1997,cheng2011} over the
wavelength ranges listed in Table~\ref{tab:C2H2 data sources}, which
extend beyond 200\,nm and are strongest shortward of 153\,nm.  A
strong absorption line of \ce{C2H2} coincides very nearly with the
hydrogen Lyman-$\alpha$ line.

The longest \ce{C2H2} photodissociation threshold, forming \ce{C2H + H}, occurs at 217\,nm and the dissociation efficiency at shorter wavelengths was studied several times \cite[e.g.,][]{okabe1983,seki1993,lauter2002,kovacs2010}, as well as the probability of forming an \ce{H2} product.
\textcite{lauter2002} convincingly determined a 100\% efficiency for H-atom formation by 121.6 and 193.3\,nm radiation after  detecting these atoms through laser-induced fluorescence, and a 100\% dissociation efficiency was assumed for all wavelengths shorter than the 217\,nm threshold.

\subsubsection{\ce{C2H6} -- ethane}
\label{sec:C2H6}
\change{
  The photoabsorption cross section of \ce{C2H6} is measured between 120 and 150\,nm by \textcite{chen2004} and their 150\,K measurement was adopted here (the temperature variation measured in this experiment was slight however). 
At shorter wavelengths, the cross section measured by \textcite{kameta1996} was used after scaling this down to match the overlapping region of \textcite{chen2004} (a factor of 0.83).
The ionisation fraction for \ce{C2H6} is also measured by \textcite{kameta1996} and adopted here.
}

\subsubsection{\ce{C3} -- tricarbon}
\label{sec:C3} 
The photoabsorption cross section of \ce{C3} is difficult to measure
due to its radical nature.  Vertical excitation energies are
calculated by \textcite{van_hemert2008} for absorption into 5
highly-excited states as well as vertical oscillator strengths.  We
combine these transitions into a photodissociation cross section after
assigning all states a Gaussian band profile of full-width
half-maximum 18\,nm.  This width was selected to match the width of
the strong \ce{C3} electronic transition near 160\,nm (corresponding
to the $1 ^1\Sigma_u^+ - 1 ^1\Sigma_g^+$ transition among the
calculations of \textcite{van_hemert2008}) observed in a
matrix-isolation experiment \cite{monninger2002}.  
Here, all bands were assumed to be 100\% dissociative, since they lie well above the \ce{C3} dissociation limit.

There is no absolute measurement of the photoionisation cross section
of \ce{C3} shorter than its 11.6\,eV threshold \cite{belau2007}.  The
onset of ionisation and its wavelength dependence is measured by
\textcite{nicolas2006}.  This wavelength dependence was adopted as a
photoionisation cross section after scaling it to the typical
molecular value of \np[cm^2]{2e-17} at 91.2\,nm.

\subsubsection{\ce{$l$-C3H} -- linear propynylidyne}
\label{sec:l-C3H}
The photodissociation of this linear molecule is dominated by the
high-lying $^2\Pi$ states around 7.8 eV. The $f$-values given in
Table~3 of \textcite{van_hemert2008} are to
simulate the \ce{$l$-C3H} spectrum, assuming $\eta^\text{d}=1$.

\subsubsection{\ce{$c$-C3H} -- cyclic propynylidyne}
\label{sec:c-C3H}
In contrast with its linear counterpart, many
different electronic states in the 5 to 7.5 eV range can contribute to
the photodissociation of cyclic C$_3$H.  The $f$-values given in
Table~4 of \textcite{van_hemert2008} are used with $\eta^\text{d}$=1.

\subsubsection{\ce{HC3H} -- propargylene}
\label{sec:HC3H}
Three isomers of \ce{H2C3} are treated in the calculations of \textcite{van_hemert2008}.
In this section, \ce{HC3H} refers to the HCCCH linear form, with the alternative linear excited isomer (\ce{H2CCC} or \ce{$l$-C3H2}) and cyclic ground isomer (\ce{$c$-C3H2}) discussed below.
The photodissociation of \ce{HC3H} is dominated by three excited ${}^3\text{A}$ levels, with vertical excitations of 4.23, 6.29, and 7.52\,eV above the ground state.

\subsubsection{\ce{$l$-C3H2} -- linear propenylidene}
\label{sec:l-C3H2}
The strongest absorptions of this molecule are into the higher
$^1A_1$ Rydberg states around 9 eV, below the ionization potential at
10.4 eV. The $f$-values given in Table~7 of \textcite{van_hemert2008} are
used, with $\eta^\text{d}$=1.  The low-lying $\tilde C$(2)$^1A_1$ state is
taken into account, even though it is not clear how efficient
predissociation is for this state. The linear HC$_3$H isomer has a
very different electronic structure from $l$-C$_3$H$_2$ with a
triplet ground state. This molecule is included as a separate species
in the database, following Table~5 of \textcite{van_hemert2008}.

\subsubsection{\ce{$c$-C3H2} -- cyclic propenylidene}
\label{sec:c-C3H2}
This molecule has only a few low-lying
dipole-allowed electronic states. The largest oscillator strengths are
found to transitions in the 9 to 11 eV range, which is above the
ionization potential of $c$-C$_3$H$_2$ at 9.15 eV. The current version
of the database, based on Table 6 of \textcite{van_hemert2008} with
$\eta^\text{d}$=1, assumes that these transitions lead to ionization rather
than dissociation.  If they would lead to dissociation, the
interstellar photodissociation rate would be increased by a factor of
2.

\subsubsection{\ce{$l$-C4} -- linear tetracarbon}
\label{sec:l-C4}
The strongest absorption occurs into the 2
$^3\Sigma_u^-$ state around 6.95 eV, which, like a similar state for
C$_3$, has a huge oscillator strength of 1.6. This state, together
with other states listed in Table~8 of \textcite{van_hemert2008}, are
included with $\eta^\text{d}$=1.  Rhombic C$_4$, which is almost isoenergetic
with $\ell$-C$_4$, is not considered and may well have a significantly
different photodissociation rate.

\subsubsection{\ce{$l$-C4H} -- butadiynyl}
\label{sec:l-C4H}
Of the dipole-allowed transitions below the ionization potential at
9.6 eV, the higher $^2\Sigma^+$ states at 7 to 9\,eV have orders of
magnitude larger oscillator strengths than other states and thus
dominate the interstellar photodissociation. The values used in Table
9 of \textcite{van_hemert2008} are used with $\eta^\text{d}$=1.

\subsubsection{\ce{$l$-C5H} -- pentynylidyne}
\label{sec:l-C5H}
\label{sec:last cross section description section}
The higher $^2\Pi$ states in the 4 to 6\,eV range
dominate the photodissociation of this molecule, using the values in
Table~10 of \textcite{van_hemert2008} with $\eta^\text{d}$=1.

\subsubsection{\ce{H2O} -- water}
\label{sec:H2O}

\begin{table}
  \caption{References and wavelength ranges of concatenated \ce{H2O} cross sections.}
  \label{tab:H2O data sources}
  \small
  \centering
  \begin{tabular}{ccl}
    \toprule
    Wavelength (nm) & Scaling & Reference \\
    \midrule
    ~~6.2~  -- ~56.6~   & 1    & \textcite{chan1993d}             \\
    ~56.6~  -- ~98.95  & 1    & \textcite{fillion2003}           \\
    ~98.95  -- 100~~~    & 1    & Linear interpolation.\myfootnotemark{1}            \\
    100~~~~ -- 107.28 & 1    & \textcite{fillion2004}           \\
    107.28  -- 108.01 & 0.5\myfootnotemark{2}  & CfA molecular database\myfootnotemark{3}\\
    108.01  -- 111.24 & 1    & \textcite{fillion2004}           \\
    111.24  -- 111.77 & 0.5\myfootnotemark{2}  & CfA molecular database\myfootnotemark{3}\\
    111.77  -- 113.9~  & 1    & \textcite{fillion2004}           \\
    113.9~~ -- 114.8~  & 1    & Linear interpolation.\myfootnotemark{1}            \\
    114.8~~ -- 123.42 & 1.09\myfootnotemark{4} & \textcite{mota2005}              \\
    123.42  -- 124.5~  & 0.91\myfootnotemark{5} & CfA molecular database\myfootnotemark{3}\myfootnotemark{6}\\
    124.5~  -- 193.9~ & 1.09\myfootnotemark{4} & \textcite{mota2005}              \\
    \bottomrule
  \end{tabular}
  \begin{minipage}{\linewidth}
    \footnotetext[1]{Linearly interpolated between surrounding cross section data.}
    \footnotetext[2]{Scaled to match the integrated overlapping cross section of \textcite{fillion2004}.}
    \footnotetext[3]{\url{www.cfa.harvard.edu/amp/ampdata/cfamols.html}}
    \footnotetext[4]{Scaled to match the integrated overlapping cross section of \textcite{chan1993d}.}
    \footnotetext[5]{Scaled to match the integrated overlapping cross section of \textcite{mota2005}.} 
    \footnotetext[6]{An apparent error in the wavelength calibration of this spectrum was corrected by adjusting the spectrum longwards by 0.05\,nm.}
  \end{minipage}
\end{table}

Measurements of the \ce{H2O} photoabsorption cross section as a function
of wavelength date back many decades (see \textcite[][Sect.~3.1.3]{van_dishoeck2013b}).
Several measurements were concatenated into a single cross section file,
as detailed in Table~\ref{tab:H2O data sources}.  Some of these data
were rescaled to agree with the integrated cross section of
more-reliable absolute measurements.

An absolutely-calibrated low-resolution cross section measured
by electron-energy loss spectroscopy \cite{chan1993d}, including the
entire ultraviolet range from its onset at 190\,nm,  was combined with several
higher-resolution direct photoabsorption measurements covering various
portions of the spectrum
\cite[][and others]{smith1981,yoshino1996,yoshino1997,parkinson2003,fillion2003,fillion2004,mota2005}.
Some of these resolve temperature-dependent rotational structure.
\textcite{fillion2003} record the photoionisation yield of \ce{H2O}
that was used to partition the experimental photoabsorption cross
section between ionisation and dissociation.

\begin{figure}
  \centering
  \includegraphics{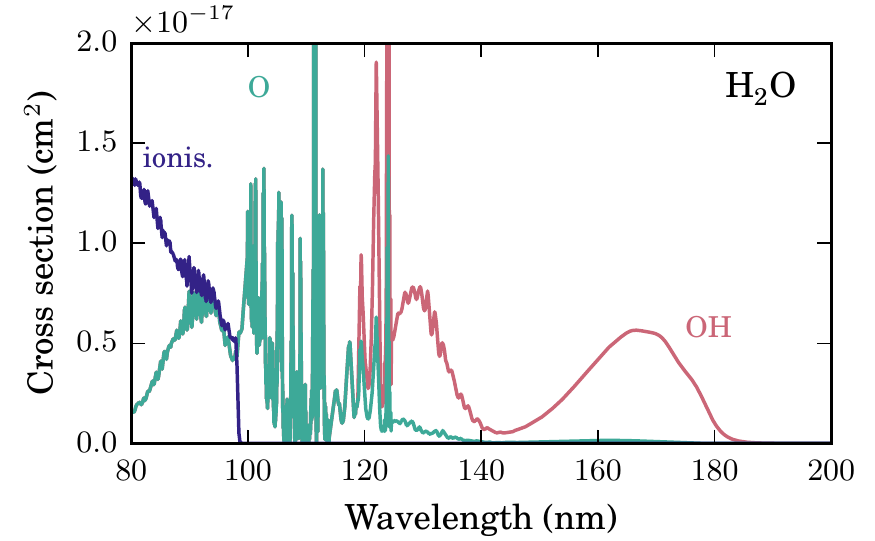}
  \label{fig:partial_cross_sections_H2O}
  \caption{Photoabsorption cross section of \ce{H2O}. Divided into channels producing O (green), OH (red), and ionisation (blue).}
\end{figure}
  The large abundance of water observed in interstellar space \cite{van_dishoeck2013b} means that its photodissociation products, either O or OH, are significant participants in the ongoing chemistry.
  A special case was made of further dividing the water photodissociation cross section into cross sections producing O or OH radicals.
  The product branching ratios are accurately measured at the Lyman-$\alpha$ wavelength \cite{slanger1982,mordaunt1994} and some information is available at other wavelengths \cite{ung1974,stief1975}.
  A coupled-states quantum mechanical wavepacket simulation of \ce{H2O} photodissociation is carried out by \textcite{van_harrevelt2008} on multidimensional potential-energy surfaces describing its ground state and two excited states.
  This calculation agrees with the experimental O and OH branching ratios at 121.6\,nm within 10\% and the calculated wavelength dependence between 118 and 146\,nm was adopted here.
This ratio was linearly extrapolated to zero at the threshold O-atom production, 177\,nm \cite{mordaunt1994}.
For wavelengths shorter than 118\,nm, an equal production of OH and H radicals was assumed. 
The resulting partial cross sections are plotted in Fig.~\ref{fig:partial_cross_sections_H2O}.
\change{\textcite{van_harrevelt2008} also calculated significant branching between the pair of products and excitations states \ce{O({}^3P) + 2H} and \ce{O({}^3D) + H2}, as well as \ce{OH(X{}^2\Pi) + H} and \ce{OH(X{}^2\Sigma) + H}.}

\subsubsection{\ce{H2O+} -- water ion}
\label{sec:H2O+}
This molecule is remarkably transparent at ultraviolet
wavelengths: calculations show that there are no dipole-allowed
dissociative electronic states below 13.6 eV so that the interstellar
H$_2$O$^+$ photodissociation rate is negligible \cite{van_dishoeck2006}.

\subsubsection{\ce{O2} -- oxygen}
\label{sec:O2}

\begin{table}
  \caption{References and wavelength ranges of concatenated \ce{O2} cross sections.}
  \label{tab:O2 data sources}
  \centering
  \begin{tabular}{cl}
    \toprule
    Wavelength (nm) & Reference \\
    \midrule
    5 -- 49  &   \parbox{0.6\linewidth}{\textcite{matsunaga1967}, MPI Mainz UV/Vis database} \\ 
    49 -- 108  & \textcite{holland1993} \\ %
    108 -- 115 & \textcite{ogawa1975}       \\
    115 -- 179 & \textcite{lu2010}             \\
    179 -- 203 & \textcite{yoshino1992}        \\
    205 -- 240 & \textcite{yoshino1988}        \\
    \bottomrule
  \end{tabular}
\end{table}

There are many laboratory studies of \ce{O2} photoabsorption. The
cross section adopted here is compiled from several sources listed in
Table~\ref{tab:O2 data sources}.  The photoionisation efficiency of
\textcite{holland1993} was used to partition the photoabsorption cross
section between dissociative and ionising branches shortwards of the
ionisation threshold at 103\,nm.

\subsubsection{\ce{O2+} -- oxygen ion}
\label{sec:O2+}
This molecule has no dissociative electronic states
below 13.6 eV \cite{honjou1978}. The only dipole-allowed states below
13.6 eV and above the O$_2^+$ dissociation limit are the A$^2\Pi_u$
and 2$^2\Pi_u$ states. Their oscillator strengths are only $f$=0.005
each and, since they are bound, $\eta^{\rm d}=0$ is assumed.

\subsubsection{\ce{H2O2} -- hydrogen peroxide}
\label{sec:H2O2}
\begin{table}
  \caption{References and wavelength ranges of concatenated \ce{H2O2} cross sections.}
  \label{tab:H2O2 data sources}
  \begin{minipage}[c]{1.0\linewidth}
    \centering
    \begin{tabular}{cl}
      \toprule
      Wavelength (nm) & Reference            \\
      \midrule
      ~65 -- 105      &\textcite{litorja1998}\myfootnotemark{1}\\
      105 -- 195      & \textcite{suto1983b} \\
      195 -- 350      &\textcite{lin1978}    \\
      \bottomrule 
    \end{tabular}
    \begin{minipage}{\linewidth}
      \footnotetext[1]{Scaled to agree with the absolutely-calibrated photoionisation cross section of \textcite{dodson2015} where they overlap.}
    \end{minipage}
  \end{minipage}
\end{table}
The combined experimental photoabsorption cross sections of \textcite{lin1978} and \textcite{suto1983b} extend from 105 to 195\,nm, with the absorption at the longest wavelength being very small (less than \np[cm^2]{e-21} and likely to be temperature dependent.
A photoionisation cross section was assembled by combining absolute and relative measurements of \ce{H2O2+} production \cite{dodson2015,litorja1998} and assuming a low production of fragmented ions.

\subsubsection{\ce{O3} -- ozone }
\label{sec:O3}
\begin{table}
  \caption{References and wavelength ranges of concatenated \ce{O3} cross sections.}
  \label{tab:O3 data sources}
  \begin{minipage}[c]{1.0\linewidth}
    \centering
    \begin{tabular}{cl}
      \toprule
      Wavelength (nm) & Reference                  \\
      \midrule
      ~53 -- ~~61     &\textcite{ogawa1958}        \\
      ~61 -- ~~93     &\textcite{berkowitz2008}    \\
      ~93 -- ~110     &\textcite{ogawa1958}        \\
      110 -- ~173     &\textcite{mason1996}        \\
      173 -- ~212     &\textcite{ackerman1971}     \\
      212 -- 1100     &\textcite{serdyuchenko2014b}\\
      \bottomrule 
    \end{tabular}
  \end{minipage}
\end{table}
The \ce{O3} photoabsorption cross section is compiled from a selection of many measurements and includes all wavelengths between 53 and 1100\,nm, with the cross section longwards of 325\,nm never exceeding \np[cm^{-2}]{1e-20}.
The references and ranges of concatenated cross sections are give in Table~\ref{tab:O3 data sources}.
The binding energy of the \ce{O3} ground state corresponds to a wavelength of 1180\,nm and all excited states decay dissociatively, even those absorbing weakly in the near-infrared \cite{grebenshchikov2007}.
The photoionisation threshold is 99\,nm and a total photoionisation cross sections is generated by \textcite{berkowitz2008} from reanalysis of the photoionisation yields of \ce{O3+}, \ce{O2+}, and \ce{O+} ions measured by \textcite{mocellin2001}.

\subsubsection{\ce{CO} -- carbon monoxide}
\label{sec:CO}

The photodissociation of the strongly bound CO molecule starts
shortward of 110 nm and occurs exclusively by predissociation. A
detailed model of the completely line-dominated CO photodissociation
cross section is constructed by \textcite{visser2009} based on
wavelengths, oscillator strengths, and predissociation probabilities
for individual lines and from a variety of experimental sources
\cite[e.g.,][]{eidelsberg1990,ubachs_etal1994,eidelsberg2006,cacciani2001}.
Synthetic spectra were then constructed by summing the profile of all
lines and including the effects of predissociation broadening, Doppler
broadening, and alternative excitation temperatures.  Visser et
al. considered all of the important isotopologues in their analysis
and proceeded to detailed calculations of photodissociation rates and
shielding-functions, \change{estimating the uncertainty of their cross sections to be 20\%.}

For the line-dominated part of the CO spectrum between 91 and 110\,nm,
a simulated photoabsorption and dissociation cross sections
was constructed from the line parameters of Visser et al.  This spectrum is shown in
Fig.~\ref{fig:cross sections CO} assuming an excitation temperature of
100\,K and a Doppler broadening of 1\,km\,s$^{-1}$.  These values are
also used in the various calculations involving CO later in the paper.
The shorter wavelength part of the spectrum, <91\,nm, is taken from
the electron-energy loss measurement of \textcite{chan1993b} that
does not resolve the CO band structure between 80 and 91.2\,nm.  The
non-dissociative part of the spectrum longer than 108\,nm is also
taken from the measurement of \textcite{chan1993b} and does not
include the full rotational structure of these bands. 

\subsubsection{\ce{CO+} -- carbon monoxide ion}
\label{sec:CO+}
Like CO, CO$^+$ has a deep potential well with a
dissociation energy of 8.34 eV. The higher excited D$^2\Pi$, G, E and
F $^2\Sigma^+$ states are likely to be {(pre-)dissociated}, however,
with oscillator strengths to the D and G states computed to be
$f$=0.01 and 0.02, respectively \cite{lavendy1993}. Similar values were assumed for the higher states, with $\eta^\text{d}$=1.

\subsubsection{\ce{CO2} -- carbon dioxide}
\label{sec:CO2}

\ce{CO2} is an important atmospheric molecule and thus well studied.
A compilation representing its photodissociation cross section is made by \textcite{huestis2010}, and was used here after being updated between 87
and 109\,nm with a more recent higher-resolution photoabsorption cross
section \cite{archer2013}. The photoionisation cross section
of \cite{shaw1995} was adopted.

\subsubsection{\ce{HCO+} -- isoformyl}
\label{sec:HCO+}

The photodissociation of this ion is studied in
detail theoretically \cite{koch1995a,koch1995b}.  The only dipole-allowed
dissociative state is the 1$^1\Pi$ state around 11.5 eV with a small
cross section.

\subsubsection{\ce{H2CO} -- formaldehyde}
\label{sec:H2CO}
\begin{table}
  \caption{References and wavelength ranges of concatenated photoabsorption \ce{H2CO} cross sections.}
  \label{tab:H2CO data sources}
  \begin{minipage}[c]{1.0\linewidth}
    \centering
    \begin{tabular}{cl}
      \toprule
      Wavelength (nm) & Reference                             \\
      \midrule
      ~6 -- 60        &\textcite{cooper1996}                  \\
      60 -- 176       &\textcite{mentall1971}\myfootnotemark{1}\\
      220 -- 375       &\textcite{meller2000}\\
      \bottomrule 
    \end{tabular}
    \begin{minipage}{\linewidth}
      \footnotetext[1]{Published absorption coefficient scaled by \np{3.9e-20} to agree with the absolutely-calibrated cross section of \textcite{suto1986} where they overlap.}
    \end{minipage}
  \end{minipage}
\end{table}

The electron-energy loss derived cross section of \textcite{cooper1996} was combined with the VUV photoabsorption spectra of \textcite{mentall1971} and \textcite{meller2000}, as listed in Table~\ref{tab:H2CO data sources}.
The weak UV absorption between 240 and 360\,nm is just shortwards of the effective threshold resulting from an internal barrier for dissociation to \ce{H2} and CO \cite{hopkins2007}.

The highly-structured \ce{H2CO} spectrum between 115 and 160\,nm consists of predissociated Rydberg series, including some showing rotational structure \cite{brint1985}, and presents the risk of cross sections deduced from under-resolved photoabsorption spectra being underestimated.
However, the integrated cross section of an absolutely-calibrated medium-resolution photoabsorption cross section \cite{suto1986} agrees with the resolution-independent electron-energy loss measurement \cite{cooper1996} within 4\%, negating this possibility.
The 225 to 360\,nm photodissociation cross section is also under resolved, but high-resolution data \cite[e.g.,][]{pope2005,ernest2012a} will not affect \ce{H2CO}'s astrophysical photodestruction properties.

The photoionisation cross section is measured near threshold by \textcite{dodson2015} and at shorter wavelengths the photoabsorption cross section was scaled by the ionisation efficiency of \textcite{cooper1996}.
This ionisation efficiency approaches 100\% at longer wavelengths than that measured by \textcite{mentall1971}, which suffers from an estimated 30\% uncertainty. 

\subsubsection{\ce{NH} -- imidogen}
\label{sec:NH}

Experimental data on photoabsorption cross sections for NH exist only
for the lower nondissociative excited states \cite{krishnamurty1969} and
via two-photon transitions \cite{de_beer1991b,clement1992}.  The direct
dissociation of NH into its first two excited repulsive curves is
calculated by \textcite{kirby1991} from ab initio
potential-energy curves and electronic transition moments, and we
adopt their theoretical cross sections.  
An additional broad feature at 100\,nm with integrated cross section
\np[cm^2\,nm]{e-16} was added to this cross section to roughly account for photoabsorption and dissociation into electronic states of higher energy than those calculated.

Predissociation of the $v=0$, 1, and 2 bound levels of the low-lying $A\,{}^3\Pi$ state contributes negligibly to the NH photodissociation rate in an ISRF \cite{kirby1991}, but are at sufficiently long wavelengths, greater than 250\,nm, that they may influence this rate in cool radiation fields.
These three bound levels were included int our cross section database without simulating their rotational structure by adopting the wavelengths of \cite{huber_herzberg1979}, band oscillator strengths (the recommended values calculated by \textcite{kirby1991}), and predissociation efficiencies.
The latter are assumed to be 0, 0, and 0.5 for the $v=0$, 1, and 2 levels, respectively, in line with their ratios of radiative and dissociative lifetimes deduced by \textcite{patel-misra1991}, and assuming rotational excitation is limited to below $J=10$, appropriate for astrophysical environments.
The combined uncertainty in the excitation and dissociation of the $A\,{}^3\Pi$ $v=2$ level is an order of magnitude.
The addition of this level increases the photodissociation rate of NH in a 4000\,K black body radiation field by 20\%.
This contribution is less than the uncertainties associated with shorter-wavelength stronger-absorbing states.

A theoretical photoionisation cross section was taken from \textcite{wang1990} that has significant magnitude, \np[cm^2]{7.4e-18}, at the Lyman-$\alpha$ wavelength.

\subsubsection{\ce{NH2} -- amidogen}
\label{sec:NH2}

There are two ab initio calculations of the dissociative
excited states, transition moments, and photodissociation cross
sections of this radical.  Here, a photodissociation cross
section file was composed, from data for the very weak
$1\,^2A''\rightarrow 2\,^2A'$ transition between 150 and 210\,nm \cite{saxon1983} added to the shorter-wavelength cross sections
due to stronger transitions to various excited states calculated by
\textcite{koch1997} based on potential energy surfaces calculated by
\textcite{vetter1996}.  An additional 2\,nm FWHM Gaussian line of
integrated cross section \np[cm^2\,nm]{e-16} was added to the
photodissociation cross section at 110\,nm to account for excitation
into states not considered in the two calculations.

A photoionisation yield of \ce{NH2} is recorded by
\textcite{gibson1985} from the ionisation threshold at 111\,nm.  We
arbitrarily adopt a photoionisation cross section of \np[cm^2]{2e-17}
at 80\,nm, in line with other molecules in the database, in order to
place this onto an absolute scale.

\subsubsection{\ce{NH3} -- ammonia}
\label{sec:NH3}

\begin{table}
  \caption{References and wavelength ranges of concatenated \ce{NH3} cross sections.}
  \label{tab:NH3 data sources}
  \begin{minipage}[c]{1.0\linewidth}
    \centering
    \begin{tabular}{cl}
      \toprule
      Wavelength (nm) & Reference \\
      \midrule
      ~~8 -- 106    &\textcite{samson1987b}\\
      106 -- 140    & \textcite{wu2007}\myfootnotemark{1}    \\
      140 -- 226    &\textcite{liang2007c} \\
      \bottomrule    
    \end{tabular}
    \begin{minipage}{\linewidth}
      \footnotetext[1]{To reduce scatter due to measurement noise this cross section was down-sampled between 106 and 115\,nm into 0.16\,nm intervals.}
    \end{minipage}
  \end{minipage}
\end{table}

The results of two high-resolution optical experiments
\cite{wu2007,liang2007c} were supplemented by an electron energy-loss
measurement \cite{samson1987b} over the wavelength ranges listed in
Table~\ref{tab:NH3 data sources} to determine a best photoabsorption
cross section for \ce{NH3}.  Since
the cross section is broad and continuous, 100\% dissociation
efficiency is assumed below the ionisation threshold.

Two measurements of the \ce{NH3} ionisation efficiency were used to
divide the cross section between photoionisation and photoabsorption
above the ionisation limit, \textcite{samson1987b} between 8 and
105\,nm and \textcite{xia1991} between 106 and 124\,nm.
\textcite{samson1987b} also determined the branching ratios of various
dissociative photoionisation products with \ce{NH3+} being the only
ion product for wavelengths longer than 79\,nm.

\begin{figure}
  \centering
  \includegraphics{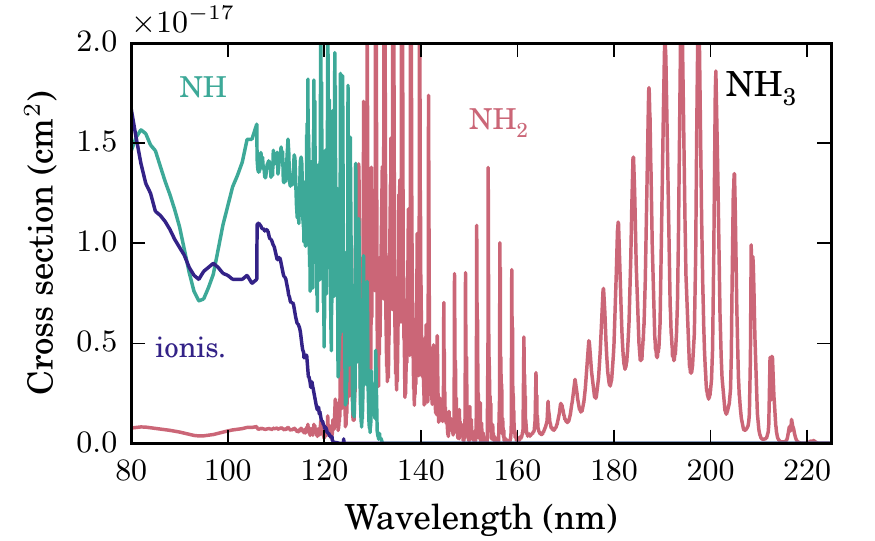}
  \label{fig:partial_cross_sections_NH3}
  \caption{Photoabsorption cross section of \ce{NH3}. Divided into channels producing NH (green), \ce{NH2} (red), and ionisation (blue).}
\end{figure}

\ce{NH3} is an abundant species in many astrophysical environments \cite[e.g.,][]{ho1983,choi2010,shinnaka2011}, and the product branching of its dissociation may have a significant affect on the total abundance of astrochemical molecules \ce{NH2} and NH.
As for \ce{H2O}, a special case was made of determining the wavelength dependence of this branching.  

A measurement of \ce{NH3} photolysis at 121.6\,nm was made by \textcite{slanger1982}, and previous experiments reviewed therein, finding a 1.89 quantum yield of H-atom production.
This requires the formation of \ce{NH + H + H} fragments with a branching ratio of 95\% and the remaining 5\% of photoabsorption events was assumed to produce \ce{NH2 + H}.
No further constraining information on these dissociation channels exists at shorter wavelengths and so the same 95 and 5\% branching was adopted.
At longer wavelengths, the observed threshold for NH fluorescence following \ce{NH3} dissociation is 132\,nm \cite{leach2005}, and 100\% \ce{NH2} production was assumed for all longer wavelengths.
The branching ratios between 121.6 and 132\,nm were linearly interpolate between experimental constraints.
The resulting partial cross sections are plotted in Fig.~\ref{fig:partial_cross_sections_NH3}.

\subsubsection{\ce{N2} -- nitrogen}
\label{sec:N2}
Detailed studies of the astrophysical photodissociation of \ce{N2} and its isotopic consequences can be found in \textcite{li2013} and \textcite{heays2014a}, and are based on a large body of work, both experimental \cite[e.g.,][]{carroll_collins1969,sprengers_etal2003,stark_etal2008,lewis_etal2008a,heays2014b} and theoretical \cite[e.g.,][]{dressler1969,stahel_etal1983,lewis_etal2005b,heays2011_thesis}.

The photoabsorption and dissociation cross sections of
\textcite{heays2014a} between 85 and 100\,nm were used.  These are the product
of a coupled-channels model \cite[e.g.,][]{gibson_lewis1996} defined
by excited-state potential-energy curves that are
experimentally-optimised \cite{heays2011b}.  These cross sections
reproduce the full ro-vibrationally resolved spectrum of the molecule,
including its temperature dependence, \change{and have an uncertainty of 10\% over the wavelength relevant to the ISRF}.  The simulation used here
assumes a thermal excitation of 100\,K and a Doppler broadening of
1\,km\,s$^{-1}$.

\ce{N2} does not absorb significantly at longer wavelengths than
100\,nm.  Some highly-excited Rydberg states appearing at wavelengths
shorter than 85\,nm are missing in the coupled-channels formulation.
The low-resolution electron-energy-loss-deduced cross sections
of \textcite{chan_etal1993} and \textcite{shaw_etal1992} were used over the
wavelength ranges $79.5<\lambda<85$ and $\lambda<79.5$ nm,
respectively.  The cross section of \textcite{chan_etal1993} was
scaled down by a factor of 0.7 in order to maintain continuity with
the adjoining measurement.

\subsubsection{NO -- nitric oxide}
\label{sec:NO}

The NO photoabsorption spectrum is well measured by
low-resolution electron-energy-loss spectroscopy
\cite{iida1986,chan1993c} and at higher resolution by photoabsorption
spectroscopy utilising He and \ce{H2} discharge sources to generate
ultraviolet continua \cite{watanabe1967}.

A further series of Fourier-transform spectroscopy measurements
\cite[][and references therein]{yoshino2006} principally employing
synchrotron radiation resolved many absorption lines between 166 and
196\,nm, and reduced them to a list of wavelengths and oscillator
strengths.  These data were used to simulate this part of the
spectrum with a line-by-line model capable of reproducing a range of
excitation temperatures.  As illustrated by various experiments, not
all lines in this range dissociate completely
\cite[e.g.,][]{brzozowski1976}.

\begin{table}
  \caption{Assumed dissociation fraction of NO excited-state vibrational levels.\myfootnotemark{1}}
  \label{tab:NO dissociation fractions}
  \small
  \centering
  \begin{tabular}{ccl}
    \toprule
    Level&Fraction&Reference\\
    \midrule
    $A(3)$\myfootnotemark{2}& 0   & \textcite{brzozowski1976} \\ %
    $B(6)$                     & 0   & \textcite{brzozowski1976} \\
    $B(7)$\myfootnotemark{3}     & 0.05 & \textcite{brzozowski1976}\\
    $B(9)$                     & 0.98& \textcite{hikida1978} \\
    $B(10)$\myfootnotemark{4}    & 1   & \\
    $B(11)$\myfootnotemark{4}    & 1   & \\
    $B(12)$\myfootnotemark{4}    & 1   & \\
    $B(14)$\myfootnotemark{4}    & 1   & \\
    $C(0)$ & $\begin{cases} 0 & J<4.5 \\ 0.9 & J\geq 4.5 \\ \end{cases}$ & \textcite{brzozowski1976}\\
    $C(1)$\myfootnotemark{5} & 1 & \\
    $C(2)$\myfootnotemark{5} & 1 & \\
    $C(3)$\myfootnotemark{5} & 1 & \\
    $D(0)$ & $\left(1 + \frac{6000}{J(J+1)}\right)^{-1}$ &\textcite{luque2000}  \\
    $D(1)$ & $\left(1 + \frac{460}{J(J+1)}\right)^{-1}$ & \textcite{luque2000} \\
    $D(2)$ & $\left(1 + \frac{290}{J(J+1)}\right)^{-1}$ & \textcite{luque2000} \\
    $D(3)$ & $\left(1 + \frac{120}{J(J+1)}\right)^{-1}$ & \textcite{luque2000} \\
    \bottomrule
  \end{tabular}
  \begin{minipage}{\linewidth}
    \footnotetext[1]{Level notation as in \textcite{yoshino2006}. Rotational angular-momentum quantum number: $J$.}
    \footnotetext[2]{Dissociates for rotational levels with $J>26$, \cite{brzozowski1976,luque2000} but this requires higher temperatures than are relevant here.}
    \footnotetext[3]{This fraction is actually an upper limit \cite{brzozowski1974,brzozowski1976}.}
    \footnotetext[4]{No measurements exist but it is reasonable to expect these levels to dissociate completely because of the large dissociation fraction of the lower-energy level $B(9)$.}
    \footnotetext[5]{Do not appear in emission \cite{brzozowski1976}.}
  \end{minipage}
\end{table}

The predissociation fraction of the various excited levels was
estimated from several studies
\cite{brzozowski1974,brzozowski1976,hikida1978,hart1987,luque2000} and
the assumed values are listed in Table \ref{tab:NO dissociation fractions} for
excited levels appearing in our line-by-line simulation.
Some of the lowest-excited rotational transitions were not observed in
the room temperature spectrum of \textcite{yoshino2006} but are
important in the interstellar medium.
The various line parameters for these transitions were extrapolated from higher levels.

The fraction of NO photoabsorption below the 134\,nm threshold
resulting in ionisation was estimated from the photoion spectroscopy
of \textcite{watanabe1967}.  The cross section plotted in
Fig.~\ref{fig:cross sections NO} and used later in various calculations
is simulated assuming a 100\,K ground state excitation and
\np[km\,s^{-1}]{1} Doppler width.

\subsubsection{CN -- cyanide radical}
\label{sec:CN}

\begin{figure}
  \centering
  \includegraphics{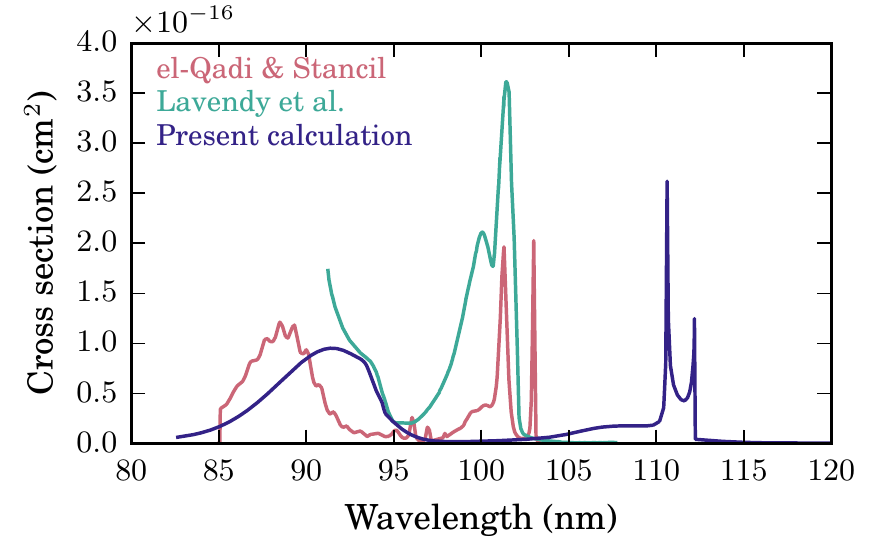}
  \caption{Photodissociation cross section for CN. Presently calculated cross section and those of \textcite{lavendy1987} and \textcite{el-qadi2013}.}
  \label{fig:compare CN cross sections}
\end{figure}

The photodissociating states of CN have not been experimentally observed.
Ab initio configuration-interaction calculations \cite{lavendy1984,lavendy1987} have identified high-lying valence states (the fourth and fifth members of ${}^2\Pi$ and ${}^2\Sigma^+$ symmetry) to be responsible for its ultraviolet photodissociation cross section, and calculated its onset to occur near 105\,nm.
A further cross section is calculated by \textcite{el-qadi2013} adopting the potential-energy curves and transition moments published by \textcite{lavendy1987} but finds an approximately factor-of-three discrepancy compared with the earlier work, as shown in Fig.~\ref{fig:compare CN cross sections}.
To resolve this, we recomputed the potential-energy curves of ${}^2\Sigma^+$ and ${}^2\Pi$ states and their transition moments with respect to photoexcitation from the ground state. 
This analysis included all states dissociating to form excited N(${}^4$S) atoms and ground state C(${}^3$P), or to lower energy limits, and some states leading to more highly-excited atoms.

Full details of the ab initio calculation using the MOLPRO software will be given in a later publication but adopts similar methods as other recent calculations of CN excited states \cite{kulik2009,shi2011} and as in \textcite{van_hemert2008}.
These calculated potential-energy curves are in good agreement with the ground state and first excited ${}^2\Pi$ and ${}^2\Sigma^+$ states of previous work \cite{lavendy1987,kulik2009}, but find the highly-excited photodissociating states to be more bound than in the previous calculations, by as much as 1\,eV.
This lowering arises from a larger basis set provided to the  ab initio potential-energy curve calculation.

A photodissociation cross section was generated from the calculated curves and transition moments by reflecting the ground state vibrational wavefunction through each excited-state potential and onto the energy axis \cite{condon1928}.
Scaling this reflection with the $R$-dependent transition moments and converting the potential-energy scale to wavelength generates the cross section in Fig.~\ref{fig:compare CN cross sections}.
The integrated calculated cross section is one quarter that determined by \textcite{lavendy1987} but in approximate agreement with the reanalysis of \textcite{el-qadi2013}, however, the increased binding of the $5\,{}^2\Pi$ state in our calculation leads to a 10\,nm longwards shift of its peak cross section, to 110\,nm.
The photodissociation rate assuming ISRF radiation is half of the value calculated assuming the cross section of \textcite{lavendy1987} (see Sect.~\ref{sec:photo rates}).
The new rate would a factor of two smaller still without the 10\,nm shift described.

There are no strong ultraviolet-absorbing excited states of CN that dissociate to form C(${}^3$P)+N(${}^4$S) ground state atoms, or the excited pair C(${}^1$D)+N(${}^4$S) (with 140\,nm dissociation threshold).
We then assumed all bound ${}^2\Pi$ and ${}^2\Sigma^+$ levels dissociate with efficiency $\eta^\text{d}=1$ for energies above the C(${}^3$P)+N(${}^2$D) dissociation limit (with a 120\,nm threshold) and all lower-energy levels have $\eta^\text{d}=0$.
It is also possible that the lower-energy bound levels are actually predissociated through spin-orbit interaction with quartet states dissociating to the C(${}^3$P)+N(${}^4$S) limit, with a corresponding 160\,nm threshold.
Even a moderate dissociation efficiency and low cross section at such long wavelengths could increase the dissociation rate of CN in a cool radiation field by an order of magnitude.

\subsubsection{\ce{HCN} -- hydrogen cyanide}
\label{sec:HCN}

\begin{table}
  \caption{References and wavelength ranges of concatenated HCN cross sections.}
  \label{tab:HCN data sources}
  \small
  \centering
  \begin{tabular}{cl}
    \toprule
    Wavelength (nm) & Reference \\
    \midrule
    ~63 -- 105      & \textcite{nuth1982}\myfootnotemark{1} \\
    105 -- 153      & \textcite{lee1980}\\
    153 -- 188      & \textcite{west1974}\\
    \bottomrule
  \end{tabular}
  \begin{minipage}{\linewidth}
    \footnotetext[1]{Scaled by a factor of \np{2.89e-20} to best agree with the integrated cross section of \textcite{lee1980} where this overlaps.}
  \end{minipage}
\end{table}

HCN absorbs strongly at wavelengths shorter than 150\,nm with a 100\%
yield of H and CN photodissociation products and two experimental cross sections for this process \textcite{lee1980,nuth1982} were combined, as listed in Table~\ref{tab:HCN data sources}.
There are weaker absorption bands longwards of 150\,nm that appear in
high-resolution spectra with broadened lineshapes suggestive of
predissociation-dominated lifetimes
\cite{herzberg1957,hsu1984,jonas1990}.
The absorption cross section of \textcite{west1974} was adopted for wavelengths greater than
157\,nm, and shows good agreement with the cross section of
\textcite{lee1980} where they overlap.  Absorption at these long
wavelengths contributes 50\% of the total HCN photodissociation rate
in a 4000\,K black-body radiation field (see Sect.~\ref{sec:photo
  rates}) but is not significant for hotter ultraviolet radiation
fields.

HCN begins to photoionise to \ce{HCN+} shortward of 92\,nm and is
assumed to have a 100\% ionisation yield by 83\,nm.  The intervening
branching to dissociation and ionisation was calculated by comparing
the wavelength dependence of the absorption cross section and a
measured photoionisation yield \cite{dibeler1968}.  

\subsubsection{\ce{HC3N} -- cyanoacetylene}
\label{sec:HC3N}
\begin{table}

  \caption{References and wavelength ranges of concatenated \ce{HC3N} cross sections.}
  \label{tab:HC3N data sources}
  \begin{minipage}[c]{1.0\linewidth}
    \centering
    \begin{tabular}{cl}
      \toprule
      Wavelength (nm) & Reference \\
      \midrule
      ~80 -- 184  & \textcite{ferradaz2009}\\
      184 -- 230 & \textcite{benilan1994} \\
      230 -- 255 & \textcite{seki1996}    \\
      \bottomrule    
    \end{tabular}
  \end{minipage}
\end{table}

A combination of three \ce{HC3N} experimental photoabsorption cross
sections provide good coverage from its threshold at 255\,nm to
80\,nm, with sources listed in Table~\ref{tab:HC3N data sources}.
Additionally, a total photoionisation cross section is recorded by
\textcite{leach2014} and here the the remaining
photoabsorption cross section was attributed to neutral photodissociation.

\subsubsection{\ce{CH3OH} -- methanol}
\label{sec:CH3OH}
\begin{table}
  \caption{References and wavelength ranges of concatenated \ce{CH3OH} cross sections.}
  \label{tab:CH3OH data sources}
  \begin{minipage}[c]{1.0\linewidth}
    \centering
    \begin{tabular}{cl}
      \toprule
      Wavelength (nm) & Reference \\
      \midrule
      ~60 -- 107  & \textcite{burton1992}\\
      107 -- 165 & \textcite{nee1985} \\
      165 -- 220 & \textcite{cheng2002}    \\
      \bottomrule    
    \end{tabular}
  \end{minipage}
\end{table}

The photoabsorption cross section of methanol is well known and the
sources of data used in our compilation are listed in
Table~\ref{tab:CH3OH data sources}.  The photoabsorption and
photoionisation cross sections measured by \textcite{burton1992} were
used to determine the dissociation versus ionisation branching ratio
shortwards of the 113\,nm ionisation limit.

\subsubsection{\ce{CH3CN} -- acetonitrile}
\label{sec:CH3CN}

The photoabsorption cross section of \textcite{eden2003}
is adopted between 140 and 182\,nm and the absorption coefficient of
\textcite{nuth1982} between 61 and 140\,nm.  The latter was multiplied
by \np{4.05e-20} to match the former over their 25\,nm overlapping
range.
The photoionisation yield measured by \textcite{schwell2008} was used
to determine partial ionisation and neutral dissociation cross
sections.

\subsubsection{\ce{CH3CHO} -- acetaldehyde}
\label{sec:CH3CHO}

The photoabsorption cross section of \ce{CH3CHO} is recorded at
high-resolution between 116 and 350\,nm by
\textcite{limao-vieira2003}.  This measurement shows a weak
long-wavelength absorption peak between 250 and 350\,nm as well as a
stronger and more structured region appearing between 184 nm and the
first ionisation limit, 121.5\,nm.

The cross section at shorter wavelengths is not well
constrained. \textcite{hurzeler1958} record the wavelength-dependent
signal of mass 44 (\ce{CH3CHO+}) and dehydrogenated mass 43
photoionisation fragments between 109 and 125\,nm.  To place this
measurement on an absolute scale the threshold \ce{CH3CHO+}
ionisation cross section calculated by \textcite{vega2010} was assumed.  We
extrapolate the photoion signal of \textcite{hurzeler1958} to shorter
wavelengths by assuming the same fall-off for both mass 44 and 43
fragments as calculated by \textcite{vega2010}.  Photoionisation 
data between 110 and 122\,nm are missing from the laboratory measurement and
a linear extrapolation to the peak of the ionisation signal at 110\,nm
is used to approximate this.  All absorption not accounted for by the
deduced photoionisation cross section is assumed to result in
dissociation.

The cross section put together in this way, and plotted in
Fig.~\ref{fig:cross sections CH3CHO}, is uncertain around the
ionisation threshold though physically plausible, and is superior to
completely neglecting wavelengths shorter than those studied by
\textcite{limao-vieira2003}.

\subsubsection{\ce{CH3NH2} -- methylamine}
\label{sec:CH3NH2}

Methylamine photoabsorbs and dissociates at wavelengths shorter than
250\,nm.  A cross section was constructed from the photoabsorption
measurements of \textcite{hubin-franskin2002} (138 to 249\,nm),
combined with their simultaneously-recorded
electron-energy-loss-derived cross section (83 to 138\,nm) and another
lower-resolution electron-energy loss measurement \cite{burton1994}
shorter than 83\,nm. Some information on branching ratios to various
products is available at a few wavelengths
\cite{michael1963,gardner1982}.

A relative photoionisation yield is measured by \textcite{hu2002}
between 121 and 135\,nm.  This was absolutely calibrated with respect
to the photoabsorption cross section by assuming a 100\% ionisation
yield at 121\,nm.  This may well be an overestimate of the yield, in
that case the photoionisation rates calculated here will be
overestimated and the non-ionising rates too low.  \textcite{hu2002}
also measured the branching of photoionisation products into
fragments.

\subsubsection{\ce{NH2CHO} -- formamide}
\label{sec:NH2CHO}

The photoabsorption cross section of \ce{NH2CHO} is measured for
wavelengths greater than 89\,nm by \textcite{gingell1997}, who
combined a direct measurement with electron energy-loss spectroscopy.
This cross section begins around 215\,nm and peaks around 168\,nm with
some resonant structure.

The ionisation threshold of \ce{NH2CHO} occurs just shortwards of the
121.3\,nm Lyman-$\alpha$ transition.  
The photoion spectra of \textcite{leach2010} was converted into
dissociation and ionisation cross sections by assuming 100\,\%
ionisation shorter than 89\,nm and scaling to match the
photoabsorption cross section of \cite{gingell1997} at this
wavelength.

\subsubsection{\ce{C2H5OH} -- ethanol}
\label{sec:C2H5OH}

A \ce{C2H5OH} photoabsorption cross section measured by
electron-energy loss spectroscopy \cite{feng2002} agrees well with
several less-complete direct photoabsorption measurements available
from the MPI Mainz UV/VIS database, and is adopted here.  All absorption events are assumed to lead to
dissociation.

An absolute ionisation cross section determined by photoion
spectroscopy \cite{cool2005} was adopted between 105\,nm and the
photoionisation threshold at 120\,nm.  For shorter wavelengths, 55 to
90.5\,nm, a cross section was calculated from the measured
photoionisation efficiency of \textcite{hatano1999b}.  The
photoionisation cross section was linearly interpolated between these
two wavelength ranges.

\subsubsection{\ce{C3H7OH} -- 1-propanol}
\label{sec:C3H7OH}

Two measurements spanning the wavelength ranges 120 to 208\,nm
\cite{salahub1971} and 30 to 120\,nm \cite{koizumi1986} were combined
into a single photoabsorption cross section.  An absolute ionisation
cross section between 125 and 107\,nm is measured by ion
mass-spectroscopy \cite{cool2005}, and a
photoionisation cross section constructed by supplementing this with a measured
photoionisation efficiency between 76 and 92\,nm \cite{hatano1999b}.
The ionisation cross section for intervening wavelengths was linearly
interpolated, and linearly extrapolated to shorter wavelengths
assuming a photoionisation fraction of 1 shortwards of 60\,nm, in line
with similar molecules \cite{hatano1999b}.

\subsubsection{\ce{OCS} -- carbonyl sulphide}
  \label{sec:OCS}
\change{
  The photoabsorption of OCS is studied extensively in view of its importance to atmospheric chemistry \cite[e.g.,][and references therein]{limao-vieira2003}. Here, the recent high-resolution measurement of its cross section by \textcite{limao-vieira2003} was adopted, and supplemented with photoabsorption and ionisation cross sections deduced form electron-energy loss spectroscopy for wavelengths shorter than 115\,nm \cite{feng2000b,feng2000c}.
}

\subsubsection{\ce{CH3SH} -- methanethiol}
  \label{sec:CH3SH}
\change{
Two experimental cross sections \cite{vagjhiani1993,tokue1987} were combined to describe \ce{CH3SH} photoabsorption from its long-wavelength threshold at 330\,nm \cite{wilson1994} to the ionisation threshold at 131\,nm \cite{morgan1995}.
For shorter wavelengths, a photodissociation cross section linearly extrapolated to zero at 100\,nm (30\,nm below the ionisation threshold) was adopted, and a photoionisation cross section using the measured efficiency for \ce{CH3SH+} production of \textcite{kutina1982} after scaling its maximum value to \np[cm^2]{5e-18}.
The various assumptions used in the shorter wavelength region were selected in a broad analogy to the photodissociation and ionisation properties of \ce{CH3OH}.
}

\subsubsection{\ce{CS} -- carbon monosulphide}
  \label{sec:CS}
\change{
  There are no absolute measurements of the CS cross section for photoabsorption into pre- or directly-dissociative excited states.
  Here, the cross section simulated and discussed by \textcite{van_dishoeck1988} was retained. 
This uses the measured wavelengths of transitions to the $B\,{}^1\Sigma^+$ states \cite{stark1987} and vertical excitation energies of higher-lying states calculated by \textcite{bruna1975}.
The strengths of these electronic transitions are estimated.
}

\subsubsection{\ce{CS2} -- carbon disulphide}
  \label{sec:CS2}
\begin{table}
  \change{
  \caption{References and wavelength ranges of concatenated \ce{CS2} cross sections.}
  \label{tab:CS2 data sources}
  \begin{minipage}[c]{1.0\linewidth}
    \centering
    \begin{tabular}{cl}
      \toprule
      Wavelength (nm) & Reference \\
      \midrule
      18 -- 78   & \textcite{wu1983}\\
      78 -- 97  & \textcite{cook1969b}\myfootnotemark{1}\\
      97 -- 105  & \textcite{carnovale1981}\myfootnotemark{2}\\
      105 -- 121   & \textcite{day1982}\\
      121 -- 193   & \textcite{sunanda2015}\\
      193 -- 205  & \textcite{xu1993}\\
      205 -- 370  & \textcite{grosch2015}\\
      \bottomrule    
    \end{tabular}
    \footnotetext[1]{Scaled by a factor of 1.27 to agree with the integrated cross section of \textcite{wu1983} for their overlapping wavelength region.}
    \footnotetext[2]{Scaled by a factor of 0.87 to agree with the integrated cross section of \textcite{wu1983} for their overlapping wavelength region.}
  \end{minipage}
  }
\end{table}

\change{
Multiple experiments were combined into a single \ce{CS2} photoabsorption cross section spanning 20 to 370\,nm, with details given in Tab.~\ref{tab:CS2 data sources}. 
The ionisation and ground-state dissociation thresholds are at 123 and 277\,nm, respectively \cite{fischer1993,okabe1972}.
A photoionisation cross section was deduced by scaling the photoionisation efficiency measured by \textcite{dibeler1967} to match the photoabsorption cross section of \textcite{wu1983} for wavelengths sufficiently shorter than the ionisation threshold that non-ionising decay is negligible.
}

\subsubsection{\ce{SO2} -- sulphur dioxide }
\label{sec:SO2}
\change{
  The photoabsorption cross section of \ce{SO2} was measured numerous times and a consolidation of these into a single spectrum between 106 and 403\,nm was generated by \textcite{manatt1993}.
This was modified by inserting a higher-resolution measured cross section  between 172 and 289\,nm, including the important photodissociating absorption bands between 170\,nm and the photodissociation threshold at 218.7\,nm \cite{becker1995}. 
A measured fluorescence yield \cite{katagiri1997} was used to estimate the dissociation fraction near this threshold. Even-higher resolution measurements of these bands are available \cite[e.g.,][]{blackie2011,endo2015} but were deemed unnecessary for our purposes.

A combination of photoionisation efficiency and short-wavelength photoabsorption cross sections measured by electron-impact and photoion spectroscopy \cite{feng1999c,holland1995} were used to determine the absorption and ionisation cross sections for wavelengths shorter than compiled by \textcite{manatt1993}.
}

\subsubsection{\ce{SH+} -- mercapto ion}
  \label{sec:SH+}
\change{
The direct photodissociation cross section of \ce{SH+} is calculated ab initio by \textcite{mcmillan2016} including transitions to several excited states.
Their calculation from the $v=0$ and $J=0$ ground-state level was adopted here and added to this additional absorption into the longer-wavelength bound-levels of the $A\,{}^3\Pi$ state, which is known to predissociate for $v\geq 1$ \cite{gustafsson1988,brites2008}.
Oscillator strength of transitions into the vibrational levels of $A\,{}^3\Pi$ were calculated here using the potential-energy curves and transition dipole moments of \textcite{mcmillan2016} and the methods of Sect.~\ref{sec:CN}, but did not include details of their rotational structure.
}

\subsubsection{\ce{SO} -- sulphur monoxide}
  \label{sec:SO}
\change{
  The predissociated $B\,{}^3\Sigma^- - X\,^3\Sigma^-$ absorption bands of SO, appearing shortward of 235\,nm, are measured by \textcite{phillips1981}.
  The short wavelength absorption between 116 and 135\,nm measured by \textcite{nee1986b} was added to this and a relative photoionisation yield shortwards of 121\,nm \cite{norwood1989} that we arbitrarily scaled to give a peak value of \np[cm^2]{5e-17}.
}

  \subsubsection{\ce{S2} -- disulphur}
  \label{sec:S2}
\change{
  The photodissociation and ionisation cross section of the \ce{S2} radical is not well known, but is important for understanding S-bearing molecular abundances evident in comets \cite{de_almeida1986}.
  Photodissociating transitions occur into the $B\,{}^3\Sigma^-_u$ state, which is predissociated for levels $v\geq 10$ \cite{kato1995}.
  Transition wavelengths for $B(v')\leftarrow X(0)$ photoabsorption bands were calculated using the molecular constants of \textcite{wheeler1998} and adopted oscillator strengths for $v'=0$ to 20 bands from the calculations of \textcite{pradhan1991}.
  These oscillator strengths to $v'=26$ were extrapolated using Franck-Condon factors calculated for the $B\leftarrow X$ transition by \textcite{smith1971}.
According to these factors, vibrational levels with $v'>26$ will contribute less than 6\% to the total absorption into the $B$ state, and the continuum absorption shorter than the 224\,nm direct dissociation limit will be weak.
The photoabsorption and photodissociation cross sections simulated here neglect rotational structure of the individual vibrational bands \cite[e.g.,][]{wheeler1998}.

  The photoionisation efficiency of \ce{S2} is measured \cite{liao1986} and a typical absolute magnitude for this was arbitrarily assumed here.
  Additional bound and unbound states exist between the $B\,{}^3\Sigma^-_u$ and ionisation continuum \cite[e.g.,][]{donovan1970,xing2013} but no quantitative information on their photoabsorption cross sections is available.  
  It is likely that the longer-wavelength absorption of $B\,{}^3\Sigma^-_u$ will dominate the photodissociation rate for \ce{S2} in most interstellar radiation fields.

Foreshadowing Sect.~\ref{sec:photo rates}, we calculate the photodissociation rate of \ce{S2} in the solar radiation field at 1\,AU distance from the sun to be \np[s^{-1}]{0.0079}, that is with a lifetime of 130\,s, which is a significant reduction from the 250\,s lifetime calculated by \textcite{de_almeida1986}.
This difference is likely due to the larger number of predissociating $B\,{}^3\Sigma^-_u$ vibrational levels considered in the present work, leading to a larger total photodissociation rate.
}

\change{
  \subsubsection{\ce{SH} -- mercapto radical}
  \label{sec:SH}
Several ab initio calculations of the SH ground and excited states have been made \cite{bruna1987,resende2001,lee_sun2001}, and there is one absolute absorption measurement, providing the oscillator strength of the predissociative $X\,^2\Pi(v''=0)$ to $A\,^2\Sigma^+(v'=0)$ transition.
This measurement was combined with calculated Franck-Condon factors for transition to higher-$v''$ $A\,^2\Sigma^+$ levels \cite{resende2001} and experimental spectroscopic constants \cite{johns1961} to estimate the cross section of $A-X$ transitions (neglecting their rotational structure).
Furthermore,  a ``reflection'' cross section to simulate continuum absorption into the lowest-energy repulsive ${}^2\Sigma^-$ state was added, using the potential-energy curves and electronic transition moment of \textcite{lee_sun2001}, and the methods described in Sect.~\ref{sec:CN}.
Higher lying bound and repulsive states were included as vertical transitions according to their calculated equilibrium energies and transition moments \cite{bruna1987}.
Finally, the unknown photoionisation cross section of SH was arbitrarily simulated by including an absorption feature of integrated cross section \np[cm^2]{5e-17} shortwards of the 119\,nm photoionisation threshold \cite{hsu1994}.
}

\change{
  \subsubsection{\ce{H2S} -- hydrogen sulphide}
  \label{sec:H2S}
  There are several measurements of the photoabsorption of \ce{H2S}, and data from \textcite{lee1987} (240 to 118.8\,nm), \textcite{xia1991} (118.8 to 106\,nm), and \textcite{feng1999} (106 to 41\,nm), was compiled here into a single cross section.
  It is likely that some of the absorption features between 120 and 160\,nm are not fully resolved by these measurements.
It was assumed that all absorption leads to dissociation or ionisation, with the branching between these two determined from the photoionisation cross sections of \textcite{xia1991,feng1999b}.
}

\subsubsection{SiO -- silicon monoxide}
\label{sec:SiO}

This molecule has a similar electronic structure to that of CO, but
with a lower dissociation energy, 8.26 eV vs.\ 11 eV.  Thus, even the
lower $^1\Sigma^+$ Rydberg states can contribute to SiO
photodissociation whereas they are bound for CO (e.g., its B and C
$v$=0 levels).  Another difference between the two molecules are quite distinct ionisation potentials: 11.4\,eV for SiO and 14\,eV for CO,
putting the latter above the Lyman-limit.  The oscillator strengths to
the 3 $^1\Sigma^+$ and 2, 3, 4 and 5 $^1\Pi$ states are taken to be
$f$=0.10, 0.32, 0.03, 0.11 and 0.10, respectively, with $\eta^{\rm
  d}=1$ as deduced by ab initio calculations \cite{van_dishoeck2006}.

\subsubsection{\ce{HCl} -- hydrogen chloride}
\label{sec:HCl}

The HCl photoabsorption cross section was adopted from the combination
of a direct measurement \cite{bahou2001} and electron-energy loss
spectroscopy \cite{brion2005} over the 5 to 135, and 135 to 230\,nm
wavelength regions, respectively.  A photoionisation cross section is
recorded by \textcite{frohlich1990}.  The HCl spectrum spans a region
of continuous absorption into the repulsive $A{}^1\Pi$ state between
135 and 250\,nm, many bound excited Rydberg and valence levels between
100 and 135\,nm, and an ionisation continuum shortwards of 97.2\,nm.

The most recent ab initio calculation of the HCl cross section by \textcite{engin2012} identifies 13 photoabsorbing electronically excited states of ${}^1\Pi$ and ${}^1\Sigma^+$ symmetry below 12.2\,eV, and is excellent in agreement with the previous computation of \textcite{van_dishoeck1982b} for those states absorbing longer than 120\,nm.
Comparing the integrated cross section of \textcite{engin2012} and the experimental data between 102 and 200\,nm leads to agreement within 4\%. 
The excited states responsible for the significant HCl absorption between 91.6 and 102\,nm are not well known. 
Many bands are observed and assigned to Rydberg-type excited states in this region \cite{ginter1981,green1991a} but without quantification of their absorption cross sections.
An independent measurement of the HCl photoabsorption spectrum below 110\,nm is warranted.

All photoabsorbing state were here assumed to be completely dissociative, ($\eta^{\rm d}=1$), despite their sometimes unbroadened linewidths.
This is based on the weakness of the fluorescence cross section from these states \cite{nee1986} and the strong coupling of Rydberg levels and unbound valence states in theoretical studies \cite[e.g.,][]{van_dishoeck1982b,alexander1998}.

Shorter than the 96\,nm ionisation limit, a small signal of \ce{H+}
or \ce{Cl+} dissociative-photoionisation products is measured
\cite{daviel1984}, as well as a small cross section for decay into
neutral products \cite{frohlich1990}.

\subsubsection{\ce{HCl+} -- hydrogen chloride ion}
\label{sec:HCl+}
Ab initio calculation of potential-energy curves for the
\ce{HCl+} ground and lowest-excited states, and the transition
moments connecting them, permitted a calculation of the molecule's
wavelength-dependent photodissociation cross section
\cite{pradhan1991}.  The dominant dissociation pathway determined by
this calculation produces H and \ce{Cl+}.

\subsubsection{AlH -- alumane}
\label{sec:AlH}
Potential-energy curves for the $A\,{}^1\Sigma^+$ and $C\,{}^1\Sigma^+$ states of AlH are calculated by \textcite{matos1987} as well as electric-dipole transition moments with respect to the $X\,{}^1\Sigma^+$ ground state.
Vertical transitions calculated from this data at the equilibrium geometry give absorption at 430 and 230\,nm with oscillator strengths of 0.0022 and 0.090, respectively.
This however represents an upper limit for photodissociation because unbroadened emission lines from the $v=0$ and 1 levels of both the $A$ and $C$ states is observed \cite{szajna2010,szajna2011}.
The predissociation of higher energy vibrational structure for both states may be possible however if they are able to tunnel through maxima predicted for both potential-energy curves \cite{matos1987,bauschlicher1988}.
Given this uncertainty, two absorption features were assumed here to represent the AlH spectrum with oscillator strengths of 0.05, one at 200\,nm to represent all photodissociating states, and another at 130\,nm to for photoionisation, the photoionisation threshold occurs at approximately 160\,nm \cite{matos1987}.
The ISRF rate is a factor of 10 smaller than previously given in the Leiden database because of the lack of dissociation via the $v=0$ and 1 levels.

\begin{centering}
    \onecolumn
    \footnotesize
    \renewcommand*{\thefootnote}{\alph{footnote}}
    \setlength{\LTcapwidth}{\linewidth}
    \begin{longtable}{ccccccccccc}
\caption{Photodissociation rates of molecules\protect\footnotemark[1]\label{tab:photodissociation rates} and parameterised dust shielding.\protect\footnotemark[2]}\\
\hline\hline

& \multicolumn{7}{c}{} & \multicolumn{3}{c}{ISRF dust shielding} \\
Species
& ISRF
& Mathis\,'83
& 4000\,K
& 10\,000\,K
& Lyman-$\alpha$
& Solar
& TW-Hydra
& $\gamma_\text{exp,ISM}$
& $\gamma_{\text{E}_2,\text{ISM}}$
& $\gamma_{\text{E}_2,\text{growth}}$
\\ 

\hline
\endfirsthead
\caption{continued.}\\
\hline\hline

& \multicolumn{7}{c}{} & \multicolumn{3}{c}{ISRF dust shielding} \\
Species
& ISRF
& Mathis\,'83
& 4000\,K
& 10\,000\,K
& Lyman-$\alpha$
& Solar
& TW-Hydra
& $\gamma_\text{exp,ISM}$
& $\gamma_{\text{E}_2,\text{ISM}}$
& $\gamma_{\text{E}_2,\text{growth}}$
\\ 

\hline
\endhead
\hline
\endfoot
\ce{H2} & 5.7E--11 & 4.6E--11 & 2.0E--16 & 3.9E--12 & -- & 3.1E--12 & 1.7E--11 & 4.18 & 3.11 & 0.45 \\
\ce{H2+} & 5.7E--10 & 3.9E--10 & 4.3E--11 & 2.0E--10 & 1.1E--09 & 8.7E--11 & 7.6E--10 & 2.78 & 1.94 & 0.44 \\
\ce{H3+} & -- & -- & -- & -- & -- & 3.9E--13 & 3.7E--12 & -- & -- & -- \\
\ce{CH} & 9.1E--10 & 6.4E--10 & 1.8E--07 & 2.2E--09 & 7.9E--12 & 2.1E--07 & 1.1E--09 & 2.12 & 1.36 & 0.32 \\
\ce{CH+} & 3.3E--10 & 2.6E--10 & 6.6E--10 & 3.8E--11 & 7.9E--12 & 9.0E--10 & 8.1E--11 & 3.54 & 2.63 & 0.45 \\
\ce{CH2} & 5.8E--10 & 3.9E--10 & 3.0E--09 & 1.3E--09 & 7.9E--12 & 3.5E--09 & 3.7E--10 & 2.35 & 1.61 & 0.31 \\
\ce{CH2+} & 1.4E--10 & 1.2E--10 & 3.7E--11 & 8.3E--11 & 7.9E--12 & 3.4E--11 & 4.7E--11 & 2.73 & 1.86 & 0.40 \\
\ce{CH3} & 6.2E--10 & 4.0E--10 & 2.6E--09 & 7.5E--10 & -- & 4.3E--09 & 2.3E--10 & 2.50 & 1.73 & 0.38 \\
\ce{CH4} & 1.5E--09 & 1.1E--09 & 1.7E--12 & 2.7E--10 & 2.8E--09 & 1.3E--10 & 2.0E--09 & 3.08 & 2.19 & 0.45 \\
\ce{CH4+} & 2.8E--10 & 1.9E--10 & 1.7E--13 & 4.6E--11 & 7.9E--12 & 3.7E--12 & 5.2E--11 & 3.11 & 2.22 & 0.45 \\
\ce{C2} & 2.4E--10 & 1.7E--10 & 4.6E--13 & 4.5E--11 & 7.8E--11 & 7.3E--12 & 9.2E--11 & 3.04 & 2.15 & 0.45 \\
\ce{C2H} & 1.6E--09 & 1.0E--09 & 4.4E--11 & 6.8E--10 & 5.1E--10 & 6.5E--11 & 8.6E--10 & 2.67 & 1.85 & 0.45 \\
\ce{C2H2} & 2.4E--09 & 1.6E--09 & 1.3E--10 & 1.1E--09 & 8.6E--09 & 4.2E--10 & 3.2E--09 & 2.64 & 1.83 & 0.45 \\
\ce{C2H4} & 3.1E--09 & 2.0E--09 & 7.5E--10 & 2.4E--09 & 3.7E--09 & 7.9E--10 & 3.5E--09 & 2.49 & 1.70 & 0.40 \\
\ce{C2H6} & 2.1E--09 & 1.5E--09 & 5.9E--12 & 4.8E--10 & 3.6E--09 & 1.6E--10 & 2.5E--09 & 2.94 & 2.07 & 0.45 \\
\ce{C3} & 5.0E--09 & 3.2E--09 & 6.5E--10 & 4.3E--09 & 2.2E--09 & 5.8E--10 & 4.2E--09 & 2.39 & 1.64 & 0.43 \\
\ce{\textit{l}-C3H} & 1.8E--09 & 1.2E--09 & 4.7E--07 & 4.4E--09 & 7.9E--12 & 4.2E--07 & 2.1E--09 & 2.08 & 1.32 & 0.31 \\
\ce{\textit{c}-C3H} & 1.1E--09 & 6.5E--10 & 1.1E--07 & 3.8E--09 & 7.9E--12 & 1.3E--07 & 1.1E--09 & 2.15 & 1.45 & 0.27 \\
\ce{HC3H} & 2.2E--09 & 1.5E--09 & 1.8E--07 & 6.5E--09 & 7.9E--12 & 2.9E--07 & 2.2E--09 & 2.15 & 1.43 & 0.29 \\
\ce{\textit{l}-C3H2} & 4.1E--09 & 2.7E--09 & 6.7E--09 & 3.7E--09 & 2.3E--09 & 7.2E--09 & 3.2E--09 & 2.51 & 1.74 & 0.40 \\
\ce{\textit{c}-C3H2} & 1.4E--09 & 8.8E--10 & 5.5E--08 & 3.3E--09 & 7.9E--12 & 3.0E--08 & 8.5E--10 & 2.26 & 1.54 & 0.31 \\
\ce{\textit{l}-C4} & 8.5E--09 & 5.7E--09 & 5.4E--08 & 2.2E--08 & 7.9E--12 & 8.4E--08 & 5.5E--09 & 2.22 & 1.52 & 0.29 \\
\ce{\textit{l}-C4H} & 3.7E--09 & 2.3E--09 & 7.4E--10 & 3.3E--09 & 7.9E--12 & 6.8E--10 & 1.2E--09 & 2.36 & 1.62 & 0.43 \\
\ce{\textit{l}-C5H} & 1.3E--09 & 9.4E--10 & 6.9E--07 & 8.0E--09 & 7.9E--12 & 9.6E--07 & 2.3E--09 & 1.76 & 1.14 & 0.22 \\
\ce{OH} & 3.8E--10 & 2.5E--10 & 1.9E--10 & 2.0E--10 & 6.4E--10 & 1.7E--10 & 5.1E--10 & 2.66 & 1.83 & 0.43 \\
\ce{OH+}& 1.3E--11   & 1.1E--11   & 4.4E--13   & 9.6E--13   & --         & 6.8E--13   & 2.9E--12& 3.97& 2.96& 0.45 \\
\ce{H2O} & 7.7E--10 & 5.3E--10 & 1.6E--10 & 4.6E--10 & 2.4E--09 & 2.3E--10 & 1.6E--09 & 2.63 & 1.80 & 0.41 \\
\ce{O2} & 7.7E--10 & 5.0E--10 & 7.5E--11 & 5.6E--10 & 3.2E--11 & 6.4E--11 & 3.7E--10 & 2.45 & 1.69 & 0.43 \\
\ce{O2+} & 3.5E--11 & 2.3E--11 & 1.4E--11 & 3.9E--11 & 4.0E--10 & 2.4E--11 & 2.4E--10 & 2.38 & 1.62 & 0.37 \\
\ce{HO2} & 6.7E--10 & 4.4E--10 & 1.8E--08 & 2.0E--09 & 7.9E--12 & 2.2E--08 & 4.8E--10 & 2.46 & 1.69 & 0.28 \\
\ce{H2O2} & 8.1E--10 & 5.3E--10 & 2.4E--09 & 5.5E--10 & 1.5E--09 & 2.9E--09 & 1.1E--09 & 2.61 & 1.80 & 0.41 \\
\ce{O3} & 1.8E--09 & 1.1E--09 & 2.4E--07 & 5.8E--09 & 4.7E--09 & 2.2E--07 & 4.1E--09 & 2.25 & 1.49 & 0.28 \\
\ce{CO} & 2.4E--10 & 2.1E--10 & 2.8E--15 & 1.8E--11 & -- & 8.0E--12 & 5.1E--11 & 3.88 & 2.88 & 0.45 \\
\ce{CO+} & 1.0E--10 & 7.1E--11 & 1.7E--13 & 2.4E--11 & 7.9E--12 & 1.4E--12 & 2.3E--11 & 2.89 & 2.04 & 0.45 \\
\ce{CO2}& 9.2E--10   & 6.8E--10   & 2.4E--12   & 1.1E--10   & 1.0E--11   & 1.7E--11   & 1.8E--10& 3.40& 2.48& 0.45 \\
\ce{HCO} & 1.1E--09 & 4.9E--10 & 5.0E--06 & 2.7E--09 & 1.2E--11 & 1.7E--08 & 7.4E--09 & 2.43 & 1.67 & 0.31 \\
\ce{HCO+} & 5.4E--12 & 3.7E--12 & 1.1E--16 & 4.9E--13 & -- & 2.8E--14 & 7.2E--13 & 3.67 & 2.68 & 0.45 \\
\ce{H2CO} & 1.4E--09 & 9.6E--10 & 4.0E--09 & 9.6E--10 & 1.5E--09 & 5.1E--09 & 1.4E--09 & 2.54 & 1.74 & 0.42 \\
\ce{NH}& 5.7E--10   & 3.8E--10   & 1.1E--11   & 2.4E--10   & 4.9E--11   & 1.7E--11   & 1.7E--10& 2.63& 1.83& 0.45 \\
\ce{NH+} & 5.3E--11 & 3.6E--11 & 1.2E--08 & 2.0E--10 & 7.4E--12 & 8.4E--09 & 5.9E--11 & 2.07 & 1.34 & 0.26 \\
\ce{NH2}& 9.5E--10   & 6.3E--10   & 5.5E--10   & 1.3E--09   & 4.8E--12   & 4.3E--10   & 4.4E--10& 2.31& 1.57& 0.35 \\
\ce{NH3} & 1.4E--09 & 9.9E--10 & 3.6E--09 & 1.7E--09 & 1.3E--09 & 4.1E--09 & 1.4E--09 & 2.61 & 1.80 & 0.36 \\
\ce{N2} & 1.7E--10 & 1.5E--10 & 3.2E--16 & 1.1E--11 & -- & 1.1E--11 & 5.2E--11 & 4.25 & 3.16 & 0.45 \\
\ce{NO} & 3.8E--10 & 2.7E--10 & 2.1E--10 & 3.1E--10 & 7.5E--11 & 2.1E--10 & 2.2E--10 & 2.56 & 1.75 & 0.40 \\
\ce{NO2} & 1.4E--09 & 9.2E--10 & 4.8E--10 & 1.1E--09 & 7.9E--12 & 4.1E--10 & 1.3E--09 & 2.50 & 1.71 & 0.40 \\
\ce{N2O} & 1.9E--09 & 1.3E--09 & 2.9E--11 & 5.1E--10 & 6.1E--10 & 7.0E--11 & 7.3E--10 & 2.81 & 1.98 & 0.45 \\
\ce{CN} & 5.2E--10 & 4.3E--10 & 2.3E--14 & 5.1E--11 & 2.1E--11 & 1.6E--11 & 1.3E--10 & 3.50 & 2.55 & 0.45 \\
\ce{HCN} & 1.6E--09 & 1.2E--09 & 5.7E--12 & 2.8E--10 & 5.3E--09 & 2.1E--10 & 3.3E--09 & 3.12 & 2.23 & 0.45 \\
\ce{HC3N} & 7.1E--09 & 4.7E--09 & 6.2E--10 & 3.5E--09 & 3.1E--09 & 9.6E--10 & 3.8E--09 & 2.59 & 1.79 & 0.45 \\
\ce{CH3OH} & 1.4E--09 & 9.5E--10 & 1.2E--10 & 5.3E--10 & 2.2E--09 & 2.1E--10 & 1.7E--09 & 2.76 & 1.92 & 0.44 \\
\ce{CH3CN} & 3.0E--09 & 2.1E--09 & 9.1E--12 & 5.4E--10 & 1.7E--09 & 1.1E--10 & 1.6E--09 & 3.07 & 2.18 & 0.45 \\
\ce{CH3SH} & 2.8E--09 & 1.8E--09 & 7.4E--09 & 2.8E--09 & 3.4E--09 & 8.5E--09 & 3.1E--09 & 2.50 & 1.72 & 0.39 \\
\ce{CH3CHO} & 2.0E--09 & 1.3E--09 & 3.8E--09 & 1.6E--09 & 3.2E--09 & 4.8E--09 & 2.7E--09 & 2.46 & 1.69 & 0.41 \\
\ce{CH3NH2} & 7.3E--10 & 4.8E--10 & 4.2E--09 & 1.3E--09 & 1.0E--15 & 6.3E--09 & 4.3E--10 & 2.37 & 1.62 & 0.34 \\
\ce{NH2CHO} & 2.7E--09 & 1.8E--09 & 2.3E--09 & 3.1E--09 & 2.9E--09 & 2.7E--09 & 3.0E--09 & 2.40 & 1.64 & 0.37 \\
\ce{C2H5OH} & 2.5E--09 & 1.7E--09 & 2.1E--10 & 9.2E--10 & 4.0E--09 & 3.7E--10 & 3.0E--09 & 2.77 & 1.93 & 0.44 \\
\ce{C3H7OH} & 4.0E--09 & 2.7E--09 & 3.8E--10 & 1.5E--09 & 8.1E--09 & 6.8E--10 & 5.8E--09 & 2.76 & 1.92 & 0.44 \\
\ce{SH} & 1.2E--09 & 8.1E--10 & 3.4E--08 & 2.8E--09 & 1.2E--09 & 4.9E--08 & 1.6E--09 & 2.40 & 1.64 & 0.32 \\
\ce{SH+} & 6.9E--10 & 5.2E--10 & 4.0E--08 & 3.2E--10 & 1.2E--11 & 4.6E--08 & 2.4E--10 & 2.83 & 1.79 & 0.40 \\
\ce{H2S} & 3.1E--09 & 2.1E--09 & 4.2E--09 & 2.2E--09 & 8.4E--09 & 6.1E--09 & 4.4E--09 & 2.64 & 1.83 & 0.41 \\
\ce{CS} & 9.5E--10 & 6.3E--10 & 5.2E--12 & 2.9E--10 & 8.4E--09 & 3.1E--10 & 4.9E--09 & 2.77 & 1.95 & 0.45 \\
\ce{CS2} & 8.8E--09 & 6.2E--09 & 5.0E--08 & 1.9E--08 & 2.7E--09 & 5.1E--08 & 6.7E--09 & 2.50 & 1.72 & 0.32 \\
\ce{OCS} & 4.7E--09 & 3.1E--09 & 1.6E--09 & 3.8E--09 & 1.1E--09 & 1.8E--09 & 3.7E--09 & 2.46 & 1.68 & 0.42 \\
\ce{S2} & 6.6E--10 & 3.7E--10 & 2.0E--07 & 4.3E--09 & -- & 2.1E--07 & 1.0E--09 & 1.90 & 1.28 & 0.21 \\
\ce{SO} & 4.2E--09 & 2.9E--09 & 8.4E--09 & 3.0E--09 & 1.7E--08 & 1.2E--08 & 1.1E--08 & 2.76 & 1.94 & 0.40 \\
\ce{SO2} & 2.4E--09 & 1.7E--09 & 3.6E--09 & 1.6E--09 & 6.1E--09 & 5.1E--09 & 4.3E--09 & 2.78 & 1.94 & 0.40 \\
\ce{SiH} & 2.7E--09 & 1.8E--09 & 1.1E--06 & 1.5E--08 & 7.9E--12 & 1.0E--06 & 5.2E--09 & 1.95 & 1.24 & 0.23 \\
\ce{SiH+}& 2.6E--09   & 2.6E--09   & 5.4E--06   & 1.1E--08   & 3.5E--09   & 4.1E--06   & 1.0E--08& 1.55& 0.94& 0.24 \\
\ce{SiO} & 1.6E--09 & 1.0E--09 & 1.4E--11 & 6.1E--10 & 7.9E--12 & 1.5E--11 & 3.5E--10 & 2.66 & 1.85 & 0.45 \\
\ce{HCl} & 1.7E--09 & 1.2E--09 & 9.4E--11 & 5.1E--10 & 1.5E--10 & 1.1E--10 & 5.6E--10 & 2.88 & 2.02 & 0.44 \\
\ce{HCl+} & 1.1E--10 & 8.5E--11 & 1.0E--12 & 2.4E--11 & 1.3E--10 & 7.9E--12 & 1.0E--10 & 3.01 & 2.12 & 0.45 \\
\ce{NaCl} & 9.5E--10 & 6.0E--10 & 6.0E--08 & 3.6E--09 & 7.9E--12 & 6.1E--08 & 8.4E--10 & 2.20 & 1.50 & 0.26 \\
\ce{PH} & 5.8E--10 & 3.8E--10 & 6.8E--11 & 4.2E--10 & 1.9E--11 & 5.9E--11 & 3.2E--10 & 2.48 & 1.71 & 0.43 \\
\ce{PH+} & 1.4E--10 & 1.1E--10 & 9.6E--08 & 4.1E--10 & 1.7E--10 & 8.0E--08 & 3.1E--10 & 1.93 & 1.16 & 0.28 \\
\ce{AlH} & 2.6E--10 & 1.9E--10 & 4.0E--09 & 1.1E--09 & -- & 5.3E--09 & 2.3E--10 & 2.42 & 1.67 & 0.26 \\
\ce{LiH} & 4.9E--09 & 3.0E--09 & 2.1E--06 & 3.3E--08 & 7.9E--12 & 2.2E--06 & 8.4E--09 & 1.81 & 1.21 & 0.21 \\
\ce{MgH} & 5.1E--10 & 3.3E--10 & 2.4E--08 & 8.8E--10 & 7.9E--12 & 3.2E--08 & 2.7E--10 & 2.30 & 1.54 & 0.36 \\
\ce{NaH} & 7.0E--09 & 4.5E--09 & 3.7E--06 & 4.6E--08 & 7.9E--12 & 4.0E--06 & 1.3E--08 & 1.77 & 1.17 & 0.21\\
\end{longtable}
\end{centering}
\begin{minipage}{\linewidth}
    \footnotetext[1]{In units of \np[s^{-1}]{}.
These rates are for unshielded atoms and molecules
exposed to the full three-dimensional interstellar radiation field,
with various wavelength dependences described
in Sect.~\ref{sec:radiation fields}.}
    \footnotetext[2]{
Dust shielding functions, $\theta$, for an infinite-slab interstellar
cloud are fit to functions of the visual extinction, $A_\text{V}$,
according to two formulae: $\theta(A_V) = \exp\left(-\gamma_{\rm exp}
A_V\right)$ and $\theta(A_V) = E_2\left(\gamma_{\text{E}_2} A_V\right)$
(where $E_2$ is the 2\textsuperscript{nd}-order exponential integral), both
assuming incident radiation at the cloud edge with the wavelength
dependence of our standard ISRF radiation field
(Eqs.~(\ref{eq:dust shielding exponential}) and (\ref{eq:dust
shielding exponential integral}) in Sect.~\ref{sec:shielding by dust}).
Values for $\gamma_{\text{E}_2}$ are
given assuming an interstellar dust size distribution (ISM), and
following the growth of dust grains in a protoplanetary disk (growth),
as described in Sect.~\ref{sec:dust grain properties}.
}
\end{minipage}
\renewcommand*{\thefootnote}{\arabic{footnote}}
\twocolumn\normalsize

\begin{centering}
    \onecolumn
    \footnotesize
    \renewcommand*{\thefootnote}{\alph{footnote}}
    \setlength{\LTcapwidth}{\linewidth}
    \begin{longtable}{ccccccccccc}
\caption{Photoionisation rates of atoms and molecules\protect\footnotemark[1]\label{tab:photoionisation rates} and parameterised dust\protect\footnotemark[2]}\\
\hline\hline

& \multicolumn{7}{c}{} & \multicolumn{3}{c}{ISRF dust shielding} \\
Species
& ISRF
& Mathis\,'83
& 4000\,K
& 10\,000\,K
& Lyman-$\alpha$
& Solar
& TW-Hydra
& $\gamma_\text{exp,ISM}$
& $\gamma_{\text{E}_2,\text{ISM}}$
& $\gamma_{\text{E}_2,\text{growth}}$
\\ 

\hline
\endfirsthead
\caption{continued.}\\
\hline\hline

& \multicolumn{7}{c}{} & \multicolumn{3}{c}{ISRF dust shielding} \\
Species
& ISRF
& Mathis\,'83
& 4000\,K
& 10\,000\,K
& Lyman-$\alpha$
& Solar
& TW-Hydra
& $\gamma_\text{exp,ISM}$
& $\gamma_{\text{E}_2,\text{ISM}}$
& $\gamma_{\text{E}_2,\text{growth}}$
\\ 

\hline
\endhead
\hline
\endfoot
\ce{H} & -- & -- & -- & -- & -- & 1.6E--12 & 5.4E--12 & 3.00 & -- & -- \\
\ce{Li} & 3.4E--10 & 2.3E--10 & 3.1E--09 & 7.1E--10 & 2.1E--10 & 4.9E--09 & 3.3E--10 & 2.45 & 1.68 & 0.32 \\
\ce{C}& 3.5E--10   & 2.6E--10   & 5.7E--15   & 2.8E--11   & --        & 1.0E--11   & 8.3E--11& 3.76& 2.77& 0.45 \\
\ce{N} & -- & -- & -- & -- & -- & 1.6E--12 & 7.8E--12 & -- & -- & -- \\
\ce{O} & -- & -- & -- & -- & -- & 1.4E--12 & 6.2E--12 & -- & -- & -- \\
\ce{Na} & 1.4E--11 & 9.1E--12 & 1.3E--10 & 1.5E--11 & 1.7E--11 & 1.4E--10 & 1.5E--11 & 2.62 & 1.81 & 0.37 \\
\ce{Mg} & 6.6E--11 & 4.3E--11 & 8.1E--12 & 5.3E--11 & 1.1E--11 & 7.3E--12 & 4.2E--11 & 2.43 & 1.67 & 0.43 \\
\ce{Al}& 4.4E--09   & 3.0E--09   & 2.1E--08   & 1.0E--08   & 1.4E--09   & 2.1E--08   & 3.7E--09& 2.36& 1.62& 0.31 \\
\ce{Si} & 4.5E--09 & 2.9E--09 & 9.3E--11 & 2.0E--09 & 5.7E--09 & 2.9E--10 & 4.4E--09 & 2.61 & 1.81 & 0.45 \\
\ce{P} & 1.9E--09 & 1.4E--09 & 1.4E--13 & 2.1E--10 & -- & 3.3E--11 & 3.6E--10 & 3.45 & 2.51 & 0.45 \\
\ce{S} & 1.1E--09 & 8.5E--10 & 8.3E--14 & 1.2E--10 & 1.1E--16 & 2.4E--11 & 2.3E--10 & 3.52 & 2.57 & 0.45 \\
\ce{Cl} & 4.7E--11 & 5.5E--11 & -- & 2.5E--12 & -- & 1.1E--11 & 4.9E--11 & 4.30 & 3.21 & 0.45 \\
\ce{K} & 3.9E--11 & 2.6E--11 & 3.7E--10 & 5.9E--11 & 3.5E--11 & 4.2E--10 & 4.0E--11 & 2.48 & 1.70 & 0.35 \\
\ce{Ca} & 3.5E--10 & 2.3E--10 & 8.0E--10 & 6.2E--10 & 1.4E--10 & 7.9E--10 & 2.9E--10 & 2.34 & 1.60 & 0.34 \\
\ce{Ti} & 2.4E--10 & 1.6E--10 & 9.8E--12 & 7.7E--11 & 4.2E--10 & 2.6E--11 & 2.9E--10 & 2.81 & 1.96 & 0.44 \\
\ce{Cr} & 1.6E--09 & 1.1E--09 & 1.7E--09 & 2.1E--09 & 1.2E--09 & 1.9E--09 & 1.4E--09 & 2.39 & 1.63 & 0.35 \\
\ce{Mn} & 3.3E--11 & 2.2E--11 & 6.8E--12 & 3.4E--11 & 7.0E--12 & 5.8E--12 & 2.6E--11 & 2.35 & 1.61 & 0.41 \\
\ce{Fe} & 4.7E--10 & 3.1E--10 & 1.2E--11 & 2.1E--10 & 7.7E--10 & 3.9E--11 & 5.9E--10 & 2.62 & 1.81 & 0.45 \\
\ce{Co} & 5.3E--11 & 3.4E--11 & 3.3E--12 & 3.7E--11 & 2.8E--11 & 3.5E--12 & 4.5E--11 & 2.47 & 1.70 & 0.45 \\
\ce{Ni} & 9.8E--11 & 6.3E--11 & 9.5E--12 & 7.5E--11 & 4.9E--11 & 9.5E--12 & 7.7E--11 & 2.43 & 1.67 & 0.44 \\
\ce{Zn}& 4.1E--10   & 2.9E--10   & 3.1E--14   & 3.8E--11   & 9.2E--12   & 2.9E--12   & 6.6E--11& 3.25& 2.35& 0.45 \\
\ce{Rb} & 2.7E--11 & 1.8E--11 & 1.6E--09 & 4.8E--11 & 2.3E--11 & 2.2E--09 & 3.0E--11 & 2.33 & 1.54 & 0.34 \\
\ce{Ca+} & 2.4E--12 & 2.0E--12 & -- & 1.7E--13 & -- & 5.2E--14 & 6.0E--13 & 4.09 & 3.04 & 0.45 \\
\ce{H-} & 1.5E--07 & 1.6E--08 & 1.6E--03 & 2.5E--07 & 7.5E--10 & 1.4E--05 & 2.3E--06 & 1.24 & 0.74 & 0.22 \\
\ce{H2} & -- & -- & -- & -- & -- & 4.9E--13 & 3.2E--12 & -- & -- & -- \\
\ce{CH} & 7.6E--10 & 5.6E--10 & 2.3E--14 & 6.8E--11 & -- & 9.6E--12 & 1.4E--10 & 3.67 & 2.70 & 0.45 \\
\ce{CH3} & 3.3E--10 & 2.3E--10 & 8.6E--14 & 4.5E--11 & 8.4E--10 & 3.6E--11 & 5.5E--10 & 3.26 & 2.36 & 0.45 \\
\ce{CH4} & 1.0E--11 & 1.2E--11 & -- & 5.5E--13 & -- & 9.3E--12 & 3.8E--11 & 4.31 & 3.21 & 0.45 \\
\ce{C2} & 4.1E--10 & 3.4E--10 & 1.4E--15 & 2.7E--11 & -- & 6.7E--12 & 8.7E--11 & 4.19 & 3.12 & 0.45 \\
\ce{C2H2} & 5.3E--10 & 4.1E--10 & 5.2E--15 & 4.0E--11 & -- & 2.4E--11 & 1.7E--10 & 3.92 & 2.91 & 0.45 \\
\ce{C2H4} & 4.1E--10 & 3.2E--10 & 1.7E--14 & 3.7E--11 & -- & 6.2E--12 & 8.1E--11 & 3.63 & 2.67 & 0.45 \\
\ce{C2H6} & 2.3E--10 & 2.0E--10 & 7.9E--16 & 1.5E--11 & -- & 2.3E--11 & 1.2E--10 & 4.17 & 3.10 & 0.45 \\
\ce{C3} & 1.4E--10 & 1.1E--10 & 9.0E--16 & 9.7E--12 & -- & 3.5E--12 & 3.3E--11 & 4.03 & 3.00 & 0.45 \\
\ce{H2O} & 2.7E--11 & 2.6E--11 & -- & 1.7E--12 & -- & 4.2E--12 & 2.2E--11 & 4.27 & 3.18 & 0.45 \\
\ce{O2} & 5.1E--11 & 4.5E--11 & 1.3E--16 & 3.4E--12 & -- & 3.4E--12 & 2.2E--11 & 4.22 & 3.14 & 0.45 \\
\ce{H2O2} & 2.5E--10 & 2.0E--10 & 3.2E--15 & 1.9E--11 & -- & 1.2E--11 & 7.7E--11 & 3.88 & 2.88 & 0.45 \\
\ce{O3} & 3.3E--11 & 3.4E--11 & -- & 2.1E--12 & -- & 7.5E--12 & 3.5E--11 & 4.28 & 3.19 & 0.45 \\
\ce{CO}& --        & --        & --        & --        & --        & 6.5E--12    & 2.5E--11 & --& --& -- \\
\ce{CO2}& --        & --        & --        & --        & --        & 5.2E--12   & 2.2E--11& --& --& -- \\
\ce{H2CO} & 4.0E--10 & 3.1E--10 & 1.1E--14 & 3.5E--11 & -- & 1.3E--11 & 1.1E--10 & 3.66 & 2.69 & 0.45 \\
\ce{NH}& 1.9E--12   & 3.0E--12   & --        & --   & --        & 2.8E--12   & 1.0E--11& 4.34& 3.24& 0.45 \\
\ce{NH2}& 1.9E--10   & 1.5E--10   & 1.7E--15   & 1.4E--11   & --        & 8.7E--12   & 5.8E--11& 3.97& 2.94& 0.45 \\
\ce{NH3} & 2.7E--10 & 2.0E--10 & 1.7E--14 & 2.8E--11 & 4.8E--11 & 9.6E--12 & 9.0E--11 & 3.49 & 2.54 & 0.45 \\
\ce{N2} & -- & -- & -- & -- & -- & 1.3E--12 & 8.8E--12 & -- & -- & -- \\
\ce{NO} & 2.6E--10 & 1.9E--10 & 7.4E--14 & 3.1E--11 & 3.3E--10 & 1.9E--11 & 2.6E--10 & 3.38 & 2.46 & 0.45 \\
\ce{NO2} & 1.5E--10 & 1.2E--10 & 4.2E--15 & 1.3E--11 & 1.7E--11 & 3.1E--12 & 4.1E--11 & 3.75 & 2.77 & 0.45 \\
\ce{N2O} & 1.7E--10 & 1.9E--10 & 1.8E--16 & 1.0E--11 & -- & 3.8E--12 & 3.7E--11 & 4.30 & 3.20 & 0.45 \\
\ce{CN} & -- & -- & -- & -- & -- & 2.0E--12 & 5.9E--12 & -- & -- & -- \\
\ce{HCN} & 4.4E--13 & 7.0E--13 & -- & 4.2E--15 & -- & 7.5E--12 & 2.8E--11 & 4.34 & 3.24 & 0.45 \\
\ce{HC3N} & 2.3E--10 & 1.8E--10 & 1.3E--15 & 1.6E--11 & -- & 1.2E--11 & 7.7E--11 & 4.07 & 3.02 & 0.45 \\
\ce{CH3OH} & 3.1E--10 & 2.5E--10 & 5.5E--15 & 2.5E--11 & -- & 1.6E--11 & 1.1E--10 & 3.78 & 2.80 & 0.45 \\
\ce{CH3CN} & 1.2E--10 & 1.1E--10 & 3.8E--16 & 7.8E--12 & -- & 1.6E--11 & 7.7E--11 & 4.19 & 3.12 & 0.45 \\
\ce{CH3SH} & 1.9E--09 & 1.3E--09 & 8.1E--13 & 2.9E--10 & 3.7E--09 & 1.6E--10 & 2.5E--09 & 3.18 & 2.29 & 0.45 \\
\ce{CH3CHO} & 8.3E--10 & 6.0E--10 & 8.3E--14 & 9.4E--11 & 1.9E--10 & 2.4E--11 & 2.8E--10 & 3.42 & 2.49 & 0.45 \\
\ce{CH3NH2} & 1.6E--09 & 1.2E--09 & 6.9E--13 & 2.3E--10 & 3.8E--09 & 1.7E--10 & 2.5E--09 & 3.21 & 2.31 & 0.45 \\
\ce{NH2CHO} & 5.2E--10 & 4.0E--10 & 2.8E--14 & 4.9E--11 & 3.0E--11 & 2.0E--11 & 1.5E--10 & 3.57 & 2.62 & 0.45 \\
\ce{C2H5OH} & 4.9E--10 & 3.9E--10 & 1.4E--14 & 4.1E--11 & -- & 2.8E--11 & 1.8E--10 & 3.72 & 2.75 & 0.45 \\
\ce{C3H7OH} & 7.7E--10 & 6.0E--10 & 2.7E--14 & 6.6E--11 & 4.4E--11 & 4.3E--11 & 2.9E--10 & 3.71 & 2.74 & 0.45 \\
\ce{SH} & 5.0E--11 & 3.7E--11 & 5.3E--16 & 3.9E--12 & -- & 8.9E--13 & 1.1E--11 & 3.92 & 2.90 & 0.45 \\
\ce{H2S} & 7.8E--10 & 5.9E--10 & 4.6E--14 & 7.8E--11 & -- & 2.6E--11 & 2.0E--10 & 3.53 & 2.58 & 0.45 \\
\ce{CS} & 2.6E--11 & 1.8E--11 & 7.3E--16 & 2.4E--12 & -- & 1.9E--13 & 3.6E--12 & 3.59 & 2.63 & 0.45 \\
\ce{CS2} & 3.6E--10 & 2.6E--10 & 2.0E--14 & 3.4E--11 & 1.6E--10 & 2.1E--11 & 2.0E--10 & 3.57 & 2.62 & 0.45 \\
\ce{OCS} & 7.7E--10 & 6.1E--10 & 1.1E--14 & 6.1E--11 & -- & 3.0E--11 & 2.2E--10 & 3.85 & 2.86 & 0.45 \\
\ce{S2} & 1.3E--10 & 9.2E--11 & 6.5E--14 & 1.8E--11 & 2.8E--10 & 1.3E--11 & 2.0E--10 & 3.27 & 2.36 & 0.45 \\
\ce{SO} & 5.3E--10 & 3.7E--10 & 3.2E--14 & 5.7E--11 & -- & 6.1E--12 & 9.8E--11 & 3.46 & 2.52 & 0.45 \\
\ce{SO2} & 1.3E--10 & 1.2E--10 & 2.7E--16 & 8.4E--12 & -- & 1.7E--11 & 8.4E--11 & 4.25 & 3.17 & 0.45 \\
\ce{HCl} & 4.5E--11 & 4.3E--11 & -- & 2.8E--12 & -- & 1.2E--11 & 5.1E--11 & 4.27 & 3.18 & 0.45 \\
\ce{AlH} & 1.5E--10 & 9.9E--11 & 1.2E--12 & 5.6E--11 & 3.4E--10 & 1.4E--11 & 2.3E--10 & 2.67 & 1.86 & 0.45\\
\end{longtable}
\end{centering}
\begin{minipage}{\linewidth}
    \footnotetext[1]{In units of \np[s^{-1}]{}.
These rates are for unshielded atoms and molecules
exposed to the full three-dimensional interstellar radiation field,
with various wavelength dependences described
in Sect.~\ref{sec:radiation fields}.}
    \footnotetext[2]{
Dust shielding functions, $\theta$, for an infinite-slab interstellar
cloud are fit to functions of the visual extinction, $A_\text{V}$,
according to two formulae: $\theta(A_V) = \exp\left(-\gamma_{\rm exp}
A_V\right)$ and $\theta(A_V) = E_2\left(\gamma_{\text{E}_2} A_V\right)$
(where $E_2$ is the 2\textsuperscript{nd}-order exponential integral), both
assuming incident radiation at the cloud edge with the wavelength
dependence of our standard ISRF radiation field
(Eqs.~(\ref{eq:dust shielding exponential}) and (\ref{eq:dust
shielding exponential integral}) in Sect.~\ref{sec:shielding by dust}).
Values for $\gamma_{\text{E}_2}$ are
given assuming an interstellar dust size distribution (ISM), and
following the growth of dust grains in a protoplanetary disk (growth),
as described in Sect.~\ref{sec:dust grain properties}.
}
\end{minipage}
\renewcommand*{\thefootnote}{\arabic{footnote}}
\twocolumn\normalsize

\section{Photodestruction due to stellar and interstellar radiation}
\label{sec:photo rates}

\subsection{Results}
\label{sec:photo results}

Photodestruction rates were calculated according
to Eq. \ref{eq:kpd} using the cross sections discussed in Sect.\ 4 and
assuming several of the radiation fields detailed in
Sect.~\ref{sec:radiation fields}, selected to represent a range of
astrochemical environments.  Tables~\ref{tab:photodissociation rates}
and \ref{tab:photoionisation rates} summarise the calculated rates for
photodissociation and photoionisation, respectively.  
The photodissociation and photoionisation rates in the ISRF have uncertainties listed in Table~\ref{tab:cross_section_properties} and described in Sect.~\ref{sec:cross section uncertainties}. 
These can be incorporated into chemical rate networks
to perform sensitivity analyses of model abundances
\cite{wakelam2012}.
For cool radiation fields, such as a 4000\,K black body or the solar
radiation field, low-lying molecular states just above the
dissociation limit dominate the photodissociation rate and uncertainty. For
example, the interstellar rate of OH is dictated by three direct and
continuous channels in the 190 to 91.2 nm range, whereas the rate assuming 4000\,K black-body radiation proceeds primarily by predissociation through the $A\,{}^2\Sigma^+$ state around 300\,nm
\cite{van_dishoeck1984a,van_dishoeck1984b}.

\change{We also computed photodestruction rates assuming the radiation field of \textcite{mathis1983} (\np[kpc]{10} Galactocentric distance). 
  These are consistently 30 to 40\% smaller than for our standard ISRF due to the differing magnitudes but similar shape of the two radiation fields over the range 100 to 300\,nm.
 For the species (about 15\%) that are primarily photodestroyed at wavelengths between the Lyman-limit and 100\,nm (e.g., dissociation for many diatomics or ionisation of many species) the Mathis 1983 rates are more similar or occasionally greater.
}

\begin{figure*}
  \centering
  \includegraphics{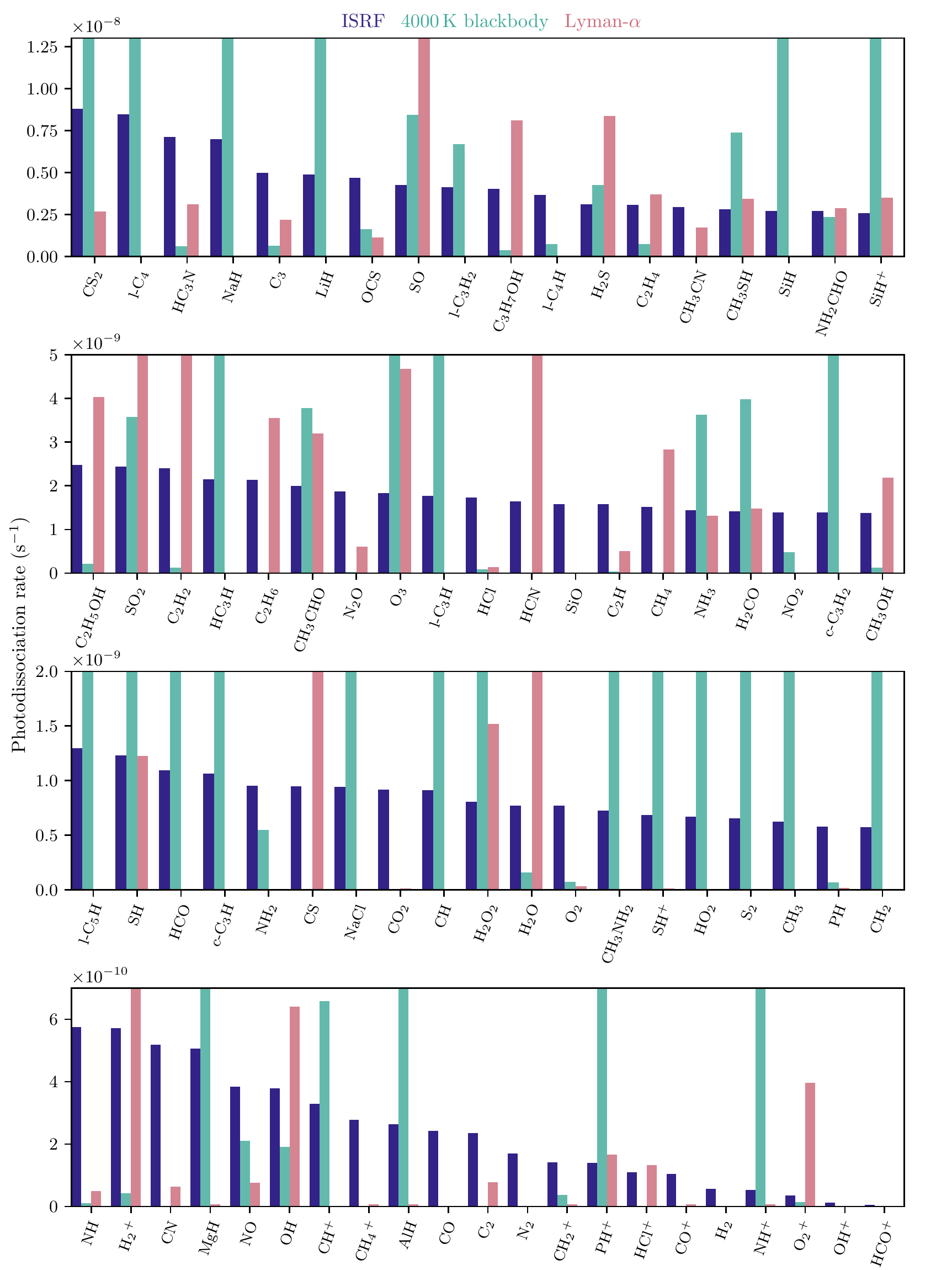}
  \caption{\change{Unshielded photodissociation rates of molecules. Rates are shown assuming three different radiation fields with ultraviolet intensity matching the standard of \textcite{draine1978}.}}
  \label{fig:compare_rates_radiation_fields_photodissociation}
\end{figure*}

\begin{figure*}
  \centering
  \includegraphics{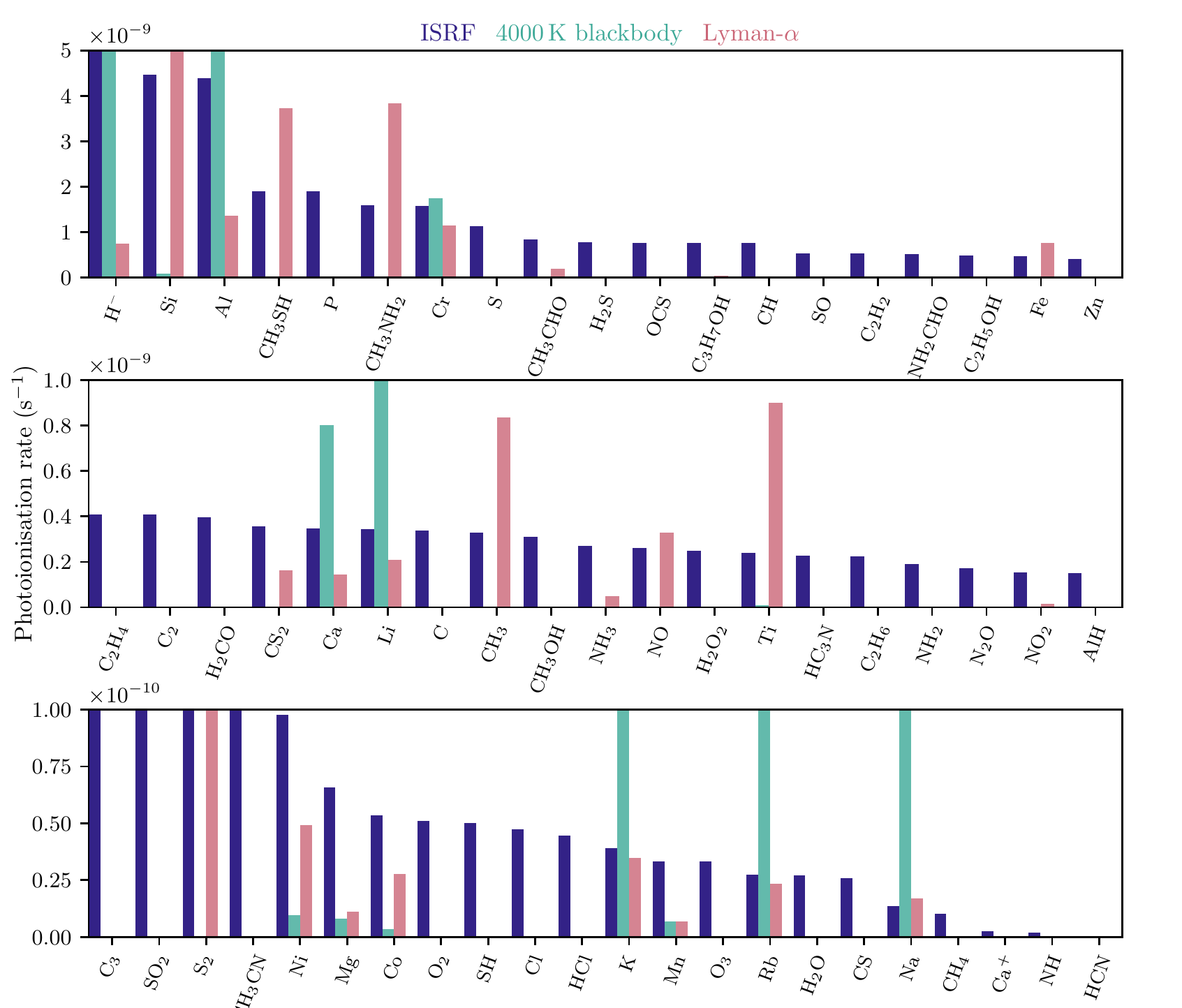}
  \caption{\change{Unshielded photoionisation rates of molecules and atoms. Rates are shown assuming different radiation fields with ultraviolet intensity matching the standard of \textcite{draine1978}.}}
  \label{fig:compare_rates_radiation_fields_photoionisation}
\end{figure*}

The photodissociation and photoionisation rates for all atoms and molecules in our database are compared in Figs.~\ref{fig:compare_rates_radiation_fields_photodissociation} and \ref{fig:compare_rates_radiation_fields_photoionisation} assuming exposure to three different wavelength-dependent radiation fields.
These are plotted in order of decreasing ISRF rate and with varying vertical scales.
The largest ISRF photodissociation rate is \np[s^{-1}]{8.5e-9} and occurs for $\ell$-C$_4$, while the smallest rate is \np[s^{-1}]{5.4e-12} for \ce{HCO+}.
These extremes correspond to lifetimes against photodissociation of 4 to 6000 years, respectively.
The lowest rates occur mostly for small ions, while neutral molecules mostly fall between 0.5 and \np[s^{-1}]{4e-9}.

The photoionisation rates of most molecules exposed to the ISRF are between 2 and 10 times smaller than their respective photodissociation rates, but can be many times smaller still.
This provides some reassurance that the dominant photodestruction pathway is included in our database for those molecules where we evaluated a photodissociation rate but neglected the corresponding photoionisation cross section and rate because of a lack of reliable information.
For two molecules, \ce{C2} and \ce{CH3NH2}, that have particularly large cross section and long wavelength thresholds for  photoionisation the ISRF photoionisation rate is actually larger than for photodissociation, by factors of 1.7 and 2.2, respectively.

The response of the various molecules to alternative radiation fields is highly variable and largely controlled by the qualitative properties of their wavelength-dependent cross sections. 
Comparative rates are plotted in Figs.~\ref{fig:compare_rates_radiation_fields_photodissociation} and \ref{fig:compare_rates_radiation_fields_photoionisation}.
We note that these are calculated assuming radiation fields with equal integrated energy intensity between 91.2 and 200\,nm, while the intensity in real astrochemical environments of interest varies by orders of magnitude.
With this normalisation, molecules with relatively long-wavelength photodissociation thresholds show massively increased rates when irradiated by a 4000\,K black body and sometimes supersede the scales of our figures by orders of magnitude.
\begin{figure}
  \centering
  \includegraphics{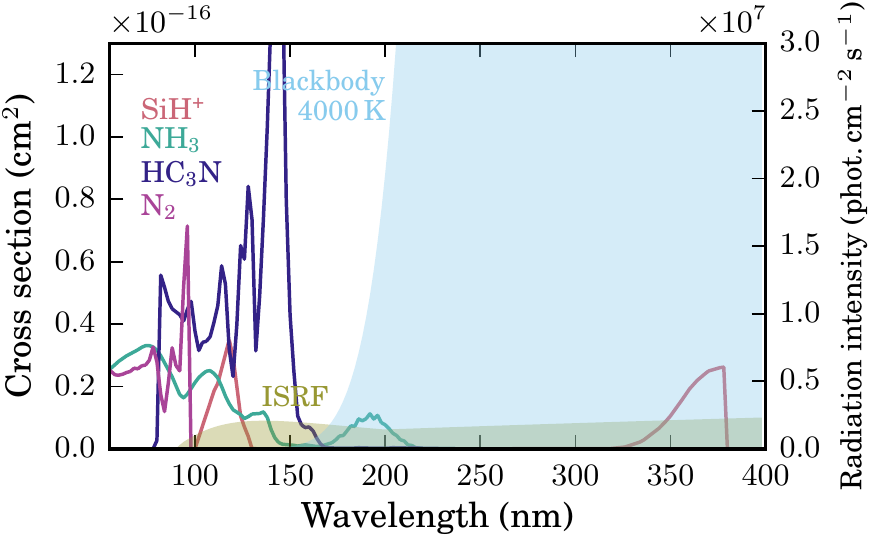}
  \caption{The photodissociation cross sections of four molecules. These are averaged into 2\,nm intervals for clarity and compared with the wavelength-dependence of two radiation fields from Sect.~\ref{sec:radiation fields}.}
  \label{fig:photodissoc_rates_radiation_field_dependence_examples}
\end{figure}
This effect is illustrated in Fig.~\ref{fig:photodissoc_rates_radiation_field_dependence_examples} for four molecules with successively shorter-wavelength dissociation thresholds, in the order \ce{SiH+}, \ce{NH3}, \ce{HC3N}, and \ce{N2}, and with respective ratios between 4000\,K and ISRF photodissociation rates of 1800, 2.5, 0.09, \np{2e-6}.
The rapid increase of black body intensity longwards of 170\,nm is responsible for this variation.
The 4000\,K photodissociation of \ce{CH3} and \ce{NH3} is dominated by absorption longer than 170\,nm into their lowest-lying electronic states.
In constrict, the very-short wavelength threshold of \ce{N2} makes it completely immune to 4000\,K radiation.
\begin{figure}
  \centering
  \includegraphics{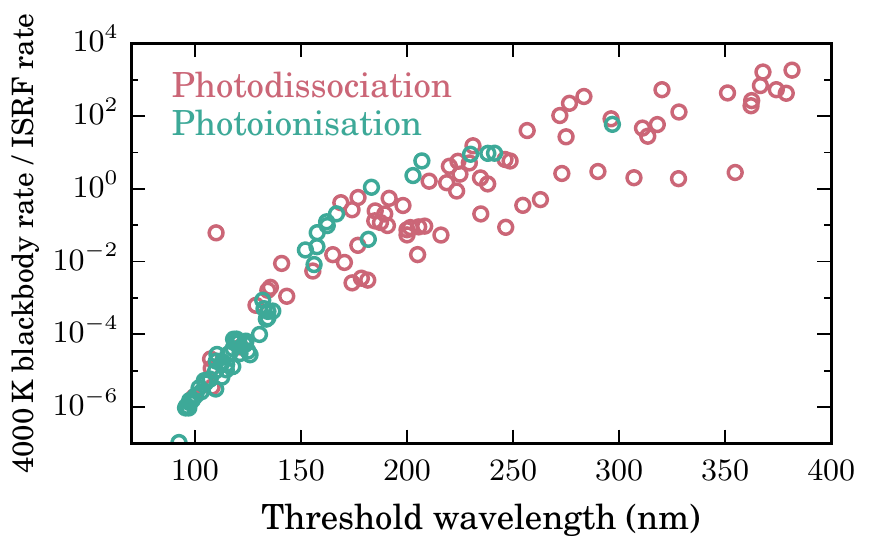}
  \caption{The ratio of photodissociation and photoionisation rates. Calculated assuming two radiation fields from Sect.~\ref{sec:radiation fields} as a function of their dissociation or ionisation thresholds.}
  \label{fig:threshold_rate_ratio_comparison}
\end{figure}
Figure~\ref{fig:threshold_rate_ratio_comparison} illustrates this point further by plotting the photodissociation or ionisation thresholds of all species versus their 4000\,K and ISRF rate ratios, showing a sharp drop shortwards of 170\,nm.
According to this figure, these ratios vary by more than an order of magnitude due to other details of each atomic or molecular cross section.

Figures~\ref{fig:compare_rates_radiation_fields_photodissociation} and \ref{fig:compare_rates_radiation_fields_photoionisation} also show rates assuming a Lyman-$\alpha$ dominated radiation field, most of which are smaller than in the ISRF case (under our normalisation scheme) or zero if there is no significant cross section at 121.6\,nm.
Most molecules and atoms have photoionisation thresholds at similar or shorter wavelengths than this (as listed in Table~\ref{tab:cross_section_properties}), explaining the general disappearance or lowering of photoionisation rates for Lyman-$\alpha$ radiation.

\begin{figure}
  \centering
  \includegraphics{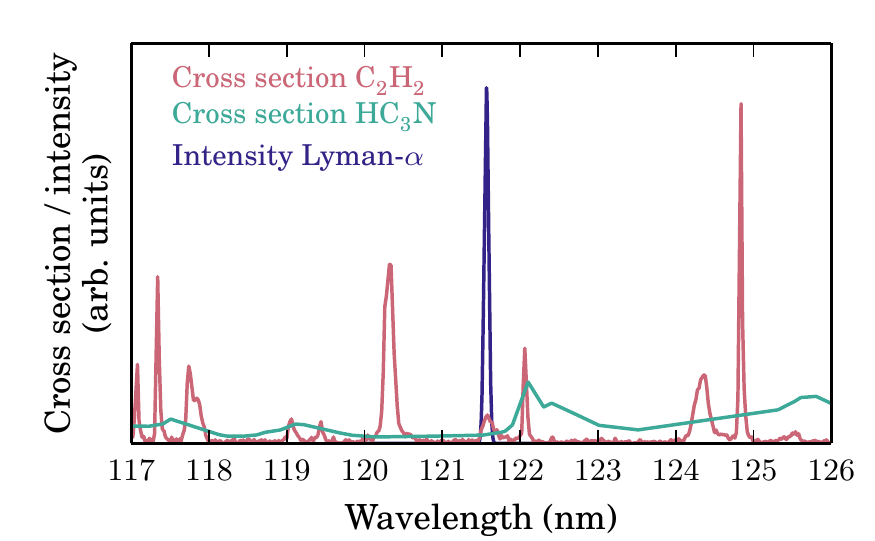}
  \caption{Comparing the wavelength dependence of the photodissociation cross sections  of \ce{C2H2} and \ce{HC3N} between 117 and 126\,nm with Lyman-$\alpha$ radiation.}
  \label{fig:C2H2_HC3N_Lyalpha}
\end{figure}
For molecules with line-dominated cross sections the Lyman-$\alpha$ cross section is sensitive to the positions of these lines.
For example, Fig.~\ref{fig:C2H2_HC3N_Lyalpha} shows why the photodissociation rate of
\ce{C2H2} increases by a factor of three when substituting the ISRF with
a Lyman-$\alpha$ emission line, while \ce{HC3N} decreases by a factor 2.
In this case the respective overlap and non-overlap of resonances is responsible.

\begin{figure*}
  \centering
  \includegraphics{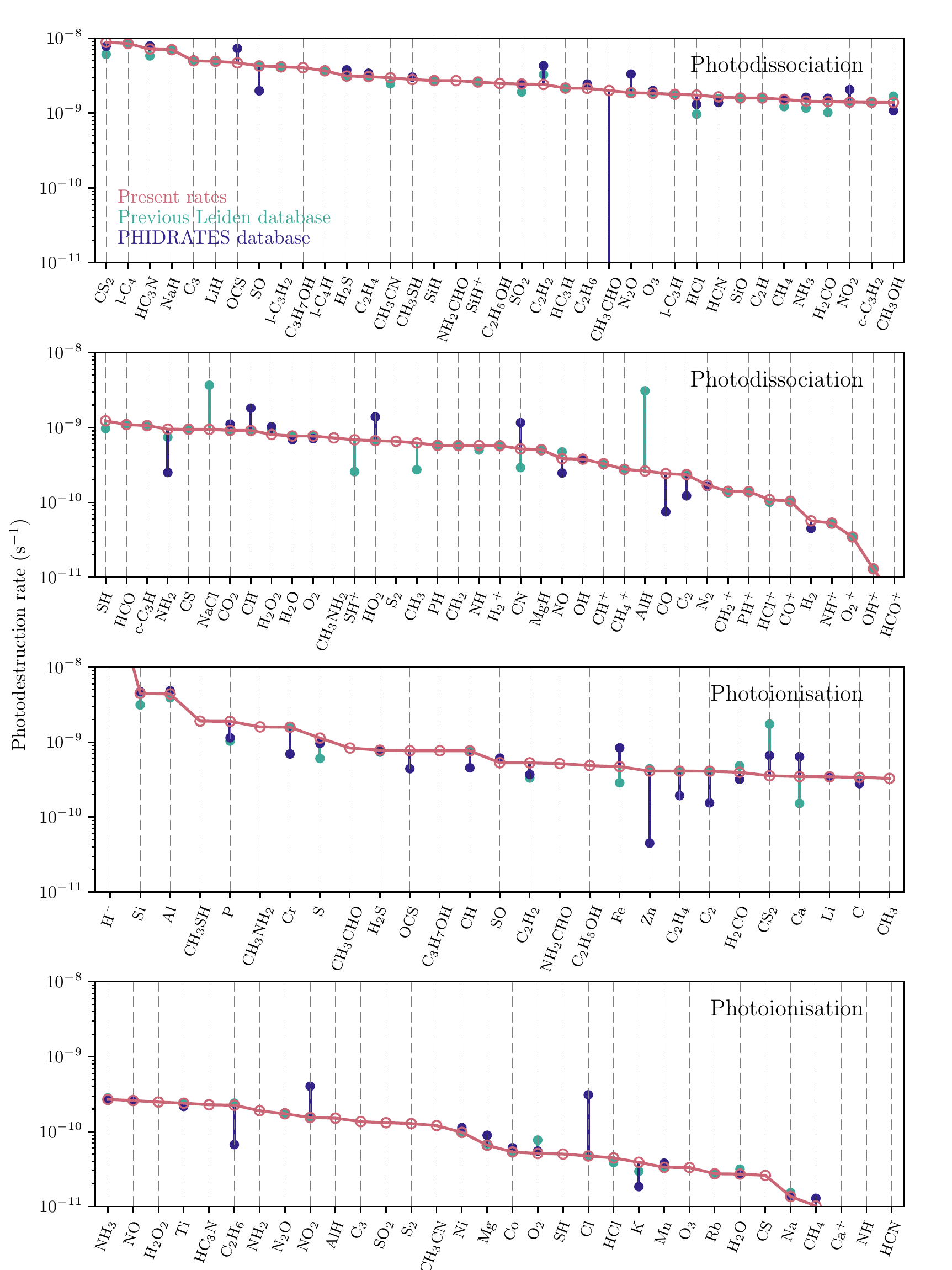}
  \caption{ ISRF photodissociation and ionisation rates for atoms and molecules. The present calculations are compared with the PHIDRATES database \cite{huebner1992,huebner2015} and the previous version of the Leiden database \cite{van_dishoeck1988,van_dishoeck2006}.}
  \label{fig:compare rates with other databases}
\end{figure*}

\subsection{Comparison with previous rates}

The new ISRF photodissociation and ionisation rates are compared with those calculated from cross sections taken from the previous version of the Leiden database \cite{van_dishoeck1988,van_dishoeck2006} and the PHIDRATES database \cite{huebner1992,huebner2015} in Fig.~\ref{fig:compare rates with other databases}.
Not all species are included in all databases.
We generally find agreement within 30\% with the molecular rates from the previous Leiden database with exceptional cases being the photodissociation of \ce{CH3}, \ce{HCl} and the photoionisation of \ce{C2H2}.
All of these processes have larger rates in the updated database, by factors between 1.6 and 2.3, and this is due to the availability of new theoretical and experimental data, especially at shorter wavelengths, leading to increased cross sections.

More disagreement follows from comparison with the PHIDRATES database, with differences spanning factors of 0.15 to 3.8, and  the extreme case of \ce{CH3CHO} photodissociation for which the PHIDRATES cross section is based on somewhat incomplete information. 
In general PHIDRATES molecular cross sections were collated originally by \textcite{huebner1992} and in previous studies, and the subsequent availability of new experimental and theoretical information explains the different ISRF photodestruction rates.

It is important to note the differing intentions of the Leiden and PHIDRATES databases.
The former considers cross sections important to interstellar chemistry, so that metastable dissociation products generally have time to decay before they can react, and the Lyman-limit at 91.2\,nm provides a hard short-wavelength limit to the radiation field in most cases.
Conversely, the PHIDRATES cross sections are intended for studying photochemistry in the solar system, including high-density environments like planetary atmospheres and cometary comae. 
Then, partial cross sections for many more dissociation and dissociative-ionisation fragments are considered, including metastable species, and with a stronger emphasis on the solar radiation field, favouring lower-lying states.

\section{Shielding functions}
\label{sec:shielding functions}

\subsection{General formulation}

Substantial molecular abundances cannot exist in the unshielded interstellar medium because of their short dissociation lifetimes.
Instead, observed molecules are found embedded inside interstellar clouds, protoplanetary disks, or similar objects that are at least partially shielded from radiation.
\change{The unattenuated photodissociation and photoionisation rates given in Sect.~\ref{sec:photo rates} must then be recalculated taking into account the intervening material, according to, for example,
\begin{multline}
  \label{eq:shielded dissoc rate}
  k = \int \sigma^\text{pd}(\lambda)
  \exp\bigg[ - \tau_\text{dust}(\lambda,N_{\ce{H}+2\ce{H2}})  \\
    - \sum_{X=\ce{H},\ce{H2},\text{self}} N_X\sigma_X^\text{abs}(\lambda) \bigg] I(\lambda)\,\text{d}\lambda.
\end{multline}

The first exponentiated term in Eq.~(\ref{eq:shielded dissoc rate}) models the attenuation of ultraviolet radiation due to dust as a function of wavelength and the column density of hydrogen nuclei (assuming this is proportional to the dust column).
The non-absorbing scattering of UV photons by dust significantly alters their radiative transfer in a shielded region so that $\tau_\text{dust}(\lambda,N_{\ce{H}+2\ce{H2}})$ is not simply proportional to column density.
Additionally, the form of this term is dependent on the dust-cloud shape and nature of the incident radiation: isotropic, normal, or otherwise.

The summation term in Eq.~(\ref{eq:shielded dissoc rate}) considers shielding by atomic and molecular species, where the most important cases are photoabsorption by H and \ce{H2}, and self shielding for a few abundant species.
The column density of species $X$ is represented by $N_X$.

The rate reduction due to dust, molecules, and atoms is characterised by a shielding function:
\begin{equation}
  \label{eq:shielding function definition}
  \theta = \frac{k}{k_0},
\end{equation}
where, $k_0$ is the photoprocess rate at the irradiated edge of the shielded region.
For an infinite-slab interstellar cloud in a region of space conforming to a standard isotropic ISRF, $k_0$ will be slightly greater than half of the rates in Tables~\ref{tab:photodissociation rates} and \ref{tab:photoionisation rates}, due to the restriction of incident radiation to $2\pi$\,sr and the occurrence of back-scattered radiation from the shielded region.
}

We calculated shielding functions according to Eqs.~(\ref{eq:shielded dissoc rate}) and (\ref{eq:shielding function definition}) for the photodissociation and ionisation of all atomic and molecular species in our database.
These results are discussed below and are also available from the Leiden database\footnote{\url{www.strw.leidenuniv.nl/~ewine/photo}} in tabulated form.
These tables contain values of $\theta$ as a function of column density for each bracketed term in Eq.~(\ref{eq:shielded dissoc rate}) treated independently.

\begin{figure*}
  \centering
  \includegraphics{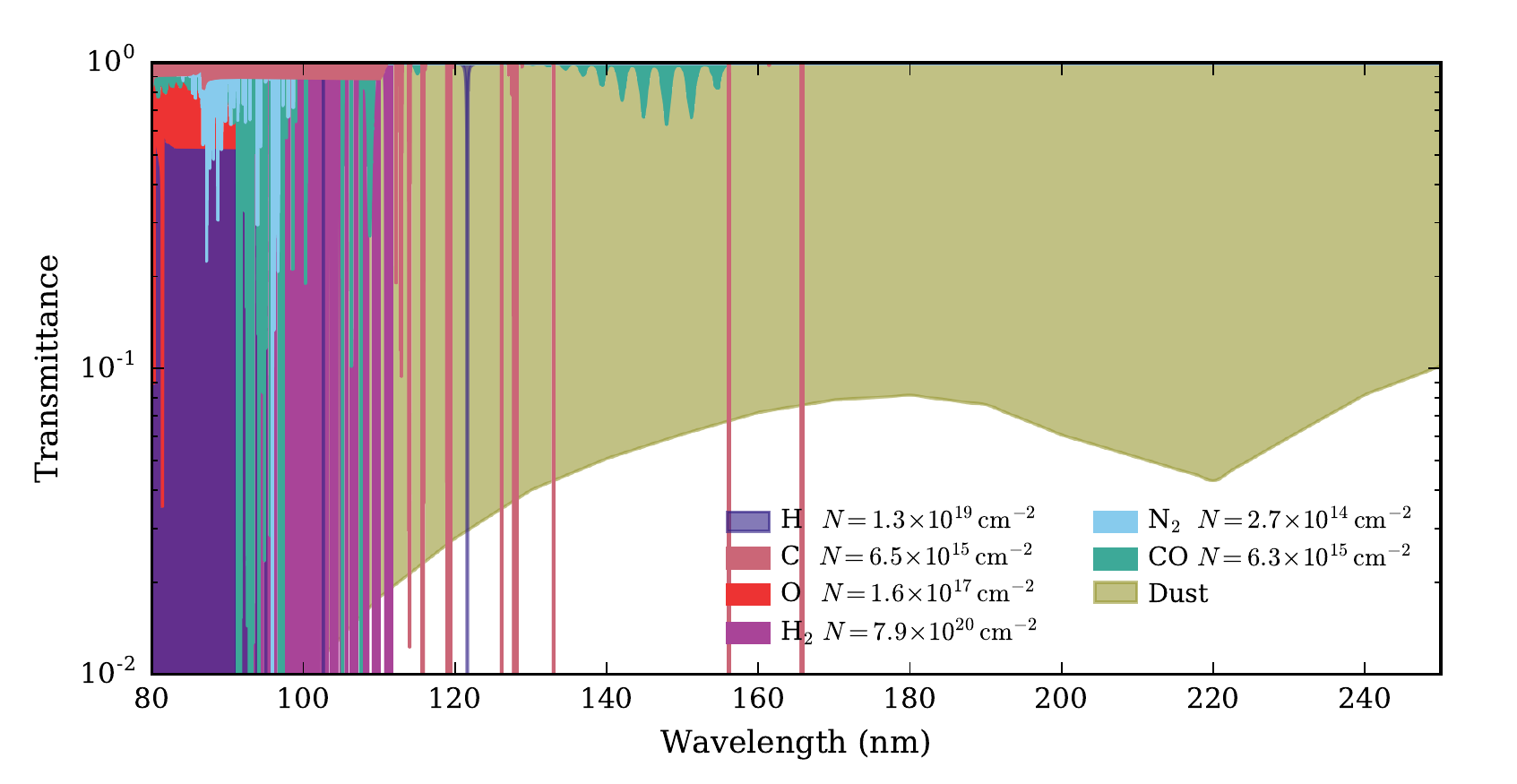}
  \caption{Possible wavelength-dependent transmittance (shaded region edges) of the highest-column density ($N$) species at $A_\text{V}=1$ depth into a semi-infinite one-sided interstellar cloud.}
  \label{fig:combination shielding}
\end{figure*}
A demonstration of the relative importance of shielding terms in Eq.~(\ref{eq:shielded dissoc rate}) is given by Fig.~\ref{fig:combination shielding}, which simulates the transmittance of ultraviolet radiation to a depth of $A_\text{V}=1$ into an interstellar cloud. 
The column densities of intervening atomic and molecular species are the result of a particular diffuse-cloud chemical model run by \textcite{heays2014a}, but their magnitudes are typical in an interstellar PDR.
For wavelengths shorter than about 130\,nm multiple sources contribute to ultraviolet shielding in this model. 
The combined shielding in this case is approximately a product of these, e.g., $\theta^\text{total} = \theta^\text{dust}\cdot\theta^{\ce{H2}}\cdot\theta^{\ce{H}}\cdots$.
This factorisation is somewhat inaccurate for multiple shielding species with overlapping and line-dominated cross sections.
Additionally, the shielding functions calculated here are appropriate for
infinite-slab geometries only.  For more sophisticated astrochemical
models the explicit simulation of ultraviolet radiative transfer using
our database of cross sections may be required.

\subsection{Shielding by dust}
\label{sec:shielding by dust}
  
\begin{figure}
  \centering
  \includegraphics{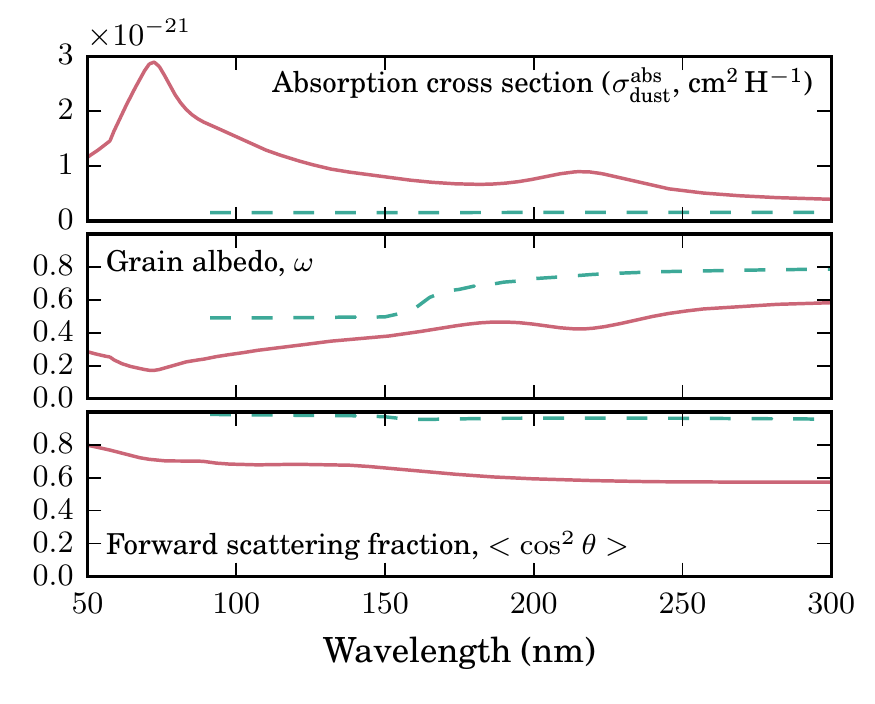}
  \caption{Solid lines: Optical properties of interstellar dust grains according to Draine et al. \cite{draine1984,draine2003b}.\protect\footnote{Model $R_\text{V}=3.1$' from \protect\url{www.astro.princeton.edu/~draine/dust/dustmix.html}} Dashed lines: Optical properties simulating grain growth in a protoplanetary disk. Upper: Photoabsorption and photoextinction cross sections. Middle: Grain albedo. Lower: Probability of forward scattering.}
  \label{fig:grain optical properties}
\end{figure}

We calculated dust-shielding-dependent photodissociation and ionisation rates using the optical properties of a dust population assuming one particular composition and size distribution.
These were taken from the mixed grain-size and composition model developed by Draine et al. \cite{draine1984,li2001,weingartner2001,draine2003a,draine2003b,draine2003c}. 
We used their ``$R_\textrm{V}=3.1$'' model published online,\footnote{\url{www.astro.princeton.edu/~draine/dust/dustmix.html}} with important optical properties plotted in Fig.~\ref{fig:grain optical properties}.
\change{We also adopt a (gas mass)/(dust mass) ratio of 124 in line with the Draine et al. dust grain model.}
The interstellar variation of these parameters and their possible effects on shielding functions is discussed in Sect.~\ref{sec:dust grain properties}.

The normal observationally-relevant extinction cross section is larger than the absorption cross section shown in Fig.~\ref{fig:grain optical properties} by a factor of $1/(1-\omega$), where $\omega$ is the grain albedo, to account for photons scattered out of the line of sight.
These photons are however still available for photodestruction and their radiative transfer through an interstellar cloud must be considered.
We did this for the case of an interstellar cloud with infinite-slab geometry and extending to large enough $A_\text{V}$ that it is effectively illuminated from one side only, \change{and assume this illumination is incident isotropically}.
The radiative transfer equations were solved according to the method of \textcite{roberge1991} and \textcite{van_dishoeck2006} that takes into account the dust absorption cross section, dust albedo, and the averaged fraction of forward-scattered photons. All of these properties are significantly wavelength dependent in the ultraviolet spectral region, as shown in Fig.~\ref{fig:grain optical properties}.
The transmission through a dust thickness corresponding to 1\,$A_\text{V}$, including scattered photons is shown in Fig.~\ref{fig:combination shielding} and is largely wavelength independent longwards of about 120\,nm. 
At shorter-wavelength dust absorption rapidly becomes more effective.

\begin{figure}
  \centering
  \includegraphics{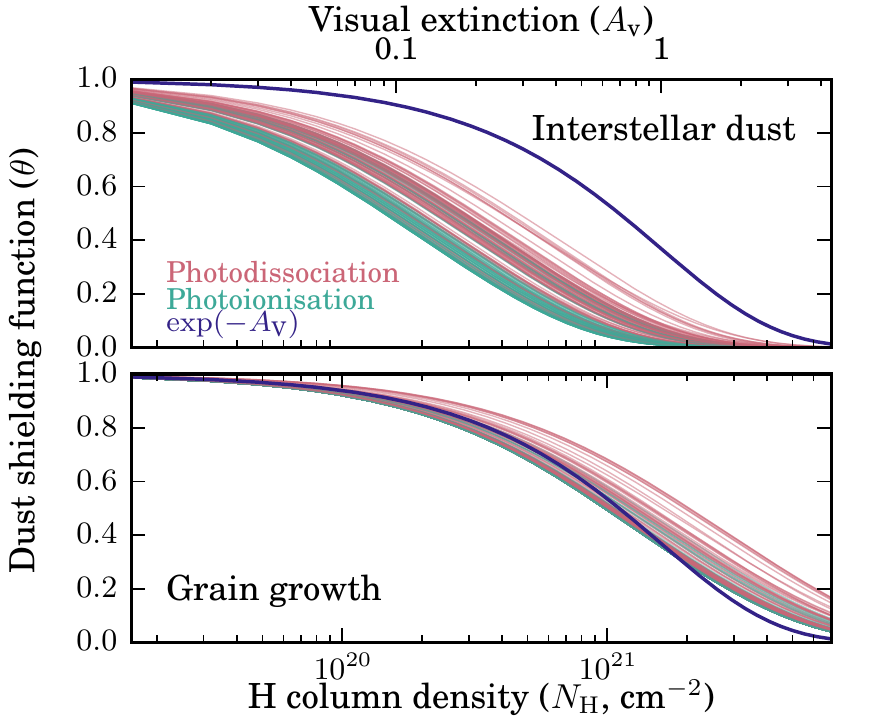}
  \caption{Shielding of photodissociation (red) and photoionisation (green) in the ISRF. For all molecules in our database by an amount of dust parameterised by H-nucleus column density. Cases are shown for interstellar dust and a population of larger dust grains (see text). An additional curve traces $\exp(-A_v)$, the visual extinction (blue).}
  \label{fig:dust shielding}  
\end{figure}
The calculated wavelength-dependent penetration of ultraviolet photons was used to calculate depth-dependent photodestruction rates for each molecule and atom in our database.
These are summarised as shielding functions in Fig.~\ref{fig:dust shielding}.
The alternative $A_V$ and H-nuclei column density scales are related by the standard proportionality \cite{savage1977}
\begin{equation}
  \label{eq:NH vs AV}
  A_V = N_{\ce{H}+2\ce{H2}}/(\np[cm^{-2}]{1.6e21})  
\end{equation}
and assuming a gas-mass to dust-mass ratio of 124.
The differences between curves arise from the varying wavelength dependencies of atomic and molecular cross sections and the dust grain optical properties.
All of the plotted shielding functions lie below the visual extinction curve, $\exp(-A_\textrm{V})$, despite the inclusion of forward scattering, which acts to increase the penetration depth.
This is because of the larger dust absorption cross section at shorter wavelengths, which also explains the faster shielding of photoionisation than photodissociation, due to the shorter-wavelength thresholds of ionisation.

\begin{figure}
  \centering
  \includegraphics{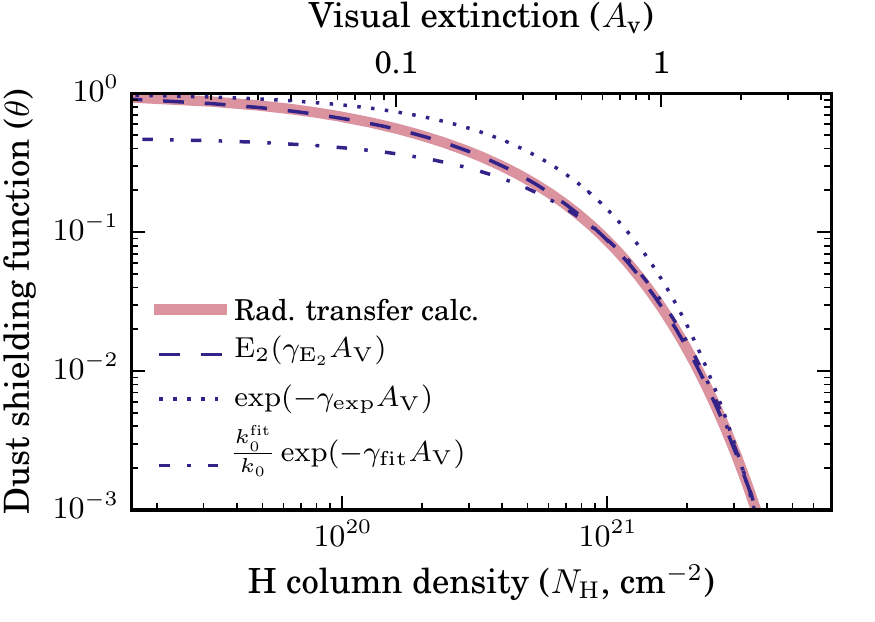}
  \caption{
    \change{
      Dust shielding function for \ce{CH4} photodissociation (solid red curve) in the ISRF.  Calculated by means of an infinite-slab radiative transfer model. Also shown are some simple parameterisations of this curve according to three formulations.
    }}
  \label{fig:dust shielding parameterisation examples}
\end{figure}

\change{
  Curves like those in Fig.~\ref{fig:dust shielding} are sometimes approximated as simple functions of $A_\text{V}$ for easier utilisation.
Either, as one-parameter exponential curves,
\begin{equation}
  \label{eq:dust shielding exponential}
  \theta(A_{\text{V}}) = \exp\left(-\gamma_\textrm{exp} A_{\text{V}}\right),
\end{equation}
\cite[e.g.,][]{van_dishoeck2006} or 2\textsuperscript{nd}-order exponential integrals,
\begin{equation}
  \label{eq:dust shielding exponential integral}
  \theta(A_{\text{V}}) =  {\textrm{E}_2}(\gamma_{\textrm{E}_2} A_\text{V}),
\end{equation}
\cite[e.g.,][]{neufeld2009,roueff2014}, or bi-exponential functions \cite[e.g.][]{roberge1991}.
We calculated values of $\gamma_{\textrm{exp}}$ and $\gamma_{\textrm{E}_2}$ in Eqs.~(\ref{eq:dust shielding exponential}) and (\ref{eq:dust shielding exponential integral}) that best fit the results of our radiative transfer model, with an example fit for the shielded photodissociation of \ce{CH4} shown in Fig.~\ref{fig:dust shielding parameterisation examples}.
Calibrated exponential-integral functions were found to reproduce the shielding effects of dust absorption within 25\% over the range of $1\geq\theta>\np{e-4}$, while the exponential function deviate by up to a factor of 3 over the same range.
Given the superior parameterisation of exponential-integral functions, we listed $\gamma_{\textrm{E}_2}$ coefficients for all species in our database in Tables~\ref{tab:photodissociation rates} and \ref{tab:photoionisation rates}, along with updated values of the normal-exponential $\gamma_{\textrm{exp}}$ parameters that are adopted in previous iterations of the Leiden database \cite{van_dishoeck2006}.

  A further reduction of shielding functions is demonstrated in Fig.~\ref{fig:dust shielding parameterisation examples}, whereby, an exponential-decay parameter $\gamma_{\rm fit}$ is adopted along with an effective unshielded rate, $k_0^{\rm fit}$, so that the depth-dependent photodestruction rates are:
\begin{equation}
  k = k_0^{\rm fit}\exp(\gamma_{\rm fit}A_{\rm{V}}),
\end{equation}
leading to shielding functions
\begin{equation}
  \theta = \frac{k_0^{\rm fit}}{k_0}\exp(\gamma_{\rm fit}A_{\rm{V}}).
\end{equation}
The values of these parameters were selected to best fit the radiative-transfer calculation between $A_{\text V}=0.1$ and 3 (a range of extinction where the details of ISRF photodestruction has most influence on PDR chemistry), and provide a better approximation of this range than Eq.~(\ref{eq:dust shielding exponential}), but a poorer fit in general than Eq.~(\ref{eq:dust shielding exponential integral}).
Fitted parameters of this modified exponential form are provided in the Leiden database for all atoms and molecules and may be useful for astrochemical codes where exponential-form dust shielding is required, but with improved accuracy over an intermediate range of $A_{\rm V}$.

The quantitative shielding discussion in this section is appropriate for use in single-sided isotropically-irradiated infinite-slab interstellar cloud models. 
The modelling of dust-shielding effects in other geometries would require specific radiative-transfer calculations.

Previously, \textcite{van_dishoeck2006} compared dust-shielding effects ad $\gamma_{\textrm{exp}}$ parameters for a range of radiation field types and dust grain properties, a point that is discussed further in Sect.~\ref{sec:dust grain properties}.}

\subsection{Shielding by H$_2$, H, and C}

\begin{figure}
  \centering
  \includegraphics{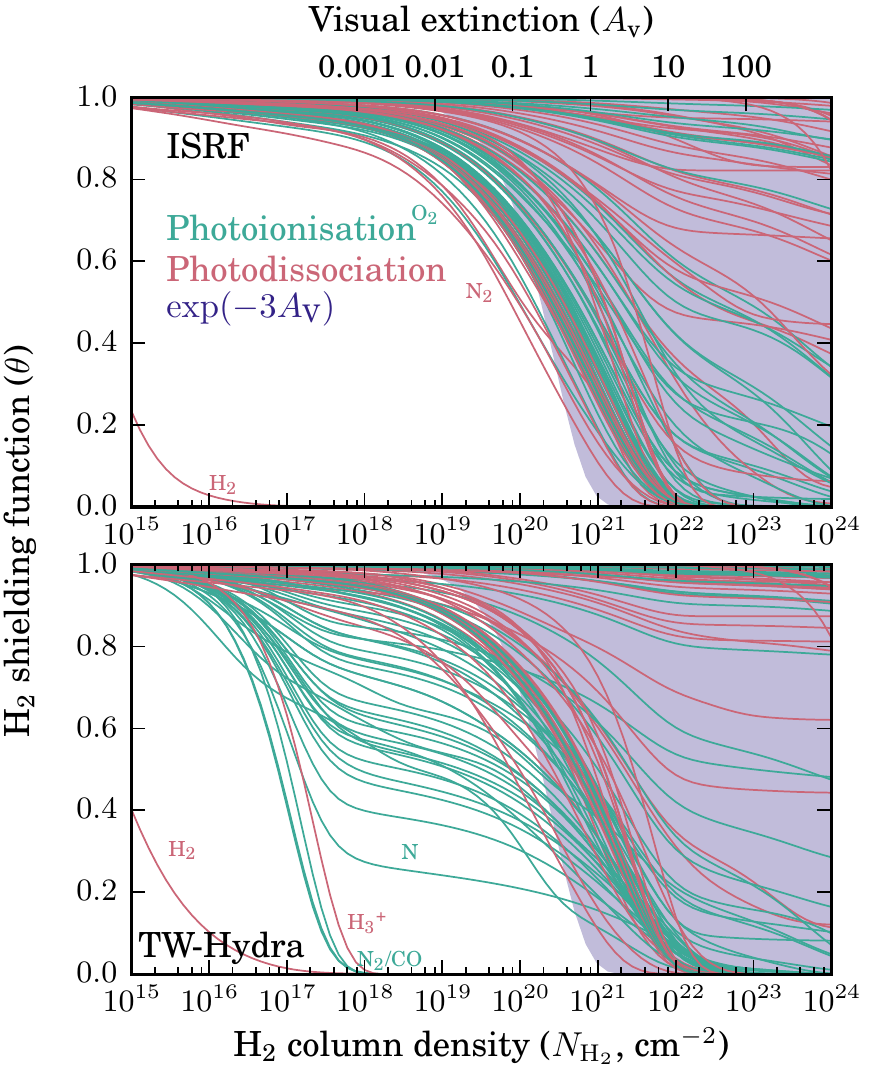}
  \caption{Shielding functions for photodissociation (red curves) and photoionisation (blue curves) by a column of \ce{H2}. Assumes an impinging ISRF or simulation of the TW-Hydra circumstellar radiation. A few extreme cases have their identity labelled. The shaded-region edge describes a typical ultraviolet-extinction curve due to interstellar dust, according to $\exp -3A_\text{V}$.}
  \label{fig:H2 shielding}
\end{figure}

The shielding of most molecules found in PDRs is dominated by dust extinction, assuming a standard amount of gas and dust.
In some cases, additional terms in Eq.~(\ref{eq:shielded dissoc rate}) must be considered.
  
\change{
  The shielding effect on our database of molecules was calculated for a one-dimensional column of \ce{H2} while neglecting scattering in the molecular lines.
  Shielding functions calculated according by means of Eqs.~(\ref{eq:shielded dissoc rate}) and (\ref{eq:shielding function definition}) are summarised in Fig.~\ref{fig:H2 shielding} for two cases: the ISRF and TW-Hydra radiation fields.
  For these calculations, the  \ce{H2} photoabsorption cross section was composed from a list of individual line parameters, as described in Sect.~\ref{sec:H2}, and assumed an excitation temperature of 100\,K and Gaussian Doppler broadening width $b=3$\,km\,s$^{-1}$.
}

Significant \ce{H2} shielding only occurs in the ISRF for column densities of more than about \np[cm^{-2}]{e20}.
Larger columns are required for some molecules that photoabsorb at wavelengths greater than 110\,nm, outside the range of \ce{H2} line absorption.
The shaded region in Fig.~\ref{fig:H2 shielding} shows a typical ultraviolet dust-shielding curve for comparison.
The influence of \ce{H2}-shielding in the ISRF is overshadowed by dust extinction in many cases, assuming a standard gas and dust mass ratio, and can safely be neglected.
In a few cases it is of comparable importance.
For instance, CO and \ce{N2}, investigated in detail by \textcite{visser2009} and \textcite{li2013}.
The extreme case of \ce{H2} shielding itself is discussed in the following section.

Figure~\ref{fig:H2 shielding} also shows \ce{H2} shielding functions assuming a radiation field simulating the TW-Hydra emission spectrum.
The resulting changes with respect to the ISRF follow from the inclusion of radiation shorter than 91.2\,nm, so that the continuum absorption of \ce{H2} can rapidly reduce the photoionisation rate of some species.
The shielding of longer-wavelength photodissociation remains dust dominated apart from the case of \ce{H3+}, whose cross section occurs entirely at short wavelengths \cite{kulander1978}.
 
\begin{figure}
  \centering
  \includegraphics{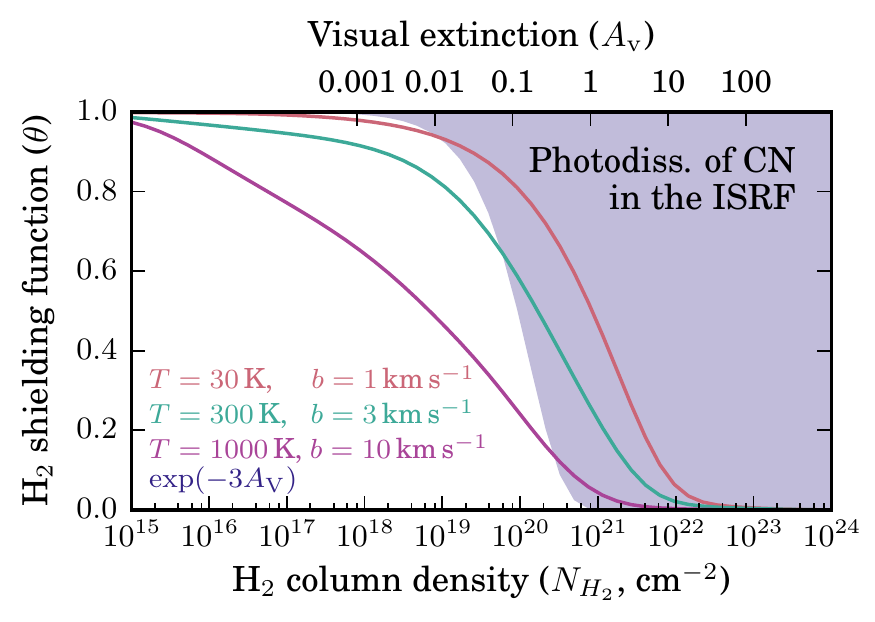}
  \caption{Shielding of CN photodissociation in the ISRF by \ce{H2} assuming various values for the \ce{H2} rotational excitation temperature, $T$, and Doppler broadening, $b$.  Also shown is a typical shielding function due to interstellar dust, $\exp(-3A_\text{V}$) (shaded region edge).}
  \label{fig:H2_shielding_various_temperatures_and_broadening}
\end{figure}
An increase in the rotational temperature and turbulent broadening of \ce{H2} increases its shielding effectiveness, due to the larger filling factor of its photoabsorption spectrum when more and broader rotational lines are included. 
This effect is quite small, for molecules and atoms that primarily absorb longwards of 110\,nm.
In some extreme cases the effect can be significant, as illustrated for CN photodissociation  in Fig.~\ref{fig:H2_shielding_various_temperatures_and_broadening}.
Here, increasing the temperature or Doppler width over astrophysically relevant ranges, from 30 to 300\,K or 1 to 3\,km\,s$^{-1}$, respectively, leads to about twice the \ce{H2} shielding between 0.1 and 3\,$A_\textrm{V}$.
Increasing these further to 1000\,K and 10\,km\,s$^{-1}$ (conditions perhaps still relevant to some interstellar shocks or in some atmospheres) results in an order of magnitude increase in \ce{H2} shielding effectiveness.
The effect of \ce{H2} temperature and broadening are studied by \textcite{visser2009} and \textcite{li2013} for the cases of CO and \ce{N2}, respectively.

Shielding by the atomic H photoionisation continuum is implicit in calculations with radiation fields having a 91.2\,nm Lyman-limit cut-off.
In principle, column-density-dependent H-shielding occurs where UV radiation includes shorter wavelengths, although this detail is often negligible in view of the greater shielding effect of \ce{H2} and dust.
Atomic-H shielding functions are discussed in Appendix~\ref{sec:additional shielding function discussion} and may be important in dust-depleted environments near the edge of a PDR.
Another species that potentially contributes to the exponential term in Eq.~(\ref{eq:shielded dissoc rate}) is atomic C \cite{rollins2012}, present near the boundary layer of a PDR, and we computed shielding functions for this species also, discussed further in the Appendix.

\change{
We did not evaluate the scattering of photons by \ce{H2} and H, whereby photoabsorption into electronically-excited levels is followed by resonant photoemission \cite[e.g.,][]{black1987}.
This is shown to significantly modify the UV spectrum in embedded regions \cite{le_petit2006}, as well as for Lyman-$\alpha$ photons scattered by atomic-H \cite{neufeld1991,bethell2011}.
For \ce{H2}, more than 80\% of absorbed UV photons are in re-emitted, mostly to ground state levels with $v>0$ and at longer wavelengths.
Ultimately, all photons will be absorbed by dust grains, but determining the influence of \ce{H2} resonant scattering on the photodestruction rates calculated here would be a worthwhile future project.} 

\subsection{Self-shielding of molecules}

\begin{figure}
  \centering
  \includegraphics{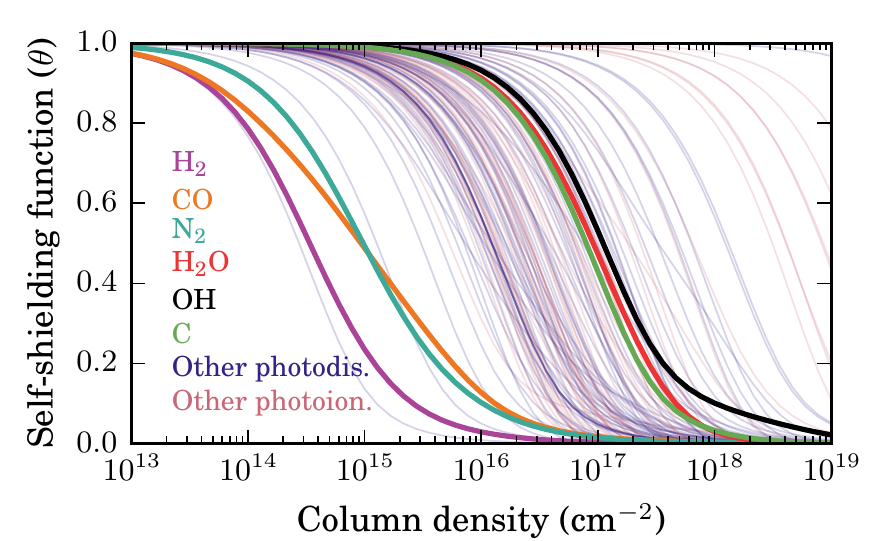}
  \caption{Self-shielding of photodissociation in the ISRF.}
  \label{fig:self shielding}
\end{figure}
Self-shielding is also important for some photodissociating species.
Shielding functions describing this phenomenon independently of other
mechanisms are plotted in Fig.~\ref{fig:self shielding}.  The three
prominent cases where self-shielding is effective at relatively low
column densities are \ce{H2}, \ce{CO}, and \ce{N2}.  The most recent
treatments of these three molecules are given in
\textcite{sternberg2014}, \textcite{visser2009}, and
\textcite{li2013,heays2014a}; respectively.  The common cause
is the line-like photoabsorption cross sections of all three
molecules, leading to almost complete attenuation of ultraviolet
radiation at their line centres for modest column densities, about
$10^{15}$\,cm$^{-2}$.  The column density of CO and \ce{N2} relative
to H in the molecule-forming region of a PDR is typically high enough
that their self-shielding competes with the simultaneous effect of
dust extinction.  

Atoms and molecules that absorb predominantly through continua may still be susceptible to self-shielding if their column densities are large enough.
Three candidates for this phenomenon are highlighted in Fig.~\ref{fig:self shielding}, \ce{H2O}, OH, and C, where the minimum  column density for effective self-shielding is approximately $10^{17}$\,cm$^{-2}$ in all cases.
Such large columns of atomic C are actually found as discussed in the previous section, and some models of the inner regions of protoplanetary disks find sufficient \ce{H2O} and OH columns \cite{bethell2009,adamkovics2014}.
These phenomena may be particularly relevant in shocked media where densities are abnormally high, for example, C in supernova remnants and \cite{white1994} protoplanetary disks \cite{tsukagoshi2015}, or \ce{H2O} in protostellar outflows \cite{mottram2014}.
All other molecular species have too low abundance or insufficiently
peaked cross sections to effectively self-shield in space.

\section{Photodestruction due to cosmic rays}
\label{sec:cosmic rays}

\subsection{Cosmic ray induced UV spectrum}

Cosmic rays penetrate deeper into an interstellar cloud, protostellar envelope, protoplanetary disk or planetary atmosphere than ultraviolet photons and ionise \ce{H2} there.
This primary process and resultant cascade of re-scattered electrons proceeds to excite further \ce{H2} and generate excited H-atoms \cite{cravens1978,gredel1995}, whose radiative decay generates a ultraviolet flux with a line-dominated structure \cite{prasad1983,gredel1987,cecchi-pestellini1992b}.  
The photolysis of molecules due to this flux is quantified several times previously \cite{sternberg1987,gredel1987,gredel1989,heays2014a} but the last major summary of rates dates back to \textcite{gredel1989}.
Here all rates are recomputed with updated cross sections. Also, the effects of grain growth, such as appropriate for protoplanetary disks, are considered.

\begin{figure}
  \centering
  \includegraphics{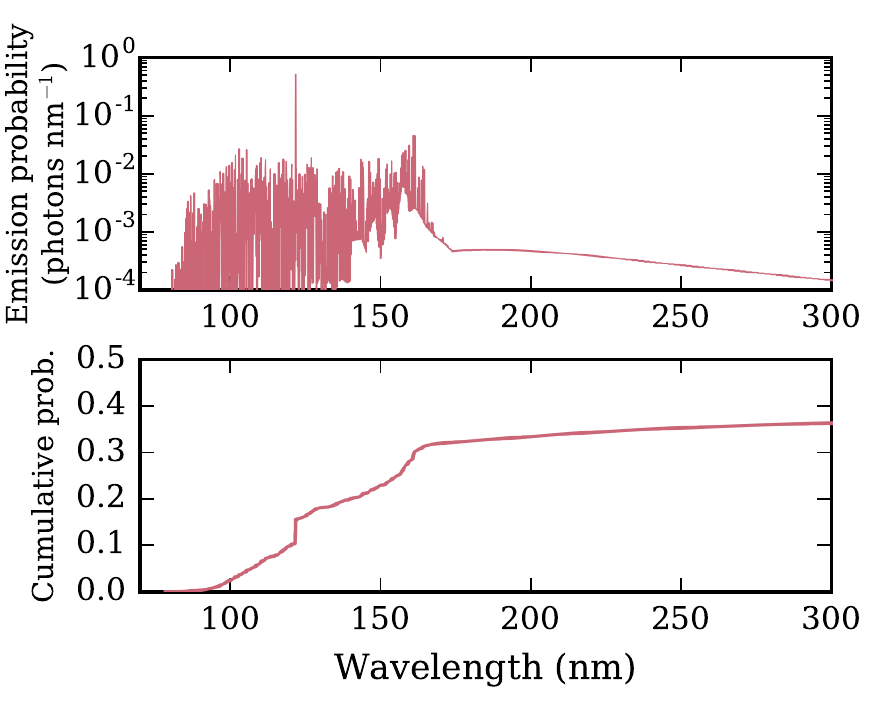}
  \caption{The cosmic-ray induced radiation field. The spectral
    density of photons generated per primary cosmic-ray ionisation
    event is shown in detail over the dominant wavelength range
    (upper) and globally (middle). These assume
    linewidths of \np[km\,s^{-1}]{1}. Lower: A cumulative integration of this
    distribution.}
  \label{fig:cosmic_ray_radiation_field}
\end{figure}

The cosmic-ray induced ultraviolet flux is modelled here as a rate of
photons generated per unit spectral density per hydrogen nucleus:
\begin{equation}
  \label{eq:cosmic ray flux}
  R(\lambda) = \zeta_{\ce{H2}} x_{\ce{H2}} P(\lambda).
\end{equation}
Here, and below, $x_{\ce{X}}= n(\ce{X}) / \left[ n(\ce{H})+2n(\ce{H2})
  \right]$ is the relative abundance of species \ce{X} with respect to
total hydrogen nuclei and $\zeta_{\ce{H2}}$ is the rate at which an H$_2$
molecule is ionised by cosmic ray collisions.
\change{In the context of diffuse interstellar clouds \cite[e.g.,][]{oka2013,indriolo2015} it is atomic-H that is being ionised, and the appropriate $\zeta$ will then be per H atom.}

We used the wavelength-dependent probability distribution of generated
photons, $P(\lambda)$, of \textcite{gredel1989}.  This distribution is plotted in
Fig.~\ref{fig:cosmic_ray_radiation_field} and consists of many \ce{H2}
and H emission lines between 80
and 170\,nm as well as some continuum emission between 122 and
300\,nm.
The modelled photoemission mostly occurs between 90
and 170\,nm, as shown by the rapid increase of the cumulative
distribution in Fig.~\ref{fig:cosmic_ray_radiation_field} over this
range, with a single step at 121.6\,nm constituting 15\% of the integrated
flux due to Lyman-$\alpha$ emission.  The cumulative distribution attains a total value of
only 0.36 because not every \ce{H2} ionisation event results in an
ultraviolet photon being generated.

The value of $\zeta_{\ce{H2}}$ is uncertain due to the unknown origin and flux
of cosmic rays and its energy spectrum, as well as the attenuation due
to matter and magnetic fields \cite{dalgarno2006,grenier2015}.
Earlier values are indirectly deduced from observations of the HD and
OH abundances \cite[e.g.,][]{hartquist1978a,van_dishoeck1986} and
\ce{H3+} abundances \cite[e.g.,][]{vanderTak00ion,hezareh2008} in
diffuse and dense interstellar clouds, and favoured a value of about
\np{3e-17}\,s$^{-1}$\,\ce{H2}$^{-1}$, although rates up to $2\times
10^{-16}$\,s$^{-1}$ are inferred for some diffuse clouds.
Subsequent observations towards more lines of sight and of other tracers
like OH$^+$ and H$_2$O$^+$ are interpreted with updated \ce{H3+}
dissociative recombination rate coefficients generally and require a
higher rate in diffuse clouds, 1 to $4\times
10^{-15}$\,s$^{-1}$\,\ce{H2}$^{-1}$
\cite{indriolo2012,rimmer2012,vaupre2014,indriolo2015}, and higher
still in the Galactic centre.  A reduction of the ionisation rate
likely occurs in dark clouds and protoplanetary disks because of the
shielding effects of the surrounding material \cite{padovani2009} or
stellar winds and magnetic fields \cite{cleeves2013,cleeves2015}.

We adopt a primary ionisation rate for Eq.~(\ref{eq:cosmic ray flux})
of $\zeta_{\ce{H2}} = \np[s^{-1}\,{\ce{H2}}^{-1}]{e-16}$. All
molecular photodissociation rates can be adapted to an alternative
$\zeta_{\ce{H2}}$ by simple scaling.

Most cosmic-ray generated photons are eventually eliminated through
absorption by dust grains inside an interstellar cloud but some excite
atoms and molecules.  The fraction of photons that lead to the
photodissociation or ionisation of an atom or molecule, X, is given
by
\begin{equation}
  \label{eq:cosmic ray absorption competition}
  p_{\ce{X}}(\lambda) = \frac{x_{\ce{X}}\sigma^\text{diss/ion}_\text{X}(\lambda)}
  {x_{\ce{dust}}\sigma^\text{abs}_{\ce{dust}}(\lambda) + \sum_j x_j \sigma^\text{abs}_{j}(\lambda)}. 
\end{equation}
Here, $\sigma^\text{diss/ion}_\text{X}$ is the photodissociation or ionisation cross section of
species X, and the denominator sums the photoabsorption cross section
of all dust and gas species. The probability of a
cosmic-ray generated ultraviolet photon being absorbed by a dust
grain, \ce{H2} molecule, or some other gas-phase species depends on
its wavelength through the various cross sections in
Eq.~(\ref{eq:cosmic ray absorption competition}).

The rate of a particular photodestruction process for species X (per X) due to cosmic-ray-induced
photons is then
\begin{equation}
  \label{eq:cosmic ray rate}
  k_\text{X} = \frac{1}{x_\mathrm{X}}\int R(\lambda)p_\mathrm{X}(\lambda)\,\text{d}\lambda.
\end{equation}
The photolysis rates calculated by \textcite{gredel1989} and
\textcite{mcelroy2013} are presented as efficiencies with the H$_2$
ionisation rate and grain albedo factored out from Eq.~(\ref{eq:cosmic ray rate}).
This is not possible if the summation terms in Eq.~(\ref{eq:cosmic ray absorption competition}) are significant \cite{gredel1987,heays2014a}, and for the sake of generality we did not make this reduction.
Our calculated rates divided by a factor of \np{2e16} are approximately comparable
with the efficiencies given in \textcite{gredel1989}, and a factor of
\np{e16} is required when comparing with \textcite{mcelroy2013}.

The  spectrum shown in
Fig.~\ref{fig:cosmic_ray_radiation_field} and used in our rate calculations assumes an
ortho-H$_2$:para-H$_2$ ratio of 0:1, that is, with \ce{H2} in its $J=0$ rotational state. 
This is appropriate for excitation temperatures corresponding to the low temperatures of a molecular cloud because only 10\% of equilibrated \ce{H2} is excited above $J=0$ at, for example, 50\,K.
Even for cases where significant quantities of super-thermally excited \ce{H2} are inferred in interstellar clouds their influence on cosmic-ray induced ultraviolet photodissociation is unlikely to be large.
This is because an altered distribution of emission lines is largely washed out in rate calculation by the integration in Eq.~(\ref{eq:cosmic ray rate}).
Indeed, after testing $J=0:J=1$ populations with the two ratios 1:0 and 1:3 we find calculated-rate differences of less than 20\% for most species in our database, in line with previous work \cite{gredel1989}.
Larger differences are found for the case of CO and \ce{N2}, which are studied in more detail previously \cite{gredel1987,heays2014a}.

\textcite{gredel1989} considered the possibility of \ce{H2} bound levels absorbing the cosmic-ray induced ultraviolet flux and subsequently re-emitting photons of the same wavelength or longer (following emission into excited ground state vibrational levels).
Their calculated rates are altered by up to 35\% by considering this phenomenon, with most species being altered by less than 5\%, and this effects is neglected here.

For the case of dust absorption, the wavelength-dependent mixed-grain
absorption cross section of Draine et al. \cite{draine1984,li2001,weingartner2001,draine2003b,draine2003c}
were used, as discussed in Sect.~\ref{sec:photo rates} and plotted in
Fig.~\ref{fig:grain optical properties}.  This absorption cross
section is somewhat different from the wavelength-independent dust
properties adopted previously \cite{gredel1987,heays2014a} that
converted an observationally-estimated dust extinction cross section,
\np[cm^{-1}\,H^{-1}]{2e-21}, to an absorption cross section by
assuming an effective grain albedo between 0 and 0.8.  It is clear
from Fig.~\ref{fig:grain optical properties} that consideration of
the wavelength dependence of dust absorption can affect the deduced
shielding of molecular photodestruction by, at most, a factor of about
two, depending on the wavelength-dependence of the molecular photoabsorption cross
section.

\begin{table*}
  \centering
  \caption{Cosmic ray photodestruction rates ($\times\np[s^{-1}]{e-16}$).}
  \label{tab:cosmic ray rates table}
  \begin{minipage}[c]{1.0\linewidth}
    \centering
  \begin{centering}
  \begin{tabular}{ccc@{\quad\qquad}ccc@{\quad\qquad}ccc}
\toprule
Species & Diss.\myfootnotemark{1} & Ion.\myfootnotemark{1} & Species & Diss. & Ion. & Species & Diss. & Ion. \\
\midrule

\ce{H} & -- & -- & \ce{C2} & 180 & 250 & \ce{NO2} & 1000 & 110 \\
\ce{Li} & -- & 250 & \ce{C2H} & 1100 & -- & \ce{N2O} & 1400 & 99 \\
\ce{C} & -- & 260 & \ce{C2H2} & 3500 & 380 & \ce{CN} & 450 & 8.3 \\
\ce{N} & -- & -- & \ce{C2H4} & 3500 & 280 & \ce{HCN} & 2000 & 14 \\
\ce{O} & -- & 2.7 & \ce{C2H6} & 2100 & 180 & \ce{HC3N} & 5900 & 160 \\
\ce{Na} & -- & 13 & \ce{C3} & 6900 & 89 & \ce{CH3OH} & 1600 & 240 \\
\ce{Mg} & -- & 110 & \ce{\textit{l}-C3H} & 3000 & -- & \ce{CH3CN} & 2600 & 97 \\
\ce{Al} & -- & 2500 & \ce{\textit{c}-C3H} & 480 & -- & \ce{CH3SH} & 2700 & 2000 \\
\ce{Si} & -- & 4200 & \ce{HC3H} & 1100 & -- & \ce{CH3CHO} & 2200 & 690 \\
\ce{P} & -- & 1500 & \ce{\textit{l}-C3H2} & 3400 & -- & \ce{CH3NH2} & 450 & 1800 \\
\ce{S} & -- & 800 & \ce{\textit{c}-C3H2} & 690 & -- & \ce{NH2CHO} & 2900 & 410 \\
\ce{Cl} & -- & 47 & \ce{\textit{l}-C4} & 1800 & -- & \ce{C2H5OH} & 2600 & 380 \\
\ce{K} & -- & 34 & \ce{\textit{l}-C4H} & 6100 & -- & \ce{C3H7OH} & 4600 & 590 \\
\ce{Ca} & -- & 270 & \ce{\textit{l}-C5H} & 170 & -- & \ce{SH} & 1100 & 34 \\
\ce{Ti} & -- & 230 & \ce{OH} & 470 & -- & \ce{SH+} & 460 & -- \\
\ce{Cr} & -- & 1200 & \ce{OH+} & 8.6 & -- & \ce{H2S} & 3400 & 620 \\
\ce{Mn} & -- & 49 & \ce{H2O} & 1000 & 23 & \ce{CS} & 1900 & 20 \\
\ce{Fe} & -- & 480 & \ce{O2} & 780 & 28 & \ce{CS2} & 5500 & 310 \\
\ce{Co} & -- & 60 & \ce{O2+} & 70 & -- & \ce{OCS} & 5200 & 560 \\
\ce{Ni} & -- & 140 & \ce{HO2} & 190 & -- & \ce{S2} & 88 & 140 \\
\ce{Zn} & -- & 180 & \ce{H2O2} & 830 & 180 & \ce{SO} & 5500 & 450 \\
\ce{Rb} & -- & 23 & \ce{O3} & 1500 & 32 & \ce{SO2} & 2700 & 110 \\
\ce{Ca+} & -- & 1.5 & \ce{CO} & 46\myfootnotemark{2} & 14 & \ce{SiH} & 620 & -- \\
\ce{H-} & -- & 1300 & \ce{CO+} & 77 & -- & \ce{SiH+} & 1200 & -- \\
\ce{H2} & -- & -- & \ce{CO2} & 600 & 8.2 & \ce{SiO} & 890 & -- \\
\ce{H2+} & 610 & -- & \ce{HCO} & 530 & -- & \ce{HCl} & 1500 & 46 \\
\ce{H3+} & -- & -- & \ce{HCO+} & 3.3 & -- & \ce{HCl+} & 97 & -- \\
\ce{CH} & 1100 & 580 & \ce{H2CO} & 1300 & 290 & \ce{NaCl} & 180 & -- \\
\ce{CH+} & 220 & -- & \ce{NH} & 370 & 7.1 & \ce{PH} & 720 & -- \\
\ce{CH2} & 290 & -- & \ce{NH+} & 22 & -- & \ce{PH+} & 90 & -- \\
\ce{CH2+} & 89 & -- & \ce{NH2} & 720 & 140 & \ce{AlH} & 54 & 150 \\
\ce{CH3} & 280 & 380 & \ce{NH3} & 1100 & 220 & \ce{LiH} & 620 & -- \\
\ce{CH4} & 1500 & 22 & \ce{N2} & 39\myfootnotemark{3} & -- & \ce{MgH} & 250 & -- \\
\ce{CH4+} & 270 & -- & \ce{NO} & 300 & 240 & \ce{NaH} & 930 & -- \\

\bottomrule
  \end{tabular}
  \footnotetext[1]{Assumes a cosmic ray ionisation rate of $\zeta_{\ce{H2}}$ = \np[s^{-1}\,\ce{H2}{^{-1}}]{e-16}. A simple scaling recovers photodestruction rates for other values of $\zeta_{\ce{H2}}$.}
  \footnotetext[2]{Assumes an abundance relative to H-nuclei of $x_{\ce{CO}}=\np{e-5}$ and significant self-shielding. Rate neglecting self-shielding: \np[s^{-1}]{9.9e-15}.}
  \footnotetext[3]{Assumes an abundance relative to H-nuclei of $x_{\ce{N2}}=\np{e-5}$ and significant self-shielding. Rate neglecting self-shielding: \np[s^{-1}]{1.2e-14}.}
\end{centering}
\end{minipage}
\end{table*}

\subsection{Results}

Photodissociation and ionisation rates in a cosmic-ray induced ultraviolet field were calculated for all molecules in our database after adopting the following set of parameters:
\begin{center}
  \begin{tabular}{r@{\hspace{0.5em}}l}
    Cosmic-ray ionisation rate ($\zeta_{\ce{H2}}$):& \np[s^{-1}\,\ce{H2}^{-1}]{e-16}\\
    Doppler broadening ($b$): & \np[km\,s^{-1}]{1} \\
    ortho-\ce{H2}:para-\ce{H2}: & 0:1 \\
    $x(\ce{H})$: & \np{e-4} \\
    $x_{\ce{N2}}$ and $x_{\ce{CO}}$: & \np{e-5} \\
 \end{tabular}
\end{center}
\change{where $x$ is the abundance relative to H-nuclei.}

The integration in Eq.~(\ref{eq:cosmic ray rate}) was performed on a
wavelength grid with 0.001\,nm (or finer) resolution, in order to
capture full details of the cross section structure of absorbing species.
The calculated photodissociation and photoionisation rates are
presented in Table~\ref{tab:cosmic ray rates table}.  
The ISRF-weighted uncertainties listed in Table~\ref{tab:cross_section_properties} provide a reasonable uncertainty estimate for these rates.

\begin{figure*}
  \centering
  \includegraphics{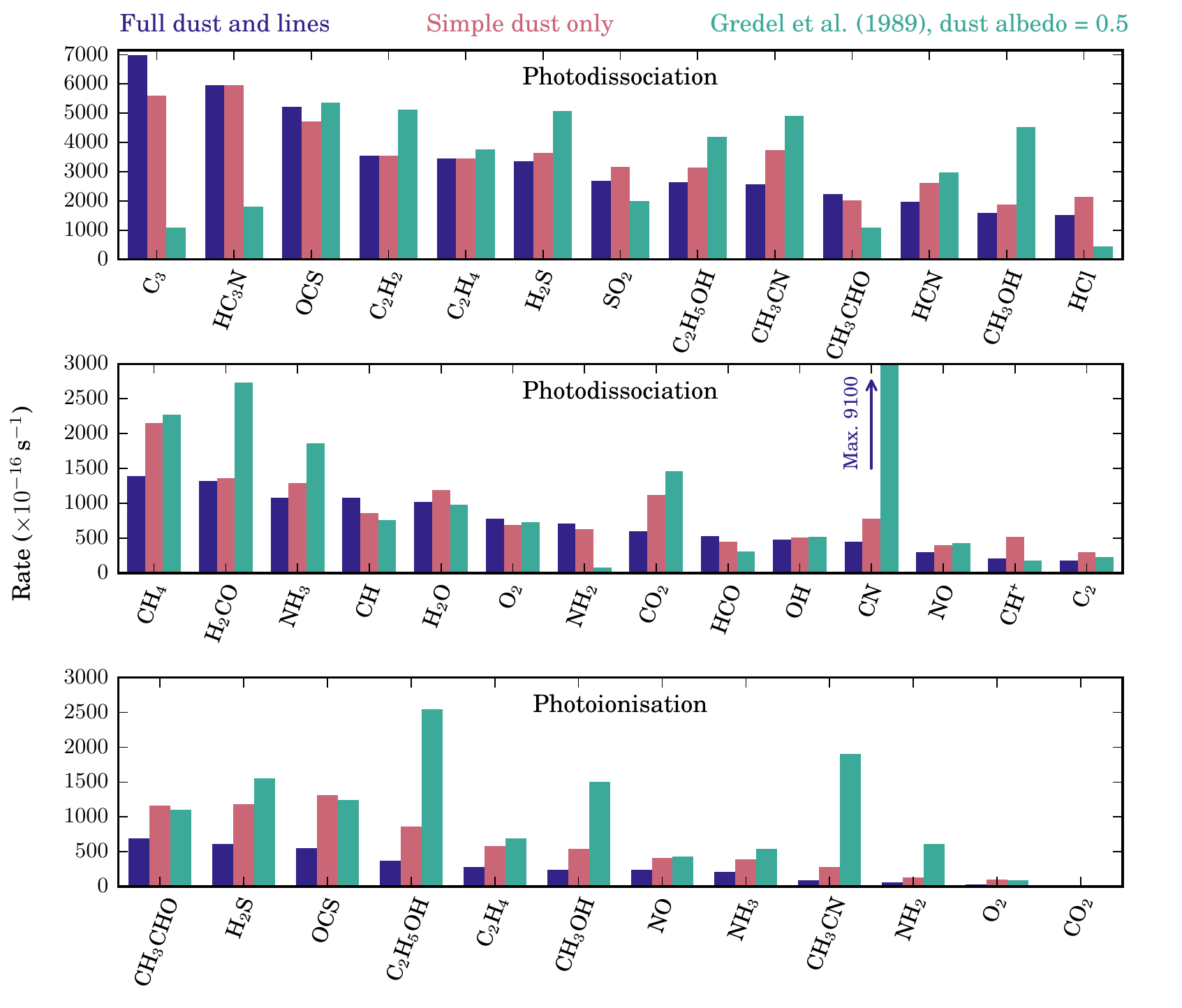}
  \caption{\change{Molecular photodissociation and ionisation rates due to cosmic-ray-induced radiation. Shown for all species common to this data base and \textcite{gredel1989}. Full dust and lines: Including all available shielding cross sections. Simple dust only: No molecular or atomic shielding, wavelength independent dust extinction cross section \np[cm^2\,\ce{H}^{-1}]{2e-21} and an albedo of 0.5. Gredel et al.: Rates computed from the efficiencies of \textcite{gredel1989} (Table 1 , column ``a'', ignoring reabsorption) following multiplication by \np{e-16}.}}
  \label{fig:selection of CR rates}
\end{figure*}

A comparison of rates for molecules common to our database and that of \textcite{gredel1989} is plotted in
Fig.~\ref{fig:selection of CR rates}.
Also shown are rates calculated while neglecting the shielding of radiation by \ce{H2} and H, and assuming a constant dust absorption cross section, \np[cm^2\,\ce{H}^{-1}]{e-21}.
These changes alter the rates in our database by factors between 0.5 and 5, the increase occurring for species absorbing mostly shortwards of 110\,nm where the \ce{H2} photoabsorption is greatest and the wavelength-dependent dust absorption cross section is also greater than its average.
These ``simple dust'' rates in
Fig.~\ref{fig:selection of CR rates} are in line with the assumption
adopted by \textcite{gredel1989}, assuming a dust grain albedo of 0.5.
Then, any further differences between rates is due to the change in our 
photodissociation cross sections relative to the previous work.

Overall, agreement is within a factor of two, with some exceptions.
The largest rate we find is for the dissociation of \ce{C3}, with a 5-fold
increase relative to \textcite{gredel1989}, due to the addition of a previously-unknown
and strong \ce{C3} absorption channel at 160\,nm as computed by
\textcite{van_hemert2008} and seen in the laboratory by \textcite{monninger2002}.
The new \ce{CH3OH} dissociation and ionisations rates are 2.5 times
smaller than that calculated by \textcite{gredel1989}, who employed
very similar cross sectional data \cite{harrison1959,salahub1971} over
the range 120 to 215\,nm.  Additional information described in
Sect.~\ref{sec:CH3OH} has permitted us to extend this cross section
shortwards to 66\,nm, without which our cosmic-ray
photodestruction rates would actually be substantially smaller.  The difference between our calculation and
Gredel et al.\ is therefore unresolved.
The new \ce{HC3N} rate is 3 times larger than that
calculated by Gredel et al.\ employing the cross section of
\textcite{connors1974}, which is nonetheless very similar to the data collected
here, which we also are unable to immediately explain.

The three-times-increased \ce{HCl} rate is due to the addition
of higher-lying photodissociating transitions (described in
Sect.~\ref{sec:HCl}) than included in the calculation of
\textcite{van_dishoeck1982b}, adopted previously.  A similar
explanation leads to the large change in the estimated \ce{NH2}
dissociation and ionisation rates.
Finally, the CN photodissociation rate was reduced by more than
10 times relative to Gredel et al., who adopted a earlier theoretical cross section \cite{lavendy1984} for this molecule that has since been updated (see Sect.~\ref{sec:CN} and \textcite{lavendy1987}).

Similar to the conclusions of Sect.~\ref{sec:shielding functions}, self-shielding of the cosmic-ray flux was only found to be important for species with highly-structured cross sections and high abundance, that is, \ce{N2} and CO 
\cite{heays2014a,gredel1987}.
The photodissociation rates of these species are reduced through self-shielding by about 50\% after assuming abundances relative to H-nuclei of \np{e-5}, as is typical for diffuse and dense interstellar clouds and prestellar cores \cite{tielens2013}, as long as the dust temperature is sufficiently high to prevent condensation of CO and \ce{N2} onto dust grains, that is greater than about 25 and 20\,K, respectively.

A relative abundance of about \np{e-7} is predicted for \ce{H2O}, OH, \ce{CO2}, \ce{NH3}, and  \ce{CH4} in some specific models of prestellar cores and dense clouds, for example, \textcite{tielens2013}, with the abundance all other species being \np{e-8} or below.
No species in our database achieves a self-shielding effect of more than 4\%  at the \np{e-7} abundance level.

\change{The assumed H, CO, and \ce{N2} abundances contribute to line-shielding of the cosmic ray induced UV flux and subsequent reduction of the photoabsorption rates of other species.
These abundances are, however, dependent on the cosmic-ray ionisation rate itself through the induced chemistry and other dynamical factors like temperature and evolutionary age.
The rates calculated here should then be considered conditional on the assumed abundances.
To test the severity of this assumption, neglecting line shielding entirely in favour of pure dust absorption increased the calculated photodissociation and ionisation rates by less than a factor of 2, apart from the self-shielding cases of \ce{N2} and CO that are treated in detail elsewhere \cite{gredel1987,heays2014a}.}

\section{Further discussion}
\label{sec:discussion}

\subsection{Effect of dust-grain properties on ultraviolet shielding}
\label{sec:dust grain properties}

The dominant absorber of ultraviolet radiation from interstellar,
stellar, or cosmic-ray sources is dust.  More radiation will be
available for gas-phase photodestruction if the dust mass is reduced.
\change{The ratio of gas-mass to dust-mass adopted in our calculations, 124, is taken in line with the Milky Way dust model of Draine et al.\footnote{\url{www.astro.princeton.edu/~draine/dust/dustmix.html}} \cite{draine2003a}}, and is somewhat larger
than the frequently-used value of 100.  This parameter is also
estimated from observations of local group Galaxies, and found to vary
between about 50 and 500 \cite{draine2007,leroy2011}, and on small
scales within the Milky Way, in one case falling to 88 and rising above 200 within one
star-forming region \cite{liseau2015}, although any such
determinations have large inherent uncertainties. Finally, low
``metallicity'' (using astronomical terminology) galaxies such as the
Magellanic Clouds have higher gas-to-dust ratios by up to an order of
magnitude \cite{roman-duval2014}.

The ultraviolet opacity of dust is also reduced by its coagulation
into larger sizes \cite{mathis1977,draine1984,van_dishoeck2006}.  This
phenomena is observed
\cite[e.g.,][]{li2003a,kessler-silacci2006,Testi14} and predicted
\cite[e.g.,][]{birnstiel2010} to occur in protoplanetary disks, at
least up to cm-sized grains, with the larger grains settling to the
midplane and drifting towards the central star. Disks also show additional complex
spatial structures seen in small and large grains being influenced by
planetary bodies or dynamical instabilities
\cite[e.g.,][]{johansen2007,andrews2011,muto2012,van_der_marel2013,pinilla2015}.

The dust shielding functions plotted in Figs.~\ref{fig:dust shielding} are shifted to higher gas column densities when their ultraviolet opacity is reduced by grain growth.
The scaling between 550\,nm visual extinction, $A_\text{V}$, and ultraviolet extinction also changes due to the wavelength-dependent variation of all dust grain optical properties when their size distribution and composition is modified \cite{mathis1977}.
\textcite{van_dishoeck2006} simulated the dust optical properties of a large-grain populated protoplanetary disk, designed to best reproduce observations of the dust disk around HD~141569A \cite{li2003a,jonkheid2006,van_dishoeck2006}.
This required the presence of dust grains as large as a few $\mu$m in diameter.
Figure \ref{fig:grain optical properties} compares the optical properties of this large-grain population with standard interstellar grains, showing a significantly-reduced ultraviolet absorption cross section (which matches interstellar grains at 550\,nm) and a somewhat increased albedo and forward scattering probability.
These changes all act to increase the penetration of photodestructive radiation through the disk.
We computed alternative shielding functions assuming a HD~141569A dust population and compare these with the interstellar dust case in Fig.~\ref{fig:dust shielding}.
The larger grains require nearly an order-of-magnitude greater column of dust mass before shielding effectively.
\change{This leads to correspondingly smaller $\gamma_{\text{E}_2}$-factors in Eq.~(\ref{eq:dust shielding exponential integral}), about 0.5 for most molecules, as listed in Tables \ref{tab:photodissociation rates} and \ref{tab:photoionisation rates}, as opposed to values of about 1.5 assuming interstellar type dust.}

Grain growth will also increase photodissociation and ionisation rates in a cosmic-ray induced UV field.
This may be significant in the midplane of a protoplanetary disk due to aggregation of dust in the high-density environment and the gravitational settling of large grains \cite{dalessio2001a}.
The reduced competitiveness of dust grains in the absorption of UV photons will then affect the chemistry of gas-phase molecules \cite{chaparro_molano2012a}.

\subsection{Effect of temperature on molecular cross sections}

Most laboratory measurements are recorded at room temperature, whereas
many theoretical calculations do not include any ground state
excitation whatsoever, simulating a low temperature of $<$10\,K with
$v$=0 and $J$=0.  Fortunately, the difference between cross sections
appropriate for interstellar or atmospheric excitation temperatures
(10 to about 1000\,K) and the available measurements and calculations
is largely obliterated by the wavelength integration of
Eq.~(\ref{eq:kpd}). At temperatures up to a few hundred K, multiple
rotational levels besides $J$=0 will be excited, in most cases
broadening the cross section slightly without changing its integrated
value \change{\cite[e.g.,][]{wu2000,miyake2011,li2013}}. At higher temperatures,
such as encountered in protoplanetary disks and exoplanet atmospheres,
vibrational levels of the ground electronic state start to be excited
but as long as $v \leq 2$ this again only results in a small
wavelength redistribution of the integrated cross section.  Such
a change in cross section shape is unlikely to affect ISRF photodestruction rates
but could significantly alter rates in a cool stellar radiation
field if the cross section shifts in or out of the maximum of the
stellar flux. This effect is particularly prominent for the case of
\ce{CO2} at long wavelengths, as demonstrated by \textcite{venot2013}.

\textcite{li2013} find the 1000\,K ISRF
photodissociation rate of \ce{N2} to be only 15\% larger than at
10\,K but a large temperature dependence for its self-shielding
and shielding by H$_2$, with factor-of-10 decreases of shielding effectiveness between 10 and 1000\,K.
A factor-of-two temperature variability of the \ce{N2} cosmic-ray induced photodissociation rate was also noted \cite{heays2014a} between 10 and 300\,K.
The temperature sensitivity of \ce{N2}, and also CO \cite{visser2009} and \ce{H2} \cite{sternberg2014}, arises from their rotational-line dominated spectra where linewidths are set mostly by the Doppler broadening.
A higher kinetic temperature leads to more Doppler broadening and a lesser cross section at the line centres, reducing the effectiveness of self-shielding.
And, a higher excitation temperature spreads the cross section over a greater number of rotational lines, also impeding self-shielding. 
A more subtle effect occurs for some predissociating molecules where the dissociation efficiency, $\eta^{\rm d}$, increases with excitation temperature (this can happen because of increased centrifugal mixing of excited states for faster rotation molecules \cite[e.g.,][]{lewis_etal2005b,eidelsberg2014}).

\subsection{Uncertainty of calculated line positions and widths}
\label{sec: uncertainty of calculated line positions and widths}

As noted in Sect. \ref{sect:theory}, theoretical cross sections are
sometimes calculated as vertical excitations from the ground to one or
more excited states, with each electronic transition  summarised by a single line
\cite[e.g.,][]{van_hemert2008}.  Then the assumption of a linewidth is required.
The ISRF photodestruction rates calculated in Sect.~\ref{sec:photo rates} are not sensitive to the
precise choice of width, however, the rates of cosmic-ray
induced processes in Sect.~\ref{sec:cosmic rays} as well as rates for
highly structured radiation fields such as Lyman-$\alpha$ are somewhat
sensitive.  For the cosmic-ray induced case, experimentation with
synthetic lines suggests that calculated rates averaged over 10 lines
and assuming 1\,nm linewidths will be accurate to within about a
factor of 2 should the real linewidths fall in the range 0.1 to
10\,nm.

\begin{figure}
  \centering
  \includegraphics{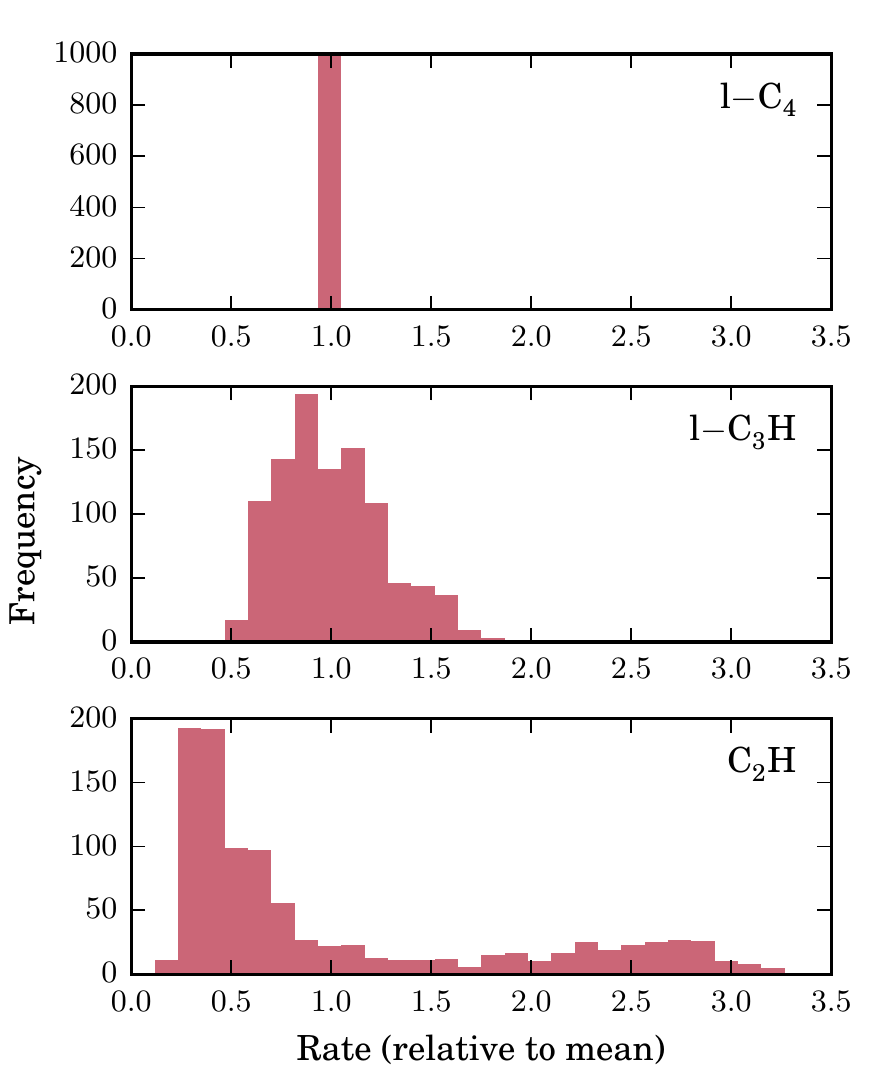} 
  \caption{Distribution of Monte Carlo simulated cosmic-ray-induced photodissociation rates of three molecules. Normalised to the mean of the distribution.}
  \label{fig:monte_carlo_example_distributions}
\end{figure}
Another test was performed to determine the sensitivity of cosmic-ray induced photodissociation rates to uncertain vertical excitation energies.
Rates were calculated for all molecules in our database that include vertical transitions (primarily those calculated by \textcite{van_hemert2008}) in a Monte Carlo simulation.
For this, the wavelength positions of all transitions were repeatedly and independently adjusted by an amount falling within a uniform probability distribution bounded by the assumed vertical-transition energy uncertainty $\pm 0.2$\,eV ($\pm 3.6$\,nm at 150\,nm).
Each line has an assumed intrinsic FWHM of 1\,nm.
In this way,  the possible range of overlap between absorption features and cosmic-ray induced emission lines was sampled.
The resultant distribution of photodissociation rates for three molecules showing quite different sensitivities is shown in Fig.~\ref{fig:monte_carlo_example_distributions}.
Line-position uncertainty does not affect the photodissociation rate for \ce{$l$-C4} because the calculated absorption for this molecule occurs longwards of 170\,nm and in the continuum part of the cosmic-ray induced spectrum.
Whereas, the photodissociation of \ce{$l$-C3H} is sensitive to the overlap of several of its transitions near to the strong \ce{H2} emission lines around 160\,nm.
The proximity of \ce{C2H} vertical transition to the Lyman-$\alpha$ wavelength means a small uncertainty in this transition energy leads to a large variation in photodissociation rate.
All other molecules tested have distributions with widths falling within the range of Fig.~\ref{fig:monte_carlo_example_distributions}, and these were considered when assigning the rate uncertainties listed in Table \ref{tab:cosmic ray rates table}.

\subsection{Effect of unresolved structure on molecular cross sections}
\label{sec:unresolved structure}

The spectral resolution of laboratory equipment used to record
photoabsorption cross sections is sometimes insufficient to completely
resolve molecular vibrational-rotational structure.  For example, our
assembly of \ce{H2O} data incorporates very high-resolution
measurements of \textcite{fillion2003,fillion2004} and taken from the CfA VUV database\footnote{\url{www.cfa.harvard.edu/amp/ampdata/cfamols.html}},
with others that do not resolve the rotational spectrum \cite{chan1993d,mota2005}.

\begin{figure}
  \includegraphics{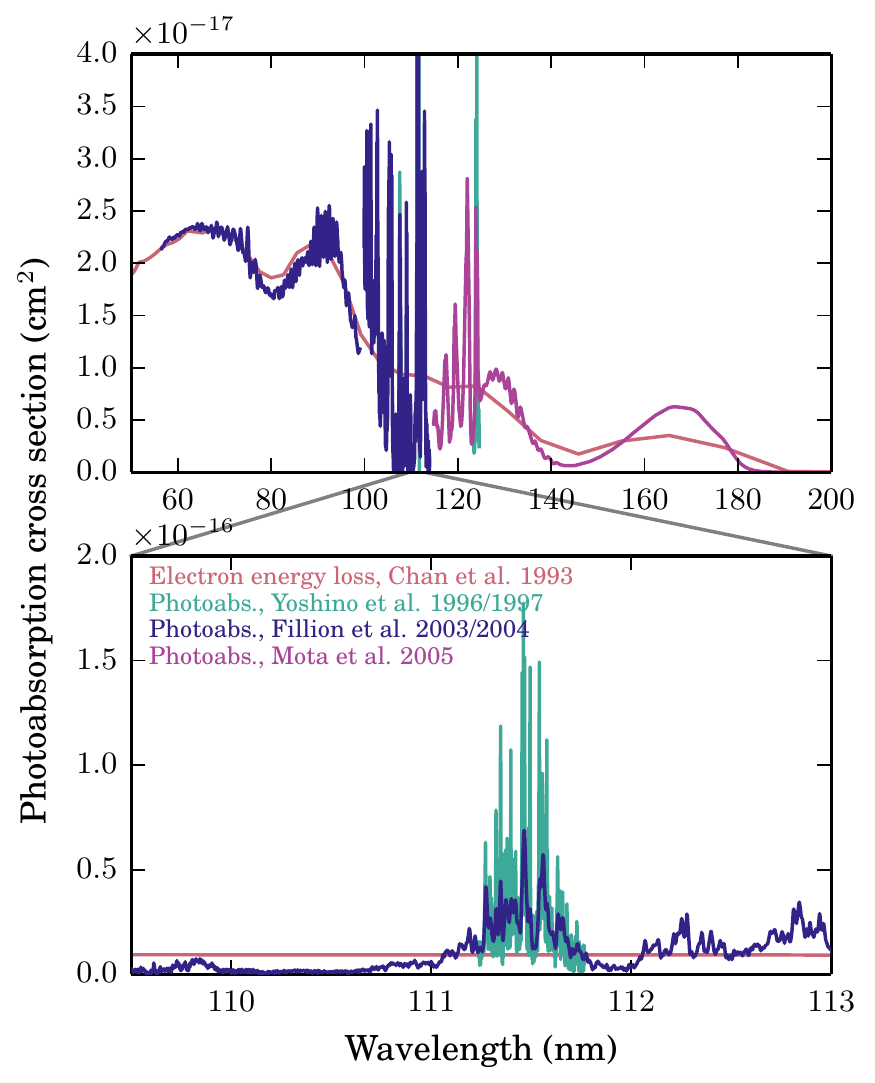}
  \caption{ \ce{H2O} photoabsorption cross section measured by low-resolution electron energy-loss spectroscopy and by direct higher-resolution photoabsorption measurements.}
  \label{fig:H2O_different_cross_sections}
\end{figure}

A comparison of \ce{H2O} spectra is plotted in
Fig.~\ref{fig:H2O_different_cross_sections}, all recorded at room
temperature.  The apparent changes between spectra are significant
despite their integrated cross sections being essentially the same.
For example, the apparent peak cross section of the vibrational band
near 111.5\,nm, whose assignment is discussed in
\textcite{fillion2004}, varies by an order of magnitude.  This
peak-value difference is still a factor of three when comparing the
two highest resolution cross sections available, CfA spectra with FWHM 0.0015\,nm and those recorded by
\textcite{fillion2004} with resolution 0.0025 to 0.005\,nm\,FWHM.

In Table~\ref{tab:H2O photodissociation rates for neglected data} we list photodissociation rates for
\ce{H2O} calculated using the highest-resolution data and neglecting this in
favour of less-resolved cross sections (which still cover the entire wavelength range).  In
most cases, any difference is negated by the smoothing effect of
wavelength integration and because the continuum part of the \ce{H2O}
cross section dominates the combined cross section of lines, despite
their large maxima.  The photodissociation rate due to Lyman-$\alpha$
dominated radiation is sharply reduced, however, when only the lowest
resolution data set is used, because of the under resolution of a
vibrational band near the Lyman-$\alpha$ wavelength.

\begin{table*}
\caption{Photodissociation rate of \ce{H2O}. These are calculated from subsets of the available experimental photodissociation cross sections. }
\label{tab:H2O photodissociation rates for neglected data}
\begin{tabular}{ccccc}
\toprule
Data sources                                                                       & Max. resolution\myfootnotemark{1} & ISRF rate\myfootnotemark{2} & Lyman-$\alpha$ rate\myfootnotemark{2} & Cosmic-ray induced              \\
(c.f. Sect.~\ref{sec:H2O} and Fig.~\ref{fig:H2O_different_cross_sections}) & (nm\,FWHM)                       & (s$^{-1}$)                 & (s$^{-1}$)                           & rate\myfootnotemark{3} (s$^{-1}$)\\
\midrule
All data                                                                           &0.0015                            &\np{7.7e-10}                &\np{2.4e-09}                          &~\np{9.8e-14}                    \\
Excluding CfA molecular database data                                              &0.0025                            &\np{7.7e-10}                &\np{2.4e-09}                          &~\np{9.8e-14}                    \\
Also excluding Fillion et al.                                                      &0.075                             &\np{8.3e-10}                &\np{2.4e-09}                          &\np{10.0e-14}                    \\
Also excluding \textcite{mota2005}                                                 &10                                &\np{8.4e-10}                &\np{1.3e-09}                          &~\np{8.4e-14}                    \\
\bottomrule
\end{tabular}
\begin{minipage}{\linewidth}
  \footnotetext[1]{The maximum resolution within the combined data set.}
  \footnotetext[2]{Calculated as in Sect.~\ref{sec:photo rates}.}
  \footnotetext[3]{Calculated as in Sect.~\ref{sec:cosmic rays}.}
\end{minipage}
\end{table*}

The photodissociation cross sections of CO, \ce{N2}, and \ce{H2} have
no continuum absorption at all and will be sensitive to insufficient
experimental resolving power.  For this reason, these molecules are
treated here with models that recreate each absorption line and its
lineshape from experimental data, but without the influence of
experimental instrumental broadening
\cite{visser2009,li2013,abgrall_etal1993}.  We can then directly test
their sensitivity to the resolution-dependent phenomenon of
self-shielding by broadening their cross sections though convolution
with Gaussian functions of increasingly greater width.

\begin{figure}
  \centering
  \includegraphics{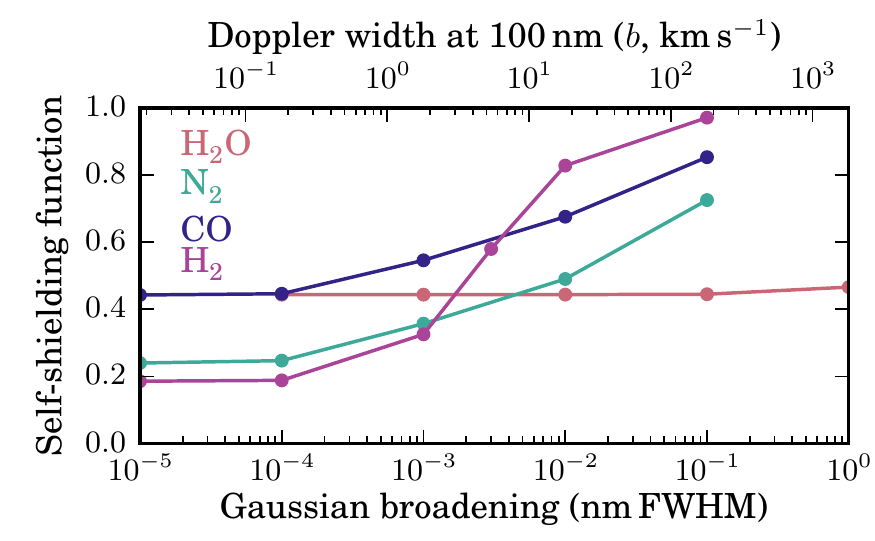}
  \caption{ISRF self-shielding factors. Shown for four molecules after convolution of their photodissociation cross sections by Gaussian functions of increasing width (equivalent Doppler widths are also shown).
    For these calculations, a column density of \np[cm^{-2}]{e15} was assumed for \ce{N2}, CO and \ce{H2}; and \np[cm^{-2}]{e17} for \ce{H2O}.
  }
  \label{fig:resolution_effect_on_self_shielding}
\end{figure}

The results of such a test are shown in
Fig.~\ref{fig:resolution_effect_on_self_shielding}, the shielding
effect decreases significantly (shielding function approaches unity)
when absorption lines of \ce{N2}, CO, and \ce{H2} are artificially broadened.
From this figure, even the use of a 0.005\,nm resolution cross section may incompletely-resolves and underestimate the self-shielding effect for a line-dominated molecular cross section spectrum.

Most species in our database are sufficiently continuum-dominated that any under-resolution will not influence their calculated photodestruction rates.
For example, we also systematically broaden the highest-resolution experimental spectrum of \ce{H2O} and find a negligible effect in Fig.~\ref{fig:resolution_effect_on_self_shielding}.
The experimentally-determined cross section of NO may benefit from the measurement of a higher resolution cross section between 130 to 165\,nm, particular if this molecule can attain a high column density in an atmosphere.
Some radical species, for example, \ce{NH2} or \ce{C2}, likely also have unquantified sharply resonant features.  

Turbulent Doppler broadening will actually reduce self-shielding in the way that experimental under-resolution mimics.
This is shown in Fig.~\ref{fig:resolution_effect_on_self_shielding} where a Doppler width of \np[km s^{-1}]{10} is significant at least for the case of \ce{H2}.
Larger Doppler widths are unlikely to occur in astrochemical environments except in locally shocked regions.

\subsection{Isotopic effects}
\label{sec:isotopic effects}

Isotopic substitutions of one or more atoms within a molecule
alter its rotational inertia and the reduced masses of its vibrational
modes \cite[e.g.,][]{herzberg1989}.  The resultant shifts in
rotational-vibrational energy levels may be different for ground and
excited states so that the wavelengths of lines in its photoabsorption
spectrum also shift.

For diatomic molecules, the largest shifts are expected when
deuterating \ce{H2} to form HD, a reduced mass (in atomic units)
increase from $(1\times 1)/(1+1)$ to $(1\times 2)/(1+2)$.  The
resulting vibrational-level shifts for the astrophysically-important
$B\,{}^1\Sigma^+_u$ and $C\,{}^1\Pi_u$ states of \ce{H2} can be as
large as \np[cm^{-1}]{1300}, equivalent to a 0.8\,nm difference
between the absorption line wavelengths of \ce{H2} and HD.  This shift
is large enough that \ce{H2} no longer shields HD, so that the
photodissociation rate of HD with depth into a cloud is much larger
than that of \ce{H2} \cite{spitzer1973,black1977}. Isotopic
substitution of a heavier element leads to smaller shifts.  For
example, the substitution of the minor isotope \ce{^{15}N} into
molecular nitrogen, \ce{N2}, leads to changes in energy levels
and absorption line wavelengths of at most \np[cm^{-1}]{200} and
0.15\,nm, respectively.  This is still more than enough to maintain
the self-shielding phenomenon for \ce{N2}, but reduce the mutual
shielding of \ce{^{14}N ^{15}N} by \ce{^{14}N2} \cite{heays2014a}.

\begin{figure}
  \includegraphics{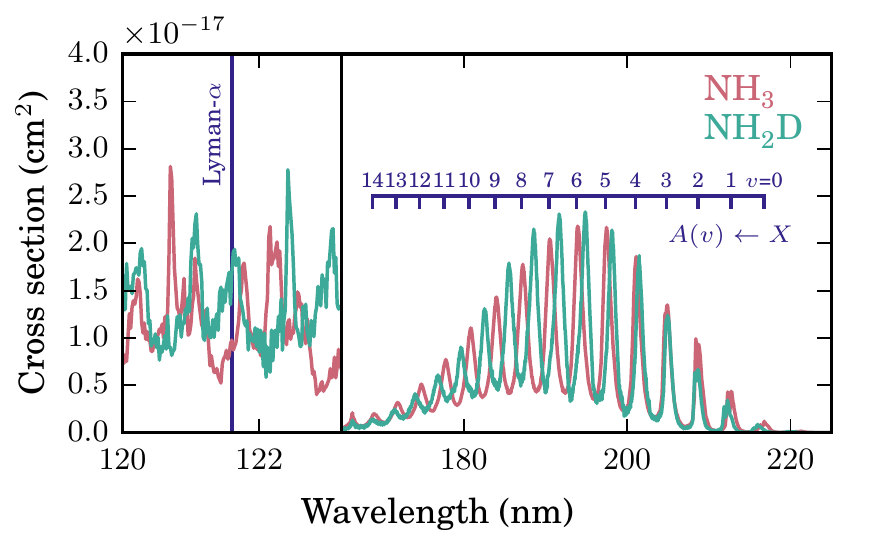}
  \caption{The long-wavelength and near-Lyman-$\alpha$ parts of the
    \ce{NH3} and \ce{NH2D} photoabsorption cross sections. The source
    of the \ce{NH3} cross section is discussed in Sect.~\ref{sec:NH3}
    and the \ce{NH2D} data are taken from two sources
    \cite{wu2007,cheng2006}. Also indicated is the wavelength of
    Lyman-$\alpha$ emission and the vibrational progression of
    absorption into the excited $\tilde A$ electronic state of \ce{NH3}.}
  \label{fig:isotopic_subsitution_cross_sections_NH3}
\end{figure}

The deuteration of \ce{NH3} to \ce{NH2D} leads to wavelength shifts of
its ${\tilde A}\leftarrow {\tilde X}$ absorption bands, shown in
Fig.~\ref{fig:isotopic_subsitution_cross_sections_NH3}, with larger shifts
for higher vibrational levels of the $\tilde A$ state.
Most important is the isotope-induced shift of higher-energy excited
states into resonance with the Lyman-$\alpha$ emission line.  The
nearly two-times larger cross section of \ce{NH2D} at 121.6\,nm leads
directly to an increased photodissociation rate of this species in a
Lyman-$\alpha$ dominated radiation field.

The Lyman-$\alpha$ resonance of \ce{C2H2} is even more critical than
for \ce{NH3}, as shown in Fig.~\ref{fig:C2H2_HC3N_Lyalpha}.  A small
isotopic shift following deuteration will lower the dissociation rate
of \ce{C2HD} in a Lyman-$\alpha$ dominated radiation field by a factor
of 10.  This reduction is in fact indicated from laboratory action
spectroscopy of \ce{C2H2} and \ce{C2HD} \cite{loffler1996,wang1997}.

The photodissociation efficiency or branching may also be sensitive to isotopic substitution.
For example, the deuteration of water to form HDO introduces the possibility of preferential branching to form OH or OD photodissociation fragments and imposing a significant influence on the ratios of \ce{H2O}, HDO, and \ce{D2O} found in interstellar space, comets, and terrestrially \cite{caselli2012,van_dishoeck2013b}.
\change{The propensity for H--OD fission is determined in laboratory measurements at a few UV wavelengths \cite[e.g.,][]{shafer1989,plusquellic1998} and theoretical calculations \cite[e.g.,][]{engel1988,zhang1989,zhou_linsen2015}, together indicating a wavelength and temperature dependent ratio of OD or OH fragments between 2 and 16 (or higher).}

\subsection{Photodissociation branching of  \ce{H2O} and \ce{NH3}}
\label{sec:H2O NH3 branching}

\begin{table*}
  \caption{Photodissociation branching of \ce{H2O} and \ce{NH3}.}
  \label{tab:partial_rates_H2O_NH3}
  \small
  \centering 
  \begin{tabular}{l@{\qquad}c@{\quad}c@{\quad}c@{\qquad}c@{\quad}c@{\quad}c@{\quad}}
    \toprule
    \smallskip
    Radiation field\myfootnotemark{1}& Rate\myfootnotemark{2}                                       & Frac.\myfootnotemark{3}                                         & Unc.\myfootnotemark{4}& Rate       & Frac.& Unc.\\
    \midrule
    \\
                                     & \multicolumn{3}{c}{\ce{H2O} $\rightarrow$ OH + H}~~~~~~      & \multicolumn{3}{c}{\ce{H2O} $\rightarrow$ O + 2H/\ce{H2}}~~~~~~ \\
    \\
    ISRF                &\np{5.9e-10}&0.77&B&\np{1.8e-10}& 0.23&B\\
    Mathis\,'83         &\np{4.0e-10}&0.75&B&\np{1.3e-10}& 0.25&B\\
    Black Body 4000\,K   &\np{1.6e-10}&0.99&A&\np{2.3e-12}& 0.01&C\\
    Black Body 10\,000\,K&\np{4.4e-10}&0.93&A&\np{3.3e-11}& 0.07&B\\
    Lyman-$\alpha$      &\np{1.8e-09}&0.74&A&\np{6.2e-10}& 0.26&A\\
    Solar               &\np{2.0e-10}&0.88&A&\np{2.7e-11}& 0.12&B\\
    TW-Hydra            &\np{1.2e-09}&0.76&A&\np{4.0e-10}& 0.24&A\\
    Cosmic-ray induced  &\np{7.6e-14}&0.77&B&\np{2.2e-14}& 0.23&B\\
    \\
                                     & \multicolumn{3}{c}{\ce{NH3} $\rightarrow$ \ce{NH2} + H}~~~~~~& \multicolumn{3}{c}{\ce{NH3} $\rightarrow$ NH + 2H/\ce{H2}}~~~~~~\\
    \\
    ISRF                &\np{8.3e-10}& 0.58&B&\np{6.1e-10}& 0.42&B\\     
    Mathis\,'83         &\np{5.6e-10}& 0.57&B&\np{4.3e-10}& 0.43&B\\
    Black Body 4000\,K   &\np{3.6e-09}& 1.00&A&\np{2.3e-13}& 0.00&C\\     
    Black Body 10\,000\,K&\np{1.6e-09}& 0.95&A&\np{9.0e-11}& 0.05&B\\     
    Lyman-$\alpha$      &\np{6.6e-11}& 0.05&B&\np{1.3e-09}& 0.95&A\\     
    Solar               &\np{4.0e-09}& 0.99&A&\np{5.6e-11}& 0.01&C\\     
    TW-Hydra            &\np{4.9e-10}& 0.35&A&\np{9.1e-10}& 0.65&A\\     
    Cosmic-ray induced  &\np{3.9e-14}& 0.39&B&\np{6.2e-14}& 0.61&B\\
    \bottomrule
  \end{tabular}
  \begin{minipage}{\linewidth}
    \footnotetext[1]{Radiation fields are defined and normalised as for Tables \ref{tab:photodissociation rates} and \ref{tab:cosmic ray rates table}.}
    \footnotetext[2]{In units of s$^{-1}$.}
    \footnotetext[3]{Estimated branching fraction between all channels producing OH and O (or \ce{NH2} and NH) regardless of excitation state or the chemical co-fragment.}
    \footnotetext[4]{Estimated rate uncertainties: accurate to within 30\% (A), a factor of 2 (B), a factor of 10 (C).}
  \end{minipage}
\end{table*}

We estimated the wavelength dependent branching ratios of \ce{H2O} and \ce{NH3} into their main astrochemically-significant photodissociation products OH and O, and \ce{NH2} and NH, respectively.
A discussion of these cross sections is given in Sects.~\ref{sec:H2O} and \ref{sec:NH3}, and the wavelength dependent partial cross sections shown in Figs.~\ref{fig:partial_cross_sections_H2O} and \ref{fig:partial_cross_sections_NH3}.
We calculated the partial photodissociation rates generating these products when exposed to various kinds of interstellar and cosmic-ray-generated radiation fields, with results given in Table~\ref{tab:partial_rates_H2O_NH3}.
The uncertainty estimates provided in this table are a combination of the overall cross section uncertainties of Table~\ref{tab:cross_section_properties} and an estimate of the branching-ratio accuracy.
Consideration was also made that minor branching channels will have larger fractional uncertainties than major ones.

The product branching ratios of both \ce{H2O} and \ce{NH3} are well known at the Lyman-$\alpha$ wavelength, reasonably constrained at longer wavelengths, but poorly known shortwards of 121.6\,nm.
Then, their ISRF and cosmic-ray-induced photodissociation rates are assigned a larger uncertainty than for other stellar radiation fields because these are approximately 40\% controlled by radiation shorter than Lyman-$\alpha$.
All other stellar rates are controlled by product branching ratios at Lyman-$\alpha$ or longwards.

The partial rates calculated here are in line with the previous version of the Leiden database for \ce{NH3} but lead to significantly greater production of atomic-O following the dissociation of \ce{H2O} in the ISRF, 23\% versus 6\%.
This is due to the product branching calculations of \cite{van_harrevelt2008} prompting the estimation of a higher O branching for wavelengths shorter than Lyman-$\alpha$.
This assumption is however quite uncertain.
The ratio of partial rates of \ce{NH3} products due to cosmic ray photodissociation is $\ce{NH2}/\ce{NH} = 0.6$ and  has decreased relative to the calculation of \textcite{gredel1989}, finding a ratio of 2.4.

\section{Updated astrochemical reaction network}

The significance of the new and updated rates listed in Tables \ref{tab:photodissociation rates}, \ref{tab:photoionisation rates}, and \ref{tab:cosmic ray rates table} are investigated by trialling a set of astrochemical models.
These single-point time-dependent gas-phase models use the integration program of \textcite{walsh2009,walsh2010} to trace the chemical evolution of a range of atoms and molecules assuming a constant temperature, density, and visual extinction; and a set of initial abundances.  
For simplicity, these models are restricted to pure gas-phase chemistry, apart from the inclusion of grain-surface-mediated \ce{H2} formation, and neglect the self-shielding of CO and \ce{N2}.

We adopt the RATE12 reaction network \cite{mcelroy2013} in a fiducial model, and then modify it by substituting or adding our newly-calculated photodissociation and photoionisation rates.
This required the updating of 111 ISRF photodissociation and ionisation rates, and 82 cosmic-ray induced rates; and adding 40 new rates for species in the network that lacked one or more photodestruction processes.
Alternative networks were also constructed that replace the ISRF rates with those appropriate for black-body fields of various temperatures, and the simulated flux from TW-Hydra, described in Sect.~\ref{sec:radiation fields}.

\begin{table}
  \caption{Physical parameters of interstellar cloud point models.}
  \label{tab:cloud_params}
  \centering
  \begin{tabular}[c]{ccc}
    \hline
    & Translucent & Dark   \\
    \hline
    Temperature (K)                                    & 100         & 10 \\
    Density (\np[cm^{-3}]{})                           & \np{e3}     & \np{e4}\\
    $A_\text{V}$ (mag.)                                & 1           & 100    \\
    $\zeta_{\ce{H2}}$ (\np[s^{-1} \ce{H2}^{-1}]{e-17}) & 1.3         & 1.3--100\\
    \hline
  \end{tabular}
\end{table}
Two kinds of models were run, with their important parameters listed in Table \ref{tab:cloud_params}.
The translucent cloud model is useful for evaluating the effect of the
updated ISRF photodestruction rates, and the dark cloud model for
isolating the effects of the new cosmic-ray induced rates.
We integrated the models until reaching chemical equilibrium, requiring about \np{e6} years of model time. 

\begin{figure*}
  \centering
  \includegraphics{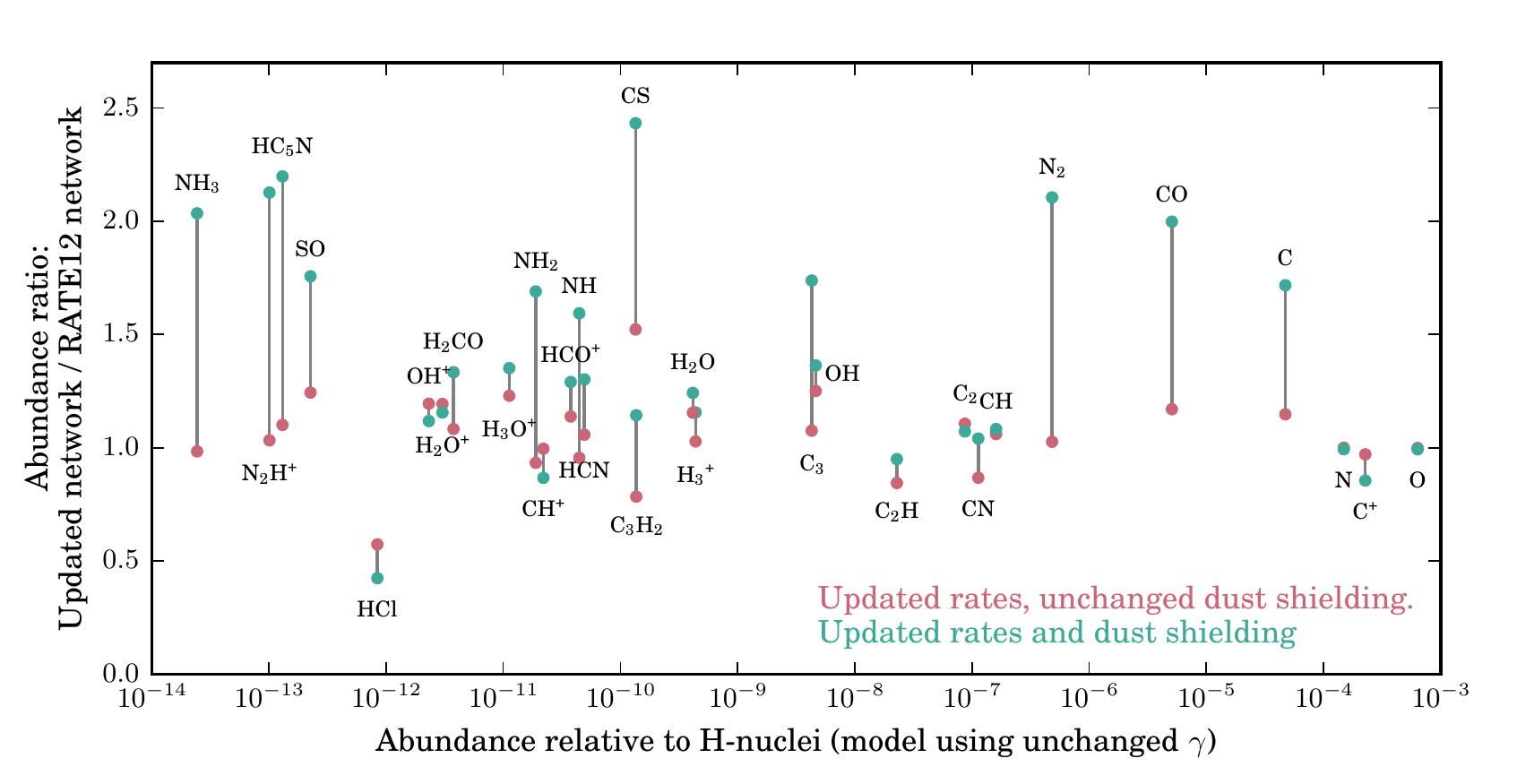}
  \caption{\change{The fractional change of chemical abundances in a translucent cloud model when substituting photodestruction rates of the RATE12 network with the newly-calculated rates.  A comparison is made with updated models including the same $\gamma_{\textrm{exp}}$ parameterisation as RATE12 for dust shielding, and with dust shielding calculated explicitly for each molecule. The plotted species are those listed in \textcite[][Table II]{tielens2013}}}.
  \label{fig:compare translucent cloud gamma}
\end{figure*}
The effects of the updated rates on model abundances at $A_\text{V}=1$ in a translucent cloud exposed to the ISRF are modest, with no abundance changes relative to the RATE12 model exceeding a factor of two.
Trialling additional pairs of models after making order-of-magnitude variations of the temperature, density, and $A_\text{V}$ in the translucent cloud model resulted in similar differences.

\change{An explicit radiative transfer (or improved parameterisation) of dust shielding is as important as the new photodestruction rates in altering the chemical model output.
The RATE12 network incorporates the $k_0\exp{(-\gamma_{\textrm{exp}} A_\text{V})}$ depth-dependent photodestruction rates of \textcite{van_dishoeck2006}.
The effect on model abundances of updating all photodestruction rates while retaining the RATE12 $\gamma_{\textrm{exp}}$ parameters is demonstrated in Fig.~\ref{fig:compare translucent cloud gamma} for a selection of atoms and molecules with $A_\text{V}=1$.
Also shown is a model with updated rates and an explicit radiative-transfer calculation simulating their reduction at a depth of 1\,$A_\text{V}$ (as in Sect.~\ref{sec:shielding functions}).
The addition of updated dust shielding leads to an increased abundance for most species by up to a factor of two after chemical equilibrium is reached.

The $\gamma_{\textrm{exp}}$ parameters used in RATE12 generally underestimate the shielding effect of a semi-infinite slab of dust at an extinction of 1, and overestimate it at higher extinction, as exampled in Fig.~\ref{fig:dust shielding parameterisation examples}.
The explicit radiative transfer calculation then results in lower photodestruction rates and generally-higher abundances of molecules and neutral atoms at 1\,$A_\text{V}$.
The disadvantages of the $\gamma_{\textrm{exp}}$ parameterisation can be largely avoided while retaining its computational efficiency by adopting the more realistic exponential-integral formulation (with $\gamma_{\textrm{E}_2}$ parameters in Sect.~\ref{sec:shielding by dust}).
}

\change{Some  species shown in Fig.~\ref{fig:compare translucent cloud gamma} are affected by the updated rates and dust shielding but are not included in our current database of  wavelength-dependent cross sections, for example, \ce{N2H+} or \ce{HC5N}.
The modelled changes are then due to chemical formation and destruction routes involving species that we updated.
The sensitivity of a chemical network to its input rates and the propagating influence of specific rates to unrelated model species is studied in the context of interstellar and atmospheric chemistry \cite[e.g.,][]{wakelam2010,loison2015}.}

\begin{figure*}
  \centering
  \includegraphics{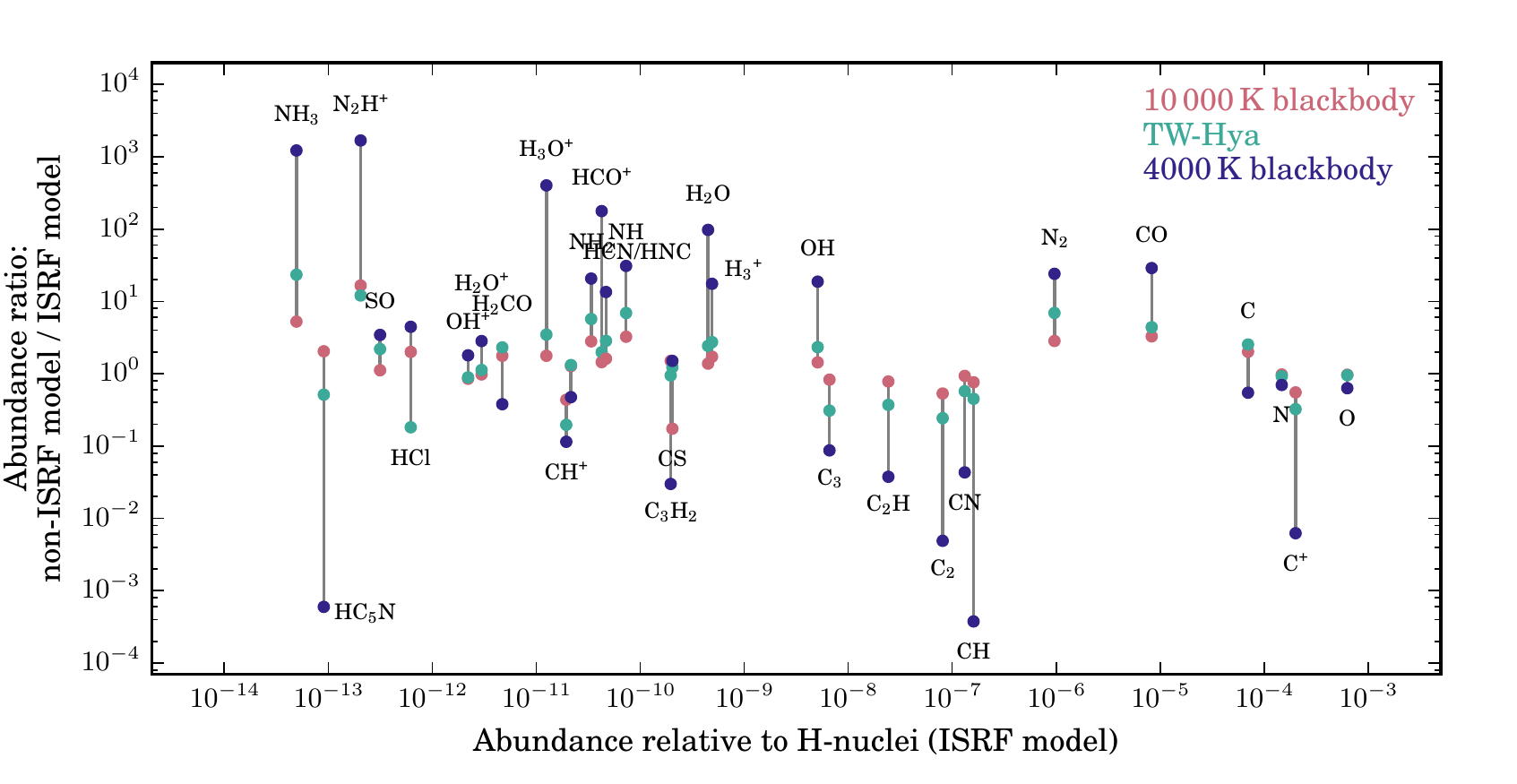}
  \caption{Comparing the chemical abundances in a translucent cloud model assuming non-ISRF radiation fields. The plotted species are those listed in \textcite[][Table II]{tielens2013}.}
  \label{fig:compare translucent cloud}
\end{figure*}
The wavelength dependence of our collected cross sections significantly alters the model output for non-ISRF ultraviolet radiation fields.
Figure~\ref{fig:compare translucent cloud} summarises the equilibrium
abundances calculated for 31 species known to compose interstellar
clouds \cite[Table~II]{tielens2013}.  The variation of these
abundances with radiation field is plotted relative to the abundances
calculated in the ISRF.  Substitution with a cooler 10\,000\,K
black-body radiation field increases the abundance of \ce{CO} and
\ce{N2} by factors of 4 and 20, respectively, because these molecules
photodissociate at relatively short wavelengths only.
The reduced occurrence of reactive C-atoms then lowers the abundance of small carbon-containing molecules by a similar factor.
On the other hand, nitrogen-containing species, for example, \ce{NH2} and \ce{N2H+}, have increased abundances.
This is due to their increased lifetime in the cooler ultraviolet flux and a formation route reliant on ion-molecule reactions with \ce{N2} and not its photodissociation \cite{walsh2015}.
Similar effects with somewhat larger magnitude follow from assuming the TW-Hydra radiation field, and with multiple-order-of-magnitude changes when assuming the extremely long-wavelength biased 4000\,K black-body field.
\change{An extreme example is the increased abundance of \ce{H3O+}, even though we did not include a direct photodestruction mechanism for this molecule in our model.
  The modelled increase is due to the decreased photoionisation of other species in the black-body radiation field, and a lowered electron abundance slowing the rate of \ce{H3O+} dissociative recombination.}

These differences are strongly-dependent on the total integrated flux
of each radiation field, which we artificially normalised as is described in
Sect.~\ref{sec:radiation fields}.  However, an alternative scaling of the integrated flux will result in similarly
divergent chemical abundances as pictured in Fig.~\ref{fig:compare
  translucent cloud} whenever photodissociation dominates molecular
destruction, due to the wavelength-dependence of atomic
and molecular cross sections: the abundances of small
carbon-bearing molecules are reduced in cool radiation fields
whereas nitrogen-containing species are enhanced.

Re-running the model while assuming weaker shielding from larger dust grains (with optical properties shown in Fig.~\ref{fig:grain optical properties}) leads to significantly more photodestruction and a reduced population of stable molecules, for example, the abundance of CO is reduced by a factor of 30, and CN by a factor of 10.
A very similar effect is achieved in our single-point steady-state model by retaining an interstellar dust opacity but reducing the model $A_\text{V}$ from 1 to 0.2.
More sophisticated 1- or multi-dimensional astrochemical models would likely distinguish between these effects, and could include the effects of \ce{H2}-shielding and self-shielding, necessary for computing a self-consistent PDR model.

\begin{figure*}
  \centering
  \includegraphics{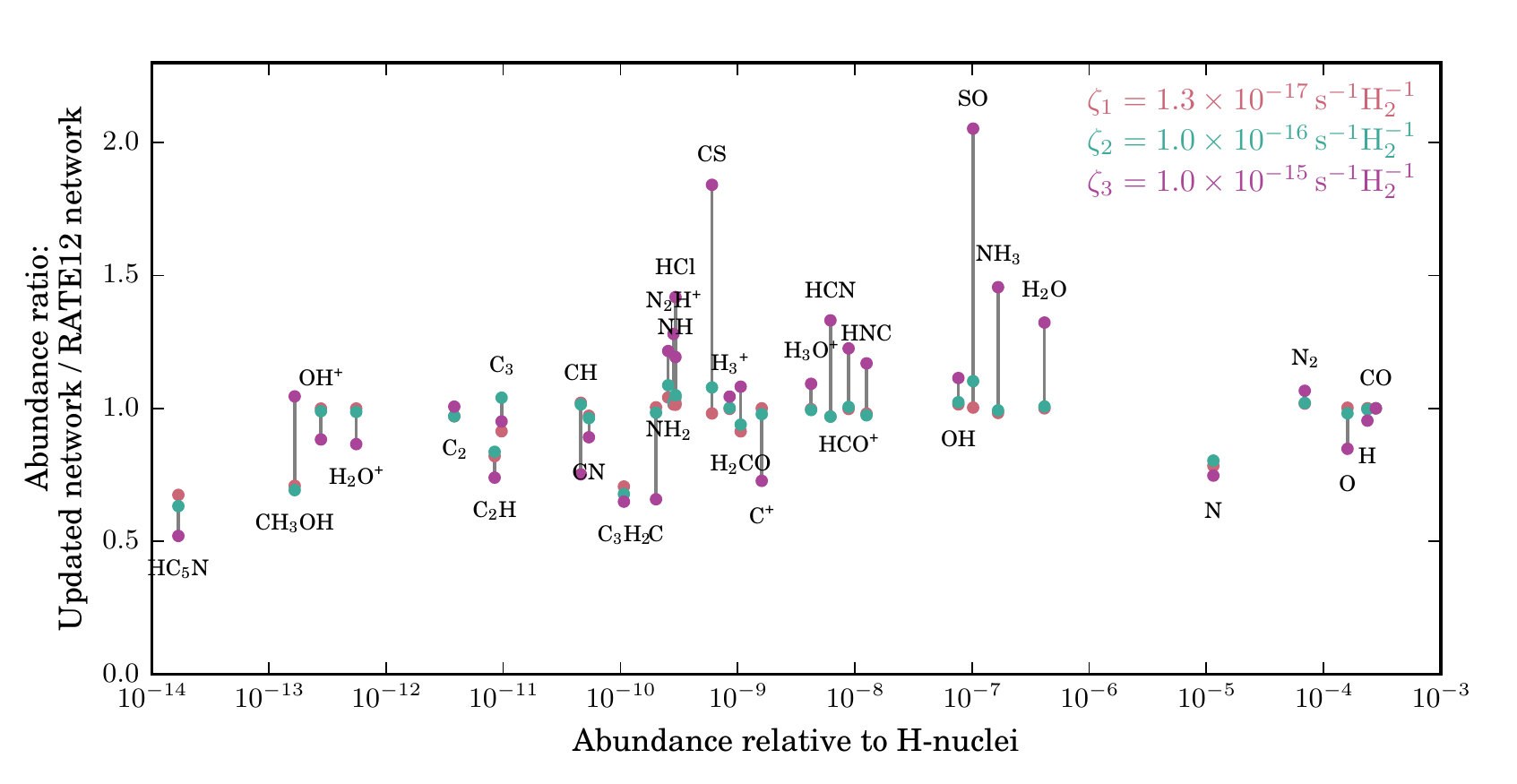}
  \caption{\change{The fractional change of chemical abundances in the dark-cloud model when substituting photodestruction rates of the RATE12 network with the newly-calculated rates.  A comparison is made for models assuming increasing ionisation rate, $\zeta$. The plotted species are those listed in \textcite[][Table II]{tielens2013}.}}
  \label{fig:compare dark cloud}
\end{figure*}
The dark cloud model was used to assess the effect of the new rates on cosmic-ray induced photodissociation and ionisation.
After comparing models using the RATE12 network and updated cosmic-ray induced photodestruction rates, and assuming a primary ionisation rate of $\zeta = \np[s^{-1} \ce{H2}^{-1}]{1.3e-17}$, no change greater than a factor of 2 was found for any species with abundance greater than \np{e-14} relative to \ce{H2}.
\change{Alternative models were run with the comic-ray induced ionisation rates of \ce{H2} and H and ultraviolet flux increased by a common factor.
Model results corresponding to $\zeta = \np{e-16}$ and \np[s^{-1} \ce{H2}^{-1}]{e-15} are also shown in Fig.~\ref{fig:compare dark cloud}.
Under the extreme model of cosmic-ray influence the effects of the updated rates calculated here are somewhat increased for some species, particularly \ce{CH3OH}, \ce{CS}, \ce{SO}, and \ce{NH3}.
Even in this case other sources of uncertainty in dark cloud chemical models may  overshadow the abundance changes engendered by our updated rates.}

\section{Summary}

  A new collection of photodissociation and photoionisation cross sections was assembled for atoms and molecules of astrochemical interest, with uncertainty estimates.
These data are used to calculate photodissociation and photoionisation rates in the ISRF and other radiation fields, including Lyman-$\alpha$ dominated radiation and a cosmic-ray induced ultraviolet flux. 
The majority of photodissociation and ionisation rates agree within 30\% when compared with other recent compilations \cite{van_dishoeck2006,huebner2015,gredel1989} (where these include comparable cross sections), with some important differences.
The reduction of these rates in shielded regions was calculated as a function of the dust, molecular and atomic hydrogen, neutral C, and self-shielding columns.
Dust opacity is generally the most important shielding effect but a comparable influence from other forms of shielding was found for some molecules, particularly if grain growth has reduced the ultraviolet absorption cross section of the dust population.

Various sensitivities of the calculated rates to the experimental and theoretical data they are derived from, or the astrophysical environments where they are applied, is given.
For most molecules, the under-resolution of resonant photoabsorption lines in experimental cross sections, small errors in the excitation electronic energies and linewidths in theoretical calculations, changes in excitation temperature, or isotopic substitution will not dramatically affect their astrochemical photodestruction.
Exceptions occur for some molecules that show a high degree of sensitivity to these details, particularly when their cross sections feature a maximum near the Lyman-$\alpha$ wavelength.

Some tests of the new rates in simple astrochemical models show sensitivity to the updated rates up to a factor of two for molecules important in translucent and dark interstellar clouds.
Additional sensitivity was shown to an improved dust-shielding parameterisation scheme that better matches the attenuation of absorbed and scattered UV light through a slab-model interstellar cloud.

The intention is to provide, along with precomputed rates and shielding functions, as detailed as possible wavelength-dependent cross sections that are suitable for use in astrochemical models of interstellar and circumstellar material that require specific treatments of photodissociation and photoionisation. 
That is, due to spatial and time variance of the ultraviolet radiation flux, temperature, density, turbulence, and dust optical properties,  which cannot be easily or comprehensively parameterised.

The cross sections and derived data for a total of 102 molecules and atoms are available from the Leiden database.\footnote{\url{www.strw.leidenuniv.nl/~ewine/photo}}
These are provided in both a binary format, explicitly including all cross section features on a dense wavelength grid, and sparser text format that is more suitable for rapid calculation in a continuum radiation field like the ISRF. 
This new database extends its previous version by the addition of 9 new species and cross sections updates for 60 more.

\begin{acknowledgements}
  Thanks to Roland Gredel for provision of his cosmic-ray induced UV spectrum, Catherine Walsh for the use of her single-point astrochemical model program, and Marc van~Hemert for ab initio calculations of CN.  
  Astrochemistry in Leiden is supported by the European Union A-ERC grant 291141 CHEMPLAN, by the Netherlands Research School for Astronomy (NOVA), by a Royal Netherlands Academy of Arts and Sciences (KNAW) professor prize, and the Dutch astrochemistry network (DAN) from the Netherlands Organisation for Scientific Research (NWO) under grant 648.000.002.
  Thanks also to the anonymous referee for many useful comments, particularly regarding the use of exponential-integral functions to represent UV-shielding by dust grains.
\end{acknowledgements}
\bibliography{article}

\begin{appendix}

\section{Shielding by atomic H and C}
\label{sec:additional shielding function discussion}

\begin{figure}
  \centering
  \includegraphics{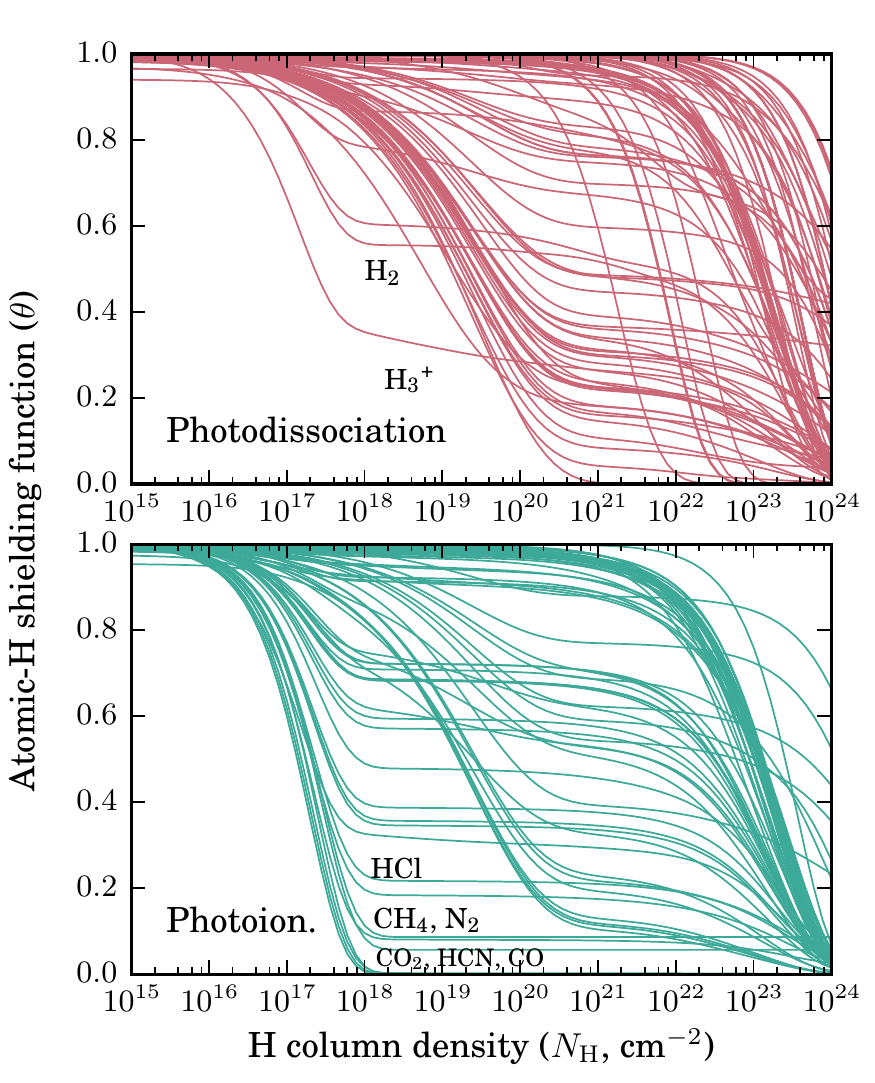}
  \caption{Shielding of photodissociation and photoionisation by
    atomic H in the simulated TW-Hydra radiation field.}
  \label{fig:H_shielding}
\end{figure}
Our simulated TW-Hydra radiation field, and other potential ultraviolet radiation environments, include shorter wavelengths than 91.2\,nm and photoprocesses there may be sensitive to shielding by atomic H.
A collection of functions for H-shielding in the TW-Hydra radiation field are shown in Fig.~\ref{fig:H_shielding}.
The photoionisation shielding by H for some molecular species (and photodissociation in the case of \ce{H3+}) is significant for H columns as low as \np[cm^{-2}]{e17}.

Photoabsorption lines of H longwards of its ionisation limit were also included and contribute to the shielding of molecules that dissociate predominantly between 91.2\,nm and the Lyman-$\alpha$ wavelength, 121.6 nm.
Of the entire series of Lyman-lines converging on the H ionisation limit, some of which are shown in Fig.~\ref{fig:combination shielding}, the Lyman-$\alpha$ transition absorbs more photons than all others combined.
It should be noted that the extinction of Lyman-$\alpha$ wavelength radiation is complicated by its re-radiation and forward scattering from dust grains \cite[e.g.,][]{bethell2011}.

\begin{figure}
  \centering
  \includegraphics{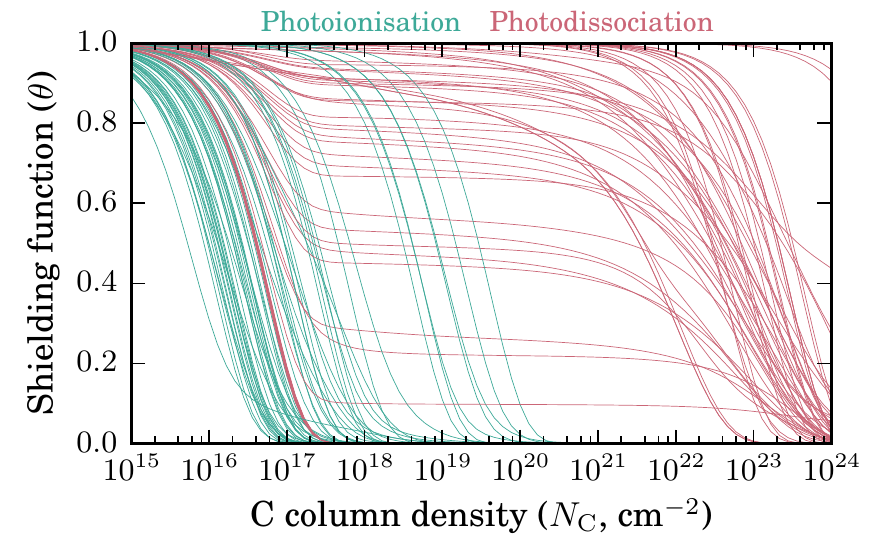}
  \caption{Shielding of photodissociation and ionisation in the ISRF by atomic C.}
  \label{fig:C_shielding}
\end{figure}
The 110\,nm-threshold ionisation continuum of C will significantly reduce the photodissociation and ionisation rates of atomic and molecular species with cross sections biased to shorter wavelengths, assuming a column density of at least \np[cm^{-2}]{e17}. Shielding functions for the ISRF are shown in Fig.~\ref{fig:C_shielding} as functions of C column density and clearly demonstrate this critical density.
The requisite amount of C is observed and modelled to exist in some kinds of photodissociation regions \cite{werner1970,frerking1989,hollenbach1991,hasegawa2003}.
For molecules with photodissociation cross sections predominantly longwards of 110\,nm an extremely unlikely C column density of at least \np[cm^{-2}]{e22} is required for shielding.
In this case, photoabsorption of non-ionising photoabsorption into excited C levels eventually provides the necessary opacity.

\end{appendix}

\end{document}